\documentclass[11pt,letterpaper]{article}

\usepackage[T1]{fontenc}
\usepackage{amsmath,amsfonts,amsthm}
\usepackage{mathtools}
\usepackage{thmtools}
\usepackage{thm-restate}
\usepackage{enumitem}
\usepackage[normalem]{ulem}

\usepackage{multirow}
\usepackage{comment}
\usepackage{complexity}
\usepackage{subcaption}
\usepackage{float}
\usepackage{nicefrac}

\usepackage[dvipsnames]{xcolor}
\usepackage[colorlinks=true,pdfpagemode=UseNone,urlcolor=RoyalBlue,linkcolor=RoyalBlue,citecolor=OliveGreen,pdfstartview=FitH,linktocpage]{hyperref}

\usepackage{dsfont}
\usepackage{xspace}
\usepackage{tikz}
\usepackage[skins]{tcolorbox}
\usepackage[nameinlink, capitalise]{cleveref}
\crefname{section}{Section}{Sections}
\crefname{subsection}{Subsection}{Subsections}
\crefname{claim}{Claim}{Claims}
\crefname{equation}{Equation}{Equations}
\crefname{observation}{Observation}{Observations}
\crefname{figure}{Figure}{Figures}
\usepackage{csquotes}
\usepackage[backend=biber, style=alphabetic, maxbibnames=10, minalphanames=3, maxalphanames=4]{biblatex}
\DeclareSourcemap{
  \maps[datatype=bibtex, overwrite]{
    \map{
      \step[fieldset=editor, null]
    }
  }
}
\DeclareSourcemap{
  \maps[datatype=bibtex]{
    \map{
       \step[fieldset=series, null]
    }
  }
}

\declaretheorem[numberwithin=section]{theorem}
\declaretheorem[sibling=theorem]{lemma}
\declaretheorem[sibling=theorem]{claim}

\declaretheorem[sibling=theorem]{claim*}
\declaretheorem[sibling=theorem]{corollary}

\declaretheorem[sibling=theorem]{remark}
\declaretheorem[sibling=theorem]{definition}

\declaretheorem[sibling=theorem]{observation}

\theoremstyle{definition}

\DeclareMathOperator*{\argmin}{arg\,min}

\usepackage{figchild}

\newcommand{\eps}{\epsilon}
\renewcommand{\epsilon}{\varepsilon}

\newcommand*{\defeq}{\stackrel{\text{def}}{=}}
\newcommand{\ind}{\mathds{1}}

\usepackage{float}
\usepackage{tikz}
\usetikzlibrary{calc} %
\usepackage{array}
\usepackage{multirow}

\usepackage[noend]{algpseudocode}
\usepackage{amsmath, amssymb}
\usepackage{amsthm}
\usepackage{comment}
\usepackage{mdframed}
\usepackage{xspace}
\usepackage{multicol}
\usepackage{tabularx}
\usepackage{makecell}

\usepackage{enumitem}

\usepackage{microtype}

\declaretheoremstyle[
    bodyfont=\normalfont  %
]{upright}

\declaretheorem[
    style=upright,  %
    refname={Algorithm,Algorithms},
    Refname={Algorithm,Algorithms},
    name={Algorithm}
]{algorithm}
\usepackage{graphicx}

\newcommand{\dist}{d}
\newcommand{\di}{\mathsf{dim}}

\newcommand{\Sf}{S_{\mathsf{free}}}
\newcommand{\Sr}{S_{\mathsf{reg}}}

\newcommand{\calF}{\mathcal{F}}
\newcommand{\calI}{\mathcal{I}}
\newcommand{\calDH}{\mathcal{D}\mathcal{H}}

\newcommand{\calB}{\mathcal{B}}
\newcommand{\calC}{\mathcal{C}}
\newcommand{\calD}{\mathcal{D}}

\newcommand{\calQ}{\mathcal{Q}}

\newcommand{\calH}{\mathcal{H}}
\newcommand{\calM}{\mathcal{M}}
\newcommand{\calP}{\mathcal{P}}

\newcommand{\calU}{\mathcal{U}}
\newcommand{\bcalU}{\widetilde{\mathcal{U}}}
\newcommand{\tmu}{\widetilde{\mu}}
\newcommand{\calR}{\mathcal{R}}

\newcommand{\calL}{\mathcal{L}}

\newcommand{\clcost}{\text{cost}}
\newcommand{\closedclcost}{\text{closedcost}}

\newcommand{\clients}{D}
\newcommand{\clientspure}{D^*_{\texttt{pure}}}
\newcommand{\clientscheap}{D^*_{\texttt{cheap}}}
\newcommand{\clientsexpensive}{D^*_{\texttt{exp}}}
\newcommand{\facilities}{F}
\newcommand{\opt}{\text{OPT}}
\newcommand{\cost}{\text{cost}}
\newcommand{\sopt}{\text{opt}}
\newcommand{\optlpfl}{\text{opt}_{\text{LP}}}

\newcommand{\Maxdist}{M}
\newcommand{\dummyset}{\Lambda}
\newcommand{\core}{C^{\text{core}}}

\newcommand{\logadaptalg}{{\sc LogAdaptiveAlgorithm}\xspace}
\newcommand{\retrieveD}{{\sc RetrieveNewCopies}\xspace}
\newcommand{\completeD}{{\sc CompleteCopies}\xspace}
\newcommand{\recoverA}{{\sc RecoverAlpha}\xspace}

\newcommand{\JMSalg}{\textsc{Greedy Algorithm}\xspace}
\newcommand{\mergealg}{\textsc{MergeSolutions}\xspace}
\newcommand{\completesol}{\textsc{CompleteSolution}\xspace}

\newcommand{\completesequence}{\textsc{CompleteSequence}\xspace}
\newcommand{\calLexp}{\calL_{\text{exp}}}
\newcommand{\bid}{\texttt{bid}}
\newcommand{\facloc}{Facility Location\xspace}

\title{A $(2{+\eps})$-Approximation Algorithm for Metric $k$-Median}

\usepackage[colorinlistoftodos,textsize=tiny,textwidth=2cm,color=green!50!gray]{todonotes}

\newlength\hwhatcwidth%
\newlength\hwhatuwidth%
\newcommand*\halfwidehat[1]{%
  \settowidth\hwhatcwidth{\ensuremath{#1}}%
  \settowidth\hwhatuwidth{\ensuremath{\mathrm{#1}}}%
  \addtolength\hwhatuwidth{-\hwhatcwidth}%
  \makebox[0pt][l]{%
    \kern\dimexpr0.25\hwhatcwidth-0.667\hwhatuwidth\relax%
    \ensuremath{\widehat{\vphantom{#1}\rule{0.5\hwhatcwidth}{0pt}}}%
  }#1%
}%

\newcommand{\avg}{\text{avg}}

\newcommand{\ho}{\hat{\opt}}

\newcommand{\apxLSkmeans}{26}

\usepackage[colorinlistoftodos,textsize=tiny,textwidth=2cm,color=green!50!gray]{todonotes}

\newcommand{\EC}{\mathsf{EC}\xspace}
\newcommand{\OPT}{{\rm OPT}\xspace}

\newcommand{\mkm}{Metric $k$-Means\xspace}
\newcommand{\ekm}{Euclidean $k$-Means\xspace}
\newcommand{\conn}{\mathrm{conn}}

\usepackage{amsthm}
\usepackage{amsfonts}
\usepackage{graphicx}
\usepackage{amsmath}
\usepackage{optidef}
\usepackage{amssymb} 
\usepackage{listings}
\usepackage{framed}
\usepackage[svgnames]{xcolor}
\usepackage{bm}
\usepackage{url}
\usepackage[colorlinks=true,pdfpagemode=UseNone,urlcolor=RoyalBlue,linkcolor=RoyalBlue,citecolor=OliveGreen,pdfstartview=FitH,linktocpage]{hyperref}
\hypersetup{hypertexnames=false}
\usepackage{multicol}
\usepackage{tabularx}
\usepackage{makecell}
\usepackage{enumitem}

\usepackage[nameinlink, capitalise]{cleveref}
\usepackage[letterpaper,top=2.5cm,bottom=2.5cm,left=3cm,right=3cm,marginparwidth=2cm]{geometry}

\usepackage{authblk}
\author[1]{Aditya Anand}
\author[2]{Moses Charikar}
\author[3]{Vincent Cohen-Addad}
\author[2]{Ruiquan Gao}
\author[4]{\\Fabrizio Grandoni}
\author[1]{Euiwoong Lee}
\author[1]{Amatya Sharma}
\author[5]{Ernest van Wijland}
\affil[1]{University of Michigan, \url{{adanand,euiwoong,amatya}@umich.edu}}
\affil[2]{Stanford University, \url{{moses, ruiquan}@cs.stanford.edu}}
\affil[3]{Google Research, \url{cohenaddad@google.com}}
\affil[4]{IDSIA, USI-SUPSI, \url{fabrizio.grandoni@gmail.com}}
\affil[5]{Université Paris-Cité, \url{ernest.vanwijland@irif.fr}}

\title{Spectral Dual Fitting for $k$-Means}
\date{}

\begin{document}

\begin{titlepage}
    \maketitle
    \thispagestyle{empty}
    \begin{abstract}
    \noindent We give a new dual fitting algorithm which gives improved approximation ratios of $3+\ln 2 + \eps\ (\approx 3.694)$ and $4.9+\eps$ for $k$-Means in (high-dimensional) Euclidean and general metrics respectively, improving upon the previously known ratios of $4+\eps$ [Charikar, Cohen-Addad, Gao, Grandoni, Lee, and van Wijland STOC'26] and $5+\eps$ [Byrka, Guo, Hu, Li, Wan, Wang FOCS'26], resp. In particular, our result for Euclidean $k$-Means breaks the hardness barrier of $1+8/e\approx 3.94$ for Metric $k$-Means. Prior to our work, no such separation between general and Euclidean metrics was known for $k$-Median, $k$-Means, or Facility Location in terms of their approximability.

    Unlike prior dual fitting approaches for $k$-Means, our new dual fitting algorithm tightly accounts for dual payments while still facilitating an effective dual feasibility analysis. We introduce a new framework that uses spectral analysis for determining the approximation factor of our algorithm.
        \thispagestyle{empty}
    \end{abstract}
\end{titlepage}

\pagestyle{empty}{
\tableofcontents
}

\newpage
\setcounter{page}{1}
\pagestyle{plain}

\section{Introduction}
Clustering is one of the most fundamental tasks in data science with wide-ranging applications across various domains, including 
unsupervised learning, text and image analysis, recommender systems, and bioinformatics.
(Metric) $k$-Means is one of the most well-studied clustering objectives.
Given a collection of $n$ clients $D$, 
a collection $\facilities$ of 
potential center locations in a metric space defined by distances $d(.,.)$, and an integer $k \in [n]$, the goal is to select $k$ centers $S \subseteq \facilities$
such that $\conn(S) := \sum_{j \in D} d^2(j,S)$ is minimized, where $d(j,S)$ is the distance of point $j$ to its nearest center in $S$. A very important special case is \emph{\ekm}, where $D$ are points in a Euclidean space, and $\facilities$ is implicitly given by any point in the Euclidean space.\footnote{Sometimes this is called the \emph{continuous} \ekm problem in the literature.}  By standard techniques~\cite{matouvsek2000approximate, de2003approximation} and losing only a $(1+\eps)$ factor in the approximation ratio for any constant $\eps > 0$, we can restrict $\facilities$ to a polynomial-time-computable subset of points. W.l.o.g., we will also assume that $D\subseteq \facilities$.

From a theoretical standpoint,  
the study of clustering problems through the lens of (polynomial-time) approximation algorithms has been extremely successful, yielding new algorithms and analysis techniques.
In particular, $k$-Means and the closely related $k$-Median (whose objective function $\sum_{j \in D} d(j, S)$ is the sum of non-squared distances) 
are two problems that have received the most attention.
For $k$-Median, there have been gradual developments over the last three decades using various algorithmic techniques, including primal LP rounding~\cite{CharikarGTS99, CL12}, primal-dual~\cite{JaiV01}, greedy algorithm with dual fitting analysis~\cite{JainMS02, JainMMSV03, cohen2023breaching}, local search~\cite{AryaGKMMP04}, and bipoint rounding~\cite{LiS13, BPRST15, GowdaPST23}. The current best approximation ratio is $(2 + \eps)$, achieved by~\cite{CGLSS25stoc}. It is known that one cannot improve the approximation ratio beyond $1 + 2/e \approx 1.73$~\cite{guha1999greedy}.

For $k$-Means, the picture has been much more elusive. While new algorithmic techniques for $k$-Median 
often translate to $k$-Means, it takes nontrivial technical effort, and the approximation ratios usually end up being much larger: this is because only approximate versions of triangle inequalities are available for squared metrics, which present technical challenges for some $k$-Median algorithms, most notably the ones based on dual fitting. 

For Metric $k$-Means, \cite{GT08} gave a $25$-approximation using local search, which was improved to a $(9 + \epsilon)$ approximation by~\cite{AhmadianNSW20} via primal-dual. 
Recently, the approximation factor was improved to $5.83 $~\cite{charikar2025kmeans} via a novel dual fitting algorithm.
For Euclidean $k$-Means, \cite{KanungoMNPSW04} gave a $(9 + \eps)$-approximation using local search, followed by improvements using primal-dual methods~\cite{AhmadianNSW20, GrandoniORSV22, CEMN22} and most recently to $4 + \eps$~\cite{CCGGLW2026kmeans} via dual fitting. The hardness for the metric case is $1 + 8/e \approx 3.94$, while for Euclidean metrics the hardness is $1.17$~\cite{Cohen-AddadS19}, and under a stronger conjecture called the Johnson Coverage Hypothesis (JCH), the hardness can be strengthened to $1.73$~\cite{cohen2022johnson}. We remark that we focus on the high-dimensional Euclidean case here, while better algorithms are known for the low-dimensional one \cite{FRS16focs,CKM16focs,C18soda}.

In this work, we give a new dual fitting based algorithm that gives significantly 
better approximation ratios for both \ekm and \mkm. Our algorithm leads to a $3+\ln 2+\eps\ (\approx 3.694)$ approximation for the Euclidean case and a $4.9 + \eps$ approximation for the metric case. Our result for Euclidean metrics is the first algorithm that gives an approximation ratio \emph{below the $3.94$-hardness barrier} for the metric case, showing that 
the optimal approximation ratios for these two versions
will be different assuming $\mathbf{P} \neq \mathbf{NP}$. Such a separation between general metrics and high-dimensional Euclidean metrics, even after 
extensive recent 
work~\cite{CEMN22, cohen2022johnson, lee2025facility}, has not been achieved for $k$-Median and Facility Location.

\begin{restatable}{theorem}{maineuclidean}
\label{thm:main_euclidean}
For any $\eps > 0$, there is a polynomial-time algorithm that yields a $(3 + \ln 2 + \eps)$-approximation for Euclidean $k$-Means. 
\end{restatable}

\begin{restatable}{theorem}{mainmetric}
\label{thm:main_metric}
For any $\eps > 0$, there is a polynomial-time algorithm that yields a $(4.9 + \eps)$-approximation for Metric $k$-Means. 
\end{restatable}

Our main technical contribution is an improved Lagrangian Multiplier Preserving (LMP) approximation for the following (Uncapacitated) Facility Location problem (with squared distances and uniform opening costs), which we simply call {\em \facloc} throughout the paper: the input consists of a set of clients $D$, a collection of centers $\facilities$, a (possibly Euclidean) metric $d$ on $D \cup \facilities$, and a {\em facility opening cost} $f$,
and the goal is to select $S \subseteq \facilities$ that minimizes $f |S| + \conn(S)$. 
In particular, $S \subseteq \facilities$ is called an LMP $c$-approximation if $c \cdot f|S|  + \conn(S) \leq 
c \cdot \OPT_{\mathrm{LP}}(f)$, where $\OPT_{\mathrm{LP}}(f)$ denotes the optimal cost of the standard LP relaxation for 
Facility Location defined in \Cref{sec:preliminaries}.
Our algorithms yield LMP $4.9$ and $(3+\ln 2)$-approximations  
for \facloc in general and Euclidean metrics, respectively.

Our improved LMP approximation is obtained via a refined dual fitting algorithm that tightly accounts for the cost incurred, while still being able to facilitate an improved one-cluster based dual feasibility analysis. We also introduce a novel spectral analysis for $k$-Means, departing from the usual factor-revealing linear programs. 
The spectral analysis provides a more accurate method to analyze the squared objective in $k$-Means compared to the previous approaches. More specifically, upper bounding the approximation factor reduces to upper bounding the largest eigenvalue of a family of matrices. The latter task is highly non-trivial, but it leads to substantially lower approximation factors compared to known methods.

This is then combined with the two-step framework
recently used in~\cite{CGLSS25stoc, 
charikar2025kmeans,
CCGGLW2026kmeans} to obtain our approximation algorithm. 
For the first step, building on our LMP approximation, we use the idea of {\em log-adaptive algorithms} to obtain a good approximate solution for $k$-Means that opens $k + O(\log n / \eps^3)$ centers. 
Let $\opt_k$ be the optimal cost for an instance of $k$-Means. 

\begin{theorem}\label{thr:mainBicriteria}
For any constant $\eps > 0$, 
there is a polynomial-time algorithm that, given an instance of $k$-Means, finds $S \subseteq \facilities$ with 
$|S| = k + O(\log n/ \epsilon^3)$ satisfying $\conn(S) \leq (\Gamma + O(\sqrt{\eps})) \cdot \opt_k$, where $\Gamma = 3+\ln 2$ for Euclidean metrics and $\Gamma = 4.9$ for general metrics. 
\end{theorem}

The second step of the framework studies {\em stable instances}.
A $k$-Means instance is $\beta$-stable if $\opt_{k-1} \geq (1 + \beta)\opt_k$, where $\opt_{k-1}$ denotes the optimal cost of $(k-1)$-Means with the same clients and centers.
We show that a slightly more careful analysis of the same algorithms as in  \cite{charikar2025kmeans, CCGGLW2026kmeans} yields the following results. 

\begin{theorem}\label{thr:mainMetricStable}
For any constants $\eps,\zeta>0$, there is a polynomial-time randomized algorithm for Metric $k$-Means that with high probability returns a solution of cost at most $(4+\eps)\opt_k$
assuming that the input instance is $(\zeta/\log n)$-stable.   
\end{theorem}
\begin{theorem}\label{thr:mainEuclideanStable}
For any constants $\eps,\zeta>0$, there is a polynomial-time randomized algorithm for Euclidean $k$-Means that with high probability returns a solution of cost at most $(2+\eps)\opt_k$ assuming that the input instance is $(\zeta/\log n)$-stable.   
\end{theorem}
These results improve the $(5+\eps)$-approximation for general metrics and the $(4+\eps)$-approximation for Euclidean metrics~\cite{charikar2025kmeans, CCGGLW2026kmeans}. 
Similar to previous results~\cite{CGLSS25stoc, 
charikar2025kmeans,
CCGGLW2026kmeans}, given \Cref{thr:mainBicriteria,thr:mainMetricStable,thr:mainEuclideanStable}, we can easily prove our main results \Cref{thm:main_euclidean} and \Cref{thm:main_metric}; given a $k$-Means instance, if it is not $\Omega(\eps^4/\log n)$-stable as a $k'$-Means instance for each $k' = k, k-1, \dots, k - O(\log n /\eps^3)$, then $\opt_{k - O(\log n /\eps^3)} \approx \opt_k$ and we can use \Cref{thr:mainBicriteria} with $k \leftarrow k - O(\log n /\eps^3)$. Otherwise, the instance is stable for some $k'$ in the above range, and we can use \Cref{thr:mainMetricStable,thr:mainEuclideanStable}. The formal proof appears in \Cref{sec:finalproof}.

\paragraph{Concurrent work.}
Concurrently and independently, Byrka, Guo, Hu, Li, Wan, and Wang~\cite{BGHLWW26focs}
gave an iterative randomized rounding framework for \(k\)-clustering with
\(\ell_p^p\)-type assignment costs. For Metric \(k\)-Means, their framework gives
a \((5+\varepsilon)\)-approximation, improving the previous \(5.83\)-approximation.
For Euclidean \(k\)-Means, they give a \((4+\varepsilon)\)-approximation and prove
an \(11/3\)-LMP approximation. Their
approach is LP-rounding based and is conceptually different from our
dual-fitting/spectral analysis. Our results have better approximation guarantees
for both settings, giving \((4.9+\varepsilon)\) for Metric \(k\)-Means and
\((3+\ln 2+\varepsilon)\) for Euclidean \(k\)-Means.

\section{Technical Overview}

We start by revisiting the dual fitting algorithm for Facility Location of the two papers by Charikar et al.~\cite{charikar2025kmeans,CCGGLW2026kmeans} for Metric and Euclidean $k$-Means, respectively. The two algorithms are very similar, except for one difference that~\cite{CCGGLW2026kmeans} lowers certain dual values which can lead to further improvements. In this overview, this difference is not important, but we will  use the stronger algorithm from~\cite{CCGGLW2026kmeans} which is still as good as the older algorithm~\cite{charikar2025kmeans} for the metric case as well.

We provide a description of their algorithm below.

\paragraph{Greedy Algorithm.}
The algorithm maintains a set $A$ of active clients, a set $S \subseteq \facilities$ of open facilities, and a partition of clients into three classes: active clients $A$, indirectly connected clients $IC$, and directly connected clients $DC$. Intuitively, active clients are not yet connected to an open facility, and they are actively growing their dual variables. Indirectly connected clients are connected to an open facility, but they do not directly contribute (a strictly positive amount) to the opening cost of that facility. On the other hand, directly connected clients are connected to an open facility and they directly contribute to the opening cost of that facility.
Initially, $A=D$, $IC=DC=S=\emptyset$, and $\alpha_j=0$ for every client $j\in D$. Fix a parameter $\rho > 1$. We let $\hat{f} = \Gamma f$ where $\Gamma$ is some constant that depends on $\rho$ to be specified later, and $f$ is the (uniform) facility cost.

As long as $A\neq\emptyset$, the algorithm increases uniformly the (\emph{dual}) values $\alpha_j$ for all clients $j\in A$ until one of the following two events occurs.
\begin{enumerate}
    \item {A facility becomes tight.}
    There exists an unopened facility $i\in \facilities\setminus S$ such that
     \begin{equation}\label{eqn:facold}
    \sum_{j\in A \cup IC}\left[\alpha_j - \rho \cdot d^2(j,i)\right]^+
    + \sum_{j\in DC}\left[\rho d^2(j,S) -  \rho d^2 (j, i)\right]^+ = \hat{f}.
    \end{equation}
    In this case, the algorithm opens facility $i$. Every client $j\in DC$ with
    $d^2(j,S)>d^2(j,i)$ is reconnected to $i$. For every client $j\in A\cup IC$, the algorithm acts as follows:
    \begin{itemize}
        \item if $\alpha_j \ge \rho \cdot d^2(j,i)$, then $j$ is connected (or reconnected) to $i$ and moved to $DC$;
        \item else, if $\alpha_j \ge d^2(j,i)$, then $j$ is connected (or reconnected) to $i$, its value $\alpha_j$ is lowered to $d^2(j,i)$, and $j$ is moved to $IC$ if it was in $A$. 
    \end{itemize}

    \item {An active client reaches its current service cost.}
    For some client $j\in A$, we have
    \[
    \alpha_j = d^2(j,S).
    \]
    In this case, the client $j$ is moved to $IC$ and connected to its closest facility in $S$.
\end{enumerate}
Intuitively, the quantities of type $[\cdot]^+$ above are \emph{bids} that client $j$ is willing to pay to open facility $i$. Via these bids, $j$ can contribute to the opening of multiple facilities. At the same time, the residual dual $\alpha_j$ of $j$ after subtracting these bids is sufficient to connect $j$ to the closest open facility. The algorithm naturally defines a notion of \emph{time}, such that at time $\theta$, all the active clients have $\alpha$-value $\theta$.

Then, they show the following two cost accounting lemmas (similar to~\cite{JainMMSV03}) for this algorithm (considering the values of $\alpha$ and $S$ at the end of the algorithm), that lead to an LMP $\Gamma$-approximation. 

\begin{lemma}[Dual Payment, for any metric]\label{lemma:payment}
$\sum_{j \in D} \alpha_j \geq \hat{f}|S| + \sum_{j \in D} d^2(j,S)$.
\end{lemma}

Note that the RHS of this inequality is the cost of the computed solution (with facility cost $\hat{f}=\Gamma\,f$). The proof of this lemma follows from the fact that a facility is only opened when it is fully paid for (by the bids), and the fact that for any client $j$, after subtracting any bid to open facilities, the remaining part of $\alpha_j$ can still pay for its connection cost $d^2(j,S)$.

\begin{lemma}[Dual Feasibility,~\cite{charikar2025kmeans,CCGGLW2026kmeans}]\label{lemma:dualfeasmetric}
Let $\rho=\sqrt{2}+1$ (resp., $\rho=2$ for Euclidean metrics).
Fix any facility $i$. Then
$$\sum_{j \in D} \left[\alpha_j - \Gamma d^2(j,i)\right]^+ \leq \hat f,$$ 
where $\Gamma=3+2\sqrt{2}$ (resp., $\Gamma=4$ for Euclidean metrics).
\end{lemma}

\Cref{lemma:dualfeasmetric} implies that $(\alpha / \Gamma)$ is dual-feasible for the Facility Location LP (see \ref{eq:lp-dual}) with opening cost $f$. Combining this with \Cref{lemma:payment}, one gets
\[
\Gamma f |S|  + \conn(S) \leq \sum_{j \in D} \alpha_j \leq \Gamma \cdot \OPT_{LP}(f), 
\]
concluding that $S$ is an LMP $\Gamma$-approximate solution.

While the dual payment lemma more or less follows directly from the algorithm description, the key challenge in~\cite{charikar2025kmeans,CCGGLW2026kmeans} has been proving dual feasibility given the complexity of the algorithm due to the presence of both IC and DC clients and the non-monotonicity of the dual variables. 

\begin{figure}[!tbh]
\centering
\begin{tikzpicture}[scale=1.0, every node/.style={font=\small}]
    \draw[->] (-8.4,0) -- (4.4,0) node[right] {$x$};

    \foreach \x in {-6,-2,0,2}
        \draw (\x,0.16) -- (\x,-0.16);

    \node[fill=red, circle, minimum size=8pt, inner sep=0pt] (c1) at (-2,0) {};
    \node[fill=red, circle, minimum size=8pt, inner sep=0pt] (c2) at (2,0) {};
    \node[fill=blue, minimum size=8pt, inner sep=0pt] (f1) at (0,0) {};
    \node[fill=blue, minimum size=8pt, inner sep=0pt] (f2) at (-6,0) {};

    \node[below=4pt] at (-6,0) {$i'$};
    \node[below=4pt] at (-2,0) {$1$};
    \node[below=4pt] at (0,0) {$i$};
    \node[below=4pt] at (2,0) {$2$};

    \node[above=4pt] at (-6,0) {$-3$};
    \node[above=4pt] at (-2,0) {$-1$};
    \node[above=4pt] at (0,0) {$0$};
    \node[above=4pt] at (2,0) {$1$};

\end{tikzpicture}
\caption{%
Barrier example for the previous dual-feasibility analysis. 
The facility $i$ is at $0$, the two clients are at $-1$ and $1$, and an already open facility $i'$ is at $-3$. 
Thus $d^2(1,i)=d^2(2,i)=1$, while $d^2(1,i')=4$ and $d^2(2,i')=16$.}
\label{fig:barrier-example}
\end{figure}
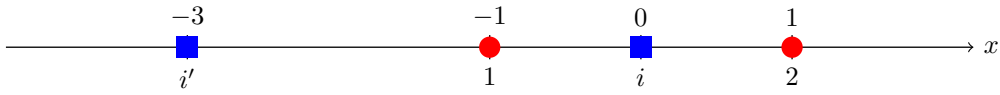

\paragraph{The barrier of previous works.} 
Our first observation is that there is a barrier for the dual feasibility analysis of the algorithm of~\cite{CCGGLW2026kmeans} for Euclidean metrics --- a very simple example shows that the LMP $4$-approximation analysis of~\Cref{lemma:dualfeasmetric} can be tight for some facility $i$. 
    As illustrated in \Cref{fig:barrier-example}, we will consider points on a line (i.e., $1$-dimensional Euclidean space).
    Let $i$ be a facility and $D^*=\{1,2\}$ located at $x_i = 0$, $x_1=-1$, $x_2=1$. Then, $d^2(1,i)+d^2(2,i)=2$.
    Let $\hat{f}=20$. The algorithm simultaneously increases $\alpha_1$ and $\alpha_2$ until $
    [\alpha_1-\rho d^2(1,i)]^+ +[\alpha_2-\rho d^2(2,i)]^+ = 
    [\alpha_1-2]^+ +[\alpha_2-2]^+=20$, i.e., when $\alpha_1=\alpha_2=12$. At this time point, suppose that client $1$ gets directly connected to the only open facility at $x_{i'}=-3$ (which was opened via bids from client $1$ and other clients not in $D^*$).
    This leads to $d^2(1,S)=4$. 
    From now on, the bid of client $1$ to facility $i$ becomes $8-2=6$. 
    We will keep increasing $\alpha_2$ until one of the following holds:
    \begin{enumerate}
        \item $\alpha_2\geq d^2(2,S)=(x_2-x_{i'})^2 = 16$. In this case, we have client 2 indirectly connected to $i'$. 
        \item $6+[\alpha_2-2]^+=20$, which means $\alpha_2=16$. In this case, we have $i$ opened and client 2 directly connected to $i$.
    \end{enumerate}
    Therefore, we have $\sum_{j\in D^*} \alpha_j = 12+16=28$. On the other hand, we have $\hat f + 4(d^2(1,i)+d^2(2,i)) = 20 + 4\cdot 2 = 28$. Hence, the dual feasibility is tight for this example, and we cannot hope for an LMP approximation ratio below $4$ using this algorithm.

\paragraph{An issue with the payments.}
Our new algorithm is inspired by the fact that while the dual feasibility can be tight, the dual payment  \cref{lemma:payment} does not tightly account for the costs: $\sum_{j} \alpha_j$ may actually be larger than the cost of the algorithm. In particular, we observe that for any client that is directly connected at the end, the algorithm always overpays the connection cost by a factor of $\rho$. To see this, let $DC^{end}$ denote the set of clients which are directly connected at the end. Let $j \in DC ^{end}$, and suppose through the execution of the algorithm, $j$ was connected to facilities $i_1, i_2...\ldots i_t$. For simplicity, assume that $j$ was directly connected to each of these facilities. Then the total bid (dual contribution, see~\Cref{eqn:facold}) 
of $j$ to these facilities is at most 
$$(\alpha_j - \rho d^2(j,i_1)) + \sum_{\ell=1}^{t-1} \rho d^2(j,i_{\ell})-\rho d^2(j, i_{\ell+1}) = \alpha_j - \rho d^2(j,i_t)~.$$
This means that after accounting for facility costs, $\alpha_j$ pays for $\rho d^2(j, i_t)$, where $\rho > 1$, leading to an overpayment of the connection cost by the factor $\rho$. Unfortunately, one cannot set $\rho = 1$ as it would not give even an LMP $O(1)$-approximation, as already observed in~\cite{charikar2025kmeans}. This means we could simply subtract $(\rho - 1 )d^2(j,i_t)$ in the dual feasibility analysis.

However, while this overpayment is not ideal, it seems challenging to account for this naturally in the analysis: dual feasibility analysis is highly \emph{local}, which cannot rely on the sequence of connections a client has: in particular, it is not clear how to meaningfully incorporate the subtraction $(\rho - 1)d^2(j,i_t)$ of client $j$, which is a \emph{global} quantity, into the local one-cluster dual feasibility analysis.
{
The previous example of \cref{fig:barrier-example} illustrates this difficulty as well; when $\alpha_2 = 16$, the events triggering the actions of ``having client 2 indirectly connected to $i'$'' and ``opening $i$ and connecting client 1 and 2 to $i$'' happen simultaneously, but depending on how the algorithm breaks the ties (which can be easily manipulated by the position of the external facility $i'$), the amount of the overpayment substantially changes. In the former case, it is $(\rho - 1) \cdot d^2(1, i') = 4$ just from client 1, and in the latter case, it is 
$(\rho - 1) \cdot (d^2(1, i) + d^2(2,i)) = 2$.}%

\paragraph{Our new algorithm with tight dual payments.}
To resolve this issue, we propose a new algorithm  (\Cref{alg:greedy}) that allows us to get the best of both worlds. Our new algorithm still has an overpayment, but we subtract this overpayment at the end, while still being able to incorporate this subtraction in the dual feasibility analysis. In our new algorithm, each client \emph{reserves} a part of its first direct connection cost, that it does not offer to other facilities for subsequent (direct) connections. The key is that clients only move to facilities where their connection cost is lower, so the first direct connection cost is an \emph{upper bound} for \emph{every} future direct connection.  

\begin{figure}[ht!]
\begin{center}
\begin{minipage}{1.0\textwidth}
\begin{mdframed}[hidealllines=true, backgroundcolor=gray!15]

    \refstepcounter{algorithm}
    \label{alg:greedy}
    \textbf{Algorithm \thealgorithm\ (Greedy Algorithm)}. \\[0.2cm]
    \textbf{Initialization:} $A=D$, $IC=DC=S=\emptyset$, and $R_j=\alpha_j=0$ for every $j\in D$. 
    
    \vspace{0.5em}

    \noindent While $A\neq\emptyset$, uniformly grow the $\alpha_j$ of every client in $A$ until: %
    \begin{enumerate}
        \item For some unopened facility $i\in \facilities \setminus S$, we have
        \[
        \sum_{j\in A \cup IC}\left[\alpha_j - \rho \cdot d^2(j,i)\right]^+
        + \sum_{j\in DC}\left[d^2(j,S) -  d^2 (j, i)\right]^+ = \hat{f}.
        \]
        
        In this case, we open facility $i$.
        For each client $j\in DC$, if $d^2(j,S)>d^2(j,i)$, we reconnect $j$ to $i$.
        For each client $j\in A\cup IC$,
        \begin{itemize}
            \item If $\alpha_j \geq \rho \cdot d^2(j,i)$, we (re)connect $j$ to $i$, move $j$ to $DC$ and set $R_j=d(j,i)$.
            \item Else, if $\alpha_j \geq d^2(j,i)$, we (re)connect $j$ to $i$, and move $j$ to $IC$ (if $j\in A$), and set $\alpha_j = d^2(j,i)$.
        \end{itemize}

        \item For some client $j\in A$, we have
        \[
        \alpha_j = d^{2}(j,S).
        \]
        In this case, we move $j$ to $IC$ and connect $j$ to the closest facility to it in $S$.
    \end{enumerate}
\end{mdframed}
\end{minipage}
\end{center}
\end{figure}

Notice that until the first direct connection, an active or IC client $j$ offers $\alpha_j - \rho d^2(j,i)$, but subsequently, instead of offering the natural $\rho d^2(j,S) - \rho d^2(j,i)$, it only offers $d^2(j,S) - d^2(j,i)$. Thus if $R_j = d(j,i_1)$ is the first direct connection distance, then $(\rho - 1)R_j^2$ is \emph{reserved} by the client and never used for any subsequent bid to facilities.  

Following the same analysis as before, consider any client $j \in DC^{end}$. If it goes through a sequence of direct connections $i_1, i_2 ...\ldots i_t$, then it offers 
$$\alpha_j - \rho d^2(j,i_1) + \sum_{\ell=1}^{t-1} d^2(j,i_{\ell}) - d^2(j, i_{\ell+1}) = \alpha_j - (\rho - 1) d^2(j,i_1) - d^2(j,i_t)~.$$
Since the connection cost at the end is only $d^2(j,i_t)$, this leads to an overpayment of $(\rho - 1)d^2(j,i_1)$, where $i_1$ is the first facility $j$ was directly connected to. But now we define $R_j = d(j,i_1)$, and subtract $(\rho - 1)R_j^2$ in the dual feasibility analysis to make this accounting exact (we will define $R_j = 0$ if $j$ is never directly connected). This turns out to be helpful in the dual feasibility analysis because we know $R_j = d(j, i_1) \geq d(j,i_2) \geq ...\ldots \geq d(j, i_t)$, as clients only move to closer facilities as the algorithm progresses. $R_j$ is thus an upper bound on all these distances, which facilitates a refined dual feasibility analysis, as discussed in greater detail in the following subsections. 

Going back to the example from \Cref{fig:barrier-example} with $\rho = 2$, 
we set $R_1 = d(1,i') = 2$ when $i'$ is open at time 12, and the new algorithm indirectly connects client $2$ to $i'$ at time 16 (which now happens regardless of the tiebreaking rule as the bid of client $1$ toward $i$ becomes smaller), leading to $R_2 = 0$. Thus we have $\sum_{j \in D^*} \alpha_j = \alpha_1 + \alpha_2 = 28 = \hat{f} + 4\bigl(d^2(1,i) + d^2(2,i)\bigr)$ just like before, but because of the subtraction of $(\rho - 1)R_1^2 = R_1^2$, the dual feasibility is no longer tight. In particular we have $\alpha_1 - R_1^2 + \alpha_2 - \hat{f} = 4 = 2\bigl(d^2(1,i) + d^2(2,i)\bigr)$, and consequently, $\sum_{j \in D^*}(\alpha_j - 2d^2(j,i) - R_j^2)^+ \leq \hat{f}$, which improves the dual feasibility factor to $2$ instead of $4$ for this example.

\paragraph{Analysis.}
From the above discussion, our modified payment and dual feasibility lemmas can be stated as follows. Let $\alpha$ and $S$ be the respective values at the end of the algorithm. The following dual payment lemma almost directly follows from the algorithm description and the discussion above.
\begin{lemma}[Dual Payment for our new algorithm, \cref{lem:approxguarantee}]\label{lem:approxguarantee-overview}
$$\sum_{j \in D} (\alpha_j - (\rho - 1)R_j^2) \geq \hat{f}|S| + \sum_{j \in D} d^2(j,S)$$ 
\end{lemma}
 The crux of our analysis is now to prove a dual feasibility lemma for a small enough constant $\Gamma$ (depending on $\rho$). In particular, we prove the following.
\begin{lemma}[Dual feasibility of our new algorithm, \cref{lem:dual-feasibility}]\label{lem:dual-feasibility-overview}%
Fix any facility $i$. Let $\rho=2.5$ (resp.,  $\rho=2$ for Euclidean metrics). Then
$$\sum_{j \in D} \left [ \alpha_j - \Gamma d^2(j,i) - (\rho - 1)R_j^2\right]^+ \leq \hat{f},$$
where $\Gamma=4.9$ (resp., $\Gamma=3+\ln(2)$ for Euclidean metrics).
\end{lemma}
Intuitively, the above dual feasibility lemma shows that the variables $\alpha'_j:=\frac{1}{\Gamma}(\alpha_j-(\rho-1)R^2_j)$ induce a dual-feasible solution for the Facility Location LP. An LMP $\Gamma$-approximation follows by the same argument as before. 

Furthermore, note that we can interpret our algorithmic idea as having smaller bids for clients who are directly connected. 
The intuition behind our new algorithmic idea (for the Euclidean case) is that these directly connected clients are ``bad'' in the dual feasibility analysis of \cite{CCGGLW2026kmeans} only when all clients and facilities are ``almost'' on a line (i.e., one-dimensional Euclidean space) and its maximum direct connection cost $R_j^2$ is close to the squared distances between it and other clients $j'$ with large dual variable $\alpha_{j'}$ (compared to $d^2(j',i)$). As we have this additional $-(\rho-1)R_j^2$ term in the final dual variable $\alpha'_j$, we can make the final dual variables of these ``bad'' directly connected clients much smaller. 

However, to apply this intuition to the analysis (for the general Euclidean cases), we confront several technical challenges: 
\begin{itemize}
    \item We need to deal with high-dimensional Euclidean spaces.
    \item We need to transform these $-(\rho-1)R_j^2$ terms into $-d^2(j,i)$ terms (because, intuitively, the dual feasibility constraint compares $\sum_{j\in D'} \alpha_j-(\rho-1)R_j^2$ with $f+\sum_{j\in D'} \Gamma d^2(j,i)$ for every $D'\subseteq D$).
    \item Finally, in the Euclidean space, \cite{CCGGLW2026kmeans} uses the pairwise distance $\frac{1}{|D'|} \sum_{j<\ell\in D'} d^2(j,\ell)$ as an intermediate quantity, which is upper bounded by $\sum_{j\in D'} d^2(j,i)$ (see, e.g., \cref{lem:opt-single-cluster}). However, this inequality may not hold if the coefficients before each $d^2(j,i)$ have an average of 1 but are ``unbalanced''. 
    After we transform the $-(\rho-1)R_j^2$ terms into the $-d^2(j,i)$ terms, we observe that the coefficients of the new terms have an average $\Omega(1)$ in the cases~\cite{CCGGLW2026kmeans}'s analysis is tight. However, the coefficients are ``unbalanced'', and some of them can even be zero (due to $R_j=0$ for the corresponding client). To prove a ratio strictly better than 4, the most significant challenge is to show that the pairwise distance terms combined with these new $-d^2(j,i)$ terms can be upper bounded by $(1-\Omega(1))\cdot \sum_{j\in D'} d^2(j,i)$. 
\end{itemize}
As an illustration on how we resolve these challenges, we will provide a framework of our spectral analysis in \cref{sec:spectral-overview}, and provide simple proofs that obtain improved ratios (which are slightly worse than our final results) on the worst cases for the analysis of~\cite{CCGGLW2026kmeans,charikar2025kmeans} in \cref{sec:euclidean-warmup,sec:metric-warmup}.

We remark that slightly smaller values of $\Gamma$ (hence slightly better approximation factors)\footnote{For example, we know that $\rho=1.96, \Gamma=3.692$ works for Euclidean metrics, and $\rho\approx 2.430,\Gamma=4.865$ works for general metrics.} can be obtained, for both the Euclidean and the metric case, with different values of $\rho$ and a substantially more technical and longer analysis. We decided to omit this for the sake of clarity (and given that the presented analysis is already very complex).
We also remark that our approach cannot yield significantly better approximation factors. Our numerical simulations suggests that our spectral analysis framework has a barrier of 3.51 for Euclidean metrics and 4.67 for general metrics. %
See the end of \cref{sec:spectral-overview} for a brief discussion of this.

\subsection{Spectral analysis: the general framework proving the dual feasibility}
\label{sec:spectral-overview}

In this part, we will introduce the general idea of our new spectral analysis of the dual feasibility, which is the key to improving the approximations for both Euclidean and Metric $k$-Means. For simplicity, we will focus on the special case where at each point in time each considered client is active or directly connected (without ever being indirectly connected). This case is already non-trivial, and allows us to illustrate the main ideas. To simplify the notation, we also assume that the final dual values of the mentioned clients are all distinct. We start with some general considerations. Then, in \cref{sec:euclidean-warmup} and \cref{sec:metric-warmup} we will detail the analysis (for the mentioned special case) in the Euclidean and metric case, respectively. The analysis for arbitrary instances is given in \cref{sec:greedy,sec:euclidean-lmp} for the Euclidean case, and in \cref{sec:greedy,sec:euclidean-lmp,sec:metric-lmp} for the metric case.

\paragraph{The general idea of the spectral analysis.}
We will consider a similar set of constraints on the dual variables $\alpha_j$ as in the original analysis of~\cite{JainMMSV03} as the starting point of our spectral analysis. Similar constraints are also used in the analysis of~\cite{charikar2025kmeans,CCGGLW2026kmeans}.

Fix a facility $i$, and let $D^* = \{j \mid \alpha_j - (\rho-1)R_j^2 \geq \Gamma d^2(j,i)\}$. We need to show that 
\begin{align}
    \label{eqn:overview-dual-feasibility-simpler}
    \sum_{j \in D^*} \alpha_j - (\rho-1)R_j^2  \leq \hat{f} + \Gamma \sum_{j\in D^*}d^2(j,i)~.
\end{align}
Let us assume by renumbering that $j \in D^*$ is sorted in increasing order of $\alpha_j$, and let us denote $D^* = [s]$. Recall that we assume that no two $\alpha_j$ are equal.

Then we consider the contributions to $i$ at time $\alpha_t - \eps$ for some arbitrarily small $\eps > 0$ for each $t \in [s]$. Observe that each client $j < t$ is already directly connected to an open facility in our special case. Let $S^t$ be the set of open facilities at time $\alpha_t - \eps$, and $r_{jt} = d(j, S^t)$ be the distance at which $j$ is connected at time $\alpha_t - \eps$.
Then, we get the following inequality.
\begin{equation}\label{eqn:facility}
\sum_{j < t} (r_{jt}^2 - d^2(j,i)) + \sum_{j \geq t} (\alpha_t - \rho d^2(j,i)) \leq \hat{f}.
\end{equation}
This inequality follows from the fact that clients $j < t$ are already (directly) connected at the time $\alpha_t - \eps$, and hence they bid $[r_{jt}^2 - d^2(j,i)]^+$ to $i$, and clients $j \ge t$ are still active, hence they bid $[\alpha - \rho d^2(j,i)]^+$ to $i$, where $\alpha$ is their current $\alpha$ value (which at time $\alpha_t - \eps$, is $\alpha_t - \eps$). 
We relax the terms of type $[x]^+$ to $x$, which can only make the LHS smaller.
Taking $\eps \to 0$ and since the algorithm ensures that at all times, every facility receives a total bid of at most $\hat{f}$, implying \Cref{eqn:facility}.

Next, we eliminate $r_{jt}^2$ via triangle inequality. Since we know that at time $\alpha_t - \eps$, $t$ was not connected, we must have \begin{align}
    \label{eqn:tri-overview}
    \alpha_t \leq d^2(t,S) \leq (d(j,t) + {r_{jt}})^2
\end{align}
See \cref{lem:p-tjl} for a formal proof. 
To get started with the spectral analysis, we will plug this lower bound on $r_{jt}$ into \cref{eqn:facility} and obtain a new inequality that upper bounds the sum of $\alpha_j-(\rho-1)R_j^2$ in terms of $\hat f$ and the sum of $d^2(j,i)$ (see \cref{lem:dual-feasibility-simpler-metric,lem:dual-feasibility-simpler-euclidean} for the formal statements we used for general and Euclidean metrics respectively). 

We will use a slightly different approach to use the inequality $\alpha_t\leq (d(j,t)+r_{jt})^2$ for the metric case and the Euclidean case. 
\begin{itemize}
    \item Consider the Euclidean space of dimension $\di$. We will assume w.l.o.g. that the facility $i$ is located at the origin of the Euclidean space and use the coordinate of the clients, $(x_j)_{j\in [s]}\in \mathbb{R}^{\di}$, to rewrite $d(j,t)=\|x_j-x_t\|$ and $d(j,i)=\|x_j\|$. Using \cref{eqn:facility,eqn:tri-overview}, some quadratic inequalities and the Cauchy-Schwarz inequality (to make every $d(\cdot,\cdot)$ appear in a quadratic form in the inequalities), we can upper bound $\sum_{j} \alpha_j - (\rho - 1)R_j^2$ by the sum of $\hat{f}$ and a quadratic form of $x=(x_j)_{j\in [s]} \in \mathbb{R}^{s \cdot \di},$ denoted as $\hat{Q}(x)$.
    Because we have $\sum_{j} d^2(j,i) = \sum_{j} \|x_j\|^2 = \| x \|_2^2$, it suffices to upper bound $\frac{\hat{Q}(x)}{\|x\|_2^2}$. As an $(s \cdot \di) \times (s \cdot \di)$ matrix, $\hat{Q}$ is a block-diagonal matrix with $\di$ many identical $s \times s$ matrices on the diagonal, so an upper bound on the maximum eigenvalue of this $s \times s$ matrix yields a bound on the approximation ratio. %
    \item In the metric case, we use the metric triangle inequality $d(j,t)\leq d(j,i) + d(t,i)$. Then, we will consider each $d(j,i)$ as a variable $a_j$, and use \cref{eqn:facility,eqn:tri-overview} and some quadratic inequalities to upper bound $\sum_{j} \alpha_j - (\rho - 1)R_j^2$ by a sum of $\hat{f}$ and a quadratic form of $(a_j)_{j\in [s]}$, denoted as $\hat{Q}(a)$. Similarly, because we have $\sum_{j} d^2(j,i) = \sum_{j} a_j^2=\|a\|^2$, it suffices to upper bound the maximum eigenvalue of $\hat{Q}(a)$ (again a quadratic form on $s$ variables) to prove the bound on our objective \cref{eqn:overview-dual-feasibility-simpler}.
\end{itemize}

Numerically, we found the following barriers for our approach:
\begin{itemize}
    \item \textbf{Euclidean case:} After the Cauchy-Schwarz step (which makes the spectral analysis work), the underlying matrix $\hat{Q}(x)$ has an eigenvalue $\geq 3.51$ in the case where there are 300 clients: the first 260 clients are immediately directly connected when they become inactive; while the last 40 clients are never directly connected. 
    \item \textbf{Metric case}: The underlying matrix $\hat{Q}(x)$ has an eigenvalue $\geq 4.67$ when the number of clients is large. More details can be found at the end of \cref{sec:metric-warmup}.
\end{itemize}

\subsection{Spectral analysis for the Euclidean case}
\label{sec:euclidean-warmup}

Let us fix $\rho=2$. Recall that we are considering the special case where all the considered clients at any point of time are either active or directly connected. To further simplify the analysis, we present a weaker bound $\Gamma\leq 3+\frac{27}{32}<3.85$ for this case.

Let $x_1, x_2 \ldots x_s$ be the vectors denoting the locations of the $s$ clients, and suppose that the facility $i$ is located at the all-zeros location, so that $d^2(j,i) = \|x_j\|^2$ for each $j \in [s]$. 
For simplicity, we will use $x=(x_j)_{j\in [s]}$ to denote the $s \cdot \di$-dimensional vector by concatenating the $s$ vectors.
In the Euclidean case, we will rewrite \cref{eqn:tri-overview} by $\alpha_t \leq (r_{jt} + ||x_j - x_t||)^2$ 
for each $t > j$. Thus we get $\alpha_t \leq r_{jt}^2 + ||x_j - x_t||^2 + 2r_{jt}||x_j - x_t||$.
Adding across all $t > j$ for a fixed $j$ and using $r_{jt} \leq R_j$, we get 
\begin{align*}
\sum_{t > j} (\alpha_t - r_{jt}^2) \leq \sum_{t > j} 2R_j||x_j - x_t|| + \sum_{t > j}||x_j - x_t||^2
\end{align*}
Subtracting $s(\rho - 1)R_j^2=sR_j^2$, 
\begin{align*}
\sum_{t > j} (\alpha_t -{r_{jt}^2}) - {sR_j^2} 
\leq \sum_{t > j} \bigl(2R_j||x_j - x_t|| + ||x_j - x_t||^2\bigr) -
{sR_j^2}
\end{align*}%
Applying the quadratic inequality $2ay - by^2 \leq \frac{a^2}{b}$, %
obtained from
$\frac{1}{b}(a-by)^2\geq 0$, we have
\begin{align}
\label{eqn:intro-euclidean-no-overbidding}
\sum_{t > j} (\alpha_t - {r_{jt}^2}) - 
{sR_j^2}
\leq \frac{\Bigl( \sum_{t > j}||x_j - x_t||\Bigr)^2}{{s}}
+ \sum_{t > j} ||x_j - x_t||^2
\end{align}

We sum~\cref{eqn:facility} over all $t\in [s]$ and the above~\cref{eqn:intro-euclidean-no-overbidding} over all $1\leq j<t\leq s$, and then divide by $s$. We get
$$\sum_{t \in [s]} \alpha_t - R_t^2\leq \hat f + Q(x)~,$$ where $Q(x)$ is a function of  $x_1, x_2 \ldots x_s$ alone:
$$Q(x) =\frac{1}{s}\left( \sum_j\frac{\bigl( \sum_{t > j}||x_j - x_t||\bigr)^2}{{s}} 
+ \sum_j\sum_{t > j} ||x_j - x_t||^2 + \sum_{t}\sum_{j < t} ||x_j||^2 + \sum_{t}\sum_{j \ge t} \rho ||x_j||^2\right)\,.$$
The first two terms come from the lower bound plugged in for $r_{jt}^2$ from above, and the last two terms come from the minus terms on the LHS of \cref{eqn:facility}.

Observe that the second term $\frac{1}{s} \sum_{j} \sum_{t>j} \|x_j-x_t\|^2$ equals to the optimal single-cluster cost for $[s]$ (see, e.g., \cref{lem:opt-single-cluster}). Hence, it is no greater than $\sum_{j} \|x_j\|^2$. 
Rearranging the third and the fourth term via $\frac{1}{s} \sum_{t} \sum_{j} \rho \|x_j\|^2 - \frac{1}{s} \sum_{t} \sum_{j<t} (\rho-1) \|x_j\|^2$, we can rewrite them as $\rho\sum_{j} \|x_j\|^2 - (\rho-1) \sum_{j} \frac{s-j}{s} \|x_j\|^2$. Using our parameter $\rho=2$, we have the following upper bound for $Q(x)$:
\begin{align*}
    Q(x) \leq \frac{1}{s^2} \sum_{j} \left(\sum\nolimits_{t>j} \|x_j-x_t\|\right)^2 - \sum_j \frac{s-j}{s} \|x_j\|^2  + 3\sum_{j} \|x_j\|^2.
\end{align*}

Now we apply Cauchy-Schwarz on the first term to
obtain $Q(x)\leq \hat{Q}(x) + 3\sum_{j} \|x_j\|^2$, where
$$\hat{Q}(x) := \sum_j \frac{s - j}{s^2} \sum_{t > j} ||x_j - x_t||^2 - \sum_{j} \frac{s-j}{s} \|x_j\|^2~. %
$$

If we show that $\hat{Q}(x) \leq \sum_j \lambda||x_j||^2$, then we obtain an LMP $(3+\lambda)$-approximation. 
This now reduces to upper bounding the largest eigenvalue of the matrix given by the quadratic form $\hat{Q}$.
We next prove that $\lambda\leq \frac{27}{32}<0.85$.

    Consider two cases for $j$: $j\geq s/4$ and $j<s/4$. 
    If $j\geq s/4$, we have $\frac{s-j}{{s^2}} \leq \frac{3}{4s}$. The total contribution of $j$ to $\hat{Q}(x)$ is then upper bounded by $\frac{3}{4s} \sum_{t>j} \|x_j-x_t\|^2$.
    If $j<s/4$, we have (in the second inequality, we use $(a-b)^2 \leq 3a^2+1.5b^2$) 
    \begin{align*}
        \frac{s-j}{s^2} \sum_{t>j} \|x_j-x_t\|^2 - \frac{s-j}{s} \|x_j\|^2 &<  \frac{1}{s} \sum_{t>j} \|x_j-x_t\|^2 - \frac{3}{4} \|x_j\|^2
        \\
        &\leq \frac{3}{4s} \sum_{t>j} \|x_j-x_t\|^2 + \frac{1}{4s} \sum_{t>j} (3\|x_j\|^2+1.5\|x_t\|^2)- \frac{3}{4} \|x_j\|^2
        \\
        &\leq \frac{3}{4s} \sum_{t>j} \|x_j-x_t\|^2 + \frac{3}{8s} \sum_{t>j} \|x_t\|^2
    \end{align*}
    The intuition behind discussing these two different cases is as follows: for small $j$, there is a large subtraction term $-\frac{s-j}{s} \|x_j\|^2$; because of this, and the flexibility of squared triangle inequalities, we can let $\|x_j\|^2$ pay more for the cost in $\sum_{t>j} \|x_j-x_t\|^2$ (and let $\|x_t\|^2$ pay less).
    
    Summing over all $j$, using the upper bound discussed above, we have \begin{align*}
        \hat{Q}(x) &\leq \frac{3}{4s} \sum_{j<t} \|x_j-x_t\|^2 + \frac{3}{8s}\sum_{j<s/4} \sum_{t>j} \|x_t\|^2 
        \\
        &\leq \frac{3}{4s} \sum_{j<t} \|x_j-x_t\|^2 + \frac{3}{32} \sum_{t}  \|x_t\|^2 
        \\
        &\leq \frac{3}{4} \sum_{j} \|x_j\|^2 + \frac{3}{32} \sum_{t} \|x_t\|^2 = \frac{27}{32} \sum_{j} \|x_t\|^2~,
    \end{align*}
    where the last inequality follows from the fact that $\frac1s\sum_{j<t} \|x_j-x_t\|^2$ is the optimal single-cluster cost for $[s]$ among all possible centers in $\mathbb{R}^{\di}$ while $\sum_{j}\|x_j\|^2$ is the cost of opening the center at the origin for $[s]$.

     In \cref{sec:euclidean-lmp},
     we generalize and refine the above derivation to the general case with both $IC$ and $DC$ clients to obtain the claimed LMP $(3+\ln 2)$-approximation for Facility Location in the Euclidean case.
     Our proof combines the above derivation with the techniques introduced in~\cite{CCGGLW2026kmeans} to deal with clients in $IC$, and it is substantially more complex.

\subsection{Spectral analysis for the metric case}
\label{sec:metric-warmup}
Now we present an overview of our spectral analysis for Metric $k$-Means in the special case when all clients are either directly connected or active (i.e., $IC$ is empty at all times). We generalize it to all possible cases in \cref{sec:metric-lmp}.

Let $a_j= d(j,i)$. 
According to \cref{eqn:tri-overview} and the metric triangle inequality, we have $\alpha_t \leq d^2(t,S_t)\leq (r_{jt}+d(j,t))^2 \leq (r_{jt}+d(j,i)+d(t,i))^2 = (r_{jt}+a_j+a_t)^2$.
Thus, we can expand the square to get the upper bound $\alpha_t - r^2_{jt} \leq 2{r_{jt}}(a_j + a_t) + (a_j + a_t)^2$.

Fix any $j$. Adding across all $t > j$,
\begin{align*}
\sum_{t > j} (\alpha_t - r^2_{jt}) &\leq \sum_{t > j}  2{r_{jt}}(a_j + a_t) + (a_j + a_t)^2  \\
&\leq 2{R_j}\Bigl((s-j)a_j + \sum_{t > j}a_t\Bigr) + \sum_{t > j} (a_j + a_t)^2,
\end{align*}
where we crucially use the fact that each $r_{jt} \leq R_j$ by the property of our algorithm. Now we subtract $s(\rho - 1)R_j^2$ for $j$, and obtain
\begin{align*}
\sum_{t > j} (\alpha_t - r_{jt}^2) - s(\rho - 1) R_j^2 &\leq 2{R_j}\Bigl((s-j)a_j + \sum_{t > j}a_t\Bigr) - s(\rho - 1)R_j^2 + \sum_{t > j} (a_j + a_t)^2  \\
&\leq \frac{1}{s(\rho - 1)}\Bigl((s-j)a_j + \sum_{t > j}a_t\Bigr)^2 + \sum_{t > j} (a_j + a_t)^2  
\end{align*}
where the last inequality is obtained by observing that the quadratic terms are controlled by $2ax - bx^2 \leq \frac{a^2}{b}$, which is equivalent to $\frac{1}{b}(a-bx)^2 \geq 0$.

If we now plug this lower bound for $r_{jt}$, and follow the same procedure of adding each inequality at time $\alpha_t - \eps$ for $t \in [s]$ and divide by $s$, we would get a bound of the form $\sum_j (\alpha_j- (\rho - 1)R_j^2) \leq \hat f + \hat{Q}(a)$ 
where $\hat{Q}(a)$ is the new quadratic form on the distances arising from this process. If the maximum eigenvalue of the quadratic form corresponding to $\hat{Q}(a)$ is $\Gamma$, then this is at most $\hat f + \Gamma a^Ta$, giving us an LMP $\Gamma$-approximation.

Now we can write the quadratic form $\hat Q (a)$ corresponding to our setup.   
\begin{align*}
\hat Q (a) = \frac{1}{s} \left( \sum_{j} \frac{1}{s(\rho - 1)}\Bigl((s-j)a_j + \sum_{t > j} a_t\Bigr)^2 + \sum_{j}\sum_{t > j} (a_j + a_t)^2 + \sum_{t} \sum_{j \geq t} \rho a_j^2 + \sum_{t} \sum_{j < t} a_j^2 \right)  
\end{align*}

The first and second terms come from substituting the lower bounds for $r_{jt}$ from the calculation above. The third and fourth terms come from the  form of the bids of the clients at time $\alpha_t - \eps$, clients $j\geq t$ are active and bid $\alpha_t - \rho a_j^2$ (hence the coefficient of $\rho$ for $a_j^2$), whereas clients $j < t$ are directly connected and bid $r_{jt}^2 - a_j^2$ (hence the coefficient of $1$ for $a_j^2$).

First, let us show via elementary analysis that when $\rho = 2$, we can bound $\Gamma \leq 5$. The first term inside the parenthesis,
$\sum_j\frac{1}{s}\bigl((s-j)a_j + \sum_{t > j} a_t\bigr)^2$, is equal to $\sum_j\frac{1}{s}\left(\sum_{t > j} (a_j + a_t)\right)^2$. Using the Cauchy-Schwarz inequality, this is at most $\sum_j\frac{s-j}{s}\sum_{t > j}(a_j + a_t)^2$, which is at most $\sum_j\frac{2(s-j)}{s}\sum_{t > j} (a_j^2 + a_t^2)$.  Again, for the middle term, we get $\sum_j\sum_{t > j} (a_j + a_t)^2 \leq \sum_j 2\sum_{t > j} (a_j^2 + a_t^2) = \sum_{j}2\sum_{t > j} (a_j^2 + a_t^2)$.

Thus with $\rho = 2$, the expression is upper bounded by

\begin{align*}
\hat{Q}(a) \leq \frac{1}{s}\left (\sum_{j} \left(2 + \frac{2(s-j)}{s}\right) \sum_{t > j}(a_j^2 + a_t^2) + \sum_t \sum_{j \geq t} 2a_j^2 + \sum_t \sum_{j < t}a_j^2\right)
\end{align*}

Now let us fix an index $j^*$ and collect coefficients of $a_{j^{*}}^2$ for the term inside the parenthesis. The total coefficient is then

$$2(s-j^*)\left(1 + \frac{s-j^*}{s}\right) + 2\sum_{j < j^*}\left(1 + \frac{s-j}{s}\right) + 2j^* + (s - j^*) = 5s - j^* - 4 + \frac{j^*(j^* + 1)}{s}$$

Dividing by the outer $s$ and using $j^* \leq s$, the total coefficient is at most $5$, which gives the desired inequality $\hat{Q}(a) \leq 5a^Ta$, and hence $\Gamma \leq 5$.

However, it is possible to do better. By running experiments for various values of $\rho$, one can observe that the largest eigenvalue of the symmetric matrix describing this quadratic form is (numerically) minimized when $\rho \approx 2.5$. With this value of $\rho$, we numerically estimate the top eigenvalue to be $4.67783$ as $s \to \infty$. It can be shown that the matrix corresponding to the quadratic form $\hat{Q}(a)$ is the sum of a diagonal and a rank one matrix, and hence it is not difficult to prove an analytic bound that reaches close to this constant for any $s$. Thus using our new algorithm with the new spectral analysis, we get an approximation ratio of $4.68$ for this special case when no client is indirectly connected.

Thus all that is left 
is to generalize this to the case when there are indirectly connected clients. Let $DC_t$ and $IC_t$ be the sets of directly and indirectly connected clients at time $\alpha_t - \epsilon$. Then the number of such valid sets\footnote{The only condition for validity is that $DC_t \subseteq DC_{t+1}$ for all $t$, since the algorithm moves clients only from $IC$ and $A$ to $DC$, while never moving from $DC$ to $IC$.} is exponential in $s$. Thus, we need a way to bound the general case directly. 

For this purpose, we write down the equations for the most general $IC$/$DC$ case, and try to solve them all at once, similarly to~\cite{charikar2025kmeans}. However, this analysis becomes more technically challenging %
as the matrices involved are no longer simple sums of diagonal and rank one matrices. %
Nevertheless, we show that the quadratic form in the general case can always be bounded by 4.9. Our proof uses the \ref{eq:collatz-wielandt}
which says that given a non-negative matrix $M$, if we find a non-zero vector $h$ such that $ \max_i \frac{(Mh)_i}{h_i}$ is at most $\Gamma$, then the top eigenvalue of $M$ is at most $\Gamma$. 

\subsection{Turning LMP approximations to bicriteria approximations}

To turn the LMP approximations to bi-criteria approximations (opening $k+O(\log n)$ centers), we employ the framework of \cite{CGLSS25stoc,charikar2025kmeans,CCGGLW2026kmeans}. The only technical improvement is as follows. The most recent work \cite{CCGGLW2026kmeans} can only turn an LMP $\Gamma$-approximation to a $\Gamma+\epsilon$ bi-criteria approximations for any constant $\Gamma\geq 4$. We remove this condition, and now this step works for any constant $\Gamma>\rho$ (which is a clear lower bound for our LMP algorithm).

\section{Preliminaries}
\label{sec:preliminaries}

For a client $j$ and a set of facilities $S$, let $d(j,S) := \min_{i\in S} d(j,i).$
For any real number $x$, let $[x]_+ := \max\{x,0\}$. For a positive integer $n$, we
write $[n] := \{1,2,\dots,n\}$, and more generally, for integers $l \le r$, we write
$[l:r] := \{l,l+1,\dots,r\}$. We use $\mathbb{R}_{>0}$ and $\mathbb{R}_{\ge 0}$ to denote
the sets of positive and non-negative reals, respectively.
\paragraph{Facility Location with uniform opening costs and squared distances.}%
In the {\em Facility Location with uniform opening costs and squared distances} problem, we are given a set of clients $D$, a set of
facilities $\facilities$, and a facility opening cost $f > 0$. The goal is to find a set $S \subseteq \facilities$
minimizing
\[
f|S| + \sum_{j\in D} d^2(j,S).
\]
We write $\OPT_{\mathrm{FL}}(f)$ for the optimum value of this problem.
We will refer to this problem simply as {\em Facility Location} throughout the paper. 

\paragraph{LP relaxation.}
The standard LP relaxation of \facloc is
\begin{equation}
\label{eq:lp-primal}
\tag{Primal FL LP}
\begin{aligned}
\min \quad & f\sum_{i\in \facilities} y_i + \sum_{j\in D}\sum_{i\in \facilities} d^2(j,i)x_{ij} \\
\text{s.t.}\quad
& \sum_{i\in \facilities} x_{ij} \ge 1 && \forall j\in D,\\
& x_{ij} \le y_i && \forall j\in D,\ i\in \facilities,\\
& x_{ij} \ge 0,\ y_i \ge 0. &&
\end{aligned}
\end{equation}
Its dual can be written as
\begin{equation}
\label{eq:lp-dual}
\tag{Dual FL LP}
\begin{aligned}
\max \quad & \sum_{j\in D} \alpha_j \\
\text{s.t.}\quad
& \sum_{j\in D} [\alpha_j - d^2(j,i)]^+ \le f && \forall i\in \facilities,\\
& \alpha_j \ge 0 && \forall j\in D.
\end{aligned}
\end{equation}
We write $\OPT_{\mathrm{LP}}(f)$ for the optimum value of the \ref{eq:lp-dual}.

\begin{definition}[LMP approximation]\label{def:lmp}
Given $\Gamma \ge 1$, an algorithm is Lagrangian Multiplier Preserving (LMP) $\Gamma$-approximation for
the \facloc problem if, for every instance with facility opening cost
$f > 0$, the output $S \subseteq \facilities$ satisfies
\[
\Gamma f \cdot |S| + \sum_{j\in D} d^2(j,S)
\;\le\;
\Gamma \cdot \OPT_{\mathrm{LP}}(f).
\]
\end{definition}

\paragraph{Bounding distances.}

Thanks to standard reductions (see, e.g., \cite{charikar2025kmeans}), in the metric case we will assume that  distances are integers in $[1,n^3/\eps]$ while losing a factor $1+O(\eps)$ in the approximation. 
\begin{lemma}[\cite{charikar2025kmeans}]
\label{lem:aspectratio}
For any constants $\eps > 0$ and $\alpha>1$, given a polynomial-time $\alpha$-approximation algorithm for $k$-Means on instances with distances in $\{1,\ldots,n^3/\eps\}$, there exists a polynomial-time $\alpha(1+O(\eps))$-approximation algorithm for Metric $k$-Means. 
\end{lemma}
In
the Euclidean case, we will assume that
non-zero 
distances are in $[1,n^2/\eps^2]$ thanks to the following reduction in \cite{CCGGLW2026kmeans}.
\begin{lemma}[\cite{CCGGLW2026kmeans}]
\label{lem:aspectratio-euclid}
For any given $\eps > 0$ and $\alpha>1$, given a polynomial-time $\alpha$-approximation algorithm for Euclidean $k$-Means on instances with
non-zero 
distances between $1$ and $n^2/\eps^2$, there exists a polynomial-time $\alpha(1+O(\eps))$-approximation algorithm for Euclidean $k$-Means on general instances. 
\end{lemma}

\paragraph{Triangle Inequalities for Squared Distances.}
We will use the following simple inequalities taken from \cite{charikar2025kmeans}. Given three distances $d,d_1,d_2$ with $d\leq d_1+d_2$, the following lemma provides a useful upper bound on $d^2$ as a function of $d_1$ and $d_2$. The next lemma provides a similar upper bound in an analogous setting where $d\leq d_1+d_2+d_3$.
\begin{lemma}[\cite{charikar2025kmeans}]\label{lem:apxTriangleInequality2}
 Let $\gamma>1$ be any constant. For any $x,y$ we have $\gamma x^2+\frac{\gamma}{\gamma-1}y^2\geq (x+y)^2$.    
\end{lemma}
\begin{lemma}[\cite{charikar2025kmeans}]\label{lem:apxTriangleInequality3}
 Let $\gamma>1$ be any constant. For any $x$, $y$, $z$, we have $\gamma x^2+\left(2+\frac{2}{\gamma-1} \right)(y^2+z^2)\geq (x+y+z)^2$.    
\end{lemma}

\section{Our Improved Greedy Algorithm}
\label{sec:greedy}
In this section, we present our improved greedy LMP approximate algorithm for the  \facloc problem with uniform {facility} cost $f$. 
As {in the} previous works \cite{charikar2025kmeans,CCGGLW2026kmeans}, our algorithm is based on the well-known greedy algorithm by \textcite{JainMMSV03}.

In our algorithm, we use parameters $\rho,\Gamma>1$, which will be defined (differently for the metric and Euclidean cases) later.
We also use $\hat{f} = \Gamma \cdot f$, where $f$ is the uniform facility opening cost. The following~\cref{alg:greedy} is our new
greedy algorithm achieving the improved LMP approximation. As standard in the area, we present the algorithm in a continuous version where there is a variable $\theta$ (interpreted as \emph{time}) that grows continuously from $0$ to $+\infty$. 
Initially, all clients are active.
We grow the $\alpha$-values of active clients at the same speed as $\theta$ grows, until one of two conditions is met: (1) a new facility becomes openable, or (2) a client's $\alpha$-value reaches its connection cost to some open facility.
It is easy to discretize this algorithm.
Here, we say one facility is openable if the total bid of the clients to this facility reaches $\hat f$, where the bid of client $j$ to a facility $i$ is defined as $[\alpha_j-\rho\cdot d^2(j,i)]^+$ when $j$ is in $A$ or $IC$, or as $[d^2(j,S)-d^2(j,i)]^+$ when $j$ is in $DC$, at the time $i$ is opened.
In addition, for each client $j \in D$, we explicitly track its first direct-connection cost, $R^2_j$. We set $R_j = d(j, i^*)$ where $i^*$ is the facility that caused $j$ to be moved to $DC$ for the first time. For a client that remains only indirectly connected or active, we keep $R_j = 0$ as it is initialized.

\begin{figure}[ht!]
\begin{center}
\begin{minipage}{1.0\textwidth}
\begin{mdframed}[hidealllines=true, backgroundcolor=gray!15]
\textbf{Algorithm \ref{alg:greedy}\ (\JMSalg), restated.} \\[0.2cm]
    \textbf{Parameters:} $\rho, \Gamma > 1$; $\hat{f} = \Gamma \cdot f$ where $f$ is the uniform facility opening cost.
    
    \vspace{0.5em}

    \textbf{Initialization:} $A=D$, $IC=DC=S=\emptyset$, and $R_j=\alpha_j=0$ for every $j\in D$. 
    
    \vspace{0.5em}

    \noindent While $A\neq\emptyset$, uniformly grow the $\alpha_j$ of every client in $A$ until:%
    \begin{enumerate}
        \item For some unopened facility $i\in \facilities \setminus S$, we have
        \[
        \sum_{j\in A \cup IC}\left[\alpha_j - \rho \cdot d^2(j,i)\right]^+
        + \sum_{j\in DC}\left[d^2(j,S) -  d^2 (j, i)\right]^+ = \hat{f}.
        \]
        
        In this case, we open facility $i$.
        For each client $j\in DC$, if $d^2(j,S)>d^2(j,i)$, we reconnect $j$ to $i$.
        For each client $j\in A\cup IC$,
        \begin{itemize}
            \item If $\alpha_j \geq \rho \cdot d^2(j,i)$, we (re)connect $j$ to $i$, move $j$ to $DC$ and set $R_j=d(j,i)$.
            \item Else, if $\alpha_j \geq d^2(j,i)$, we (re)connect $j$ to $i$, set $\alpha_j$ to be $d^2(j,i)$, and move $j$ to $IC$ (if $j\in A$). %
        \end{itemize}

        \item For some client $j\in A$, we have
        \[
        \alpha_j = d^{2}(j,S).
        \]
        In this case, we move $j$ to $IC$ and connect $j$ to the closest facility to it in $S$.
    \end{enumerate}
\end{mdframed}
\end{minipage}
\end{center}
\end{figure}

\begin{theorem}
\label{thr:greedyLMP}
\cref{alg:greedy} is a LMP $\Gamma$-approximation for the {\facloc} problem with uniform {facility} cost $f$ if 
\begin{itemize}
    \item $d$ is a (general) metric, $\rho=2.5$ and $\Gamma=4.9$; or
    \item $d$ is a Euclidean metric, $\rho=2$ and $\Gamma=3+\ln(2)$.
\end{itemize}
\end{theorem}

\begin{observation}
    \label{lem:intermediate-alpha-values}
    At any time during the algorithm, for any $j\in IC$, we have $\alpha_j=d^2(j,S)$.
\end{observation}

{Because} we freeze $\alpha_j$ when a client $j$ becomes directly connected, we can show that the $\alpha$ values are non-increasing after connection. 
For some intermediate time $\theta$, we will use $\alpha_j(\theta)$ to denote the $\alpha$ value of client $j\in D$ {at that time}.  
Further, for each $j\in D$, we define $\hat{\alpha}_j$ to be its $\alpha$ value at the first time when $j$ is connected to some facility.
Let $(\alpha^*_j)_{j\in D}$ be the final vector of $\alpha$-values. 
\begin{observation}
    \label{lem:alpha-decreasing-after-connection}
    For any client $j\in D$, $\alpha_j(\theta)$ is non-increasing in $\theta\in [\hat{\alpha}_j, +\infty)$. %
\end{observation}
\begin{proof}
    This follows {from} the fact that we only modify $\alpha_j$ {after time $\hat{\alpha}_j$} when some facility $i$ is opened and $\alpha_j>d^{2}(j,i)$. Then the resulting value of $\alpha_j$ is {lowered to} $d^{2}(j,i)$.
\end{proof}

\begin{corollary}
    \label{cor:alpha-vs.-hat-alpha}
    For each client $j$, we have $\alpha^*_j\leq \hat{\alpha}_j$.
\end{corollary}
Our goal is to show that $(\alpha^*_j - (\rho-1)R_j^2)_{j\in D}$ can characterize the total cost of \cref{alg:greedy} and can be scaled to a feasible dual solution to the {\ref{eq:lp-dual}}. %

First, we prove that the sum of $(\alpha^*_j - (\rho-1)R_j^2)_{j\in D}$ is an upper bound of the total cost. The proof relies on the following observation: for each client $j$ that is finally directly connected, its $\alpha$-value can be written as the sum of its final connection cost $d^2(j,S)$, $\rho-1$ times its first direct-connection cost $R_j^2$, and the bids of $j$ to the open facilities.

\begin{lemma}[Dual Payment, {\Cref{lem:approxguarantee-overview} restated, {replacing $\alpha$ by $\alpha^*$}} ]\label{lem:approxguarantee}
At the end of the execution of \Cref{alg:greedy},%
\[
\sum_{j\in D} d^2(j,S) + |S|\hat{f} \leq \sum_{j\in D}(\alpha^*_j - (\rho-1)R_j^2).
\]
\end{lemma}
\begin{proof}
    First, observe that it is sufficient to prove that at any point in the execution, we have
    \begin{equation}
    \label{eq:induc-cost-of-alg-metric}
    \sum_{j\in DC} (\alpha_j - (\rho-1)R_j^2) \geq \sum_{j\in DC} d^2(j,S) + \sum_{i\in S} \hat{f}.
    \end{equation}
    Indeed, at the end of the execution, $A=\emptyset$, and for every $j\in IC$, we have $\alpha_j^*= d^2(j,S)$ and $R_j = 0$ (since $j$ never became directly connected). Thus, their contribution to the left-hand side is exactly their connection cost, making \eqref{eq:induc-cost-of-alg-metric} imply the lemma.
    
    To prove this, we proceed by induction and show that \eqref{eq:induc-cost-of-alg-metric} is maintained throughout the execution.

    The inequality is initially true since $DC=\emptyset$ and $S = \emptyset$. Furthermore, since whenever the $\alpha$-value of a client $j$ reaches $d^2(j, S)$ it stops growing and moves to $IC$, no client is added to $DC$ except when a new facility is opened (i.e., added to $S$).
    
    We now consider what happens when we open a facility $i$, i.e., add it to $S$. 
    Let $(\alpha, S,$ $A, IC, DC)$ be the state right before opening $i$, and set $DC' = \{ j \in A\cup IC : \alpha_j \geq \rho d^2(j, i) \}$, and $X=\{j\in DC:d^2(j, S)>d^2(j, i)\}$.
     The change of cost of the right-hand side of~\eqref{eq:induc-cost-of-alg-metric} is exactly
    \begin{align*}
     \hat{f} &+   \sum_{j\in DC'} d^2(j, i) + \sum_{j \in X} (d^2(j, i) - d^2(j, S))\,.
    \end{align*}
    Since the algorithm decided to open $i$, we know that the total offer reaches $\hat{f}$, so we have
    \begin{align*}
      \hat{f} &\leq \sum_{j\in DC'}(\alpha_j - \rho d^2(j, i)) + \sum_{j\in X}(d^2(j, S)- d^2(j, i))\,.
    \end{align*}
    Substituting this bound into our expression, we get that the change of cost of the right-hand side is at most
    \begin{align*}
        \sum_{j\in DC'} (\alpha_j - \rho d^2(j, i) + d^2(j, i)) &= \sum_{j\in DC'} (\alpha_j - (\rho-1) d^2(j, i)).
    \end{align*}
    Because each client $j \in DC'$ is becoming directly connected for the first time, the algorithm sets its penalty variable $R_j = d(j, i)$, meaning $R_j^2 = d^2(j, i)$. Therefore, the change of the right-hand side is at most $\sum_{j\in DC'} (\alpha_j - (\rho-1)R_j^2)$.
    Because we do not change the $\alpha$-value or $R_j$ value after a client is moved to $DC$, this is exactly the change of the left-hand side of~\eqref{eq:induc-cost-of-alg-metric}. The invariant is thus maintained, completing the proof.
\end{proof}

Second, we prove that $\big((\alpha^*_j - (\rho-1)R_j^2)/\Gamma\big)_{j\in D}$ can be scaled to a dual feasible solution %
{for an appropriate choice of}
$\rho, \Gamma$. Since the proof of this part is very long, we will first present a general framework for both the metric case and Euclidean case in \cref{subsec:dual-feasibility-framework}, where we establish a set of constraints for the variables $(\alpha^*_j)_{j\in D}$ and $(R_j)_{j\in D}$. Then, we will use the constraints to establish the following lemma for the metric case and the Euclidean case separately in \cref{sec:metric-lmp,,sec:euclidean-lmp}, using significantly different techniques. 

\begin{lemma}[Dual Feasibility, {\Cref{lem:dual-feasibility-overview} restated,} {replacing $\alpha$ by $\alpha^*$}]
    \label{lem:dual-feasibility}
    For any facility $i\in \facilities$, we have 
    $$\sum_{j\in D} \left[\alpha^*_j-(\rho-1)R_j^2 - \Gamma d^2(j,i)\right]^+\leq  {\hat f~},$$
    where
    \begin{itemize}
        \item if $d$ is a (general) metric, we choose $\rho=2.5, \Gamma=4.9$; or
        \item if $d$ is a Euclidean metric, we choose $\rho=2, \Gamma = 3+\ln(2)$.
    \end{itemize}
\end{lemma}

\begin{proof}[Proof of \cref{thr:greedyLMP}]

Consider the standard \ref{eq:lp-dual} for \facloc problem {with $f=\hat f/\Gamma$} and define
\[
\widetilde{\alpha}_j
\;:=\;
\frac{\alpha_j^*-(\rho-1)R_j^2}{\Gamma}
\qquad \forall j\in D.
\]
By \Cref{lem:dual-feasibility}, $(\widetilde{\alpha}_j)_{j\in D}$ is dual feasible. Hence, by weak duality,
\[
\sum_{j\in D}\widetilde{\alpha}_j
\;\le\;
{\OPT_{\mathrm{LP}}(f)}
\]
Multiplying by $\Gamma$, we obtain
\[
\sum_{j\in D}\bigl(\alpha_j^*-(\rho-1)R_j^2\bigr)
\;\le\;
\Gamma\cdot \OPT_{\mathrm{LP}}(f).
\]
On the other hand, \Cref{lem:approxguarantee} gives
\[
\hat f \cdot |S| + \sum_{j\in D} d^2(j,S)
\;\le\;
\sum_{j\in D}\bigl(\alpha_j^*-(\rho-1)R_j^2\bigr).
\]
Using the fact that $\hat f=\Gamma f$, we conclude that
\[
\Gamma f \cdot |S| + \sum_{j\in D} d^2(j,S)
\;\le\;
\Gamma\cdot \OPT_{\mathrm{LP}}(f).
\]
The claim follows from \Cref{def:lmp}.
\end{proof}

\subsection*{Dual feasibility: the analysis framework}
\label{subsec:dual-feasibility-framework}
Next, we show the analysis framework to establish \cref{lem:dual-feasibility}.  We will use $[n]$ to denote the set $\{1,2,\dots, n\}$ and $[l:r]$ to denote the set $\{l, l+1,\dots, r\}$. From now on, we fix any facility $i\in \facilities$. 
Let $D^*=\{j\in D:\alpha_j^*-(\rho-1)R^2_j > \Gamma\cdot d^2(j,i)\}$.
Suppose that $|D^*|=s$ and we index them by $[s]$ and order them in the increasing order of the time they become inactive.

That is, we have $\hat{\alpha}_1\leq \hat{\alpha}_2 \leq \dots \leq \hat{\alpha}_{s}$. Then, our goal (which will be formally shown at the end of this subsection) becomes proving 
\begin{align}
    \label{eqn:new-goal-dualf}
    \sum_{j\in [s]} (\alpha^*_j-(\rho-1)R_j^2) \leq \hat f + \Gamma \cdot \sum_{j\in [s]} d^2(j,i)~.
\end{align}

Consider the time $\hat{\alpha}_t-\epsilon$ for each client $t$, where $\epsilon>0$ is sufficiently small so that neither condition in the algorithm is satisfied between time $\hat{\alpha}_t-\epsilon$ (inclusive) and $\hat{\alpha}_t$ (exclusive).
For simplicity, we define $DC^t$ as the set of directly connected clients in $[t-1]$ (i.e., $DC\cap [t-1]$) at time $\hat{\alpha}_t-\epsilon$, and $IC^t$ as the set of indirectly connected or active clients in $[t-1]$ (i.e., $(A\cup IC)\cap [t-1]$) at time $\hat{\alpha}_t-\epsilon$. 
Moreover, we use $S^t$ to denote the set $S$ at time $\hat{\alpha}_t-\epsilon$.
Because the sets $IC, DC, A$ are disjoint and the set $DC$ only grows, we have the following observation.
\begin{observation}
    \label{obv:disjointness-and-monotonicity}
For each $t\in [s]$, we have $DC^t\cap IC^t=\emptyset$ and $DC^t\cup IC^t = [t-1]$.
For each $t\in [s-1]$, we have $DC^t\subseteq DC^{t+1}$.
\end{observation}

Suppose that $r_{jt}$ (for $t\in [s]$ and $j\in DC^t$) is the distance of client $j\in DC^t$ to its closest opened facility at time $\hat{\alpha}_t-\epsilon$ (i.e., $d(j,S^t)$). One simple observation is that $r_{jt}\leq R_j$ for each $j\in DC^t$, as the set $S$ only grows in the algorithm.
\begin{observation}
    \label{obv:distance-to-opened-facility}
    For each $t\in [s]$ and $j\in DC^t$, we have $r_{jt}\leq R_j$. 
\end{observation}
Next, we will establish two sets of constraints for the variables $(\alpha^*_j)_{j\in [s]}$, $(r_{jt})_{1\leq j<t\leq s}$, and $(R_j)_{j\in [s]}$. The first set of constraints is as follows. The proof of the first set of constraints relies on the fact that for each client $t\in [s]$, its $\alpha$-value at time $\hat{\alpha}_t-\epsilon$ is strictly less than its distance to any open facility.
\begin{lemma}
  \label{lem:p-tjl}
  For each $t\in [s], j\in DC^t, \ell\in IC^t\cup\{t\}$, we have the following constraint:
  \begin{align}
      \label{eqn:p-tjl}
      & \alpha^*_{\ell} \leq (r_{jt}+d(j,\ell))^2~.
      \tag{$p_{t,j,\ell}$}
  \end{align}
\end{lemma}
\begin{proof}
    Consider any client $t\in [s]$.
    Consider the time $\hat{\alpha}_t-\epsilon$.
    Consider any client $\ell\in IC^t\cup\{t\}$. As $\ell\leq t$, we have $\hat{\alpha}_{\ell}\leq \hat{\alpha}_t$ by our assumption. 
    Next, we prove that $\alpha^*_{\ell}\leq d^2(\ell,S^t)$ by discussing two cases:
    \begin{itemize}
        \item If $\hat{\alpha}_{\ell}<\hat{\alpha}_{t}$ (since $\epsilon$ is sufficiently small, we have $\hat{\alpha}_{\ell}\leq \hat{\alpha}_t-\epsilon$), then $\ell$ is indirectly connected at time $\hat{\alpha}_{t}-\eps$. Because of \cref{lem:intermediate-alpha-values,,lem:alpha-decreasing-after-connection}, we have $\alpha^*_{\ell}\leq \alpha_{\ell}(\hat{\alpha}_t - \eps) = d^2(\ell,S^t)$.
        \item Otherwise (i.e., if $\hat{\alpha}_{\ell}=\hat{\alpha}_t$), we can prove $\hat{\alpha}_t-\epsilon < d^2(\ell,S^t)$ by contradiction. Suppose that $\hat{\alpha}_t-\epsilon \geq d^2(\ell,S^t)$. Then, $\ell$ will be connected to some opened facility using the second condition {in \cref{alg:greedy}} at time $\hat{\alpha}_t-\epsilon$, contradicting that $\ell$ is active at time $\hat{\alpha}_{t}-\eps$. 
        Since $\epsilon$ can be arbitrarily small here, we have $\hat{\alpha}_{\ell} = \hat{\alpha}_{t} \leq d^2(\ell,S^t)$. Because of \cref{cor:alpha-vs.-hat-alpha}, we get $\alpha^*_{\ell}\leq d^2(\ell,S^t)$.
    \end{itemize}  
    Because the triangle inequality holds for the metric $d$, we have $d(\ell,S^t)\leq d(j,S^t)+d(j,\ell)$. Therefore, $\alpha^*_{\ell}\leq d^2(\ell,S^t)\leq (r_{jt}+d(j,\ell))^2$, completing the proof.
\end{proof}

Our second set of constraints is as follows. The proof of the second set of constraints relies on the fact that for each client $t\in [s]$, the total bid to the fixed facility $i$ at time $\hat{\alpha}_t-\epsilon$ is strictly less than $\hat{f}$, as otherwise $t$ will be connected to $i$ at time $\hat{\alpha}_t-\epsilon$ and become inactive, contradicting the definition of $\hat{\alpha}_t$.
\begin{lemma}
  \label{lem:beta-t}
For each $t\in [s]$, we have the following constraint:
\begin{align}
    \label{eqn:beta-t}
    (s-t+1) \cdot \alpha^*_t  + \sum_{\ell \in IC^t} \alpha^*_\ell + \sum_{j\in DC^t} r_{jt}^2 \leq \hat{f} + \sum_{\ell \in IC^t \cup [t:s]} \rho \cdot d^2(\ell,i) + \sum_{j\in DC^t} d^2(j,i)~.
    \tag{$\beta_t$}
\end{align}
\end{lemma}
\begin{proof}
    Consider any client $t\in [s]$.
    Consider the time $\hat{\alpha}_t-\epsilon$. 
Note that we assume that $\alpha^*_t > (\rho-1)R^2_t + \Gamma \cdot d^2(t,i)$, where $\rho, \Gamma$ are some constants strictly greater than $1$. Because of \cref{cor:alpha-vs.-hat-alpha} 
and because $\epsilon$ is sufficiently small, we have $\hat{\alpha}_t-\epsilon \geq \alpha^*_t - \epsilon > d^2(t,i)$.
If facility $i$ is opened at time $\hat{\alpha}_t-\epsilon$, $t$ should be connected to some facility at time $\hat{\alpha}_t-\epsilon$, contradicting our assumption that $\hat{\alpha}_t$ denotes the time client $t$ is connected for the first time. 
Therefore, the total bid to $i$ at this time stamp should be strictly less than $\hat{f}$, i.e., (recall that $r_{jt}=d(j,S^t)$ for each $j\in DC^t$)
\begin{align*}
    \sum_{j\in DC^t} \left[r_{jt}^2 - d^2(j,i)\right]^+ 
    + \sum_{j\notin DC^t} \left[\alpha_j(\hat{\alpha}_t-\epsilon) - \rho \cdot d^2(j,i)\right]^+ < \hat{f}~.
\end{align*}

If $j\geq t$, we have $\alpha_j(\hat{\alpha}_t-\epsilon) = \hat{\alpha}_t - \eps \geq \alpha^*_t - \eps$. Otherwise,  if $j<t$ and $j\notin DC^t$, we have either $j$ being indirectly connected (i.e., $\hat{\alpha}_j\leq \hat{\alpha}_t-\epsilon$), or active (i.e., $\hat{\alpha}_j=\hat{\alpha}_t$ and $\alpha_j(\hat{\alpha}_t-\epsilon)=\hat{\alpha_t}-\epsilon$). 
If $j$ is indirectly connected, because of \cref{lem:alpha-decreasing-after-connection}, we have $\alpha_j^*\leq \alpha_j(\hat{\alpha}_t-\epsilon)$. 
Otherwise, because of \cref{cor:alpha-vs.-hat-alpha}, we have $\alpha_j^* - \epsilon \leq \hat{\alpha}_j - \epsilon = \hat{\alpha}_t - \epsilon = \alpha_j(\hat{\alpha}_t-\epsilon)$.
Because $\epsilon$ is sufficiently small, we have 
\begin{align*}
    \sum_{j\in DC^t} \left[r_{jt}^2 - d^2(j,i)\right]^+ 
    + \sum_{j\in IC^t} \left[\alpha^*_j - \rho \cdot d^2(j,i)\right]^+ + \sum_{j\geq t} [\alpha^*_t - \rho \cdot d^2(j,i)]^+ \leq \hat{f}.
\end{align*}
Since $[x]^+\geq x$, we can complete the proof by rearranging the terms in the above inequality.
\end{proof}

Finally, we only need to prove the following lemmas to complete the proof of \cref{thr:greedyLMP}. Because we will use the weighted version of \cref{lem:dual-feasibility-simpler-metric,,lem:dual-feasibility-simpler-euclidean} in %
{the log-adaptive algorithm and}
{walking between two solutions later,}
we will directly claim the weighted generalization. To establish \cref{lem:dual-feasibility}, we only need to consider the special case of them where $w(j)=1$ for each $j\in [s]$.

\begin{restatable}{lemma}{DualFeasibilitySimplerEuclidean}
    \label{lem:dual-feasibility-simpler-euclidean}
    Let $\rho=2$.
    Let $\hat{f}$ be any non-negative real number.
    Let $([s]\cup \{i\},d)$ be a general metric.
    Let $w:[s]\to \mathbb{R}_{>0}$ be any {positive} weight function.
    Let $(IC^t)_{t\in [s]}$ and $(DC^t)_{t\in [s]}$ be two families of sets satisfying:
    \begin{enumerate}[itemsep=0em,topsep=0.5em,label=(\alph*)]
        \item For each $t\in [s]$, $DC^t\cap IC^t=\emptyset$ and $DC^t\cup IC^t = [t-1]$.
        \item For each $t\in [s-1]$, $DC^t\subseteq DC^{t+1}$.
    \end{enumerate}
    Suppose that the variables $(\alpha^*_j)_{j\in [s]}, (R_j)_{j\in [s]}, (r_{jt})_{1\leq j<t\leq s}$ {satisfy} the following constraints
    \begin{enumerate}[itemsep=0em,topsep=0.5em,label=(\arabic*)]
        \item For any $t \in [s], j\in DC^t$, we have $r_{jt}\leq R_j$. \label{constr:euclidean-first}
        \item For each $t\in [s], j\in DC^t$ and $\ell\in IC^t\cup\{t\}$, we have $\alpha^*_{\ell} \leq (r_{jt}+d(j,\ell))^2$. \label{constr:euclidean-second}
        \item For each $t\in [s]$, we have \label{constr:euclidean-third}
    $$\sum_{\ell\in[t:s]} w(\ell) \alpha^*_t  + \sum_{\ell \in IC^t} w(\ell) \alpha^*_\ell + \sum_{j\in DC^t} w(j) r_{jt}^2 \leq \hat{f} + \sum_{\ell \in IC^t \cup [t:s]} \rho \cdot w(\ell) d^2(\ell,i) + \sum_{j\in DC^t} w(j) d^2(j,i).$$
    \end{enumerate}
    Then, the following holds:
    \begin{align}
        \label{eqn:dual-feasibility-simpler-euclidean}
        \sum_{j\in [s]} w(j) \cdot (\alpha^*_j - (\rho-1)R_j^2) \leq \hat{f} + \Gamma \cdot \sum_{j\in [s]} w(j) \cdot d^2(j,i)~,
    \end{align}
    {where $\Gamma=5$ if $d$ is a general metric, and $\Gamma=3+\ln 2$ if $d$ is a Euclidean metric.}
\end{restatable}

\begin{restatable}{lemma}{DualFeasibilityGeneralMetric}
    \label{lem:dual-feasibility-simpler-metric}
    Let $\rho=2.5$ and $\Gamma=4.9$.
    Let $\hat{f}$ be any non-negative real number. 
    Let $([s]\cup \{i\},d)$ be a metric.
    Let $w:[s]\to \mathbb{R}_{>0}$ be any {positive} weight function. %
    Let $(IC^t)_{t\in [s]}$ and $(DC^t)_{t\in [s]}$ be two families of sets satisfying:
    \begin{enumerate}[itemsep=0em,topsep=0.5em, label = (\alph*)]
        \item For each $t\in [s]$, $DC^t\cap IC^t=\emptyset$ and $DC^t\cup IC^t = [t-1]$.
        \item For each $t\in [s-1]$, $DC^t\subseteq DC^{t+1}$.
    \end{enumerate}
    Suppose that the variables $(\alpha^*_j)_{j\in [s]}, (R_j)_{j\in [s]}, (r_{jt})_{t\in [s], j\in DC^t}$ {satisfy} the following constraints
    \begin{enumerate}[itemsep=0em,topsep=0.5em,label=(\arabic*)]
        \item For any $t\in [s], j\in DC^t$, we have $r_{jt}\leq R_j$. 
        \label{constr:metric-first}
        \item For each $t\in [s], j\in DC^t$, we have $\alpha^*_{t} \leq (r_{jt}+d(j,t))^2$. \label{constr:metric-second}
        \item For each $t\in [s]$, we have  \label{constr:metric-third}
    $$\sum_{\ell\in[t:s]} w(\ell)\alpha^*_t  + \sum_{\ell \in IC^t} w(\ell)\alpha^*_\ell + \sum_{j\in DC^t} w(j) r_{jt}^2 \leq \hat{f} + \sum_{\ell \in IC^t \cup [t:s]} \rho w(\ell) d^2(\ell,i) + \sum_{j\in DC^t} w(j) d^2(j,i).$$
    \end{enumerate}
    Then, the following holds:
    \begin{align}
        \label{eqn:dual-feasibility-simpler-metric}
        \sum_{j\in [s]} w(j) \cdot (\alpha^*_j - (\rho-1)R_j^2) \leq \hat{f} + \Gamma \cdot \sum_{j\in [s]} w(j) \cdot d^2(j,i)~.
    \end{align}
\end{restatable}

\begin{remark}
    The constraint \ref{constr:metric-second}
    of \cref{lem:dual-feasibility-simpler-metric} is a weaker version of the corresponding constraint in \cref{lem:dual-feasibility-simpler-euclidean}: we don't require the constraint for each $\ell\in IC^t\cup\{t\}$ (instead, we only require it for $\ell=t$). 
\end{remark}

\begin{proof}[Proof of \cref{lem:dual-feasibility}]
    \cref{obv:disjointness-and-monotonicity,obv:distance-to-opened-facility,lem:p-tjl,,lem:beta-t} imply that the variables $(\alpha^*_j)_{j\in [s]}, (R_j)_{j\in [s]}, (r_{jt})_{t\in [s], j\in DC^t}$ satisfy the constraints in \cref{lem:dual-feasibility-simpler-metric} and \cref{lem:dual-feasibility-simpler-euclidean}. Therefore, we can apply \cref{lem:dual-feasibility-simpler-metric} {in the metric case} and \cref{lem:dual-feasibility-simpler-euclidean} {in the Euclidean case} to get the desired result.
\end{proof}

\section{Establishing the LMP Approximations}
\label{sec:euclidean-lmp}

In this section, we will prove \cref{lem:dual-feasibility-simpler-euclidean}.
For convenience, we restate the lemma below.
Using the proof of \cref{lem:dual-feasibility-simpler-euclidean}, we will also prove a generalization of it, \cref{lem:dual-feasibility-simpler-euclidean-special}, which will be used in the later sections for the log-adaptive and robust analysis of the greedy algorithm for Euclidean metrics.
Throughout this section, we will assume that $\rho=2$.
For simplicity, in the rest of this section, we will define $\Phi=\sum_{j\in [s]} w(j) \cdot d^2(j,i)$.

Readers can assume $w(j)=1$ for all $j\in [s]$ in the first reading, which is enough for understanding the main ideas and establishing the LMP approximations.

\DualFeasibilitySimplerEuclidean*

\subsection{Proof Roadmap}
To combine the constraints in the lemma, we will create a set of non-negative parameters $p_{tj\ell}$ 
for each $t\in [s], j\in DC^t, \ell\in IC^t \cup \{t\}$ such that $\sum_{\ell\in IC^t \cup\{t\}} p_{tj\ell}=1$ for each $t\in [s]$ and each $j\in DC^t$. 
\footnote{These parameters are corresponding to the parameters $\phi^{(t)}_{j,\ell}$ in \cite{CCGGLW2026kmeans}. We are using a slightly different notation here.}

Using these parameters, we can lower bound the LHS of the third constraint \ref{constr:euclidean-third} by adding the following quantity which is no greater than 0, according to the second constraint \ref{constr:euclidean-second}:
\begin{align*}
0\geq \sum_{j\in DC^t} w(j)& \cdot \sum_{\ell\in IC^t\cup\{t\}} p_{tj\ell}(\alpha^*_\ell - (r_{jt}+d(j,\ell))^2) \\
&= \sum_{j\in DC^t} w(j) \cdot \left(-r_{jt}^2 + \sum_{\ell\in IC^t\cup\{t\}} p_{tj\ell} \left(\alpha^*_\ell - 2r_{jt}\cdot d(j,\ell) - d^2(j,\ell)\right)\right)
\end{align*}
Therefore, we can lower bound the LHS of the third constraint \ref{constr:euclidean-third} by the following quantity, where we use $r_{jt}\leq R_j$ in the second inequality:
\begin{align*}
  &\text{LHS of constraint \ref{constr:euclidean-third}} \\
  &~\geq \sum_{\ell\in [t:s]} w(\ell)\cdot \alpha^*_t  + \sum_{\ell \in IC^t} w(\ell) \cdot \alpha^*_\ell + \sum_{j\in DC^t} w(j) \cdot \sum_{\ell\in IC^t\cup\{t\}} p_{tj\ell} \left(\alpha^*_\ell - 2r_{jt}\cdot d(j,\ell) - d^2(j,\ell)\right)\\
  &~\geq \sum_{\ell\in [t:s]} w(\ell)\cdot \alpha^*_t  + \sum_{\ell \in IC^t} w(\ell)\cdot \alpha^*_\ell + \sum_{j\in DC^t} w(j) \cdot \sum_{\ell\in IC^t\cup\{t\}} p_{tj\ell} \left(\alpha^*_\ell - 2R_j\cdot d(j,\ell) - d^2(j,\ell)\right).
\end{align*}
Putting this inequality back to the third constraint \ref{constr:euclidean-third}, we have
\begin{align*}
  & \sum_{\ell\in [t:s]} w(\ell) \alpha^*_t  + \sum_{\ell \in IC^t} w(\ell) \alpha^*_\ell + \sum_{j\in DC^t} w(j)  \sum_{\ell\in IC^t\cup\{t\}} p_{tj\ell} \alpha^*_\ell \\
  \leq& \hat{f} + \rho\sum_{\ell \in IC^t \cup [t:s]}   w(\ell)  d^2(\ell,i) + \sum_{j\in DC^t} w(j)  d^2(j,i) + \sum_{j\in DC^t} w(j)  p_{tj\ell} \sum_{\ell\in IC^t\cup \{t\}} 2R_j d(j,\ell) +  d^2(j,\ell) \\
  =& \hat{f} + 2\Phi -  \sum_{j\in DC^t}   w(j) d^2(j,i) + \sum_{j\in DC^t} w(j)  p_{tj\ell} \sum_{\ell\in IC^t\cup \{t\}} 2R_j d(j,\ell) +  d^2(j,\ell)
  \tag{$\rho=2$}
\end{align*}

Further, we create another set of non-negative parameters $\beta_t$ for each $t\in [s]$.\footnote{These parameters are corresponding to the parameters $\beta_t$ in \cite{CCGGLW2026kmeans}.} The goal is to ensure that $\sum_{t\in [s]} \beta_t \leq 1$ and the following inequality holds:
\begin{align*}
  \sum_{t\in [s]} \beta_t \cdot \left(\sum_{\ell \in [t:s]} w(\ell) \alpha^*_t  + \sum_{\ell \in IC^t} w(\ell) \alpha^*_\ell + \sum_{j\in DC^t} w(j) \sum_{\ell\in IC^t\cup\{t\}} p_{tj\ell} \alpha^*_\ell\right) 
  = \sum_{j \in [s]}  w(j) \alpha^*_j ~.
\end{align*}
In this way, we can have the following upper bound:
\begin{align}
  &\sum_{j\in [s]} w(j)\left(\alpha^*_j - (\rho-1)R_j^2\right) \leq \hat{f} + 2 \Phi 
  \notag
  \\
  \label{eqn:lmp-dual-feasibility-goal-2}
  &~~
  \underbrace{- \sum_{j\in [s]} w(j) R_j^2 + \sum_{t\in [s]} \beta_t \cdot \left(\sum_{j\in DC^t} w(j)p_{tj\ell}  \left(\sum_{\ell\in IC^t\cup \{t\}} 2R_j d(j,\ell) + d^2(j,\ell)\right) -  w(j) d^2(j,i)\right)
   }_{\text{the extra cost } \EC}.
\end{align}

To obtain \cref{lem:dual-feasibility-simpler-euclidean}, it suffices to explicitly define the parameters $(p_{tj\ell})_{t\in [s], j\in DC^t, \ell\in [s]}$, $(\beta_{t})_{t\in IC^t\cup\{t\}}$ and prove the extra cost terms after $\hat{f} + 2\Phi$ in the RHS of \cref{eqn:lmp-dual-feasibility-goal-2} is no greater than $(\Gamma-\rho)\cdot \Phi$, where $\Gamma$ depends on whether the underlying metric is metric or Euclidean according to \cref{lem:dual-feasibility-simpler-euclidean}.

\newcommand{\suf}{\mathsf{suf}}
\subsection{Defining the parameters}
In this subsection, we will define the parameters $(p_{tj\ell})_{t\in [s], j\in DC^t, \ell\in [s]}$ and $(\beta_{t})_{t\in [s]}$ explicitly.
These parameters are essentially the same as those in \cite{CCGGLW2026kmeans}, but we write them in an explicit formula and prove some new properties here\footnote{The definitions in \cite{CCGGLW2026kmeans} follow a recursion way. Properties here are either based on the explicit form or only implicitly proved in \cite{CCGGLW2026kmeans}.}.
For the further establishment of the LMP approximation guarantee for this paper, we will also define $P_{j\ell}$, which indicates how much $d^2(j,\ell)$ will be charged in the extra cost term $\EC$, and $d^+_j$, which indicates how much the extra cost is reduced by $d^2(j,i)$ (see \cref{lem:extra-cost-term-upperbound} for a formal statement).
After defining these parameters, we will prove \cref{eqn:dual-feasibility-simpler-euclidean} in next subsection.

For simplicity, we will use $\suf(j)=\sum_{\ell \geq j} w(\ell)$ to denote the suffix sum of $w(j)$ for each $j\in [s]$.
For each $t\in [s]$ and each $u\in [s]$, we consider 
\begin{align*}
  \pi_t(u) = \prod_{j\in DC^t: j<u} \frac{\suf(j)}{\suf(j+1)}~,
\end{align*}
where if $[u-1]\cap DC^t=\emptyset$, we simply let $\pi_t(u)=1$. 
We define the following coefficients for each $t\in [s]$ and each $j\in [s]$:
\begin{align}
  \label{eqn:euclidean-eta-tl}
  \eta_{t\ell} = \begin{cases}
    \suf(t) \cdot \pi_t(t) & \text{if } \ell=t~,\\
    \hfil w(\ell) \cdot \pi_t(\ell) \hfil & \text{if } \ell\in IC^t~,\\
    \hfil 0 \hfil & \text{otherwise}~,
  \end{cases}
\end{align}

First, we present the \textbf{definitions of the parameters $p_{tj\ell}$} 
for each $t\in [s]$, $j\in DC^t$, and $\ell \in IC^t \cup \{t\}$, so that $\sum_{\ell \in IC^t \cup \{t\}} p_{tj\ell} =1$ for each $t\in [s]$ and each $j\in DC^t$. For simplicity, we define $p_{tj\ell}=0$ if $\ell \notin IC^t\cup\{t\}$. The definition of $p_{tj\ell}$ will follow the rule below,
\begin{align}
  \label{eqn:euclidean-p-tjl}
  p_{tj\ell} \propto \begin{cases}
    0 & \text{if } \ell \le j \text{ or } \ell \notin IC^t\cup\{t\}~,\\
    \eta_{t\ell} & \text{if } \ell>j \text{ and } \ell \in IC^t \cup \{t\}~.
  \end{cases}
\end{align}
Alternatively, we can define for each $t\in [s]$ and each $j\in DC^t$, the suffix sum of $\eta_{t\ell}$ as follows:
\begin{align}
    \label{eqn:euclidean-sigma}
  \sigma_t(j) = \sum_{\ell\geq j} \eta_{t\ell}~,
\end{align}
For any $t\in[s]$, $j\in DC^t$, and those $\ell \in [j+1:t] \cap(IC^t \cup \{t\})$, we can rewrite $p_{tj\ell } = \eta_{t\ell}/\sigma_t(j+1)$ (or, $\eta_{t\ell}/\sigma_t(j)$ because $\eta_{tj}=0$).
For simplicity, we assume that $p_{tj\ell}=0$ if $j\notin DC^t$. 
It is clear that $\sum_{\ell\in IC^t \cup \{t\}} p_{tj\ell} =1$ for each $t\in [s]$ and each $j\in DC^t$. 
It is straightforward that we have $p_{tj\ell}\geq 0$.
{Further}, we have the following properties:
\begin{observation}
  \label{obv:sigma-vs-pi}
  For each $t\in [s]$ and each $j\leq t$, we have $\sigma_t(j) = \suf(j) \cdot \pi_t(j)$. Moreover, if $j\in DC^t$, we have $\frac{w(j)}{\sigma_t(j)} = \frac{1}{\pi_t(j)} - \frac{1}{\pi_t(j+1)}$.
\end{observation}
\begin{proof}
  We first prove the first part by induction on $j$. For the base case $(j=t)$, we have $\sigma_t(t) = \eta_{tt} = \suf(t) \cdot \pi_t(t)$. For the inductive step, we have $\sigma_t(j) = \eta_{tj} + \sigma_t(j+1) = \eta_{tj}+ \suf(j+1) \cdot \pi_t(j+1)$. If $j\in DC^t$, we have $\pi_t(j) = \frac{\suf(j+1)}{\suf(j)} \cdot \pi_t(j+1)$. Because $\eta_{tj}=0$,  $\sigma_t(j) = \suf(j) \cdot \pi_t(j)$. Otherwise (if $j\notin DC^t$), we have $\pi_t(j) = \pi_t(j+1)$, and thus $$\sigma_t(j) = \eta_{tj} + \suf(j+1)\pi_t(j) = w(j) \pi_t(j) + \suf(j+1) \pi_t(j)  = \suf(j) \cdot \pi_t(j)~.$$
  
  For the second part, it is straightforward to establish the following equality:
  \begin{align*}
    \frac{1}{\pi_t(j)} - \frac{1}{\pi_t(j+1)} = \frac{1}{\pi_t(j)} - \frac{\suf(j+1)}{\suf(j)} \cdot \frac{1}{\pi_t(j)} = \frac{w(j)}{\suf(j+1) \cdot \pi_t(j)} = \frac{w(j)}{\sigma_t(j)}~. & \hfill \qedhere
  \end{align*}
\end{proof}
\begin{corollary}
  \label{cor:sum-w-over-sigma}
  For each $t\in [s]$ and each $\ell \in IC^t \cup \{t\}$, we have $\sum_{j<\ell: j\in DC^t} \frac{w(j)}{\sigma_t(j)} = 1 - \frac{1}{\pi_t(\ell)}$.
\end{corollary}

\begin{lemma}[essentially, in the proof of \cite{CCGGLW2026kmeans}]
  \label{lem:justify-p-tjl}
  We have the following equality for any $(\alpha_j)_{j\in [s]}$ and any $t\in [s]$:
  \begin{align}
    \label{eqn:justify-p-tjl}
    \sum_{\ell \in [t:s]} w(\ell) \alpha^*_t  + \sum_{\ell \in IC^t} w(\ell) \alpha^*_\ell + \sum_{j\in DC^t} w(j) \sum_{\ell\in IC^t\cup\{t\}} p_{tj\ell} \alpha^*_\ell
    = \sum_{j \in [s]}  \eta_{tj} \alpha^*_j ~.
  \end{align}
\end{lemma}

\begin{proof}
  For each $t\in [s]$, we have 
  \begin{align*}
    &\sum_{\ell \in [t:s]} w(\ell) \alpha^*_t  + \sum_{\ell \in IC^t} w(\ell) \alpha^*_\ell + \sum_{j\in DC^t} w(j) \sum_{\ell\in IC^t\cup\{t\}} p_{tj\ell} \alpha^*_\ell
    \\
    =\ & \suf(t) \alpha^*_t  + \sum_{\ell \in IC^t} w(\ell) \alpha^*_\ell + \sum_{j\in DC^t} w(j) \sum_{\ell\in IC^t\cup\{t\}: \ell>j} \frac{\eta_{t\ell}}{\sigma_t(j)} \cdot \alpha^*_\ell
    \\
    =\ & \suf(t) \alpha^*_t  + \sum_{\ell \in IC^t} w(\ell) \alpha^*_\ell + \sum_{\ell\in IC^t\cup\{t\}} \alpha^*_\ell \cdot \eta_{t\ell} \cdot \sum_{j\in DC^t: j<\ell} \frac{w(j)}{\sigma_t(j)}
    \\
    =\ & \suf(t) \alpha^*_t  + \sum_{\ell \in IC^t} w(\ell) \alpha^*_\ell + \sum_{\ell\in IC^t\cup\{t\}} \alpha^*_\ell \cdot \eta_{t\ell} \cdot \left(1-\frac{1}{\pi_t(\ell)}\right) \tag{\cref{cor:sum-w-over-sigma}}
    \\
    =\ & \sum_{\ell \in IC^t \cup \{t\}} \eta_{t\ell} \alpha^*_\ell + \suf(t) \alpha^*_t  + \sum_{\ell \in IC^t} w(\ell) \alpha^*_\ell - \sum_{\ell\in IC^t\cup\{t\}} \frac{\eta_{t\ell}}{\pi_t(\ell)} \cdot \alpha^*_\ell 
    \\
    =\ & \sum_{\ell \in IC^t \cup \{t\}} \eta_{t\ell} \alpha^*_\ell~. \tag{by definition of $\eta_{t\ell}$ in \cref{eqn:euclidean-eta-tl}}
  \end{align*}
  Because we have $\eta_{t\ell}=0$ if $\ell \notin IC^t \cup \{t\}$, we can rewrite the LHS of \cref{eqn:justify-p-tjl} as $\sum_{t\in [s]} \beta_t \cdot \sum_{\ell\in [s]} \eta_{t\ell} \alpha^*_\ell$.
\end{proof}

Next, we present \textbf{the definitions of the parameters $\beta_t$}
for each $t\in [s]$ by the following recurrence relation (note that $\eta_{jj}>0$ for any $j\in [s]$):
\begin{align}
  \label{eqn:Euclidean-beta-recurrence}
  \beta_j \cdot \eta_{jj}  = w(j) -  \sum_{t>j} \beta_t \cdot \eta_{tj}.
\end{align}
The definitions are in the same spirit as the definitions of the parameters $\beta_t$ in the metric case (\cref{lem:focsweights}). 
Note that because we have $\eta_{tj}=0$ for any $t<j$, we have 
\begin{observation}
  \label{obv:beta-eta-sum}
  For each $j\in [s]$, we have $\sum_{t\in [s]} \beta_t \eta_{tj} = w(j)$.
\end{observation}
We will defer the proof that $\beta_j\geq 0$ to \cref{lem:beta-non-negative} and the proof that $\sum_{t\in [s]} \beta_t \leq 1$ to \cref{cor:beta-sum-euclidean}. Using the above definition and \cref{lem:justify-p-tjl}, we can show that
\begin{corollary}
  \label{lem:justify-beta-p-tjl}
  We have the following equality for any $(\alpha_j)_{j\in [s]}$:
  \begin{align}
    \label{eqn:justify-beta-p-tjl}
    \sum_{t\in [s]} \beta_t \cdot \left(\sum_{\ell \in [t:s]} w(\ell) \alpha^*_t  + \sum_{\ell \in IC^t} w(\ell) \alpha^*_\ell + \sum_{j\in DC^t} w(j) \sum_{\ell\in IC^t\cup\{t\}} p_{tj\ell} \alpha^*_\ell\right) 
    = \sum_{j \in [s]}  w(j) \alpha^*_j ~.
  \end{align}
\end{corollary}

Finally, for simplicity, we \textbf{construct the matrix} $P$, in which, for any $j<\ell$, $P_{j\ell}$ equals the coefficient (divided by $2w(j)$) of $d(j,\ell)$ in the extra cost term $\EC$ in \cref{eqn:lmp-dual-feasibility-goal-2}. 
We define $P_{j\ell}=0$ for any $j\geq \ell$.
Intuitively, we can think of $P_{j\ell}$ as how much we will charge $d^2(j,\ell)$ in the extra cost term $\EC$ (see \cref{lem:extra-cost-term-upperbound} for a formal statement).
Because we have $p_{tj\ell}\neq 0$ only when $j\in DC^t$, $\ell\in IC^t\cup\{t\}$ and $j<\ell$, we have following simple expression for $P_{j\ell}$:
\begin{align*}
  P_{j\ell} = \sum_{t\in [s]} \beta_t \cdot p_{tj\ell}~.
\end{align*}
We also \textbf{define} $d^+_j = \sum_{\ell>j} P_{j\ell}$. 
Intuitively, this indicates how much the extra cost is reduced by $d^2(j,i)$ (see \cref{lem:extra-cost-term-upperbound} for a formal statement).

Further, we present a simple characterization of $d^+_j$ in terms of $\{\beta_t\}$, which will be used in the next subsection to upper bound the extra cost term $\EC$ in \cref{eqn:lmp-dual-feasibility-goal-2} by $ \Phi $.
We define $t_j\geq j$ as the threshold such that for any $t>t_j$, we have $j\in DC^t$, and for any $j<t\le t_j$, we have $j\notin DC^t$. 
\begin{observation}
  \label{obv:dplus-vs.-sum-beta}
  For each $j\in [s]$, we have $d^+_j= \sum_{t: j\in DC^t} \beta_t = \sum_{t>t_j} \beta_t$. 
\end{observation}
\begin{proof}
  By definition, we have
  \begin{align*}
    d^+_j = \sum_{\ell>j} P_{j\ell} = \sum_{\ell>j} \sum_{t\in [s]} \beta_t \cdot p_{tj\ell} = \sum_{t\in [s]: j\in DC^t} \beta_t \cdot \sum_{\ell>j} p_{tj\ell} = \sum_{t>t_j} \beta_t~. & \hfill  \qedhere
  \end{align*}
\end{proof}

\subsection{\texorpdfstring{Proof of \cref{lem:dual-feasibility-simpler-euclidean}}{Proof of lem-dual-feasibility-simpler-euclidean}}
\subsubsection{\texorpdfstring{Step 1: Bounding the extra cost term via $P_{j\ell}$ and $d^+_j$}{Step 1: Bounding the extra cost term via Pjell and d+j}}
Using above definitions, we can rewrite the extra cost term in \cref{eqn:lmp-dual-feasibility-goal-2}:
\begin{lemma}
  \label{lem:extra-cost-term-upperbound}
  We have the following upper bound for the extra cost term $\EC$ in \cref{eqn:lmp-dual-feasibility-goal-2}:
  \begin{align*}
    \EC \leq \sum_{j<\ell} w(j) P_{j\ell} d^2(j,\ell) + \sum_{j\in [s]} w(j) d^+_j \sum_{\ell>j} P_{j\ell} d^2(j,\ell) - \sum_{j\in [s]} w(j) d^+_j  d^2(j,i)~.
  \end{align*}
\end{lemma}
\begin{proof} According to our definition of $P_{j\ell}$ and $d^+_j$, we can rewrite the extra cost term $\EC$ in \cref{eqn:lmp-dual-feasibility-goal-2} as follows:
\begin{align*}
  \EC \leq& - \sum_{j\in [s]}  w(j) R_j^2 + \sum_{j<\ell} w(j) P_{j\ell} \cdot \left(2R_j\cdot d(j,\ell) + d^2(j,\ell)\right)  - \sum_{j\in [s]} w(j) d^+_j \cdot d^2(j,i) \\
  \leq& \sum_{j<\ell} w(j) P_{j\ell} d^2(j,\ell) + \sum_{j\in [s]} w(j) \left(-R_j^2 + \sum_{\ell>j} P_{j\ell} \cdot 2R_jd(j,\ell)\right)  - \sum_{j\in [s]} w(j) d^+_j d^2(j,i)
  \\
  \leq& \sum_{j<\ell} w(j) P_{j\ell} d^2(j,\ell) + \sum_{j\in [s]} w(j) \left(\sum_{\ell>j} P_{j\ell} d(j,\ell)\right)^2 - \sum_{j\in [s]} w(j) d^+_j d^2(j,i)
  \tag{Applying $-R_j^2 + 2bR_j \leq b^2$ for any $b\in \mathbb{R}$}
  \\
  \leq& \sum_{j<\ell} w(j) P_{j\ell} d^2(j,\ell) + \sum_{j\in [s]} w(j) d^+_j \sum_{\ell>j} P_{j\ell} d^2(j,\ell) - \sum_{j\in [s]} w(j) d^+_j  d^2(j,i)
  \tag{Cauchy-Schwarz inequality}
\end{align*}
\end{proof}

\subsubsection{\texorpdfstring{Step 2: Bounding the Laplacian matrix of $P$}{Step 2: Bounding the Laplacian matrix of P}}
The following lemma upper bounds the Laplacian matrix of $P$ by the identity matrix, which is essentially a key lemma in \cite{CCGGLW2026kmeans} to establish the LMP 4-approximation for Euclidean $k$-means. For the sake of completeness, we prove this lemma in \cref{app:euclidean-lmp}.
\begin{restatable}[essentially, \cite{CCGGLW2026kmeans}]{lemma}{STOCKeyLemma}
  \label{lem:stoc-key-lemma}
  For any $w:[s]\to \mathbb{R}_{> 0}$ and any $x_1,\dots, x_s\in \mathbb{R}$, we have
  $$\sum_{j<\ell} w(j) P_{j\ell} (x_j-x_\ell)^2 \leq \sum_{j\in [s]} w(j) x_j^2~.$$
\end{restatable}
As a corollary, we have the following properties of the matrix $P$, which give bounds for the first term of \cref{lem:extra-cost-term-upperbound}.
\begin{corollary}
  \label{cor:stoc-key-lemma-metric}
  For any metric $([s]\cup \{i\},d)$, we have 
  \begin{align}
    \label{eqn:stoc-key-lemma-metric}
    \sum_{j<\ell} w(j) P_{j\ell} d^2(j,\ell) \leq 2 \sum_{j\in [s]} w(j) d^2(j,i)~.
  \end{align}
  In particular, if $d$ is Euclidean, we have an improved upper bound:
  \begin{align*}
    \sum_{j<\ell} w(j) P_{j\ell} d^2(j,\ell) \leq \sum_{j\in [s]} w(j) d^2(j,i)~.
  \end{align*}
\end{corollary}
\begin{proof}
  Let $a_j = d(j,i)$ for each $j\in [s]$. By the triangle inequality, we have $d(j,\ell) \leq a_j + a_\ell$ for any $j<\ell$. Therefore, we have
  \begin{align*}
    \sum_{j<\ell} w(j) P_{j\ell} d^2(j,\ell) \leq \sum_{j<\ell} w(j) P_{j\ell} (a_j+a_\ell)^2 \leq 2 \sum_{j<\ell} w(j) P_{j\ell} (a_j^2 + a_\ell^2)~.
  \end{align*}
  For each $j\in [s]$, consider the vector $x=(x_1,\dots, x_s)$ such that $x_j = a_j$ and $x_\ell = 0$ for all $\ell \neq j$. By \cref{lem:stoc-key-lemma}, we have $\sum_{\ell<j} w(\ell)P_{\ell j} a_j^2 + \sum_{\ell>j} w(j)P_{j\ell} a_j^2 \leq w(j)a_j^2$. Therefore, we have obtained the desired \cref{eqn:stoc-key-lemma-metric}.

  In particular, if $d$ is Euclidean, we can w.l.o.g. assume that $i$ is located at the origin $0^{\di}$ and $j$ is located at $v_j\in \mathbb{R}^{\di}$ for each $j\in [s]$. Then we can rewrite $\sum_{j<\ell} w(j) P_{j\ell} d^2(j,\ell)$ as $\sum_{t\in [\di]} \sum_{j<\ell} w(j) P_{j\ell} (v_{jt}-v_{\ell t})^2$. By \cref{lem:stoc-key-lemma}, we have $\sum_{j<\ell} w(j) P_{j\ell} (v_{jt}-v_{\ell t})^2 \leq \sum_{j\in [s]} w(j) v_{jt}^2$ for each $t\in [\di]$. Therefore, we have $$\sum_{j<\ell} w(j) P_{j\ell} d^2(j,\ell) \leq \sum_{t\in [\di]} \sum_{j\in [s]} w(j) v_{jt}^2 = \sum_{j\in [s]} w(j) d^2(j,i)~.$$
  Hence we have completed the proof.
\end{proof}

Another direct corollary of this lemma is that the sum of $P$ on any column $\ell$ is upper bounded by $w(\ell) (1-d^+_\ell)$.
\begin{corollary}
  \label{cor:bounded-degrees}
  For any $\ell\in [s]$, we have $\sum_{j<\ell} w(j) P_{j\ell} \leq w(\ell) (1-d^+_\ell)$.
\end{corollary}
\begin{proof}
  Consider the vector $x=e_{\ell}$ which is the indicator vector for $\ell$. By \cref{lem:stoc-key-lemma}, we have $\sum_{j<\ell} w(j)P_{j\ell} + \sum_{\ell<\ell'} w(\ell)P_{\ell \ell'} \leq w(\ell)$. Because $d^+_\ell = \sum_{\ell'>\ell} P_{\ell \ell'}$, we have $\sum_{j<\ell} w(j) P_{j\ell} \leq w(\ell) (1-d^+_\ell)$.
\end{proof}

\subsubsection{\texorpdfstring{Step 3: A key property on the conditional column sum of $P$}{Step 3: A key property on the conditional column sum of P}}
The third key property is an upper bound for the conditional column sum of $P$ in terms of $d^+_j$ and $w(\ell)$, which will be used to upper bound the second and the third term in \cref{lem:extra-cost-term-upperbound}.
Intuitively, consider the $\ell$-th column (corresponding to client $\ell$) in the matrix $P$. The conditional column sum of $P$ is the sum of all rows $j$ in this column whose corresponding client $j$ has $d^+_j>\theta$.
For a better understanding, $d^+_j$ can be interpreted as a parameter indicating whether client $j$ becomes directly connected early in the process of the greedy algorithm (according to \cref{obv:dplus-vs.-sum-beta}). If $d^+_j>d^+_{j'}$ for other client $j'$, it means that client $j$ becomes directly connected earlier than $j'$.
Hence, the conditional column sum of $P$ can be interpreted as the sum of the coefficients of a set of clients $j<\ell$ that become directly connected at some time point before $\ell$ becomes directly connected in the process of the greedy algorithm.
The reason we only need to take care of those clients $j$ is because we only charge $d^2(j,\ell)$ for those clients $j$ in the upper bound of the extra cost term $\EC$ in \cref{lem:extra-cost-term-upperbound} (i.e., $P_{j'\ell}=0$ for other clients $j'$).
\begin{lemma}
  \label{lem:sum-threshold-wPij-upperbound}
    For any $\theta \in (0,1)$ and any $\ell\in [s]$, we have
    \begin{align*}
        \sum_{j<\ell} \ind(d^+_j > \theta) \cdot w(j) P_{j\ell} \leq w(\ell) (1-\theta). 
    \end{align*}
\end{lemma}
\begin{remark}
  We shall remark that if $\theta<d^+_\ell$, this lemma has a weaker bound than \cref{cor:bounded-degrees}. However, if $\theta\geq d^+_\ell$, this lemma gives a stronger bound than \cref{cor:bounded-degrees}.
\end{remark}

To prove this lemma, we start with the following simple observation, which is a direct consequence of the definition of $\pi_t(j)$.
\begin{observation}
  For any $j\in [s]$, $\pi_t(j)$ is non-decreasing in $t$.
\end{observation}
\begin{proof}
  This observation follows the definition of $\pi_t(j)$ and the fact that $DC^t\subseteq DC^{t+1}$ for any $t\in [k-1]$.
\end{proof}
\begin{corollary}
  \label{cor:sum-w-over-sigma-monotone-restricted-to-DCu}
  Suppose $u\leq t$. Then, for any $\ell \in [s]$, we have $$\sum_{j<\ell: j\in DC^u} \frac{w(j)}{\sigma_t(j)} \leq \sum_{j<\ell: j\in DC^u} \frac{w(j)}{\sigma_u(j)}~.$$
\end{corollary}
\begin{proof}
  According to \cref{obv:sigma-vs-pi}, we have $\sigma_t(j)=\suf(j)\cdot \pi_t(j)$ and $\sigma_u(j)=\suf(j)\cdot \pi_u(j)$. Because $\pi_t(j)$ is non-decreasing in $t$, we have $\pi_t(j)\geq \pi_u(j)$, and thus $\frac{w(j)}{\sigma_t(j)} \leq \frac{w(j)}{\sigma_u(j)}$ for each $j\in DC^u$. Therefore, we have $\sum_{j<\ell: j\in DC^u} \frac{w(j)}{\sigma_t(j)} \leq \sum_{j<\ell: j\in DC^u} \frac{w(j)}{\sigma_u(j)}$.
\end{proof}

Our first intermediate lemma is an upper bound for the conditional column sum of $P$ restricted to a subset of clients $DC^u$ for any $u\in [s]$.
\begin{lemma}
  \label{lem:sum-Pij-by-DCu-upperbound}
  Fix any $u\in [s]$. For any $\ell \in [s]$, we have
  \begin{align*}
    \sum_{j<\ell: j\in DC^u} w(j) P_{j\ell} \leq w(\ell) \cdot \left(1-\frac{1}{\pi_{u}(\ell)}\right)~.
  \end{align*}
\end{lemma}
\begin{proof}
  Using the definition of $P_{j\ell}$, we have
  \begin{align*}
    \sum_{j<\ell: j\in DC^u} w(j) P_{j\ell} &= \sum_{j<\ell: j\in DC^u} w(j) \sum_{t\in [s]} \beta_t \cdot p_{tj\ell} = \sum_{t\in [s]} \beta_t \cdot \sum_{j<\ell: j\in DC^u} w(j) p_{tj\ell}~.
  \end{align*}

  First, we consider the case when $t<u$. In this case, we have $DC^t\subseteq DC^u$. Since by our definition, $p_{tj\ell}=0$ if $j\notin DC^t$, we have $\sum_{j<\ell: j\in DC^u} w(j) p_{tj\ell} = \sum_{j<\ell: j\in DC^t} w(j) p_{tj\ell}$. Therefore, for $t<u$, we have the following equation, where the second equality is by our definition of $p_{tj\ell}$ and the final equality is by \cref{cor:sum-w-over-sigma}:
  \begin{align*}
    \sum_{j<\ell: j\in DC^u} w(j) p_{tj\ell} &= \sum_{j<\ell: j\in DC^t} w(j) p_{tj\ell} = \sum_{j<\ell: j\in DC^t} w(j) \frac{\eta_{t\ell}}{\sigma_t(j)} = \eta_{t\ell} \left(1-\frac{1}{\pi_t(\ell)}\right)~.
  \end{align*}
  Since $\pi_{t}(\ell)$ is non-decreasing in $t$, we have $\sum_{j<\ell: j\in DC^u} w(j) p_{tj\ell} \leq \eta_{t\ell} (1-\frac{1}{\pi_{u}(\ell)})$ for $t<u$.

  Second, we consider the case when $t\geq u$. 
  Using our definition of $p_{tj\ell}$ and \cref{cor:sum-w-over-sigma}, we have
  \begin{align*}
    \sum_{j<\ell: j\in DC^u} w(j) p_{tj\ell} = \sum_{j<\ell: j\in DC^u} w(j) \cdot \frac{\eta_{t\ell}}{\sigma_t(j)} = \eta_{t\ell} \sum_{j<\ell: j\in DC^u} \frac{w(j)}{\sigma_t(j)} &\leq \eta_{t\ell} \sum_{j<\ell: j\in DC^u} \frac{w(j)}{\sigma_u(j)} 
    \\
    &= \eta_{t\ell} \left(1-\frac{1}{\pi_u(\ell)}\right)~.
  \end{align*}

  In conclusion, using the above discussions, the recurrence relation for $\beta_t$ (\cref{eqn:Euclidean-beta-recurrence}) and the fact that $\eta_{t\ell}=0$ for $t<\ell$, we have 
  \begin{align*}
    \sum_{j<\ell: j\in DC^u} w(j) P_{j\ell} = \sum_{t\in [s]} \beta_t \sum_{j<\ell: j\in DC^u} w(j) p_{tj\ell} \leq \sum_{t\in [s]} \beta_t \eta_{t\ell} \left(1-\frac{1}{\pi_u(\ell)}\right) = w(\ell) \left(1-\frac{1}{\pi_u(\ell)}\right) .
  \end{align*}
\end{proof}

Next, we show the first deferred proof that $\beta_u\geq 0$ for any $u\in [s]$.
\begin{lemma}
  \label{lem:beta-non-negative}
For any $u\in [s]$, we have $\beta_u\geq 0$.
\end{lemma}
\begin{proof}
  First, we show an upper bound for the following term:
\begin{align*}
    \sum_{t>u} \beta_t \cdot \eta_{tu} \leq \sum_{t>u} \beta_t \cdot w(u) \pi_t(u) &= \sum_{t>u} \beta_t \cdot \frac{w(u) \sigma_t(u)}{\suf(u)} \tag{\cref{obv:sigma-vs-pi}}\\
    &= \frac{w(u)}{\suf(u)} \cdot \sum_{t>u} \beta_t \cdot \sum_{j\geq u} \eta_{tj}
    \\
    &= \frac{w(u)}{\suf(u)} \cdot \sum_{j\geq u} \sum_{t>u} \beta_t \cdot \eta_{tj}
    \\
    &= \frac{w(u)}{\suf(u)} \cdot \left(\sum_{j>u} w(j) + \sum_{t>u} \beta_t \eta_{tu}\right) \tag{\cref{eqn:Euclidean-beta-recurrence}}
    \\
    &=w(u) \cdot \frac{\suf(u+1)}{\suf(u)} + \frac{w(u)}{\suf(u)} \cdot \sum_{t>u} \beta_t \cdot \eta_{tu}. 
  \end{align*}
Therefore, $\frac{\suf(u+1)}{\suf(u)} \sum_{t>u} \beta_t \cdot \eta_{tu} \leq w(u) \cdot \frac{\suf(u+1)}{\suf(u)}$, and thus $\sum_{t>u} \beta_t \cdot \eta_{tu} \leq w(u)$. 

Recall \cref{eqn:Euclidean-beta-recurrence}, where we have $\beta_{u}\eta_{uu} = w(u) - \sum_{t>u} \beta_t \cdot \eta_{tu}$. Because $\eta_{uu}>0$, we have $\beta_u\geq 0$.
\end{proof}

We show a lemma that could imply the second deferred proof that $\sum_{t\in [s]} \beta_t \leq 1$. This lemma is more general in the sense that it also gives an upper bound for the tail sum of $\beta_t$, which is useful for establishing \cref{lem:sum-threshold-wPij-upperbound}, the focus of this part.
\begin{lemma}
  For any $u\in [s]$, we have $\sum_{t\geq u} \beta_t \leq \frac{1}{\pi_u(u)}$.
\end{lemma}
\begin{proof}
  For this lemma, we will prove a stronger statement that for any $u\in [s]$, we have
  \begin{align}
    \label{eqn:stronger-beta-pi-sum}
    \sum_{t\geq u} \beta_t \cdot \pi_t(u) = 1~.
  \end{align}
  Then, because $\pi_t(u)$ is non-decreasing in $t$, we have $\sum_{t\geq u} \beta_t \leq \frac{1}{\pi_u(u)}$.

  To prove the stronger statement, we will use \cref{obv:sigma-vs-pi}:
  \begin{align*}
    \sum_{t\geq u} \beta_t \cdot \pi_t(u) &= \sum_{t\geq u} \beta_t \cdot \frac{\sigma_t(u)}{\suf(u)} \tag{\cref{obv:sigma-vs-pi}}\\
    &= \frac{1}{\suf(u)} \cdot \sum_{t\geq u} \beta_t \cdot \sum_{j\geq u} \eta_{tj}
    \\
    &= \frac{1}{\suf(u)} \cdot \sum_{j\geq u} \sum_{t\geq u} \beta_t \cdot \eta_{tj}
    \\
    &= \frac{1}{\suf(u)} \cdot \sum_{j\geq u} w(j)=1. \tag{by \cref{eqn:Euclidean-beta-recurrence}}
  \end{align*}
\end{proof}
\noindent
By choosing $u=1$ for \cref{eqn:stronger-beta-pi-sum} in the proof, we have the following corollary as we have $\pi_t(u)=1$ for $t\in [s]$.
\begin{corollary}
  \label{cor:beta-sum-euclidean}%
  We have $\sum_{t\in [s]} \beta_t = 1$.
\end{corollary}
\noindent
This further implies the following corollary because of \cref{obv:dplus-vs.-sum-beta}.
\begin{corollary}
  \label{cor:dplus-upperbound}
  For each $j\in [s]$, we have $d^+_j \leq 1$.
\end{corollary}

Finally, we are ready to present the proof of the lemma in this part.

\begin{proof}[Proof of \cref{lem:sum-threshold-wPij-upperbound}]
    Suppose that $u^*$ is the largest index $u$ such that $\sum_{t\geq u} \beta_t > \theta$. Such index $u$ always exists because of \cref{cor:beta-sum-euclidean}.  Then, according to \cref{obv:dplus-vs.-sum-beta}, we have $d^+_j > \theta$ if and only if $t_j<u^*$, i.e., $j\in DC^{u^*}$. %
    Therefore, we have
    \begin{align*}
      \sum_{j<\ell} \ind(d^+_j > \theta) \cdot w(j) P_{j\ell} &= \sum_{j<\ell: j\in DC^{u^*}} w(j) P_{j\ell} \\
      &\leq w(\ell)\left(1 - \frac{1}{\pi_{u^*}(\ell)}\right) \tag{\cref{lem:sum-Pij-by-DCu-upperbound}}
      \\
      &\leq w(\ell)\left(1-\frac{1}{\pi_{u^*}(u^*)}\right) 
      \\
      &\leq w(\ell)\left(1-\sum_{t\geq u^*} \beta_t\right) < w(\ell)(1-\theta)~,
    \end{align*}
    where the second inequality is using two facts: (i) $\pi_{u^*}(\ell)$
    is non-decreasing in $\ell$, so $\pi_{u^*}(\ell) \leq \pi_{u^*}(u^*)$ for $\ell \leq u^*$; and (ii) we have $DC^{u^*} \subseteq [u^*-1]$ and so $\pi_{u^*}(\ell)=\pi_{u^*}(u^*)$ for $\ell\geq u^*$.
\end{proof}

\subsubsection{Step 4(a): Proof for the Euclidean case}
\label{subsubsec:euclidean-lmp-final-proof}
In this part, we will prove \cref{lem:dual-feasibility-simpler-euclidean} for the Euclidean case.
We will consider $\Gamma = 3+\ln 2$ here.
Without loss of generality, we assume the facility $i$ is located at $x_i=0^{\di}$ 
and the corresponding $s$ clients are located at $x_1,\dots, x_s$ respectively.
According to \cref{eqn:lmp-dual-feasibility-goal-2,,lem:extra-cost-term-upperbound}, we have 
\begin{align*}
  &\sum_{j\in [s]} w(j) \left(\alpha^*_j - (\rho-1)R_j^2\right) 
  \\
  &\quad \leq \hat{f} + 2\Phi + \sum_{j<\ell} w(j) P_{j\ell} d^2(j,\ell) + \sum_{j\in [s]} w(j) d^+_j \sum_{\ell>j} P_{j\ell} d^2(j,\ell) - \sum_{j\in [s]} w(j) d^+_j  d^2(j,i)
  \\
  &\quad = \hat{f} + 2\Phi + \sum_{j<\ell} w(j) P_{j\ell} \|x_j-x_{\ell}\|^2 + \sum_{j\in [s]} w(j) d^+_j \sum_{\ell>j} P_{j\ell} \|x_j-x_{\ell}\|^2 - \sum_{j\in [s]} w(j) d^+_j  \|x_j\|^2
\end{align*}
Because of \cref{cor:stoc-key-lemma-metric}, the first term after $\hat{f} + 2\Phi$ can be upper bounded by $\Phi$.
Next, we will focus on upper bounding the remaining term (shown below) by $\ln(2)\cdot \Phi$. %
$$\sum_{j\in [s]} w(j) d^+_j \cdot \sum_{\ell>j} P_{j\ell} \cdot \|x_j-x_\ell\|^2 - \sum_{j\in [s]} w(j) d^+_j \cdot \|x_j\|^2~.$$

Using \cref{lem:sum-threshold-wPij-upperbound}, for any $\theta\in [0,1]$, we have 
\begin{align*}
    \sum_{j<\ell} w(j)&d^+_j P_{j\ell} \|x_j-x_\ell\|^2 - \sum_{j\in [s]} w(j) d^+_j \|x_j\|^2 
    \\
    &\leq \theta \sum_{j<\ell} w(j)P_{j\ell}\cdot \|x_j-x_\ell\|^2 + \sum_{j<\ell} [d^+_j-\theta]^+ \cdot w(j)P_{j\ell} \|x_j-x_\ell\|^2 - \sum_{j\in [s]} w(j) d^+_j \|x_j\|^2
    \\
    &\leq \theta \cdot \Phi  + \underbrace{\sum_{j<\ell} [d^+_i-\theta]^+ \cdot w(j)P_{j\ell} \|x_j-x_\ell\|^2 - \sum_{j\in [s]} w(j)d^+_j \|x_j\|^2}_{S_2} \tag{\cref{lem:stoc-key-lemma}}
\end{align*}
For the term $S_2$, we derive the following upper bound: %
\begin{align*}
  S_2&=\sum_{j\in [s]: d^+_j>\theta} \sum_{\ell>j} (d^+_j-\theta) w(j)P_{j\ell} \cdot  \|x_j-x_\ell\|^2 - \sum_{j\in [s]} w(j) d^+_j \|x_j\|^2 
  \\
  &\leq \sum_{j\in [s]: d^+_j>\theta} \sum_{\ell>j} (d^+_j-\theta) w(j)P_{j\ell} \cdot  \left(\frac{\|x_j\|^2}{d^+_j-\theta} + \frac{\|x_\ell\|^2}{1-d^+_j +\theta}\right) - \sum_{j\in [s]} w(j)d^+_j \|x_j\|^2
  \tag{\cref{lem:apxTriangleInequality2}}
  \\
  &\leq \sum_{j\in [s]: d^+_j>\theta} w(j)\left(\sum_{\ell>j} P_{j\ell}\right) \|x_j\|^2 + \sum_{j\in [s]: d^+_j>\theta} \sum_{\ell>j} \frac{d^+_j-\theta}{1-d^+_j+\theta} \cdot w(j)P_{j\ell} \|x_\ell\|^2 - \sum_{j\in [s]} w(j)d^+_j \|x_j\|^2
  \\
  &\leq \sum_{j\in [s]: d^+_j>\theta} \sum_{\ell>j} \frac{d^+_j-\theta}{1-d^+_j+\theta} \cdot w(j)P_{j\ell} \|x_\ell\|^2
  \tag{definition of $d^+_j$}
  \\
  &= \sum_{\ell\in [s]} \|x_\ell\|^2 \cdot \sum_{j<\ell: d^+_j>\theta} \frac{d^+_j-\theta}{1-d^+_j+\theta} \cdot w(j)P_{j\ell}~.
\end{align*}
Let $g(y) = \frac{y-\theta}{1-y+\theta}$ for any $y\in [\theta,1]$. This is easy to verify that $g(\theta)=0$. Therefore, we have $g(y) = \int_{\theta}^y g'(z) dz$.
Hence, we can further upper bound $S_2$ by
\begin{align*}
  S_2 &\leq  \sum_{\ell \in [s]} \|x_\ell\|^2 \cdot \sum_{j<\ell: d^+_j>\theta} \int_{\theta}^{d^+_j} g'(z)\cdot w(j)P_{j\ell} dz
  \\
  &= \sum_{\ell \in [s]} \|x_\ell\|^2 \cdot \int_{z=\theta}^1 g'(z) \cdot \sum_{j<\ell: d^+_j>z} w(j)P_{j\ell} dz~.
  \\
  &\leq \sum_{\ell \in [s]} \|x_\ell\|^2 \cdot \int_{z=\theta}^1 g'(z) \cdot w(\ell)(1-z)dz \tag{\cref{lem:sum-threshold-wPij-upperbound}}
  \\
  &= \sum_{\ell \in [s]} w(\ell)\|x_\ell\|^2 \cdot \int_{z=\theta}^1 g'(z) (1-z)dz
\end{align*}
Since $g(z)=0$ when $z=\theta$ and $1-z=0$ when $z=1$, we have 
\begin{align*}
  \int_{z=\theta}^1 g'(z) \cdot (1-z)dz = \int_{z=\theta}^1 g(z)dz = \int_{z=\theta}^1 \frac{z-\theta}{1-z+\theta} dz = \theta-1-\ln(\theta)~.
\end{align*}
We have $S_2\leq (\theta-1-\ln(\theta)) \cdot \Phi$. Therefore, we have 
\begin{align*}
\sum_{j<\ell} w(j)d^+_j P_{j\ell} \|x_j-x_\ell\|^2 - \sum_{j\in [s]} w(j) d^+_j \|x_j\|^2  &\leq (\theta + \theta -1 - \ln(\theta))\cdot \Phi \\
&= (2\theta -1 - \ln(\theta)) \cdot \Phi~.
\end{align*}
If we choose $\theta = 1/2$, we get our desired bound that 
$$\sum_{j<\ell} w(j)d^+_j P_{j\ell} \|x_j-x_\ell\|^2 - \sum_{j\in [s]} w(j) d^+_j \|x_j\|^2 \leq \ln(2) \cdot \Phi~.$$

\subsubsection{Step 4(b): Proof for the general metric case}
In this part, we will prove \cref{lem:dual-feasibility-simpler-euclidean} for the general metric case.
We will consider $\Gamma = 5$ here. For simplicity, we use $a_j=d^2(j,i)$ for each $j\in [s]$ in this part. According to \cref{eqn:lmp-dual-feasibility-goal-2,,lem:extra-cost-term-upperbound}, we have 
\begin{align*}
  &\sum_{j\in [s]} w(j) \left(\alpha^*_j - (\rho-1)R_j^2\right) 
  \\
  &\quad \leq \hat{f} + 2\Phi + \sum_{j<\ell} w(j) P_{j\ell} d^2(j,\ell) + \sum_{j\in [s]} w(j) d^+_j \sum_{\ell>j} P_{j\ell} d^2(j,\ell) - \sum_{j\in [s]} w(j) d^+_j  d^2(j,i)
\end{align*}
First, because of \cref{cor:stoc-key-lemma-metric}, we can upper bound the first term after $\hat{f}+2\Phi$ by $2\Phi$.
Next, we upper bound the remaining term, i.e., 
\begin{align}
  \label{eqn:remaining-term-general-metric}
  \sum_{j\in [s]} w(j) d^+_j \sum_{\ell>j} P_{j\ell} d^2(j,\ell) - \sum_{j\in [s]} w(j) d^+_j  d^2(j,i)
\end{align}
by $\Phi$ for the general metric case.

Using the definition of $a_j$, we have
\begin{align*}
  \cref{eqn:remaining-term-general-metric} &\leq \sum_{j\in [s]} w(j) d^+_j \sum_{\ell>j} P_{j\ell} \cdot 2(a_j + a_\ell) - \sum_{j\in [s]} w(j) d^+_j  a_j
  \tag{definition of $a_j$ and \cref{lem:apxTriangleInequality2}}
  \\
  &= 2\sum_{j<\ell} w(j) d^+_j P_{j\ell} \cdot a_\ell + 2\sum_{j<\ell} w(j) d^+_j P_{j\ell} \cdot a_j - \sum_{j\in [s]} w(j) d^+_j  a_j
  \\
  &= 2\sum_{j<\ell} w(j) d^+_j P_{j\ell} \cdot a_\ell + \sum_{j\in [s]} w(j) (2(d^+_j)^2 - d^+_j)  a_j
  \tag{definition of $d^+_j$}
  \\
  &= 2\sum_{j<\ell} w(j) \left(\int_{\theta=0}^{1} \ind(d^+_j>\theta) d\theta\right) \cdot P_{j\ell} a_\ell + \sum_{j\in [s]} w(j) (2(d^+_j)^2 - d^+_j)  a_j
  \\
  &= 2\sum_{\ell\in [s]} \left(\int_{\theta=0}^{1} \sum_{j<\ell}\ind(d^+_j>\theta) w(j) P_{j\ell} d\theta\right) \cdot a_\ell + \sum_{j\in [s]} w(j) (2(d^+_j)^2 - d^+_j)  a_j
\end{align*}
Note that for any $\ell\in [s]$, the coefficient of $a_\ell$ in the first term can be upper bounded as follows: 
\begin{align*}
  \int_{\theta=0}^{1} \sum_{j<\ell}\ind(d^+_j>\theta) w(j) P_{j\ell} d\theta &\leq \int_{\theta=0}^{d^+_\ell} \sum_{j<\ell}  w(j) P_{j\ell} d\theta + \int_{\theta=d^+_\ell}^1 \sum_{j<\ell} \ind(d^+_j>\theta) w(j) P_{j\ell}  d\theta
  \\
  &\leq d^+_\ell \cdot w(\ell) (1-d^+_\ell) + \int_{\theta=d^+_\ell}^1 w(\ell)(1-\theta) d\theta \tag{\cref{cor:bounded-degrees,,lem:sum-threshold-wPij-upperbound}}
  \\
  &= w(\ell) \left(d^+_\ell (1-d^+_\ell) + \frac{(1-d^+_\ell)^2}{2}\right)
\end{align*}
Therefore, we can further upper bound \cref{eqn:remaining-term-general-metric} by (the final inequality is because $d^+_j\in [0,1]$ for each $j\in [s]$):
\begin{align*}
  \cref{eqn:remaining-term-general-metric} &\leq \sum_{j\in [s]} w(j) \cdot \left(2d^+_j(1-d^+_j) + (1-d^+_j)^2+2(d^+_j)^2 - d^+_j\right) \cdot a_j
  \\
  &= \sum_{j\in [s]} w(j) \cdot \left(1-d^+_j+(d^+_j)^2\right)\cdot a_j \leq \Phi~.
\end{align*}

\subsection{\texorpdfstring{A generalization of \cref{lem:dual-feasibility-simpler-euclidean} for the Euclidean case}{A generalization of lem:dual-feasibility-simpler-euclidean for the Euclidean case}}
In this subsection, we will prove a generalization of \cref{lem:dual-feasibility-simpler-euclidean},
where we only have some approximate version of the constraints. 
We will show that the same guarantee can be achieved up to a small additive factor. 
Later, we will use this generalization to obtain the bicriteria approximation guarantee for the Euclidean case. 
Because the proof is highly technical with dealing with the small additive factor and mostly follows the spirit of the earlier derivations, we defer it to \Cref{app:dual-feasibility-simpler-euclidean-special}.

\begin{restatable}{lemma}{DualFeasibilityEuclideanSpecial}
  \label{lem:dual-feasibility-simpler-euclidean-special}
  Let $\rho=2$ and $\Gamma=3+\ln(2)$.
    Let $\hat{f}$ be any non-negative real number.
    Let $\eps\geq 0$ be any sufficiently small non-negative real number. 
    Let $([s]\cup \{i\},d)$ be a Euclidean metric.
    Let $w:[s]\to \mathbb{R}_{>0}$ be any non-negative weight function.
    Let $a\in [s]$ be any positive integer such that $a\leq s$.
    Let $(IC^t)_{t\in [s]}$ and $(DC^t)_{t\in [s]}$ be two families of sets satisfying:
    \begin{enumerate}[itemsep=0em,topsep=0.5em,label=(\alph*)]
        \item For each $t\in [s]$, $DC^t\cap IC^t=\emptyset$ and $DC^t\cup IC^t = [t-1]$.
        \item For each $t\in [s-1]$, $DC^t\subseteq DC^{t+1}$.
        \item $[a+1:s-1] \subseteq IC^s$.
    \end{enumerate}
    Suppose that the variables $(\alpha^*_j)_{j\in [s]}, (R_j)_{j\in [s]}, (r_{jt})_{t\in [s], j\in DC^t}$ {satisfy} the following constraints
    \begin{enumerate}[itemsep=0em,topsep=0.5em,label=(\arabic*)]
        \item For any $t\in [s], j\in DC^t$, we have $r_{jt}\leq R_j$. 
        \label{constr:euclidean-special-first}
        \item For each $t\in [a], j\in DC^t$ and $\ell\in IC^t\cup\{t\}$, we have $\alpha^*_{\ell} \leq (r_{jt}+d(j,\ell))^2$. \label{constr:euclidean-special-second}
        \item For each $t\in [a]\cup \{s\}$, we have \label{constr:euclidean-special-third} %
    $$\sum_{\ell\in[t:s]} w(\ell) \alpha^*_t  + \sum_{\ell \in IC^t} w(\ell)\alpha^*_\ell + \sum_{j\in DC^t} w(j) r_{jt}^2 \leq \hat{f} + \sum_{\ell \in IC^t \cup [t:s]} \rho w(\ell)d^2(\ell,i) + \sum_{j\in DC^t} w(j) d^2(j,i).$$
        \item There is a set $L\subseteq [a]\cap IC^s$ such that: \label{constr:euclidean-special-second-approximate}
        \begin{itemize}
          \item $\sum_{\ell\in L} w(\ell) \leq \epsilon \cdot \sum_{\ell\in [a+1:s]} w(\ell)$;
        \item for any $j\in DC^s, \ell\in (IC^s\cup\{s\})\setminus L$, we have $\alpha^*_{\ell} \leq (1+\epsilon) \cdot (r_{js}+d(j,\ell))^2$;
        \item for any $j\in DC^s, \ell\in L$ and $\ell'\in [a+1:s]$, we have $\alpha^*_{\ell} \leq 
        (r_{js}+d(j,\ell'))^2$.
        \end{itemize}
    \end{enumerate}
    Then, the following holds:
    \begin{align}
        \label{eqn:dual-feasibility-simpler-euclidean-special}
        \sum_{j\in [s]} w(j) \cdot (\alpha^*_j - (\rho-1)R_j^2) \leq \hat{f} + (\Gamma+10\epsilon) \cdot \sum_{j\in [s]} w(j) \cdot d^2(j,i)~.
    \end{align}
\end{restatable}

\begin{remark}
  This is a generalization of \cref{lem:dual-feasibility-simpler-euclidean}, because \cref{lem:dual-feasibility-simpler-euclidean} is a special case when $\epsilon=0, a=s-1$ and $L=\emptyset$.  
\end{remark}

\section{\texorpdfstring{An Improved LMP $4.9$-Approximation for the Metric Case}{An Improved the LMP 4.9-Approximation for the Metric Case}}
\label{sec:metric-lmp}

In this section, we will prove \cref{lem:dual-feasibility-simpler-metric}, restated below for convenience.
\DualFeasibilityGeneralMetric*

\subsection{Unweighted}

\providecommand{\sdist}{a}
\providecommand{\bcoef}{\eta}
\providecommand{\wcoeff}{\beta}

In this subsection, we prove \Cref{lem:dual-feasibility-simpler-metric} in the special case $w(\ell)=1$ for all $\ell\in[s]$. Define $ \sdist_t \defeq d(t,i) $ for all $ t\in[s].$ {The final proof of this lemma for the unweighted case was obtained by GPT 5.2/5.4 Pro/Thinking, after several rounds of exchange with the authors implementing various theoretical ideas and experiments, and revised by the authors.}

As briefly discussed in the overview in {\Cref{sec:spectral-overview}}, the proof has four steps. First, from the inequalities at time $\alpha_t-\eps$ given in condition~\ref{constr:metric-third} of \Cref{lem:dual-feasibility-simpler-metric}, 
we derive a family of constraints $\eqref{eq:Pk-unweighted}$ by using the triangle inequality to lower bound the terms involving $r_{jt}$. Second, we choose coefficients $(\wcoeff)_{t\in[s]}$ so that, in the weighted combination of these constraints, the coefficient of each $\alpha_t^*$ is exactly $1$. Third, after collecting terms, we obtain a quadratic form in the distances to the facility $i$. Finally, we bound the top eigenvalue of the corresponding matrix by the Collatz--Wielandt Formula, which yields the claimed constant.

\paragraph{Step 1: From $\eqref{eq:Ik-unweighted}$ to $\eqref{eq:Pk-unweighted}$.}
For each $t\in[s]$, given $w(t)=1$ and $a_t\defeq d(t,i)$ for all $t\in[s]$, rewriting and rearranging the condition~\ref{constr:metric-third} gives
\begin{align}
    \tag{Condition~\ref{constr:metric-third}}
     &\sum_{j\in[t:s]} \alpha^*_t  + \sum_{j \in IC^t} \alpha^*_j + \sum_{j\in DC^t} w(j) r_{jt}^2 \leq \hat{f} + \sum_{j \in IC^t \cup [t:s]} \rho d^2(j,i) + \sum_{j\in DC^t} w(j) d^2(j,i).\\
    \label{eq:Ik-unweighted}
    \tag{$I_t$}
    \implies & \sum_{j\in DC^t}(r_{jt}^2-\sdist_j^2)
    +
    \sum_{j\in IC^t}(\alpha_j^*-\rho\sdist_j^2)
    +
    \sum_{j\ge t}(\alpha_t^*-\rho\sdist_j^2)
    \le \hat f.
\end{align}
Fix $t\in[s]$. For each $j\in DC^t$, conditions~\ref{constr:metric-first}--\ref{constr:metric-second} and the triangle inequality imply
\[
    \alpha_t^* 
    \le (r_{jt}+d(j,t))^2
    \le (r_{jt}+\sdist_j+\sdist_t)^2
    \le r_{jt}^2 + (\sdist_j+\sdist_t)^2 + 2(\sdist_j+\sdist_t)R_j.
\]
Rearranging and using $2\sdist_j\sdist_t\le \sdist_j^2+\sdist_t^2$, we obtain
\[
    r_{jt}^2-\sdist_j^2
    \ge
    \alpha_t^* - 2\sdist_t^2 - 3\sdist_j^2 - 2(\sdist_j+\sdist_t)R_j.
\]
Substituting into \Cref{eq:Ik-unweighted} gets us
\begin{align}
    &\sum_{j\in DC^t}\alpha_t^*
    +
    \sum_{j\in IC^t}\alpha_j^*
    +
    \sum_{j\ge t}\alpha_t^*
    - \hat f
    \notag\\
    &\qquad\le
    \sum_{j\in DC^t}\bigl(2\sdist_t^2+3\sdist_j^2+2(\sdist_t+\sdist_j)R_j\bigr)
    +
    \sum_{j\in IC^t}\rho\sdist_j^2
    +
    \sum_{j\ge t}\rho\sdist_j^2.
    \label{eq:Pk-unweighted}
    \tag{$P_t$}
\end{align}

\paragraph{Step 2: Choosing coefficients.} 
We now choose nonnegative coefficients $\wcoeff_1,\ldots,\wcoeff_s$ and multiply each inequality $\eqref{eq:Pk-unweighted}$ with $\wcoeff_t$ and sum over all  $t\in[s]$ such that the coefficient of each $\alpha_t^*$ is exactly $1$. This is very similar to the proof of how coefficients are derived in \cite{charikar2025kmeans}. %

\begin{lemma}
    \label{lem:focsweights}
    For each $t\in [s]$, define
    \[
        m_t\defeq |DC^t|,
        \qquad
        c_t\defeq s-|IC^t|=s-t+1+m_t,
        \qquad
        g_t\defeq \sum_{u\ge t} \wcoeff_u,
    \]
    \[
        \bcoef_t\defeq \sum_{u>t:\,t\in DC^u} \wcoeff_u,
        \qquad
        A_t\defeq m_t \wcoeff_t.
    \]
    Then there exist nonnegative coefficients $\wcoeff_1,\dots,\wcoeff_s$ such that, for every $t\in[s]$,
    \begin{equation}
        \label{eq:alphatcoeff}
        \wcoeff_t c_t + \sum_{u>t:\,t\in IC^u} \wcoeff_u = 1,
    \end{equation}
    and moreover,
    \[
        0\le \bcoef_t\le g_t,
        \qquad
        \sum_{t=1}^s \wcoeff_t = 1,
        \qquad
        A_t+g_t \le 1.
    \]
\end{lemma}

\begin{proof}
    Define the coefficients recursively, for $t=s,s-1,\dots,1$, by
    \begin{equation}
        \label{eq:focsreccur}
        \wcoeff_t=
        \frac{1-\sum_{u>t:\,t\in IC^u} \wcoeff_u}{c_t}.
    \end{equation}
    Since \Cref{eq:alphatcoeff} is triangular, this is the unique solution of the system. We verify that the resulting coefficients are nonnegative.

    Fix $t\in \{0\}\cup [s]$ and consider the suffix $\{\alpha^*_{t+1},\dots,\alpha^*_s\}$. In the final weighted combination, its total coefficient is $s-t$. On the other hand, for each $u>t$, the contribution of the $u$'th inequality \Cref{eq:Pk-unweighted} to this suffix is
    \[
        c_u + |IC^u\cap [t+1,u-1]|.
    \]
    Since $IC^u$ and $DC^u$ partition $[u-1]$, we have
    \begin{align*}
        c_u + |IC^u\cap [t+1,u-1]|
        &= (s-u+1)+m_u +(u-t-1)-|DC^u\cap [t+1,u-1]| \\
        &= s-t + |DC^u\cap [1,t]| \\
        &\ge s-t+m_t,
    \end{align*}
    where we used $DC^t\subseteq DC^u$ and $DC^t\subseteq [t-1]$. Summing over $u>t$ gives
    \begin{equation}\label{eq:suffixbeta-unweighted}
        (s-t+m_t)\sum_{u>t} \wcoeff_u \le s-t,
    \end{equation}
        
    and hence
    \[
        \sum_{u>t} \wcoeff_u \le \frac{s-t}{s-t+m_t}.
    \]
    Therefore,
    \[
        \sum_{u>t:\,t\in IC^u} \wcoeff_u \le \sum_{u>t} \wcoeff_u \le \frac{s-t}{s-t+m_t}<1,
    \]
    so \Cref{eq:focsreccur} indeed yields $\wcoeff_t\ge 0$.

    The inequality $0\le \bcoef_t\le g_t$ is immediate from the definitions.

    To prove $\sum_{t=1}^s \wcoeff_t=1$, sum the coefficients of all $\alpha^*$-variables in the weighted combination of \Cref{eq:Pk-unweighted}. Each inequality contributes total coefficient $s$, while by construction each $\alpha_u^*$ has coefficient $1$. Hence
    \[
        s\sum_{t=1}^s \wcoeff_t = s,
    \]
    so $\sum_{t=1}^s \wcoeff_t=1$, and in particular $g_t\le 1$.

    Finally, fix $t\in[s]$ and consider the suffix $\{\alpha_t^*,\dots,\alpha_s^*\}$. Its total coefficient in the final weighted combination is $s-t+1$. For each $u\ge t$, the contribution of the $u$'th inequality to this suffix is
    \[
        c_u + |IC^u\cap [t,u-1]|.
    \]
    Again using that $IC^u$ and $DC^u$ partition $[u-1]$, we obtain
    \begin{align*}
        c_u + |IC^u\cap [t,u-1]|
        &= (s-u+1)+m_u+(u-t)-|DC^u\cap [t,u-1]| \\
        &= s-t+1 + |DC^u\cap [1,t-1]| \\
        &\ge s-t+1+m_t = c_t.
    \end{align*}
    Summing over $u\ge t$ yields $c_t g_t\le s-t+1$, so
    \begin{equation}\label{eq:suffixg-unweighted}
        g_t\le \frac{s-t+1}{c_t}.
    \end{equation}
    Since \Cref{eq:alphatcoeff} implies $c_t \wcoeff_t\le 1$, we also have $\wcoeff_t\le 1/c_t$. Therefore,
    \[
        A_t+g_t = m_t \wcoeff_t + g_t \le \frac{m_t}{c_t} + \frac{s-t+1}{c_t}=1.
    \]
\end{proof}

\paragraph{Step 3: The quadratic form} We next compute the quadratic form produced by the weighted sum of $\eqref{eq:Pk-unweighted}_{t\in[s]}$.

\begin{lemma}
    \label{lem:quadraticform}
    For each $t\in[s]$ and each $x\in\mathbb{R}^s$, define  $U_t(x) \defeq \sum_{u:\,t\in DC^u} \wcoeff_u x_u.$
    Then the weighted sum of all inequalities \Cref{eq:Pk-unweighted} over $t\in[s]$ implies
    \[
        \sum_{t\in[s]} \alpha_t^*-(\rho-1)R_t^2 - \hat f
        \le
        \rho \sum_{t\in[s]} \sdist_t^2 + Q_\rho(\sdist),
    \]
    where, for every $x\in\mathbb{R}^s$, 
    \begin{equation}
        \label{eq:Qrho-first}
        Q_\rho(x)
        =
        \sum_{t\in[s]} \bigl(2A_t+(3-\rho)\bcoef_t\bigr)x_t^2
        +
        \frac{1}{\rho-1}\sum_{t\in[s]} \bigl(U_t(x)+\bcoef_t x_t\bigr)^2.
    \end{equation}
\end{lemma}

\begin{proof}
    By \Cref{eq:alphatcoeff}, the weighted sum of the left-hand sides of $\eqref{eq:Pk-unweighted}_{t\in[s]}$ is $\sum_{t\in[s]}\alpha_t^*-\hat f$. Let $\mathcal R$ denote the weighted sum of the right-hand sides. Then
    \[
        \sum_{t\in[s]}\alpha_t^*-\hat f \le \mathcal R.
    \]
    We now collect the terms in $\mathcal R$. 

    First,
    \[
        \sum_{t\in[s]} \wcoeff_t\sum_{j\in DC^t} 2\sdist_t^2
        = \sum_{t\in[s]} 2A_t\sdist_t^2.
    \]
    Second,
    \[
        \sum_{t\in[s]} \wcoeff_t\sum_{j\in DC^t} 3\sdist_j^2
        = \sum_{t\in[s]} 3\bcoef_t\sdist_t^2.
    \]
    Third,
    \[
        \sum_{t\in[s]} \wcoeff_t\sum_{j\in DC^t} 2(\sdist_t+\sdist_j)R_j
        = 2\sum_{t\in[s]} R_t\bigl(U_t(\sdist)+\bcoef_t\sdist_t\bigr).
    \]
    Finally, for each fixed $t$, the sets
    \[
        \{u\in[s]: t\in IC^u\},
        \qquad
        \{u\in[s]: t\in DC^u\},
        \qquad
        \{u\in[s]: u\le t\}
    \]
    partition $[s]$. Since $\sum_{t=1}^s \wcoeff_t=1$ by \Cref{lem:focsweights}, it follows that
    \[
        \sum_{u:\,t\in IC^u} \wcoeff_u + \sum_{u\le t} \wcoeff_u = 1-\bcoef_t,
    \]
    and hence
    \[
        \sum_{u\in[s]} \wcoeff_u\left(\sum_{j\in IC^u}\rho\sdist_j^2 + \sum_{j\ge u}\rho\sdist_j^2\right)
        = \rho\sum_{t\in[s]} (1-\bcoef_t)\sdist_t^2.
    \]
    Therefore,
    \[
        \mathcal R
        =
        \sum_{t\in[s]} 2A_t\sdist_t^2
        + \sum_{t\in[s]} 3\bcoef_t\sdist_t^2
        + 2\sum_{t\in[s]} R_t\bigl(U_t(\sdist)+\bcoef_t\sdist_t\bigr)
        + \rho\sum_{t\in[s]} (1-\bcoef_t)\sdist_t^2.
    \]
    Equivalently,
    \[
        \mathcal R
        =
        \rho\sum_{t\in[s]} \sdist_t^2
        + \sum_{t\in[s]} \bigl(2A_t+(3-\rho)\bcoef_t\bigr)\sdist_t^2
        + 2\sum_{t\in[s]} R_t\bigl(U_t(\sdist)+\bcoef_t\sdist_t\bigr).
    \]
    Hence
    \[
        \sum_{t\in[s]} \alpha_t^*-(\rho-1)R_t^2 - \hat f
        \le
        \rho\sum_{t\in[s]} \sdist_t^2
        + \sum_{t\in[s]} \bigl(2A_t+(3-\rho)\bcoef_t\bigr)\sdist_t^2
        + \sum_{t\in[s]} \Bigl(2R_t\bigl(U_t(\sdist)+\bcoef_t\sdist_t\bigr)- (\rho-1)R_t^2\Bigr).
    \]
    As in the warm-up, we get rid of the mixed term $R$ using
    \[
        2Rx-(\rho-1)R^2 \le \frac{x^2}{\rho-1} \qquad (R,x\in\mathbb{R}).
    \]
    Applying this with $x=U_t(\sdist)+\bcoef_t\sdist_t$ for each $t$, we obtain
    \[
        \sum_{t\in[s]} \alpha_t^*-(\rho-1)R_t^2 - \hat f
        \le
        \rho\sum_{t\in[s]} \sdist_t^2
        + \sum_{t\in[s]} \bigl(2A_t+(3-\rho)\bcoef_t\bigr)\sdist_t^2
        + \frac{1}{\rho-1}\sum_{t\in[s]} \bigl(U_t(\sdist)+\bcoef_t\sdist_t\bigr)^2.
    \]
    This is exactly the claimed inequality.
\end{proof}

\paragraph{Step 4: Matrix Form} It remains to bound $Q_\rho$. 

\begin{lemma}
    \label{lem:matrix}
    Let $K$ be the unique symmetric matrix such that $Q_\rho(x)=x^\top Kx$ for all $x\in\mathbb{R}^s$. Then, for every $t\in[s]$ and every $x\in\mathbb{R}_{\ge 0}^s$,
    \[
        (Kx)_t
        =
        \Bigl(2A_t+(3-\rho)\bcoef_t+\frac{\bcoef_t^2}{\rho-1}\Bigr)x_t
        + \frac{\bcoef_t}{\rho-1}U_t(x)
        + \frac{\wcoeff_t}{\rho-1}\sum_{u\in DC^t} \bigl(U_u(x)+\bcoef_u x_u\bigr).
    \]
    Moreover, $K$ is entrywise nonnegative for all $1<\rho\le 3$. Consequently,
    \[
        Q_\rho(x)\le \lambda_{\max}(K)\|x\|_2^2
        \qquad\text{for all }x\in\mathbb{R}_{\ge 0}^s.
    \]
\end{lemma}

\begin{proof}
    Since $Q_\rho(x)=x^\top Kx$ and $K$ is symmetric, we have
    \[
        \frac{\partial Q_\rho(x)}{\partial x_t}=2(Kx)_t.
    \]
    Differentiating \Cref{eq:Qrho-first}, we obtain
    \begin{align*}
        (Kx)_t
        &= \bigl(2A_t+(3-\rho)\bcoef_t\bigr)x_t
        + \frac{\bcoef_t}{\rho-1}\bigl(U_t(x)+\bcoef_t x_t\bigr)
        + \frac{\wcoeff_t}{\rho-1}\sum_{u\in DC^t} \bigl(U_u(x)+\bcoef_u x_u\bigr) \\
        &= \Bigl(2A_t+(3-\rho)\bcoef_t+\frac{\bcoef_t^2}{\rho-1}\Bigr)x_t
        + \frac{\bcoef_t}{\rho-1}U_t(x)
        + \frac{\wcoeff_t}{\rho-1}\sum_{u\in DC^t} \bigl(U_u(x)+\bcoef_u x_u\bigr).
    \end{align*}
    Every coefficient in this expression is nonnegative when $1<\rho\le 3$, so $K$ is entrywise nonnegative. The final claim follows from the definition $\lambda_{\max}(K)$.
\end{proof}

We now pick $\rho=5/2$ and replace $\bcoef_t$ by $g_t$, since $\bcoef_t\leq g_t$ (\Cref{lem:focsweights}). For every $t\in[s]$ and every $x\in\mathbb{R}_{\ge 0}^s$, \Cref{lem:matrix} gives
\begin{equation}
    \label{eq:finalkentry}
    (Kx)_t
    \le
    \Bigl(2A_t + \frac12 g_t + \frac23 g_t^2\Bigr)x_t
    + \frac23 g_t U_t(x)
    + \frac23 \wcoeff_t\sum_{u\in DC^t} U_u(x)
    + \frac23 \wcoeff_t\sum_{u\in DC^t} g_u x_u.
\end{equation}

The next lemma applies the Collatz--Wielandt {Formula} \cite{meyer2000matrix} 

\begin{lemma}
    \label{lem:function}
    Fix $\tau=\sqrt[3]{12}$. Then
    \[
        \lambda_{\max}(K)\le \max_{1\le x\le \tau} F(x),
    \]
    where
    \begin{equation}
        \label{eq:function}
        F(x)=
        \frac{48x^6-47\tau x^5-357x^3+(564\tau-396)x^2+4665}{2178}.
    \end{equation}
\end{lemma}

\begin{proof}
    We choose $h\in\mathbb{R}_{\ge 0}^s$ by
    \[
        h_t = (1+11g_t)^{-2/3} \qquad (t\in[s]).
    \]
    By the min--max Collatz--Wielandt Formula \cite[Theorem 8.3.3]{meyer2000matrix} for non-negative symmetric matrix $K\geq 0$, with spectral radius $\rho(K)$, we have
    \begin{equation}\label{eq:collatz-wielandt}
        \lambda_{\max}(K) = \rho(K) = \min_{x\in \mathbb{R}_{\geq 0}^s} \max_{t\in[s], x_t\neq 0} \frac{(Kx)_t}{x_t} \le \max_{t\in[s]} \frac{(Kh)_t}{h_t} \tag{Collatz--Wielandt Formula}
    \end{equation}

    \begin{claim}
        \label{lem:suffixintegrabound}
        For every $t\in[s]$,
        \begin{align}
            \frac{2g_tU_t(h)}{3h_t}
            &\le
            \frac{2}{11}g_t\Bigl((1+11g_t)-(1+11g_t)^{2/3}\Bigr),\label{eq:lambda1}\\
            \frac{2\wcoeff_t}{3h_t}\sum_{u:\,t\in DC^u} g_uh_u
            &\le
            A_t\cdot \frac{2}{3}(1+11g_t)^{2/3}\tau^{-2},\label{eq:lambda2}\\
            \frac{2\wcoeff_t}{3h_t}\sum_{u:\,t\in DC^u} U_u(h)
            &\le
            A_t(1+11g_t)^{2/3}\cdot \frac{2}{11}(\tau-1).\label{eq:lambda3}
        \end{align}
        Consequently,
        \begin{equation}
            \label{eq:lambda}
            \frac{(Kh)_t}{h_t} \le G(g_t),
        \end{equation}
        where
        \begin{align}
            G(g)
            \defeq{}&
            2(1-g)+\frac12 g+\frac23 g^2
            + \frac23(1-g)(1+11g)^{2/3}\tau^{-2}
            \notag\\
            &+ \frac{2}{11}(1-g)(1+11g)^{2/3}(\tau-1)
            + \frac{2}{11}g\Bigl((1+11g)-(1+11g)^{2/3}\Bigr).
            \label{eq:scalar-majorant-g}
        \end{align}
    \end{claim}

    \begin{proof}
        Since the sets $DC^u$ are monotone in $u$, the set $\{u\in[s]: t\in DC^u\}$ is a suffix of $\{t+1,\dots,s\}$. Let $u_t$ be its first element. Let $u_t=s+1$ if $t\notin DC^u$ for every $u$. Then
        \[
            U_t(h)=\sum_{u\ge u_t} \wcoeff_u h_u = \sum_{u\ge u_t} (g_u-g_{u+1})(1+11g_u)^{-2/3}.
        \]
        Since the function $z\mapsto (1+11z)^{-2/3}$ is decreasing, this sum is bounded by the corresponding integral:
        \begin{equation}
            \label{eq:ukh}
            U_t(h)
            \le \int_0^{\bcoef_t} (1+11z)^{-2/3}\,dz
            \le \frac{3}{11}\Bigl((1+11\bcoef_t)^{1/3}-1\Bigr).
        \end{equation}
        Together with $\bcoef_t\le g_t$, this gives \Cref{eq:lambda1}.

        For \Cref{eq:lambda3}, \Cref{eq:ukh} implies
        \[
            U_u(h)
            \le
            \frac{3}{11}\Bigl((1+11\bcoef_u)^{1/3}-1\Bigr)
            \le
            \frac{3}{11}(\tau-1),
        \]
        since $\bcoef_u\in[0,1]$. Therefore,
        \[
            \frac{2\wcoeff_t}{3h_t}\sum_{u:\,t\in DC^u} U_u(h)
            \le
            \frac{2\wcoeff_t}{3}(1+11g_t)^{2/3}\cdot m_t\cdot \frac{3}{11}(\tau-1)
            =
            A_t(1+11g_t)^{2/3}\cdot \frac{2}{11}(\tau-1).
        \]

        For \Cref{eq:lambda2}, the function $g\mapsto g(1+11g)^{-2/3}$ is increasing on $[0,1]$, so
        \[
            g_u h_u = g_u(1+11g_u)^{-2/3} \le 12^{-2/3}=\tau^{-2}.
        \]
        Hence
        \[
            \sum_{u:\,t\in DC^u} g_u h_u \le m_t\tau^{-2},
        \]
        and therefore
        \[
            \frac{2\wcoeff_t}{3h_t}\sum_{u:\,t\in DC^u} g_uh_u
            \le
            \frac{2\wcoeff_tm_t}{3}(1+11g_t)^{2/3}\tau^{-2}
            =
            A_t\cdot \frac{2}{3}(1+11g_t)^{2/3}\tau^{-2}.
        \]

        Combining \Cref{eq:finalkentry,eq:lambda1,eq:lambda2,eq:lambda3} and using $A_t+g_t\le 1$ from \Cref{lem:focsweights}, we obtain
        \begin{align*}
            \frac{(Kh)_t}{h_t}
            &\le
            \Bigl(2A_t+\frac12 g_t+\frac23 g_t^2\Bigr)
            + \frac{2g_tU_t(h)}{3h_t}
            + \frac{2\wcoeff_t}{3h_t}\sum_{u:\,t\in DC^u} U_u(h)
            + \frac{2\wcoeff_t}{3h_t}\sum_{u:\,t\in DC^u} g_uh_u \\
            &\le
            2A_t+\frac12 g_t+\frac23 g_t^2
            + \frac{2}{11}g_t\Bigl((1+11g_t)-(1+11g_t)^{2/3}\Bigr) \\
            &\qquad + A_t(1+11g_t)^{2/3}\cdot \frac{2}{11}(\tau-1)
            + A_t\cdot \frac{2}{3}(1+11g_t)^{2/3}\tau^{-2} \\
            &\le G(g_t).
        \end{align*}
    \end{proof}

    Finally, we simplify the expression in \Cref{eq:lambda} by substituting $g=(x^3-1)/11$ and simplifying, using $\tau^{-2}=\tau/12$. This gives
    \[
        \frac{(Kh)_t}{h_t} \le G(g_t)=F(x),
    \]
    where
    \[
        F(x)=
        \frac{48x^6-47\tau x^5-357x^3+(564\tau-396)x^2+4665}{2178}.
    \]
\end{proof}

The remaining task is to prove an upper bound on $F(x)$ as follows:

\begin{lemma}
    \label{lem:functionupperbound}
    $
        \max_{x\in[1,\tau]} F(x) < \frac{12}{5}.
    $
\end{lemma}

\begin{proof}
    This is deferred to \Cref{sec:prooffunctionupperbound}.
\end{proof}

We can now finish the proof of the unweighted version of \Cref{lem:dual-feasibility-simpler-metric}. Using \Cref{lem:quadraticform} with $\rho=5/2$, we obtain
\begin{align*}
    \sum_{t\in[s]} \alpha_t^*-(\rho-1)R_t^2 - \hat f
    &\le \frac52 \sum_{t\in[s]} \sdist_t^2 + Q_{5/2}(\sdist) \tag{\Cref{lem:quadraticform}} \\
    &\le \frac52 \sum_{t\in[s]} \sdist_t^2 + \lambda_{\max}(K)\sum_{t\in[s]} \sdist_t^2 \tag{\Cref{lem:matrix}} \\
    &\le \frac52 \sum_{t\in[s]} \sdist_t^2 + \left(\max_{x\in[1,\tau]}F(x)\right)\sum_{t\in[s]} \sdist_t^2 \tag{\Cref{lem:function}} \\
    &\le \frac52 \sum_{t\in[s]} \sdist_t^2 + \frac{12}{5}\sum_{t\in[s]} \sdist_t^2 \tag{\Cref{lem:functionupperbound}} \\
    &= 4.9\sum_{t\in[s]} \sdist_t^2.
\end{align*}
This proves \Cref{lem:dual-feasibility-simpler-metric} in the unweighted case, i.e., $w(\ell)=1$ for all $\ell\in[s]$.

\subsection{Weighted}

\providecommand{\sdist}{a}
\providecommand{\bcoef}{\eta}
\providecommand{\wcoeff}{\beta}
\providecommand{\swt}{W}

We now prove \Cref{lem:dual-feasibility-simpler-metric} in its full generality for any weight function $w:[s]\to\mathbb{R}_{> 0}$. %

To extend the argument to the weighted setting, the overall structure is the same as in the unweighted case: given the condition (3) of \Cref{lem:dual-feasibility-simpler-metric}, we derive inequalities $\eqref{eq:Pt-weighted-short}_{t\in[s]}$, choose coefficients so that the $\alpha_t^*$ terms sum up to $w(t)$ for each $t\in[s]$, rewrite the resulting bound as a quadratic form, and then upper bound its top eigenvalue. The new issue is the bookkeeping. The cardinalities of clients are replaced by weighted masses, the normalization of the coefficients changes accordingly. After introducing the appropriate weighted analogues of the unweighted parameters and rescaling by $\sqrt{w(t)}$, however, the same spectral argument goes through.

Since the proof is a weighted analogue of the unweighted subsection, we mention only the differences in the proof of the coefficients and in the quadratic form, and defer the full proof to \Cref{sec:dual-feasibility-weighted-full}.

Define

\[
    \sdist_t \defeq d(t,i),
    \qquad
    W(S)\defeq \sum_{j\in S}w(j),
    \qquad
    \swt_t\defeq W([t:s]),
    \qquad
    M_t\defeq W(DC^t).
\]

For each $t\in[s]$, condition~(3) gives
\begin{equation}
    \label{eq:Ik-weighted-short}
    \tag{$I_t^w$}
    \sum_{j\in DC^t} w(j)(r_{jt}^2-\sdist_j^2)
    +
    \sum_{j\in IC^t} w(j)(\alpha_j^*-\rho\sdist_j^2)
    +
    \sum_{j\ge t} w(j)(\alpha_t^*-\rho\sdist_j^2)
    \le \hat f.
\end{equation}
Exactly as in the unweighted case, for every $j\in DC^t$,
\[
    r_{jt}^2-\sdist_j^2
    \ge
    \alpha_t^* - 2\sdist_t^2 - 3\sdist_j^2 - 2(\sdist_j+\sdist_t)R_j.
\]
Substituting into \Cref{eq:Ik-weighted-short} yields
\begin{align}
    &M_t\alpha_t^*
    +
    \sum_{j\in IC^t} w(j)\alpha_j^*
    +
    \swt_t\alpha_t^*
    - \hat f
    \notag\\
    &\qquad\le
    2M_t\sdist_t^2
    + \sum_{j\in DC^t} w(j)\bigl(3\sdist_j^2+2(\sdist_t+\sdist_j)R_j\bigr)
    + \sum_{j\in IC^t}\rho w(j)\sdist_j^2
    + \sum_{j\ge t}\rho w(j)\sdist_j^2.
    \label{eq:Pt-weighted-short}
    \tag{$P_t^w$}
\end{align}

\begin{lemma}
    \label{lem:focsweights-weighted-short}
    Define
    \[
        c_t\defeq \frac{\swt_t+M_t}{w(t)},
        \qquad
        g_t\defeq \sum_{u\ge t}\wcoeff_u,
        \qquad
        \bcoef_t\defeq \sum_{u>t:\,t\in DC^u}\wcoeff_u,
        \qquad
        A_t\defeq \frac{M_t\wcoeff_t}{w(t)}.
    \]
    Then there exist nonnegative coefficients $\wcoeff_1,\dots,\wcoeff_s$ such that, for every $t\in[s]$,
    \begin{equation}
        \label{eq:alphatcoeff-weighted-short}
        \wcoeff_t c_t + \sum_{u>t:\,t\in IC^u}\wcoeff_u = 1,
    \end{equation}
    and moreover,
    \[
        0\le \bcoef_t\le g_t,
        \qquad
        \sum_{t=1}^s \wcoeff_t=1,
        \qquad
        A_t+g_t\le 1.
    \]
\end{lemma}

The proof is the same as that of \Cref{lem:focsweights}, with cardinalities replaced by weighted masses. The only changes are the following two weighted analogues of Inequalities \ref{eq:suffixbeta-unweighted}, \ref{eq:suffixg-unweighted}
\[
    (\swt_{t+1}+M_t)\sum_{u>t}\wcoeff_u\le \swt_{t+1},
    \qquad
    (\swt_t+M_t)g_t\le \swt_t.
\]

\begin{lemma}
    \label{lem:quadraticform-weighted-short}
    For $x\in\mathbb R^s$, define $U_t(x)\defeq \sum_{u:\,t\in DC^u}\wcoeff_u x_u.$
    Then
    \[
        \sum_{t\in[s]} w(t)\bigl(\alpha_t^*-(\rho-1)R_t^2\bigr)-\hat f
        \le
        \rho\sum_{t\in[s]} w(t)\sdist_t^2 + Q_\rho^w(\sdist),
    \]
    where
    \begin{equation}
        \label{eq:Qrho-weighted-short}
        Q_\rho^w(x)
        =
        \sum_{t\in[s]} w(t)\bigl(2A_t+(3-\rho)\bcoef_t\bigr)x_t^2
        +
        \frac{1}{\rho-1}\sum_{t\in[s]} w(t)\bigl(U_t(x)+\bcoef_t x_t\bigr)^2.
    \end{equation}
\end{lemma}

\begin{proof}
With the coefficients chosen as above, the same counting as in the unweighted proof gives
\[
    \sum_{t\in[s]} w(t)\alpha_t^* - \hat f \le \mathcal R,
\]
where
\[
    \mathcal R
    =
    \rho\sum_{t\in[s]} w(t)\sdist_t^2
    + \sum_{t\in[s]} w(t)\bigl(2A_t+(3-\rho)\bcoef_t\bigr)\sdist_t^2
    + 2\sum_{t\in[s]} w(t)R_t\bigl(U_t(\sdist)+\bcoef_t\sdist_t\bigr).
\]
Subtracting $(\rho-1)\sum_t w(t)R_t^2$ and applying $2Rx-(\rho-1)R^2\le x^2/(\rho-1)$ pointwise yields the claim.
\end{proof}

\begin{lemma}
    \label{lem:matrix-weighted-short}
    Let
    \[
        \widetilde x\defeq (\sqrt{w(1)}x_1,\dots,\sqrt{w(s)}x_s),
    \]
    and let $B$ be the symmetric matrix such that
    \[
        Q_\rho^w(x)=\widetilde x^\top B\widetilde x.
    \]
    Then, for every $t\in[s]$,
    \[
        \frac{(B\widetilde x)_t}{\sqrt{w(t)}}
        =
        \Bigl(2A_t+(3-\rho)\bcoef_t+\frac{\bcoef_t^2}{\rho-1}\Bigr)x_t
        + \frac{\bcoef_t}{\rho-1}U_t(x)
        + \frac{\wcoeff_t}{(\rho-1)w(t)}\sum_{u\in DC^t}w(u)\bigl(U_u(x)+\bcoef_u x_u\bigr).
    \]
    In particular, for $\rho=5/2$ and $x\in\mathbb R_{\ge 0}^s$,
    \[
        \frac{(B\widetilde x)_t}{\sqrt{w(t)}}
        \le
        \Bigl(2A_t+\tfrac12 g_t+\tfrac23 g_t^2\Bigr)x_t
        + \tfrac23 g_tU_t(x)
        + \frac{2\wcoeff_t}{3w(t)}\sum_{u\in DC^t}w(u)U_u(x)
        + \frac{2\wcoeff_t}{3w(t)}\sum_{u\in DC^t}w(u)g_ux_u.
    \]
\end{lemma}

\begin{proof}
This is obtained by expanding \Cref{eq:Qrho-weighted-short}, exactly as in the proof of \Cref{lem:matrix}.
\end{proof}

\begin{lemma}
    \label{lem:function-weighted-short}
    Let
    \[
        h_t\defeq (1+11g_t)^{-2/3},
        \qquad
        \widetilde h_t\defeq \sqrt{w(t)}\,h_t.
    \]
    Then
    \[
        \lambda_{\max}(B)\le \max_{1\le x\le \tau}F(x)<\frac{12}{5}.
    \]
\end{lemma}

\begin{proof}
Since $B$ is symmetric and entrywise nonnegative, \ref{eq:collatz-wielandt} gives
\[
    \lambda_{\max}(B)\le \max_{t\in[s]}\frac{(B\widetilde h)_t}{\widetilde h_t}.
\]
The three inequalities used in the proof of \Cref{lem:function} are unchanged, except that $|DC^t|$ is replaced by $M_t/w(t)$, which is exactly absorbed by the definition of $A_t$. Hence the same functions $G(g)$ and $F(x)$ are obtained, and the final inequality follows from \Cref{lem:functionupperbound}.
\end{proof}

We can now conclude exactly as in the unweighted case:
\begin{align*}
    &\sum_{t\in[s]} w(t)\bigl(\alpha_t^*-(\rho-1)R_t^2\bigr)-\hat f \\
    &\qquad\le \frac52\sum_{t\in[s]} w(t)\sdist_t^2 + Q_{5/2}^w(\sdist) \tag{\Cref{lem:quadraticform-weighted-short}} \\
    &\qquad\le \frac52\sum_{t\in[s]} w(t)\sdist_t^2 + \lambda_{\max}(B)\sum_{t\in[s]} w(t)\sdist_t^2 \tag{\Cref{lem:matrix-weighted-short}} \\
    &\qquad\le 4.9\sum_{t\in[s]} w(t)\sdist_t^2. \tag{\Cref{lem:function-weighted-short}}
\end{align*}
This proves \Cref{lem:dual-feasibility-simpler-metric}.

\section{Log-Adaptive Algorithm for Weighted Facility Location}
\label{sec:log_adaptivity}

Let $\epsilon \in (0, 10^{-100})$ be a sufficiently small constant. We choose $\delta = \epsilon/2$. In weighted facility location, each client $j \in D$ has a weight $w(j) > 0$ and the objective is to minimize $\sum_{j \in D} w(j) \cdot d^2(j, S) + |S|f$.

We present the modified greedy algorithm for weighted facility location that works in $O(\log n / \epsilon^3)$ phases.

\begin{definition}[openable]
\label{def:openable}
A facility $i$ is \emph{openable} with respect to the algorithm's state $(\alpha, S, \theta)$ if there are increased dual values $(\tau_j)_{j \in A}$ of active clients that satisfy the following conditions:
\begin{itemize}
    \item Only nearby clients are increased:
    $\tau_j = \alpha_j$ for every $j \in A \setminus B(i, \sqrt{\epsilon\theta})$.
    
    \item Nearby clients are only increased slightly: 
    $\alpha_j \le \tau_j \le \min\{(1-\delta) d^2(j,S), (1+\epsilon^3)\theta\}$ for every $j \in A \cap B(i, \sqrt{\epsilon}\theta)$.
    
    \item Facility $i$ is paid for: 
    \begin{align*}
    (1-\delta)\sum_{j \in DC} w(j) \cdot [d^2(j,S) - d^2(j,i)]^+ 
    &+ \sum_{j \in IC} w(j) \cdot [\alpha_j - \rho(1-\delta)d^2(j,i)]^+ \\
    &+ \sum_{j \in A} w(j) \cdot [\tau_j - \rho(1-\delta)d^2(j,i)]^+ \ge \hat{f}.
    \end{align*}
    
    \item Dual feasibility: for every facility $i_0$:
    \begin{align*}
    \sum_{j \in DC} w(j) \cdot [(1-\delta)d^2(j,S) - d^2(j,i_0)]^+ 
    &+ \sum_{j \in IC} w(j) \cdot [\alpha_j - \rho d^2(j,i_0)]^+ \\
    &+ \sum_{j \in A} w(j) \cdot [\tau_j - \rho d^2(j,i_0)]^+ \le \hat{f}.
    \end{align*}
\end{itemize}
\end{definition}

The openability can be checked by solving a linear program.

\begin{lemma}
    There is a polynomial-time algorithm that, given the state of the algorithm and a facility $i$, either outputs $(\tau_j)_{j \in A}$ that satisfy all conditions of openability or certifies that $i$ is not openable.
\end{lemma}
\begin{proof}
Fix $i$ and $(\alpha, S, A, IC, DC, \theta)$ and consider a linear program with variables $(\tau_j)_{j \in A}$. We claim that each of the four bullets in \Cref{def:openable} can be expressed as linear constraints. The first two bullets are immediate. Indeed, as $\tau_j = \alpha_j$ for every $j \in A \setminus B(i, \sqrt{\epsilon \theta})$, we can treat them as constants and only have $(\tau_j)_{j \in A \cap B(i, \sqrt{\epsilon \theta})}$ as the true variables. 

Then, the third bullet, 
\begin{align*}
    (1-\delta)\sum_{j \in DC} w(j) \cdot [d^2(j,S) - d^2(j,i)]^+ 
    &+ \sum_{j \in IC} w(j) \cdot [\alpha_j - \rho(1-\delta)d^2(j,i)]^+ \\
    &+ \sum_{j \in A} w(j) \cdot [\tau_j - \rho(1-\delta)d^2(j,i)]^+ \ge \hat{f}.
\end{align*}
becomes a linear inequality in the true variables, since the only terms involving the true variables are $w(j)\cdot [\tau_j - \rho(1 - \delta) d^2(j,i)]^+$ for $j \in A \cap B(i, \sqrt{\epsilon \theta})$. For those $j$, $[\tau_j - \rho(1 - \delta) d^2(j,i)]^+$ is indeed equal to $(\tau_j - \rho(1 - \delta)d^2(j,i))$ (without $[\cdot]^+$) because $\epsilon$ is sufficiently small, $d^2(j,i)\leq \eps\theta$ and $\tau\geq \theta$. 

Finally, for the fourth bullet, we can use an additional variable $z'_{i_0,j}\geq 0$ for each $j\in D$. We add the constraint $z'_{i_0,j}\geq (1-\delta)d^2(j,S)-d^2(j,i_0)$ for $j\in DC$, the constraint $z'_{i_0,j} \geq \alpha_j - \rho  d^2(j,i_0)$ for $j\in IC$ and the constraint $z'_{i_0,j}\geq \tau_j - \rho d^2(j,i_0)$ for $j\in A$. In the linear program, we can rewrite the corresponding fourth bullet as 
$$\sum_{j\in D} w(j) z'_{i_0,j} \leq \hat{f}~.$$
\end{proof}

\Cref{alg:logadaptive} is our log-adaptive algorithm. It works in $O(\log n/\eps^3)$ phases where, in the $i$th phase, every active client $j$ has $\alpha_j = (1+\eps^3)^{i} = \theta$ and the algorithm opens openable facilities in an arbitrary order by increasing $\alpha_j$'s to $\tau_j$'s.
However, when $i$ is open, the only scenario $j$ with $\tau_j > \alpha_j$ is when $j \in B(i, \sqrt{\eps \theta})$, so $j$ becomes immediately put into $DC$ after $i$ is open.
Unlike \cref{alg:greedy}, we don't lower the $\alpha$-values for clients in $IC$ immediately after a facility is opened. Instead, we only lower the $\alpha$-values in the second stage of each phase.

\begin{figure}[ht!]
    \begin{center}
    \begin{minipage}{1.0\textwidth}
    \begin{mdframed}[hidealllines=true, backgroundcolor=gray!15]
    
\begin{algorithm}
\label{alg:logadaptive}

\vspace{0.5em}

\textbf{Parameters:} $\rho, \Gamma > 1, \eps\in (0,10^{-100})$; $\hat{f} = (\Gamma + 201\sqrt{\eps}) \cdot f$ where $f$ is the uniform facility opening cost.
    
\vspace{0.5em}

\textbf{Initialization:} Set $\theta = 1$, $S = \emptyset$, $A = D$, $DC = \emptyset$, $IC = \emptyset$, and $\alpha_j = \theta$ for every $j \in D$. \\
While $A \neq \emptyset$:
\begin{enumerate}
    \item \textbf{While} there is an unopened facility $i$ that is openable:
    \begin{enumerate}
        \item Compute $(\tau_j)_{j \in A}$ so that $i$ is openable with $(\tau_j)_{j \in A}$.
        \item Add $i$ to $S$. Set $\alpha_j \gets \tau_j$ for every $j \in A \cap B(i, \sqrt{\epsilon \theta})$.
        \item For every $j \in A\cup IC$:
        \begin{itemize}
            \item If $\alpha_j \ge \rho(1-\delta)d^2(j,i)$, move $j$ to $DC$ and set $R_j=\sqrt{1-\delta}d(j,i)$. 
            \item Otherwise, if $j \in A$ and $\alpha_j \ge (1-\delta)d^2(j,i)$, move $j$ to $IC$.
        \end{itemize}
    \end{enumerate}
    \item Perform the following step and move to the next phase:
    \begin{itemize}
        \item For every $j \in IC$, set $\alpha_j \gets (1-\delta)d^2(j,S)$.
        \item For every remaining $j \in A$, set $\alpha_j \gets \min\{(1+\epsilon^3)\theta, (1-\delta)d^2(j,S)\}$ and move it to $IC$ when $\alpha_j = (1-\delta)d^2(j,S)$.
        \item Update $\theta \gets (1+\epsilon^3)\theta$.
    \end{itemize}
\end{enumerate}
\end{algorithm}

\end{mdframed}
\end{minipage}
\end{center}
\end{figure}

For each client $j \in D$, we explicitly track its first direct-connection cost, $R_j$. We set $R_j = \sqrt{1-\delta}d(j, i^*)$ where $i^*$ is the facility that caused $j$ to be moved to $DC$ for the first time. For a client that remains only indirectly connected or active, we set $R_j = 0$.

In the rest of this subsection, let $(\alpha^*_j)_{j\in D}$ be the final dual solution produced by the algorithm.

\subsection{Dual Payment}

\begin{lemma} \label{lem:approx_guarantee_logadaptive}
At the end of the log-adaptive algorithm, the final dual solution $\alpha^*$ satisfies:
$$ \sum_{j \in D} w(j) \cdot (\alpha_j^* - (\rho-1)R_j^2) \ge (1-\delta) \sum_{j \in D} w(j) \cdot d^2(j,S) + \sum_{i \in S} \hat{f} $$
\end{lemma}
\begin{proof}
First, let us show that it suffices to prove that at the end of the algorithm, we have
$$ \sum_{j \in DC} w(j) \cdot \alpha_j \ge \sum_{j \in DC} w(j) \cdot \left[ (1-\delta) d^2(j,S) + (\rho-1)R_j^2 \right] + \sum_{i \in S} \hat{f} $$
Indeed, at the end of the algorithm, $A = \emptyset$. For every $j \in IC$, we have $\alpha_j \ge (1-\delta)d^2(j,S)$ by the construction of the update steps in Stage 2. Because $R_j = 0$ for $j \in IC$, their contribution naturally satisfies the required bound. Therefore, proving the $DC$ client inequality implies the lemma.

To this end, we prove by induction that at any point in the execution of the algorithm, the $DC$ invariant holds. The base case is at the beginning of the algorithm, where $DC = \emptyset$ and $S = \emptyset$. Next, we show that this invariant is preserved in each stage of the algorithm.

\textbf{Stage 1.} In each step of this stage, we open a facility $i$, i.e., adding it to $S$. Suppose that $(\alpha, S, A, IC, DC, \theta)$ denotes the state right before opening $i$. Let $A'$ and $IC'$ be the clients that are moved to $DC$ from $A$ and $IC$ respectively by the opening of $i$. This means that $A' = \{j \in A : \tau_j \ge \rho(1-\delta)d^2(j,i)\}$ and $IC' = \{j \in IC : \alpha_j \ge \rho(1-\delta)d^2(j,i)\}$. 

For these newly directly connected clients, their initial direct-connection parameter is exactly $R_j = {\sqrt{1-\delta}}d(j,i)$. Therefore, their combined contribution to the right-hand side of our invariant is ${w(j)}\cdot((1-\delta)d^2(j,i) + (\rho-1)R _j^2) = {w(j)}\cdot((1-\delta)d^2(j,i) + (\rho-1)(1-\delta) d^2(j,i))$. 
Notice that since $\rho > 1$ and $\delta > 0$, we can bound this contribution:
\begin{align*}
    w(j) \cdot ((1-\delta)d^2(j,i) + (\rho-1)(1-\delta) d^2(j,i)) &= w(j)(1-\delta)d^2(j,i) [1 + (\rho-1)] \\
    &= w(j) \rho(1-\delta) d^2(j,i)
\end{align*}

Let $X = \{j \in DC : d^2(j,i) < d^2(j,S)\}$ denote the clients in $DC$ whose distance to $i$ is smaller than their distance to $S$. For clients in $X$, their fixed penalty $R_j$ does not change, meaning the change in cost on the right-hand side is strictly driven by the distance dropping to $d^2(j,i)$.

Using the bound established above, the change of cost of the right-hand side is at most:
$$ \hat{f} + \sum_{j \in A' \cup IC'} w(j) \cdot \rho(1-\delta)d^2(j,i) - \sum_{j \in X} w(j) \cdot (1-\delta)(d^2(j,S) - d^2(j,i)) $$

Since $i$ is fully paid according to the third bullet of the openability condition, we also have the capacity bound:
$$ \hat{f} \le \sum_{j \in A'} w(j) \cdot (\tau_j - \rho(1-\delta)d^2(j,i)) + \sum_{j \in IC'} w(j) \cdot (\alpha_j - \rho(1-\delta)d^2(j,i)) + \sum_{j \in X} w(j) \cdot (1-\delta)(d^2(j,S) - d^2(j,i)) $$

Substituting this capacity bound back into our change of cost, the distance and capacity terms cancel out, and we get that the change of the right-hand side is at most:
$$ \sum_{j \in A'} w(j) \cdot \tau_j + \sum_{j \in IC'} w(j) \cdot \alpha_j $$
Because we do not change the $\alpha$-value after a client is moved to $DC$, this is exactly the change of the left-hand side of our invariant. Thus, the inequality holds through Stage 1.

\textbf{Stage 2.} No facility is open at this stage, and no client is added to $DC$ either. Because $DC$ is unaltered and no new $\hat{f}$ is introduced, the invariant remains trivially preserved through the continuous updates of $A$ and $IC$ clients.
\end{proof}

\subsection{Dual Feasibility}

\begin{lemma}
\label{lem:dual-feasibility-logadaptive}
We have for any $i\in\mathcal{F}$, 
\begin{equation}
    \sum_{j\in D}w(j) \cdot [\alpha_j^* - (\rho-1)R_j^2 - (\Gamma+201\sqrt{\eps}) d^2(j,i)]^+\leq \hat f,
\end{equation}
where
    \begin{itemize}
        \item if $d$ is a (general) metric, we choose $\rho=2.5, \Gamma=4.9$; or
        \item if $d$ is a Euclidean metric, we choose $\rho=2, \Gamma = 3+\ln(2)$.
    \end{itemize}
\end{lemma}

\begin{claim}[No over-bidding] \label{clm:no_overbidding}
At any point in the algorithm, for every facility $i_0 \in \mathcal{F}$, we have
$$ \sum_{j \in DC} w(j) \cdot [(1-\delta)d^2(j,S) - d^2(j,i_0)]^+ + \sum_{j \in A \cup IC} w(j) \cdot [\alpha_j - \rho d^2(j,i_0)]^+ \le \hat{f}. $$
\end{claim}
\begin{proof}
Consider an arbitrary facility $i_0$. 

\textbf{Stage 1:} Suppose that a facility $i'$ is opened with dual increases $(\tau_j)_{j \in A}$. Because $i'$ is openable, the proposed state satisfies the fourth bullet of \Cref{def:openable} (dual feasibility) for any $i_0 \in \mathcal{F}$:
$$ \sum_{j \in DC} w(j) \cdot [(1-\delta)d^2(j,S) - d^2(j,i_0)]^+ + \sum_{j \in IC} w(j) \cdot [\alpha_j - \rho d^2(j,i_0)]^+ + \sum_{j \in A} w(j) \cdot [\tau_j - \rho d^2(j,i_0)]^+ \le \hat{f}. $$

Let us analyze how the left-hand side of our actual state constraint differs from this openability bound after $i'$ is added to $S$ and clients are processed. First, observe that for every active client $j \in A$, its updated dual value effectively becomes $\tau_j$. This is because Step 1(b) explicitly sets $\alpha_j \gets \tau_j$ for $j \in A \cap B(i', \sqrt{\epsilon\theta})$, and for active clients outside this ball, the first openability condition guarantees $\tau_j = \alpha_j$. Thus, going into Step 1(c), the duals are $\tau_j$ for $A$ and remain $\alpha_j$ for $IC$.

For a client $j$ that was already directly connected ($DC$) before opening $i'$, its contribution can only decrease, since the distance $d^2(j, S \cup \{i'\}) \le d^2(j, S)$.

For a client $j$ that was active ($A$) or indirectly connected ($IC$) and now moves to $DC$, it does so precisely because its updated dual value satisfies the condition in Step 1(c): $\tau_j \ge \rho(1-\delta)d^2(j,S \cup \{i'\})$ (if it was from $A$) or $\alpha_j \ge \rho(1-\delta)d^2(j,S \cup \{i'\})$ (if it was from $IC$). Since $\rho > 1$ and the function $[x]^+$ is non-negative and non-decreasing, we can strictly bound its new actual contribution. For a client moving from $A$:
\begin{align*}
[(1-\delta)d^2(j,S \cup \{i'\}) - d^2(j,i_0)]^+ &\le \left[ \frac{\tau_j}{\rho} - d^2(j,i_0) \right]^+ \le \left[ \tau_j - \rho d^2(j,i_0) \right]^+.
\end{align*}
The identical algebraic bound holds for a client moving from $IC$ using $\alpha_j$. In both cases, the client's new connection contribution is strictly bounded by its corresponding dual term in the openability bound.

Finally, for the active or indirectly connected clients that do not move to $DC$, their actual dual values are exactly $\tau_j$ and $\alpha_j$, respectively. Whether an active client remains in $A$ or moves to $IC$, its contribution is evaluated using its dual variable, resulting in $w(j)[\tau_j - \rho d^2(j,i_0)]^+$. This perfectly matches its term in the openability bound. Therefore, every term in the updated state is less than or equal to its corresponding term in the openability bound, guaranteeing the total sum remains $\le \hat{f}$.

\textbf{Stage 2:} In the second stage, we have the following steps: decreasing the $\alpha$-values of clients in $IC$ to $(1-\delta)d^2(j,S)$, and increasing the $\alpha$-values of clients in $A$ to $\min\{(1+\epsilon^3)\theta, (1-\delta)d^2(j,S)\}$. 

The first step can only decrease the left-hand side of the claim. Hence, it suffices to show that the continuous increase in the second step does not violate the claim. Assume towards contradiction that at the end of the phase, if we increase $\alpha_j$ of every $j \in A$ to $\min\{(1+\epsilon^3)\theta, (1-\delta)d^2(j,S)\}$, the claim is violated for a non-empty subset of facilities $F'$.

We select a ``minimal'' such counter-example: let $\tau' \le (1+\epsilon^3)\theta$ be the smallest value such that if we set $\alpha'_j := \min\{\tau', (1-\delta)d^2(j,S)\}$, it satisfies
$$ \sum_{j \in A} w(j)[\alpha'_j - \rho d^2(i,j)]^+ + \sum_{j \in IC} w(j)[\alpha_j - \rho d^2(i,j)]^+ + \sum_{j \in DC} w(j)[(1-\delta)d^2(j,S) - d^2(i,j)]^+ = \hat{f} $$
for some $i \in F'$. We remark that $\tau' \ge \theta$ since the constraint was strictly satisfied at the end of Phase 1. Furthermore, $i \notin S$ is not yet open; if it were open, there would be no active client $j \in A$ for which $\alpha'_j - \rho d^2(i,j)$ is strictly positive (since $\alpha'_j \le (1-\delta)d^2(j,S) \le (1-\delta)d^2(j,i) < \rho d^2(j,i)$). Hence, the increase in $\alpha$-values cannot cause $i$ to violate the constraint if it were already opened, which contradicts $i \in F'$.

We now show that this facility $i$, combined with dual values $\tau_j = \alpha'_j$ for $j \in A \cap B(i, \sqrt{\epsilon\theta})$ and $\tau_j = \alpha_j$ for $j \in A \setminus B(i, \sqrt{\epsilon\theta})$, satisfies the conditions of openability. In other words, $i$ is an openable facility that was not yet opened, which contradicts the proper completion of the first stage's while-loop. 

We verify the conditions of openability one-by-one. The first two bullets (only nearby clients increased, and only increased slightly) are directly satisfied by our choice of $\tau_j \le \tau'$ and the definition of $\alpha'_j$. For the third bullet (facility $i$ is paid for):
\begin{align*}
&\sum_{j \in A} w(j) \cdot [\tau_j - \rho(1-\delta)d^2(i,j)]^+ + \sum_{j \in IC} w(j) \cdot [\alpha_j - \rho(1-\delta)d^2(i,j)]^+ \\
&+ \sum_{j \in DC} w(j) \cdot [(1-\delta)d^2(j,S) - (1-\delta)d^2(i,j)]^+ \\
&\ge \sum_{j \in A \cap B(i, \sqrt{\epsilon\theta})} w(j) \cdot [\alpha'_j - \rho d^2(i,j)]^+ + \sum_{j \in A \setminus B(i, \sqrt{\epsilon\theta})} w(j) \cdot [\alpha'_j - \rho d^2(i,j)]^+ \\
&+ \sum_{j \in IC} w(j) \cdot [\alpha_j - \rho d^2(i,j)]^+ + \sum_{j \in DC} w(j) \cdot [(1-\delta)d^2(j,S) - d^2(i,j)]^+ \\
&\ge \hat{f}.
\end{align*}
Here, for the subset outside the ball $A \setminus B(i, \sqrt{\epsilon\theta})$, we used the fact that $\alpha_j - \rho(1-\delta)d^2(i,j) \ge (1+\epsilon^3)\alpha_j - \rho d^2(i,j) \ge \alpha'_j - \rho d^2(i,j)$ when $d^2(i,j) \ge \epsilon\theta$ (recalling that $\alpha_j = \theta$ for all active clients). 

It remains to verify the fourth bullet of openability (dual feasibility). For any facility $i_0 \in \mathcal{F}$, we evaluate the dual feasibility constraint using our constructed $(\tau_j)_{j \in A}$. Because $\tau_j \le \alpha'_j$ for all $j \in A$, and the function $[x]^+$ is non-decreasing, we can explicitly bound the left-hand side of the dual feasibility constraint:
\begin{align*}
&\sum_{j \in DC} w(j)[(1-\delta)d^2(j,S) - d^2(j,i_0)]^+ + \sum_{j \in IC} w(j)[\alpha_j - \rho d^2(j,i_0)]^+ + \sum_{j \in A} w(j)[\tau_j - \rho d^2(j,i_0)]^+ \\
&\le \sum_{j \in DC} w(j)[(1-\delta)d^2(j,S) - d^2(j,i_0)]^+ + \sum_{j \in IC} w(j)[\alpha_j - \rho d^2(j,i_0)]^+ + \sum_{j \in A} w(j)[\alpha'_j - \rho d^2(j,i_0)]^+ \\
&\le \hat{f}.
\end{align*}
The first inequality holds by direct substitution of the upper bound $\tau_j \le \alpha'_j$. The second inequality follows strictly from the minimality of $\tau'$: by definition, $\tau'$ was chosen as the \emph{smallest} value at which the total sum (evaluated with $\alpha'_j$) reaches $\hat{f}$ for \emph{some} facility ($i \in F'$). Therefore, for any arbitrary facility $i_0$, this sum cannot strictly exceed $\hat{f}$. Since all four conditions hold, $i$ is openable, which is a contradiction.
\end{proof}

Next, we fix a facility $i$ and show that the dual-feasibility for $i$ is satisfied at the end of the algorithm.
We say that facility $i$ is frozen if there exists some open facility $i_0$ such that 
$$d(i,i_0) \le 2\sqrt{\eps\theta}~.$$ 

\paragraph{Modified $\alpha$-values.} For analysis purposes, we will introduce a new modification of the $\alpha$-values for each client $j$ in terms of the fixed facility $i$. We will denote this modified $\alpha$-value as $\alpha^c_j$.
This modified $\alpha^c$-value will be used to write the analog of the constraints in \Cref{lem:p-tjl,lem:beta-t}. 

The modifications happen only when we open a facility $i_0$ that does not freeze $i$, i.e., when $d(i,i_0) > 2\sqrt{\epsilon\theta}$.
More specifically, at the stage with time $\theta$, if we open such a facility $i_0$ that does not freeze $i$, we keep $\alpha^c_j$ as the value before opening $i_0$. That is, we don't increase $\alpha^c_j$ to $\tau_j$ if $j\in B(i_0,\sqrt{\epsilon\theta})$ (step 1b) for such $i_0$. In other steps (step 1b for a facility $i_0$ that freezes $i$ or step 2), we still modify $\alpha^c_j$ in the same way as  we modify $\alpha_j$. 

In the rest of this subsection, we will consider $\alpha^c$ instead of $\alpha^*$ in writing the constraints. 
To ensure that the $\alpha^*$-value can ensure dual feasibility, we will show that $\alpha_j^c$ and $\alpha_j^*$ are close to each other for every client $j$.

\begin{lemma}
\label{lem:alphaclose}
    For any client $j\in D^*$, we have $\alpha^*_j-\alpha^c_j \leq \eps^2 \cdot d^2(j,i)$.
\end{lemma}
\begin{proof}
    According to the definition of $\alpha^c_j$, the only case where $\alpha^c_j\neq \alpha^*_j$ is if at some time $\theta$, client $j$ becomes directly connected to a facility $i_0$ that does not freeze $i$. Suppose that $\alpha_j$ is the $\alpha$-value of $j$ before opening $i_0$ and $\tau_j$ is the $\alpha$-value of $j$ after opening $i_0$. Because we will no longer change the $\alpha$-value of a client if it is in $DC$, we have $\alpha^c_j=\alpha_j$ and $\alpha^*_j=\tau_j$. Note that $\tau_j-\alpha_j\leq \eps^3\theta$ according to \Cref{def:openable}. On the other hand, if $\tau_j>\alpha_j$, we have $d(j,i_0)\leq \sqrt{\eps\theta}$. Since $i_0$ does not freeze $i$, we know $d(i,i_0) > 2\sqrt{\eps\theta}$. By the triangle inequality, we have $d(j,i) \ge d(i,i_0) - d(j,i_0) > \sqrt{\eps\theta}$. Squaring both sides implies $d^2(j,i) > \eps\theta$, which rearranges to $\theta < \frac{1}{\eps} d^2(j,i)$. Substituting this into our dual difference yields $\alpha^*_j - \alpha^c_j = \tau_j - \alpha_j \le \eps^3 \theta < \eps^2 d^2(j,i)$.
\end{proof}

Now, we assume that 
\begin{align}
    \label{eqn:Dstar}
    D^*= \{j\in D: \alpha^c_j - (\rho-1)R^2_j > \Gamma d^2(j,i)\}~.
\end{align} Then, it suffices to show that 
\begin{align*}
\sum_{j \in D^*} w(j) \cdot (\alpha_j^c - (\rho-1)R^2_j)  \le \hat{f} + (\Gamma+200\sqrt{\epsilon})  \cdot \sum_{j \in D^*} w(j) \cdot d^2(j,i)~.
\end{align*}
For simplicity, let us number the clients in $D^*$ from $1$ to $s=|D^*|$ in the order they become inactive, breaking ties by order of increasing $\alpha^c$- value. We will use $D^*$ and $[s]$ interchangeably.

\paragraph{Establishing the constraints.}

Next, for each client $t\in [s]$:
\begin{itemize}
    \item If $t$ becomes inactive strictly before $i$ is frozen, we define $DC^t$ as the set of clients in $[t-1]$ that are directly connected at the beginning of the stage $t$ becomes inactive, $IC^t = [t-1]\setminus DC^t$, and $S^t$ as the set of opened facilities at the beginning of the stage $t$ becomes inactive.
    \item Else, we define $DC^t$ as the set of clients in $[t-1]$ that are directly connected right before $t$ becomes inactive, $IC^t = [t-1] \setminus DC^t$, and $S^t$ as the set of opened facilities right before $t$ becomes inactive.
\end{itemize}

\begin{claim}
\label{clm:notfrozen}
For any $t \in D^*$ that becomes inactive strictly before $i$ becomes frozen,
\[
\sum_{j \in DC^t} w(j) [(1-\delta)d^2(j,S^t) - d^2(j,i)]^+ 
+ \sum_{j \in IC^t} w(j) [\alpha_j^c - \rho d^2(j,i)]^+ 
+ \sum_{j \ge t} w(j) [\alpha_t^c - \rho d^2(j,i)]^+ \le \hat{f}.
\]
\end{claim}
\begin{proof}
Let us analyze the situation depending on the stage in which $t$ becomes inactive. Since the claim assumes $t$ becomes inactive strictly before $i$ becomes frozen, any facility $i_0$ opened at or before the time $t$ becomes inactive cannot freeze $i$.

\paragraph{Stage 1.} Suppose $t$ becomes inactive because of the opening of some facility $i_0$. Because $t$ becomes inactive before $i$ is frozen, $i_0$ does not freeze $i$. By the definition of the modified duals, $\alpha^c$ does not incorporate the Step 1b increases associated with opening a non-freezing facility $i_0$. Specifically, for any client $j$, its modified value $\alpha^c_j$ is bounded by its actual dual value $\alpha_j$ at the beginning of the stage
$i_0$ is opened.

At the beginning of the stage
where $i_0$ is opened, let $\alpha$ denote the current dual values. We have $\alpha_j = \theta$ for all active clients, including $t$ and all $j > t$. Therefore, the modified dual for $t$ is exactly $\alpha^c_t = \theta$. By applying \Cref{clm:no_overbidding} (no over-bidding) for facility $i$ 
at the beginning of the stage
$i_0$ is opened, we establish:
$$\sum_{j \in DC^t} w(j) [(1-\delta)d^2(j, S^t) - d^2(j,i)]^+ + \sum_{j \in IC^t} w(j) [\alpha_j - \rho d^2(j,i)]^+ + \sum_{j \geq t} w(j) [\alpha_j - \rho d^2(j,i)]^+ \le \hat{f}.$$

For any $j \in IC^t$, the modified dual value $\alpha^c_j$ never exceeds the true dual value $\alpha_j$ at the beginning of the stage $i_0$ was opened, so $\alpha^c_j \le \alpha_j$. For the active clients $j \geq t$, we have $\alpha_j = \theta = \alpha^c_t$. Substituting these into the inequality, and noting that the function $[v]^+$ is non-decreasing with respect to $v$, we obtain the desired bound:
$$\sum_{j \in DC^t} w(j) [(1-\delta)d^2(j,S^t) - d^2(j,i)]^+ + \sum_{j \in IC^t} w(j) [\alpha_j^c - \rho d^2(j,i)]^+ + \sum_{j \geq t} w(j) [\alpha_t^c - \rho d^2(j,i)]^+ \le \hat{f}.$$

\paragraph{Stage 2.} Suppose that $t$ becomes inactive due to the continuous update of the $\alpha$-values in Stage 2 (at the end of a phase). In Stage 2, the modified duals $\alpha^c$ track the actual duals $\alpha$ exactly, meaning $\alpha^c_t = \alpha_t \le (1+\epsilon^3)\theta$.

By \Cref{clm:no_overbidding}, the over-bidding constraint is satisfied for facility $i$ at the exact moment $t$ becomes inactive:
$$\sum_{j \in DC} w(j) [(1-\delta)d^2(j, S) - d^2(j,i)]^+ + \sum_{j \in A \cup IC} w(j) [\alpha_j - \rho d^2(j,i)]^+ \le \hat{f}.$$

Notice that Stage 2 does not alter the sets $DC$ and $IC \cup A$. Therefore, at this moment, $DC^t \subseteq DC$ and $IC^t \subseteq IC \cup A$. We compare the contributions of each client in this inequality to the terms in our claim:
\begin{itemize}
    \item For $j \in DC^t \subseteq DC$, the contribution is identically $[(1-\delta)d^2(j,S^t) - d^2(j,i)]^+$.
    \item For $j \in IC^t$, we have $\alpha^c_j \le \alpha_j$ since dual values are non-increasing once a client is in $IC$, and our modification $\alpha^c$ strictly lowers or matches $\alpha$. Thus, the term for $j$ is upper bounded by the corresponding term in the over-bidding claim.
    \item For $j > t$, these clients either remain in $A$, or becomes inactive at the same stage as $t$ but have $\alpha^c_j\geq \alpha^c_t$. In the first case, we have $\alpha_t=(1+\eps^3)\theta$, and thus $\alpha^c_t \leq \alpha_t \leq \alpha_j$ because $\alpha^c$-value is non-increasing after becoming inactive.
    In the second case, we have $\alpha^c_t \leq \alpha^c_j \leq \alpha_j$.
    Thus, $[\alpha^c_t - \rho d^2(j,i)]^+ \le [\alpha_j - \rho d^2(j,i)]^+$.
\end{itemize}
Summing these bounds confirms the inequality holds.
\end{proof}

\begin{claim}
\label{clm:frozenstay}
    If $i$ is frozen by $i_0$ with $(\tau_j)_{j\in A}$, then, for every $j\in D^*\cap A$, $j$ becomes inactive when opening $i_0$ and $\alpha_j^c\leq\tau_j$.
\end{claim}
\begin{proof}
    Consider $j\in A$. First, note that if $\tau_j\geq (1-\delta)d^2(j,i_0)$, then $j$ is removed from $A$ when $i_0$ is opened (moving to either $IC$ or $DC$), and so $\alpha_j^c\leq\tau_j$ since $\alpha$-values are non-increasing after a client becomes inactive.

    In the other case, when $\alpha_j\leq \tau_j<(1-\delta)d^2(j,i_0)$, where $\alpha_j$ is the $\alpha$-value of $j$ before $i_0$ is open, we show that $j\notin D^*$. Consider the (future) time right after $j$ is removed from the active clients. The $\alpha$-value of $j$ then (which upper bounds the final $\alpha_j^c$), cannot be strictly greater than $(1-\delta)d^2(j,i_0)$ whether it is increased in stage 1 or in stage 2, since we have $i_0\in S$ at that time. By the metric triangle inequality:
    $$d(i, j)\geq d(j,i_0) - d(i_0,i) = \left(1-\frac{d(i_0, i)}{\sqrt{\theta}}\cdot \frac{\sqrt{\theta}}{d(j,i_0)}\right)d(j,i_0)$$
    where $\theta$ is the value when $i_0$ was open. As $i_0$ freezes $i$, we have $d(i_0,i)\leq 2\sqrt{\epsilon\theta}$, and we know $\theta=\alpha_j < (1-\delta)d^2(j,i_0)$. Plugging in those bounds to the above inequality yields:
    \begin{align*}
        d(i, j) &> \left(1 - 2\sqrt{\epsilon(1-\delta)}\right)d(j,i_0)~.
    \end{align*}
    Because $\epsilon \in (0, 10^{-100})$ is sufficiently small, the term $2\sqrt{\epsilon(1-\delta)}$ is strictly less than $1$, meaning the distance remains strictly positive. Squaring both sides gives:
    $$d^2(i,j) > \left(1 - 2\sqrt{\epsilon(1-\delta)}\right)^2 d^2(j,i_0)~.$$
    
    Since $\Gamma > 2$, we have $\Gamma \left(1 - 2\sqrt{\epsilon(1-\delta)}\right)^2 > 1 > 1-\delta$. Thus:
    $$ \Gamma d^2(i,j) > (1-\delta)d^2(j,i_0) \ge \alpha_j^c~.$$
    Hence, $\alpha_j^c < \Gamma d^2(i, j)$, which implies $j$ cannot be in $D^*$.
\end{proof}

\begin{claim}
\label{clm:3rdconstraint}
For any $t\in D^*$,
\[
\sum_{j\in DC^t}w(j)((1-\delta)d^2(j, S^t) - d^2(j, i)) + \sum_{j\in IC^t}w(j)(\alpha_j^c - \rho d^2(j, i)) + \sum_{j \geq t}w(j)(\alpha_t^c-\rho d^2(j,i)) \le \hat{f}.
\]
\end{claim}
\begin{proof}
    Consider the point in the execution at which $t$ becomes inactive. As established by \Cref{clm:frozenstay}, if a facility $i_0$ freezes $i$, every active client $j \in D^* \cap A$ becomes inactive during the exact step $i_0$ is opened. Therefore, no client in $D^*$ can become inactive \emph{after} $i$ is frozen. This leaves us with exactly two exhaustive cases for $t \in D^*$:

    \paragraph{Case 1: $t$ becomes inactive strictly before $i$ is frozen.}
    By \Cref{clm:notfrozen},
    \[
    \sum_{j \in DC^t} w(j) [(1-\delta)d^2(j,S^t) - d^2(j,i)]^+ 
    + \sum_{j \in IC^t} w(j) [\alpha_j^c - \rho d^2(j,i)]^+ 
    + \sum_{j \ge t} w(j) [\alpha_t^c - \rho d^2(j,i)]^+ \le \hat{f}.
    \]
    Dropping the $[\cdot]^+$ completes the proof for this case.

    \paragraph{Case 2: $t$ becomes inactive exactly when $i$ is frozen.}
    Suppose $t$ becomes inactive due to the opening of a facility $i_0$ that freezes $i$. Because all clients in $D^* \cap A$ become inactive at this step (\Cref{clm:frozenstay}), and our ordering of $1, \dots, s$ in $D^*$ is determined by the time they become inactive, it must be that all clients $j \ge t$ in $D^*$ become inactive simultaneously at this step. Furthermore, because ties are broken by order of increasing $\alpha^c$-value, we are guaranteed that $\alpha_t^c \le \alpha_j^c$ for all $j \ge t$. 

    Now, consider the fourth bullet of openability (dual feasibility) for $i_0$ evaluated at the fixed facility $i$, just before $i_0$ is opened:
    \[
    \sum_{j \in DC} w(j) [(1-\delta)d^2(j,S) - d^2(j,i)]^+ 
    + \sum_{j \in IC} w(j) [\alpha_j - \rho d^2(j,i)]^+ 
    + \sum_{j \in A} w(j) [\tau_j - \rho d^2(j,i)]^+ \le \hat{f}.
    \]
    
    Thus, letting $A_{before}$ and $IC_{before}$ be the sets $A$ and $IC$ just before $i_0$ is opened, we have exactly $S = S^t$, $DC^t \subseteq DC$, and $IC^t = (IC^t \cap IC_{before}) \cup (IC^t \cap A_{before})$. Note also that $t \in IC^t \cap A_{before}$, and all $j > t$ belong to $A_{before}$.

    We drop the non-negative $[\cdot]^+$ operators from the openability constraint and restrict the sums to our subsets of interest to establish a strict lower bound on the left-hand side:
    \begin{itemize}
        \item For $j \in DC^t \subseteq DC$, we drop the positive part. Because $S = S^t$, we exactingly have: 
        $[(1-\delta)d^2(j,S) - d^2(j,i)]^+ \ge (1-\delta)d^2(j,S^t) - d^2(j,i)$.
        
        \item For $j \in IC^t \cap IC_{before}$, the modified dual never exceeds the true dual prior to freezing, so $\alpha_j^c \le \alpha_j$. Thus:
        $[\alpha_j - \rho d^2(j,i)]^+ \ge \alpha_j - \rho d^2(j,i) \ge \alpha_j^c - \rho d^2(j,i)$.
        
        \item For $j \in IC^t \cap A_{before}$, these clients are active and in $D^*$. By \Cref{clm:frozenstay}, we have $\alpha_j^c \le \tau_j$. Evaluated in the summation over $A$, we have:
        $[\tau_j - \rho d^2(j,i)]^+ \ge \tau_j - \rho d^2(j,i) \ge \alpha_j^c - \rho d^2(j,i)$.
        (Notice that combining this with the previous bullet perfectly reconstructs the $\alpha_j^c - \rho d^2(j,i)$ term for \emph{all} $j \in IC^t$).
        
        \item For $j \ge t$, these clients are in $D^* \cap A_{before}$. By \Cref{clm:frozenstay}, $\alpha_j^c \le \tau_j$. Applying our tie-breaking rule, $\alpha_t^c \le \alpha_j^c \le \tau_j$. Evaluated in the $A$ sum, we have:
        $[\tau_j - \rho d^2(j,i)]^+ \ge \tau_j - \rho d^2(j,i) \ge \alpha_t^c - \rho d^2(j,i)$.
    \end{itemize}
    
    Because all subsets are disjoint and strictly contained within the original openability sums, summing these lower bounds yields:
    \[
    \sum_{j \in DC^t} w(j) ((1-\delta)d^2(j,S^t) - d^2(j,i)) 
    + \sum_{j \in IC^t} w(j) (\alpha_j^c - \rho d^2(j,i)) 
    + \sum_{j \ge t} w(j) (\alpha_t^c - \rho d^2(j,i)) \le \hat{f}.
    \]
    This matches the required inequality, completely resolving the case.
\end{proof}

For any client $t$ and $j\in DC^t$, define
\[
r_{jt}:= \sqrt{1-\delta}d(j,S^t).
\]
Next, we will prove \cref{lem:dual-feasibility-logadaptive} for the general metric setting and the Euclidean setting separately based appropriate choices of $\rho$ and $\Gamma$.

\subsubsection*{General metric}

Now, we will show \Cref{lem:dual-feasibility-logadaptive} for $\rho = 2.5$ and $\Gamma = 4.9$. To this end, let's show that $(IC^t), (DC^t), (\alpha_j^c), (R_j), (r_{jt})$ satisfy all the constraints of \Cref{lem:dual-feasibility-simpler-metric}, restated below for convenience.

\DualFeasibilityGeneralMetric*

By definition of $(IC^t)_t$ and $(DC^t)_t$, the two constraints on them hold. Let $j\in DC^t$. Since $R_j$ is the distance to the first facility that $j$ directly connects to, and $d(j, S^t)$ is non-increasing in $t$, we have $r_{jt}\le R_j$ for any $t>j$. Now, let $t\in [s]$, we have:
\begin{align*}
    \alpha_t^c &\le \alpha_t^*\\
    &\le (1-\delta)d^2(t, S^t)\\
    &\le (\sqrt{1-\delta}d(t, j) + \sqrt{1-\delta}d(j, S^t))^2\\
    &\le (r_{jt} + d(j, t))^2.
\end{align*}
\Cref{clm:3rdconstraint} gives the last constraint.

Hence, by \Cref{lem:dual-feasibility-simpler-metric}, we have:
\[
\sum_{j\in [s]}w(j)\cdot (\alpha_j^c - (\rho-1)R_j^2)\le \hat{f} + \Gamma\cdot \sum_{j\in [s]}w(j)\cdot d^2(j,i).
\]
This completes the proof of \Cref{lem:dual-feasibility-logadaptive} for $\rho=2.5$ and $\Gamma = 4.9$.

\subsubsection*{Euclidean metric}

Now, we will show \Cref{lem:dual-feasibility-logadaptive} for $\rho=2$, $\Gamma=3+\ln 2$, and $d$ a Euclidean metric. Let $a$ be the index of the last client that becomes inactive strictly before $i$ is frozen.

First, let's prove the following claim:

\begin{claim}
\label{clm:2nd-constraint-euclidean}
For each $t\in [a]$, $j\in DC^t$ and $\ell\in IC^t \cup \{t\}$, we have:
\[
    \alpha_{\ell}^c \le (r_{jt} + d(j, \ell))^2.
\]
\end{claim}
\begin{proof}
    Consider any client $t\in [a]$. Let $P_t$ be the phase in which $t$ becomes inactive. 
    Consider the state of the algorithm at the exact beginning of phase $P_t$. By definition, $S^t$ is the set of open facilities at this time.
    
    Consider any client $\ell\in IC^t\cup\{t\}$. As $\ell\leq t$, $\ell$ becomes inactive in phase $P_t$ or in an earlier phase. 
    Next, we prove that $\alpha^c_{\ell}\leq (1-\delta)d^2(\ell,S^t)$ by discussing two cases:
    \begin{itemize}
        \item If $\ell$ becomes inactive strictly before phase $P_t$, then $\ell$ is indirectly connected at the beginning of phase $P_t$. Because of the Stage 2 continuous updates at the end of the previous phase, we have $\alpha^c_{\ell} \le \alpha_{\ell} = (1-\delta)d^2(\ell,S^t)$.
        
        \item Otherwise (i.e., if $\ell$ becomes inactive during phase $P_t$), we can prove $\alpha^c_\ell \le (1-\delta)d^2(\ell,S^t)$ by contradiction. Suppose that $\alpha^c_\ell > (1-\delta)d^2(\ell,S^t)$. Let $\theta$ be the active dual parameter at the beginning of phase $P_t$. Since $\ell$ is active at this time, $\theta  = \alpha^c_\ell > (1-\delta)d^2(\ell, S^t)$. Then, $\ell$ would have been moved to $IC$ during the Stage 2 update of the previous phase (when the threshold $(1+\epsilon^3)\theta_{prev} = \theta$ exceeded the connection cost), contradicting that $\ell$ is still active at the beginning of phase $P_t$. 
        Therefore, $\alpha^c_{\ell} \le \theta \le (1-\delta)d^2(\ell,S^t)$.
    \end{itemize}  
    
    Because the triangle inequality holds for the Euclidean metric $d$, we have $d(\ell,S^t)\leq d(j,S^t)+d(j,\ell)$. Taking the square root of our established bound, we get:
    \begin{align*}
        \sqrt{\alpha^c_\ell} &\le \sqrt{1-\delta} d(\ell, S^t) \\
        &\le \sqrt{1-\delta} (d(j, S^t) + d(j, \ell)) \\
        &= r_{jt} + \sqrt{1-\delta}d(j, \ell)
    \end{align*}
    
    Since $\delta > 0$, we have $\sqrt{1-\delta} < 1$. Because metric distances are non-negative, replacing the scaled distance with the full distance preserves the inequality:
    $$ \sqrt{\alpha^c_\ell} \le r_{jt} + d(j, \ell) $$
    
    Squaring both sides yields $\alpha^c_{\ell}\leq (r_{jt}+d(j,\ell))^2$, completing the proof.
\end{proof}

We will make several cases according to how $i$ gets frozen.

\paragraph{Case 1: All clients in $D^*$ become inactive strictly before $i$ is frozen.} In this case, let's show that $(IC^t), (DC^t), (\alpha_j^c), (R_j), (r_{jt})$ satisfy all the constraints of \Cref{lem:dual-feasibility-simpler-euclidean}, restated below for convenience.

\DualFeasibilitySimplerEuclidean*

Since in this case $a=s$, \Cref{clm:2nd-constraint-euclidean} implies constraint (2) is satisfied. \Cref{clm:3rdconstraint} implies constraint (3) is satisfied. Hence, we can apply \Cref{lem:dual-feasibility-simpler-euclidean}, and get:

\[
    \sum_{j\in [s]}w(j)\cdot(\alpha_j^c-(\rho-1)R_j^2)\le \hat{f}+\Gamma\cdot \sum_{s\in [s]}w(j)\cdot d^2(j,i),
\]
which completes the proof in this case.

\paragraph{Case 2: Some clients in $D^*$ become inactive when $i$ is frozen.} In this case, we assume that $i_0$ freezes $i$. 
Let $IC^F$ be the set of indirectly connected clients right before $i_0$ is opened, and
\[
    L:= \{j\in IC^F:d(j,i)>\sqrt{\eps\theta}\}.
\]
Note that $IC^F$ is different with $IC^s$ in the way that $IC^F$ does not include active clients right before $i_0$ is opened, i.e., $j\notin IC^F$ for any $j\in [a+1:s]$ while $IC^s=IC^F \cup [a+1:s-1]$.

We will further discuss two subcases to prove \cref{lem:dual-feasibility-logadaptive}.

\paragraph{Subcase 2(a): $[a+1:s]\cap B(i_0,\sqrt{\eps\theta})=\emptyset$ or $\sum_{j\in L}w(j)>\eps\cdot\sum_{j\in [a+1:s]}w(j)$.} We redefine $DC^s$ and $IC^s$ as the respectively directly connected and active or indirectly connected client at the beginning of the stage $s$ becomes inactive. First, let's bound the total increment of $\alpha_j^c$ when $i_0$ is opened.

\begin{claim}
\label{clm:increment-bound}
The total increment of $\alpha^c_j$ for $j\in D^*$ when $i_0$ is opened is at most $\epsilon \sum_{j\in D^*} w(j) d^2(j,i)$.
\end{claim}
\begin{proof}
    If $[a+1:s] \cap B(i_0, \sqrt{\eps\theta}) = \emptyset$, or if no active client in $[a+1:s]$ strictly increases its $\alpha^c$-value when $i_0$ is opened, the total increment for all $j \in D^*$ is exactly $0$. Because weights and squared distances are non-negative, the bound $0 \le \epsilon \sum_{j\in D^*} w(j) d^2(j,i)$ holds trivially.
    
    Next, we consider the case where $\sum_{j\in L}w(j)>\eps\cdot\sum_{j\in [a+1:s]}w(j)$.

    Let $U = [a+1:s]$ denote the set of clients in $D^*$ that become inactive exactly when $i$ is frozen (i.e., when $i_0$ is opened). For any client $j \in [1:a]$, $j$ became inactive strictly before this step. By the algorithm's rules, the dual values of inactive clients are never increased during Stage 1. Therefore, the total increment of $\alpha^c_j$ across all $j \in D^*$ is exactly the sum of the increments for the active clients in $U$.

    Right before $i_0$ is opened, every $j \in U$ is active, meaning its dual value is exactly $\alpha_j = \theta$. When $i_0$ is opened, its modified dual value $\alpha^c_j$ is set to $\tau_j$. By the second bullet of openability (\Cref{def:openable}), the maximum increased value is bounded by $\tau_j \le (1+\epsilon^3)\theta$. Thus, for every $j \in U$, the increment is bounded by $\epsilon^3 \theta$.
    
    Summing this over all $j \in U$, the total increment for $D^*$ is bounded by:
    $$ \sum_{j \in U} w(j)(\alpha^c_j - \alpha_j) \le \epsilon^3 \theta \sum_{j \in U} w(j). $$

    By our Case 2(a) assumption, we have $\sum_{j\in L}w(j) > \epsilon \sum_{j\in U}w(j)$, which rearranges to:
    $$ \sum_{j \in U} w(j) < \frac{1}{\epsilon} \sum_{j \in L} w(j). $$

    Substituting this weight bound into our total increment yields:
    $$ \sum_{j \in U} w(j)(\alpha^c_j - \alpha_j) < \epsilon^3 \theta \left( \frac{1}{\epsilon} \sum_{j \in L} w(j) \right) = \epsilon^2 \theta \sum_{j \in L} w(j). $$

    Recall the definition of the subset $L = \{j \in IC^F : d(j,i) > \sqrt{\epsilon\theta}\}$. For every client $j \in L$, we have $d^2(j,i) > \epsilon\theta$, which algebraically rearranges to $\theta < \frac{1}{\epsilon} d^2(j,i)$. Substituting this upper bound for $\theta$ into our increment expression, we obtain:
    $$ \sum_{j \in U} w(j)(\alpha^c_j - \alpha_j) < \epsilon^2 \sum_{j \in L} w(j) \left( \frac{d^2(j,i)}{\epsilon} \right) = \epsilon \sum_{j \in L} w(j) d^2(j,i). $$
    
    Finally, note that $L \subseteq IC^F \subseteq [1:a] \subset D^*$. Because client weights $w(j)$ are strictly positive and squared metric distances $d^2(j,i)$ are non-negative, the sum over the subset $L$ is trivially bounded by the sum over the entire set $D^*$:
    $$ \epsilon \sum_{j \in L} w(j) d^2(j,i) \le \epsilon \sum_{j \in D^*} w(j) d^2(j,i). $$
    
    This completes the proof.
\end{proof}
Apply \cref{clm:no_overbidding} at the beginning of the stage $s$ becomes inactive, this lemma guarantees that 
\begin{align*}
    w(s) (\alpha^c_s-\rho d^2(s,i)) + \sum_{\ell\in IC^s} w(\ell)(\alpha^c_\ell -\rho d^2(\ell,i)) + \sum_{j\in DC^s} w(j) (r^2_{js} - d^2(j,i)) \\
    \leq \hat f + \eps \sum_{j\in D^*} w(j)d^2(j,i).
\end{align*}
Note that because we redefine $IC^s, DC^s, S^s$ according to the sets $IC,DC,S$ at the beginning of the stage $s$ becomes inactive, it's easy to see that \cref{clm:2nd-constraint-euclidean} generalizes for $t=s$.
Thus, %
we can apply \Cref{lem:dual-feasibility-simpler-euclidean-special} (restated, see below), replacing $\hat{f}$ with $\hat{f} + \eps\sum_{j\in D^*}w(j)d^2(j,i)$ and using $L=\emptyset$ and $\eps'=0$ (the $\eps$-parameter in the lemma). We get:
\[
    \sum_{j\in [s]}w(j)\cdot (\alpha_j^c - (\rho-1)R_j^2)\le \hat f + (\Gamma + \eps)\cdot\sum_{j\in [s]}w(j)\cdot d^2(j,i).
\]
This completes the proof in this case.

\paragraph{Subcase 2(b): $[a+1:s]\cap B(i_0,\sqrt{\eps\theta})\neq \emptyset$ and $\sum_{j\in L}w(j)\le \eps\cdot\sum_{j\in [a+1:s]}w(j)$.}
In this case, we define $DC^s$ and $IC^s$ as respectively the directly connected and the active or indirectly connected client right before $s$ becomes inactive.
Let's show that $a$, $L$, $(IC^t), (DC^t), (\alpha_j^c), (R_j), (r_{jt})$ satisfy all the constraints of \Cref{lem:dual-feasibility-simpler-euclidean-special}, restated below for convenience.

\DualFeasibilityEuclideanSpecial*

Just like for the other cases, constraints (a), (b), (c), (1), (2) and (3) hold. Remains to prove that constraint \ref{constr:euclidean-special-second-approximate} is satisfied. Because $L\subseteq IC^F$, $IC^F\subseteq [a]$ and $IC^F\subseteq IC^s$, we have $L\subseteq [a]\cap IC^s$. Then, by proving the following claim, we establish constraint \ref{constr:euclidean-special-second-approximate}. Note that because $IC^F \setminus L \cup [a+1:s] = IC^s \setminus L$, the second and third bullet of the following claim together implies the second bullet of constraint \ref{constr:euclidean-special-second-approximate}. The fourth bullet of the following claim implies the third bullet of constraint \ref{constr:euclidean-special-second-approximate}.

\begin{claim}
\label{clm:4th-constraint-euclidean}
The following hold:
\begin{itemize}
    \item $\sum_{\ell\in L} w(\ell) \leq \epsilon \cdot \sum_{\ell\in [a+1:s]} w(\ell)$;
    \item for any $j\in DC^s, \ell\in IC^F\setminus L$, we have $\alpha^c_{\ell} \leq (1+20\sqrt{\epsilon}) \cdot (r_{js}+d(j,\ell))^2$;
    \item for any $j\in DC^s$ and $\ell \in [a+1:s]$, we have $\alpha^c_{\ell} \leq (r_{js}+d(j,\ell))^2$;
    \item for any $j\in DC^s, \ell\in L$ and $\ell'\in [a+1:s]$, we have $\alpha^c_{\ell} \leq (r_{js}+d(j,\ell'))^2$.
\end{itemize}
\end{claim}
\begin{proof}
    We prove the four bullets of the claim one by one.
    
    \paragraph{First bullet.} By the definition of Case 2(b), the weight bound is our direct assumption. Substituting the dummy variable $\ell$ for $j$ yields the first bullet immediately.

    \paragraph{Second bullet.} Consider any $j\in DC^s$ and $\ell\in IC^F\setminus L$. Let $\theta$ be the active dual parameter right before $i_0$ is opened. Because $\ell \notin L$, its distance to $i$ satisfies $d(\ell, i) \le \sqrt{\epsilon\theta}$. Because $i_0$ freezes $i$, we know $d(i, i_0) \le 2\sqrt{\epsilon\theta}$. By the metric triangle inequality, $d(\ell, i_0) \le 3\sqrt{\epsilon\theta}$. 
    
    To bridge the distance to the open facilities $S^s$, we instantiate an active client $\ell' \in [a+1:s] \cap B(i_0, \sqrt{\epsilon\theta})$, which is guaranteed to exist by the Subcase 2(b) definition. Because $\ell'$ is active, we strictly know $\theta \le (1-\delta)d^2(\ell', S^s)$, which implies $d(\ell', S^s) \ge \sqrt{\theta}$.
    
    Applying the triangle inequality across $j, \ell, \ell',$ and $S^s$ yields the core distance relationship:
    $$d(\ell, S^s) \ge d(\ell', S^s) - d(\ell, \ell') \ge \sqrt{\theta} - 4\sqrt{\epsilon\theta}~.$$
    
    By applying the triangle inequality to $j, \ell$, and $S^s$ and scaling by $\sqrt{1-\delta}$, we directly obtain (because $\delta=\eps/2$):
    $$(r_{js} + d(j, \ell))^2 \ge \theta \left( \sqrt{1-\delta} - 4\sqrt{\epsilon} \right)^2 \geq \theta\left(1-\eps/2-4\sqrt{\eps}\right)^2>\theta(1-9\sqrt{\eps})~.$$
    
    Squaring both sides and applying the Taylor expansion $(1 - 9\sqrt{\epsilon})^{-1} \le 1 + 10\sqrt{\epsilon}$ (valid for our sufficiently small $\epsilon$) provides the exact upper bound for $\theta$:
    $$\theta \le (1 + 10\sqrt{\epsilon}) (r_{js} + d(j, \ell))^2~.$$
    
    Because $\ell \in IC^F$, it is indirectly connected right before $i_0$ opens, meaning its modified dual value is strictly bounded by the active parameter: $\alpha^c_{\ell} \le \theta$. Substituting our geometric bound yields the final inequality:
    $$\alpha^c_{\ell} \le (1+20\sqrt{\epsilon}) (r_{js} + d(j, \ell))^2~.$$

    \paragraph{Third bullet.} Consider any $j\in DC^s$ and $\ell\in [a+1:s]$. Because $\ell$ is active right before $i_0$ is opened, its maximum possible modified dual value is dictated by the second bullet of the openability condition (\Cref{def:openable}), which strictly enforces $\tau_\ell \le (1-\delta)d^2(\ell, S^s)$. Thus, we bypass the parameter $\theta$ entirely:
    $$\alpha^c_\ell \le \tau_\ell \le (1-\delta)d^2(\ell, S^s)~.$$
    
    Applying the triangle inequality to $j, \ell$, and $S^s$, we have $d(\ell, S^s) \le d(j, S^s) + d(j, \ell)$. Multiplying by $\sqrt{1-\delta}$ and substituting $r_{js} = \sqrt{1-\delta}d(j, S^s)$ yields:
    $$\sqrt{1-\delta}d(\ell, S^s) \le r_{js} + \sqrt{1-\delta}d(j, \ell) \le r_{js} + d(j, \ell)~.$$
    
    Squaring both sides and substituting into our openability bound gives exactly:
    $$\alpha^c_\ell \le (1-\delta)d^2(\ell, S^s) \le (r_{js} + d(j, \ell))^2~,$$
    completing the proof of the third bullet.

    \paragraph{Fourth bullet.} Consider any $j\in DC^s$, $\ell\in L$, and $\ell'\in [a+1:s]$. Because $\ell'$ is active right before $i_0$ is opened, we strictly have $\theta \le (1-\delta)d^2(\ell', S^s)$.
    
    Applying the triangle inequality to $j, \ell',$ and $S^s$, we have $d(\ell', S^s) \le d(j, S^s) + d(j, \ell')$. Multiplying by $\sqrt{1-\delta}$ yields:
    $$\sqrt{1-\delta}d(\ell', S^s) \le r_{js} + \sqrt{1-\delta}d(j, \ell') \le r_{js} + d(j, \ell')~.$$
    
    Squaring both sides provides the geometric upper bound for $\theta$:
    $$\theta \le (1-\delta)d^2(\ell', S^s) \le (r_{js} + d(j, \ell'))^2~.$$
    
    Finally, because $\ell \in L \subseteq IC^F$, it is strictly indirectly connected. According to the algorithm, its dual value is never increased during Stage 1 updates, strictly ensuring $\alpha^c_{\ell} \le \theta$. Combining this with our geometric upper bound gives the exact final relation:
    $$\alpha^c_{\ell} \le \theta \le (r_{js} + d(j, \ell'))^2~,$$
    completing the proof of the fourth bullet.
\end{proof}

Thus, we can apply \Cref{lem:dual-feasibility-simpler-euclidean-special}. We get:
\[
    \sum_{j\in [s]}w(j)\cdot(\alpha_j^c-(\rho-1)R_j^2)\le \hat{f} + (\Gamma + 200\sqrt{\eps})\cdot\sum_{j\in [s]}w(j)\cdot d^2(j,i).
\]
This completes the proof.

\section{\texorpdfstring{Approximation with $O(\log n / \eps^3)$ Extra Centers}{Approximation with O(log n / eps3) Extra Centers}}
In this section, we present bi-criteria approximations with $O(\log n / \eps^3)$ extra centers, as claimed in \cref{thr:mainBicriteria} (restated in \cref{thm:robust}). 
For simplicity, we will assume that $\rho\in [2,3]$ in the rest of this section. 

\subsection{Pre-processing}
For this section, we will consider the case where the distances between clients and facilities are lower bounded by $1$, thanks to the following variant of \cref{lem:aspectratio-euclid}:
\begin{lemma}
\label{lem:aspectratio-euclid2}
For any given $\eps > 0$ and $\alpha>1$, given a polynomial-time $\alpha$-approximation algorithm for Euclidean $k$-Means on instances with
distances between $1$ and $2n^3/\eps^3+1$, there exists a polynomial-time $\alpha(1+O(\eps))$-approximation algorithm for Euclidean $k$-Means on general instances. 
\end{lemma}
\begin{proof}
Given \cref{lem:aspectratio-euclid}, it suffices to prove the following argument:
\begin{quote}
    For any given $\eps > 0$ and $\alpha>1$, given a polynomial-time $\alpha$-approximation algorithm for Euclidean $k$-Means on instances with
    distances between $1$ and $2n^3/\eps^3+1$, there exists a polynomial-time $\alpha(1+O(\eps))$-approximation algorithm for Euclidean $k$-Means on instances with
    non-zero distances between $1$ and $n^2/\eps^2$. 
\end{quote}

Suppose that we have a Euclidean $k$-Means instance with non-zero distances between $1$ and $n^2/\eps^2$ (namely, the old instance), in which $x:D\cup F\to \mathbb{R}^{\di}$ denotes the coordinate of each point. 
Now, we index $D\cup F$ by $1, 2, \dots, |D|+|F|$ arbitrarily. 
We construct the following instance (namely, the new instance) by assigning each point at $x':D\sqcup F \to \mathbb{R}^{\di+|D|+|F|}$, where $x'(i) = (\frac{2n}{\eps} \cdot x(i), e_i)$ for each $i\in D\sqcup F$, and $e_i\in \mathbb{R}^{|D|+|F|}$ denotes the indicator vector, which has a value of 1 on coordinate $i$ and a value of 0 on all other coordinates.
Note that we will separate two copies for $i\in D\cap F$ in the new instance.
Let $d$ and $d'$ denote the Euclidean metric induced by the old and new instances, respectively.
Consider any $i\in F, j\in D$. We have 
\begin{align}
\label{eqn:aspectratio-euclid2}
    d'^2(i,j) = \frac{4 n^2}{\eps^2} \|x(i)-x(j)\|^2 + \|e_i-e_j\|^2 = \frac{4n^2}{\eps^2} d^2(i,j) + 2~.
\end{align}
It is easy to see that $d'(i,j)\in [1, \frac{2n}{\eps} d(i,j)+1]\subseteq [1, 2n^3/\eps^3+1]$.

Now, we assume that we have an $\alpha$-approximation on the new instance. We first run a constant-factor approximation $k$-Center on the old instance and decide if the optimal value for the old instance $OPT$ on $k$-Means is $0$. If it's not, it is guaranteed that $OPT$ is at least 1. According to \cref{eqn:aspectratio-euclid2}, the optimal value of the new instance $OPT'$ is $4n^2/\eps^2 \cdot OPT+2n\geq 4n^2/\eps^2 + 2n$. Then we run the $\alpha$-approximation on the new instance, getting a solution with value $ALG'$ on the new instance such that $ALG'\leq \alpha \cdot OPT'$. We obtain the solution $ALG$ on the old instance with the same set of open facilities and the same connections for the clients. Again, because of \cref{eqn:aspectratio-euclid2}, we have $ALG'=4n^2/\eps^2 \cdot ALG + 2n$. This implies, on the old instance, we have 
\begin{align*}
    \frac{ALG}{OPT} = \frac{(\eps^2/4n^2) \cdot (ALG'-2n)}{(\eps^2/4n^2) \cdot (OPT'-2n)} = \frac{ALG'-2n}{OPT'-2n}= \frac{ALG'-2n}{OPT'} \cdot \frac{OPT'}{OPT'-2n} \leq \alpha \dot (1+\eps^2/(4n^2))~.
\end{align*}
This completes the proof of this lemma.
\end{proof}

We shall remark here that under this assumption, our log-adaptive~\cref{alg:logadaptive} still runs in $O(\log n/\eps^3)$ phases.

\subsection{Walking Between Two Solutions: Setup}
\label{sec:walk-setup}

This subsection is an adaptation of Subsection 5.2 of \cite{CCGGLW2026kmeans} to our new bidding strategy.

Given the description of the log-adaptive algorithm, we present our final bicriteria approximation \mergealg. It results in a solution that opens $k+O(\log n)$ centers by maintaining two partial solutions that lead to opening at least $k$ and at most $k$ centers respectively, and \emph{gradually merging them} into a single solution that opens exactly $k+O(\log n)$ facilities, exploiting the $O(\log n)$ adaptivity of the algorithm. 

Before detailing the merging procedure \mergealg, it is helpful to observe that we may generalize the analysis to make it more ``robust''. We introduce this more robust analysis in this subsection, where we also define the notation that will be used later to describe \mergealg. 

Starting from the two solutions obtained in the standard way whose facility costs differ by at most $\eta := 2^{-n}$, 
the two solutions we maintain will have an exponentially small difference $\eta$ in one of their parameters. 
(This parameter will be either the facility cost $f$, some distance related to a \emph{free facility} that will be introduced later, or some probability that a client is in $IC$ or $DC$.)
For this reason, when we merge the two solutions, it may occur that some facilities in one solution are almost completely paid for, rather than fully paid for, or that the dual constraint of some facilities is slightly violated. To address this, we generalize the definition of ``openable'' to ``$\eta$-openable'', where the only difference is an $O(n\eta)$ slack in the third and fourth conditions.

Note that, in the algorithm, each inactive client $j$ is always either directly connected to some facility $i^*$ or indirectly connected to some facility $i^*$, where $i^*$ is given by $i^* = \argmin_{i\in S} d^2(j,i)$.
In the rest of this section, we will maintain this invariant, but allow some flexibility to its connection type and $\alpha$-value. 
Specifically, we allow each client $j\in D$ to have a set of copies $D_j$, with each $\tilde{j}\in D_j$ defined by (i) the connection type $ct(\tilde{j})$, i.e., whether it is in $A$, $IC$ or $DC$; (ii) the $\alpha$-value $\tilde{\alpha}(\tilde{j})$; (iii) the probability mass $\mu(\tilde{j})$. The probability mass should satisfy $\sum_{\tilde{j}\in D_j} \mu(\tilde{j}) = 1$. 
In particular, for each client $j$, $D_j$ has two possibilities:
\begin{itemize}
\item $D_j$ has only one copy that is of type $A$ and with probability mass $1$;
\item $D_j$ has some copies of type $DC$ and at most one copy of type $IC$.
\end{itemize}
In short, each $D_j$ contains at most one copy that is of type $IC$ or $A$.
We will say $j$ is active if the first bullet holds, and inactive if the second bullet holds.
In this way, we define the $\alpha^I$-value of a client as the $\alpha$-value of its active copy or the indirectly connected copy. If the client has no such copy, we say its $\alpha^I$-value is $0$ (as this value will not be used).
We remark this $\alpha^I$-value is different from the $\alpha$-values of the copies, and will only be used when we merge two solutions (\Cref{subsec:merging_algorithm}). 
Moreover, we will treat each copy $\tilde{j}$ as a normal weighted client located at the same point as $j$ in the metric space.
When we compute the total bid, we average out over their different copies.

Further, we consider $\calD$ as \emph{the collection of (succinct) copies} by dropping all the $\alpha$-values of directly connected copies from $\cup_{j\in D} D_j$. 
This collection will be used to define the algorithm that merges two solutions.
In this way, we can still deduce $\alpha^I$ values for each client and this set can be independent of the $\alpha$-values of directly connected copies, which are not necessary in defining the algorithm.

Equivalently, for each inactive client $j$, we consider its connection as a discrete probability distribution over (i) the connection type $ct_j$, i.e., whether it is in $IC$ or $DC$; (ii) the $\alpha$-value of the client $\alpha_j$. 
In particular, in the support of this distribution, there can be only one possible $\alpha$-value for indirect connects, i.e., there exists at most one value $\alpha^I$ such that $\Pr[(ct_j, \alpha_j) = (IC, \alpha^I)] > 0$.
When we compute the total bid, we average out their contributions. Adapting to our new dual fitting approach, the expected contribution of client $j$ to facility $i$ now becomes:
\begin{align*}
\mathbb{E}\left[(1-\delta)[d^2(j,S) - d^2(j, i)]^+ \mid ct_j=DC\right] \cdot \Pr[ct_j=DC] \hspace{2em}
\\
+ \mathbb{E}\left[[\alpha_j - \rho(1-\delta)d^2(j, i)]^+ \mid ct_j=IC\right] \cdot \Pr[ct_j=IC],
\end{align*}
where the probability and the expectations are all taken over the discrete distribution. Letting $\pi(j)=\Pr[ct_j=DC]$ and letting $\alpha^I_j$ be the only possible $\alpha$-value for indirect connects of $j$, we can rewrite the above as:
\begin{align*}  
    \pi(j)\cdot (1-\delta)[d^2(j,S) - d^2(j, i)]^+ + (1-\pi(j))\cdot [\alpha^I_{j} - \rho(1-\delta)d^2(j, i)]^+ 
\end{align*}
For each active client $j$, the connection distribution is a distribution with one element in the support having type $ct_j=A$. Therefore, we can define $\pi(j)=0$ and let $\alpha^I_j$ be the $\alpha$-value of $j$. In the rest of this section, we will use both points of view interchangeably.

Given the time $\theta = (1+\epsilon^3)^{p-1}$, the sequence $\mathcal{H}=(H_1, \dots , H_p)$ of the sequences of facilities opened by each stage 1 of the algorithm that was executed so far, we easily can obtain the set of open facilities $S$. 
Further given the collection of copies $\calD$ for the clients at the beginning of the phase, we can deduce the $\alpha$-value for each indirectly connected or active copy. Then, we can use the following process \retrieveD, which is the third step of stage 1 of our modified log-adaptive algorithm, to obtain the current collection of copies:

\begin{mdframed}[hidealllines=true, backgroundcolor=gray!15]
\vspace{-5mm}
\paragraph{\retrieveD($\calD, H_p$)}\ \\
For each copy $j\in \calD$ that is active or indirectly connected:\\[-0.5em]
\begin{itemize}[topsep=0.5em,itemsep=0.2em]
    \item Let $\alpha^I$ be its $\alpha$-value and $j$ be the client it corresponds to.
    \item If $\alpha^I \geq \rho(1-\delta)d^2(j, H_p)$, we modify the copy to be directly connected and drop its $\alpha$-value. 
    \item Else if $\alpha^I \geq (1-\delta)d^2(j, H_p)$, we modify the copy to be indirectly connected and keep its $\alpha$-value unchanged.
\end{itemize}
\end{mdframed}

Given the output of the procedure, we can deduce all current copies for each client, and their corresponding connection types, probability mass and $\alpha$-values. 
Based on these variables, we have a new definition of $\eta$-openability tailored for our new dual fitting approach. 

\begin{definition}[$\eta$-openable]
    \label{def:robustopenable}
Consider the time $\theta=(1+\epsilon^3)^{p-1}$. Let $\mathcal{H}=(H_1,\dots, H_{p-1}, H_p)$ be the sequence of sequences of opened facilities so far and let $\calD$ be the collection of copies for all clients at the beginning of the phase $p$.
Then sets $S, A, IC, DC$ can be deduced from $\calH$ and \retrieveD$(\calD, H_p)$ defined as above.

For each copy $j$, we can also deduce its probability mass $\mu(j)$, and its $\alpha$-value $\alpha_j$ if $j\in IC\cup A$.
We say that a facility $i\in F$ is \emph{$\eta$-openable} (with respect to $\theta$, $\mathcal{H}$ and $\calD$) if there are $(\tau_j)_{j\in A}$ that satisfy the following conditions. 

\begin{itemize}
    \item Only nearby clients are increased:
    \[
    \tau_j=\alpha_j\text{ for every }j\in A\setminus B(i, \sqrt{\epsilon \theta}).
    \]

    \item Nearby clients are only increased slightly:
    \[
    \alpha_j\leq\tau_j\leq \min\{(1-\delta)d^2(j,S), (1+\epsilon^3)\theta\} \text{ for every } j\in A\cap B(i,\sqrt{\epsilon\theta}).
    \]

    \item Facility $i$ is almost paid for:
    \begin{align*}
    (1-\delta)\sum_{j\in DC} \mu(j) \cdot [d^2(j,S) - d^2(j,i)]^+ + \sum_{j\in IC} \mu(j) \cdot [\alpha_j - \rho(1-\delta)d^2(j,i)]^+ \hspace{2em} \\
    + \sum_{j\in A} [\tau_j - \rho(1-\delta)d^2(j,i)]^+ \geq \hat f - 8n\eta.
    \end{align*}

    \item Dual feasibility almost holds: for every facility $i_0$,
    \begin{align*}
    \sum_{j\in DC} \mu(j) \cdot [(1-\delta)d^2(j,S) - d^2(j,i_0)]^+ + \sum_{j\in IC} \mu(j) \cdot [\alpha_j - \rho d^2(j,i_0)]^+ \hspace{2em} \\
    + \sum_{j\in A} [\tau_j - \rho d^2(j,i_0)]^+ \leq \hat f+8n\eta. 
    \end{align*}
\end{itemize}
\end{definition}

\begin{remark}
    \label{remark:robustopenable}
    Equivalently, the third bullet can be rewritten in the point of view of the clients (instead of the copies): Suppose $A$ is the set of active clients and $\alpha^I_j$ is the $\alpha^I$-value for client $j$, and $\pi(j)$ is the probability that $j\notin A$ is directly connected. Then, the four bullets can be rewritten as
    \begin{itemize}
        \item Only nearby clients are increased:
    \[
    \tau_j=\alpha^I_j\text{ for every }j\in A\setminus B(i, \sqrt{\epsilon \theta}).
    \]

    \item Nearby clients are only increased slightly:
    \[
    \alpha^I_j\leq\tau_j\leq \min\{(1-\delta)d^2(j,S), (1+\epsilon^3)\theta\} \text{ for every } j\in A\cap B(i,\sqrt{\epsilon\theta}).
    \]
    \item Facility $i$ is almost paid for:
    \begin{align*}
    \sum_{j\notin A} \bigg( \pi(j)\cdot (1-\delta) \cdot [d^2(j,S) - d^2(j,i)]^+ + (1-\pi(j))\cdot [\alpha^I_j - \rho(1-\delta)d^2(j,i)]^+\bigg)  \hspace{2em} \\
    + \sum_{j\in A} [\tau_j - \rho(1-\delta)d^2(j,i)]^+ \geq \hat f - 8n\eta.
    \end{align*}
    \item Dual feasibility almost holds: for every facility $i_0$,
    \begin{align*}
    \sum_{j\notin A} \bigg( \pi(j)\cdot \left[(1-\delta)d^2(j,S) - d^2(j,i_0)\right]^+ + (1-\pi(j))\cdot \left[\alpha^I_{j} - \rho d^2(j,i_0)\right]^+ \bigg) \hspace{2em}\\
    + \sum_{j\in A} [\tau_j - \rho d^2(j,i_0)]^+ \leq \hat f + 8n\eta.  
    \end{align*}
    \end{itemize}
\end{remark}

Similar as \cite{CGLSS25stoc,charikar2025kmeans,CCGGLW2026kmeans}, we introduce the concept of \emph{free facilities}. When we merge the two solutions, we introduce free facilities, whose opening costs are not necessarily paid. We will introduce at most $O(1)$ such free facilities in each phase. However, we do not restrict the number of free copies a regular facility may have. Let $\Sf \subseteq \mathcal{F}$ denote the multiset of free facilities that have been opened.

Moreover, we augment the metric space for the free facilities. For each free copy $\tilde{i} \in \Sf$ of a regular facility $i \in \mathcal{F}$, we define a parameter $u(\tilde{i}) \ge 0$, which acts as the squared distance overhead of the free copy. Therefore, for any point $x$ in the metric space (either a facility or a client), the distance squared is given by $d^2(\tilde{i}, j) = u(\tilde{i}) + d^2(i, j)$. 

We further introduce the following terminology: as mentioned, we refer to the facilities in $\Sf$ as \emph{free facilities}, the facilities in $\mathcal{F}$ and $\Sr = S - \Sf$ as \emph{regular} facilities, and when we simply mention ``facilities'', we are referring to their union.

Note that if $d$ is Euclidean, we can have one additional dimension, which equals $\sqrt{u(\tilde{i})}$ for any free facility $\tilde{i}$ and equals $0$ for any regular facility or client, to maintain the Euclidean property.

Finally, we will use the letters $i$ and $h$ to denote facilities. Specifically, we use $i \in \mathcal{F}$ to denote a regular facility and use $\tilde{i}$ to denote a free copy of $i$. When referring to a facility that can be either free or regular, we use the letter $h$.

We now introduce the definition of valid sequences, which aims to capture the sequence of facilities that are opened during the while-loop of stage 1 of the log-adaptive algorithm.
Let us say $\calH'=(H_1',\dots,H_p')$ is a super-sequence of $\calH=(H_1,\dots,H_p)$ and denote it $\calH'\supseteq \calH$ if they have a common prefix of length $p-1$, i.e., for all $i<p$, $H_i'=H_i$, and $H_p'$ is a super-sequence of $H_p$. Given two sequences $\calH_1=(H_1,\dots,H_r)$ and $\calH_2=(H_{r+1}, \dots,H_l)$, we define $\calH_1\oplus\calH_2=(H_1,\dots,H_l)$ as the concatenation of the two sequences.

\begin{definition}[$\eta$-valid sequence] 
    Consider the time $\theta$. Let $\calH$ be the sequence of sequences of opened facilities and $\calD$ be the collection of copies at the beginning of the phase. 
A sequence $\langle h_1, \ldots, h_\ell\rangle$ of facilities is $\eta$-valid (with respect to $\theta, \calD$ and $\calH$), if the following conditions hold. 
\begin{itemize}
    \item Each $h_t$ is either free or $\eta$-openable with respect to the time $\theta$, a super-sequence $\calH' \supseteq \calH \oplus (\langle h_1, \ldots, h_{t-1}\rangle)$ of the sequence of sequences of opened facilities, where $\oplus$ is just the concatenation, and the collection $\calD$.
    \item  The sequence is maximal in terms of (0-)openability, i.e., there is no (0-)openable facility with respect to the time $\theta$,  $\calH \oplus (\langle h_1, \ldots, h_\ell \rangle)$, and the collection $\calD$.
\end{itemize}
\label{def:valid_sequence}
\end{definition}

We consider the following procedure \completeD to return the collection of copies after opening a sequence of facilities $H_p$ during the first stage of the phase.
This procedure is identical to executing stage 2 of the log-adaptive algorithm on the collection of copies.  

\begin{mdframed}[hidealllines=true, backgroundcolor=gray!15]
\vspace{-5mm}
\paragraph{\completeD$(\theta, \calD, \calH=(H_1,\dots, H_p))$}\ \\

\noindent
Set $S=\cup_{r\in [p]} H_r$ and $\calD'=\text{\retrieveD}(\calD, H_p)$. For each copy $j\in \calD'$:\\

\begin{itemize}
    \item If $j$ is of type $IC$, set $\alpha_j\leftarrow (1-\delta)d^2(j,S)$.
    \item If $j$ is of type $A$, set $\alpha_j\leftarrow \min\{(1+\epsilon^3)\theta, (1-\delta)d^2(j, S)\}$, and change its connection type to $IC$ when $\alpha_j=(1-\delta)d^2(j,S)$.
\end{itemize}

\end{mdframed}

\begin{definition}[Valid collection of copies]
    \label{def:valid-collection}
    Consider the time $\theta=(1+\epsilon^3)^{p-1}$. Let $\calH= (H_1,\dots, H_p)$ be the sequence of sequences of opened facilities and $\calD$ be the collection of copies after the first $p-1$ phases. 
    A collection of copies $\calD'$ is valid (with respect to $\theta, \calH, \calD$), if $\calD'$ can be obtained by the following procedure.
    \begin{quote}
        Suppose $\calD''=\text{\completeD}(\theta, \calD, \calH)$.
        Then, we can obtain $\calD'$ by the following process: for each copy $j$ that is indirectly connected or active in $\calD$ but is directly connected in $\calD''$, if $\rho(1-\delta)d^2(j,S)\leq \alpha_{j}\leq \rho(1-\delta)d^2(j,S) + 4\eta$ (where $\alpha_j$ is the $\alpha$-value of $j$ in $\calD$), then it can be split into a directly connected copy with an arbitrarily smaller probability mass, while the remaining probability mass is given to an indirectly connected copy with $\alpha$-value $\alpha'_j=(1-\delta)d^2(j,S)$.
    \end{quote}
    In particular, if $\calD'=\calD''$, we say that $\calD'$ is perfectly valid.
\end{definition}

We remark that 
\begin{itemize}
    \item The log-adaptive algorithm generates $0$-valid sequences in the first stage of each phase. Specifically, it generates sequences $\calH = (H_1, H_2, \ldots, H_L)$ and a history of the collections of copies $\calDH = (\calD_0, \calD_1, \ldots, \calD_L)$, where each $H_p$ is $0$-valid (with respect to $\theta = (1+\epsilon^3)^{p-1}$, $\calD_{p-1}$ and $\calH$) and contains only regular facilities. 
    \item The log-adaptive algorithm only maintains a single copy for each inactive client, i.e., each inactive client $j$ is either directly connected to some facility $i$ with probability $1$ or indirectly connected to some facility $i$ with probability $1$. Therefore, each collection of copies $\calD_p$ maintained by the log-adaptive algorithm is perfectly valid (with respect to $\theta=(1+\epsilon^3)^{p-1}$, $\calH$ and $\calD_{p-1}$).
    \item The set $S$ of opened facilities is such that $A = \emptyset$ at the final time $\theta$.
\end{itemize}

For a general sequence of sequences $\calH = (H_1, H_2, \ldots, H_L)$ and a history of collection of copies $\calDH = (\calD_0, \calD_1, \ldots, \calD_L)$, let us say they consist of $\eta$-valid sequences and valid collections of copies if, for every $p\in [L]$,
\begin{itemize}
\item $H_p$ is an $\eta$-valid sequence with respect to $\theta = (1+\epsilon^3)^{p - 1}$, $\calH$, $\calD_{p-1}$; and
\item $\calD_p$ is a valid collection of copies with respect to $\theta = (1+\epsilon^3)^{p-1}$, $\calH$, $\calD_{p-1}$.
\end{itemize} 
It is easy to see from the definitions that the following claim holds.

\begin{claim}
    \label{claim:alpha_values_robust}
    Suppose that $\calH, \calDH$ consist of $\eta$-valid sequences and valid collections of copies. For each $p\in [L]$, letting $S=\cup_{r\in [p]} H_r$, then we have 
    \begin{itemize}
        \item If $j\in \calD_p$ is indirectly connected, then $\alpha_j = (1-\delta)d^2(j, S)$;
        \item If $j\in \calD_p$ is active, then $\alpha_j=(1+\epsilon^3)^p$ and $(1+\epsilon^3)^p<(1-\delta)d^2(j,S)$.
    \end{itemize}
\end{claim}

Furthermore, we say that $\calH, \calDH$ form a \emph{solution} with open facilities $S =\cup_{r\in [L]} H_r$ if there are no active copies in $\calD_L$.
Otherwise, we say that $\calH, \calDH$ form a partial solution.

The results in \cref{sec:log_adaptivity}, namely \cref{lem:approx_guarantee_logadaptive,lem:dual-feasibility-logadaptive}, 
can thus be stated as: if $\calH, \calDH$ is a solution of $0$-valid sequences consisting of regular facilities, then the set $S$ of opened facilities satisfies
\[
    \sum_{j\in D} w(j) \cdot d^2(j, S) \leq \frac{\Gamma}{1-\delta} \cdot \left( \optlpfl(f) - f \cdot |S| \right)\,.
\]
We now allow an additive $8n\eta$ slack in the third bullet (facility paying) and fourth bullet (dual feasibility) of the openability (Definition \ref{def:robustopenable}). 
In addition, in the second stage of each phase, we allow a directly connected copy $j$ that satisfies $\rho(1-\delta)d^2(j,S) \leq \alpha_j < \rho(1-\delta)d^2(j,S) + (1+\rho)\eta$ to split into two copies with arbitrary probability mass: one is identical, while the other is indirectly connected with $\alpha$-value $\alpha_j=(1-\delta)d^2(j,S)$. 
All these modifications lead to the following generalization.

\begin{restatable}{theorem}{thmrobust}

Consider a solution $\calH, \calDH$ of $\eta$-valid sequences and valid collections of copies. Let $S$ be the opened facilities. Further, let $\Sf$ and  $\Sr = S - \Sf$ be the free copies and regular facilities, respectively. Then,
we have
\[
    \sum_{j\in D} d^2(j, S) \leq \frac{\Gamma+201\sqrt{\eps} + 20n\eta}{1-\delta} \cdot \left( \optlpfl(f) - |\Sr| \cdot f\right)\,.
\]
where
    \begin{itemize}
        \item if $d$ is a (general) metric, we choose $\rho=2.5, \Gamma=4.9$; or
        \item if $d$ is a Euclidean metric, we choose $\rho=2, \Gamma = 3+\ln(2)$.
    \end{itemize}
\label{thm:robust}
\end{restatable}

In the next subsection, we will merge two solutions into one with $|\Sr| =k$ and $|\Sf| =O(\log n/\epsilon^3)$. To our help, we will use two basic routines defined as follows: 
\begin{itemize}
    \item $\completesol_{f,u}(\calH, \calDH)$ takes a partial solution $\calH, \calDH$ as input and returns a solution by running the while-loop of the log-adaptive algorithm using $f,u$ and the existing copies induced by the last collection in $\calDH$.

    In particular, we also consider a slightly different input for this routine. If $\calH = (H_1, \ldots, H_{p})$ and $\calDH = (\calD_0, \ldots, \calD_{p-1})$, where the first $p-1$ sequences of $\calH$ form a partial solution with $\calDH$, and $H_p$ is $\eta$-valid. Then, we will first generate $\calD_p$ by \completeD$(\theta=(1+\epsilon^3)^{p-1}, \calD_{p-1}, (H_1,\dots, H_p))$, append it to the end of $\calDH$ (so that the resulting $\calH,\calDH$ form a partial solution), and then start from phase $p+1$ of the log-adaptive algorithm.
    
    \item $\completesequence_{f,u}(\calH = (H_1, \ldots, H_{p-1}), H_p = \langle h_1, \ldots, h_\ell\rangle, \calDH=(\calD_0,\dots,\calD_{p-1}))$ where $\calH, \calDH$ form a partial solution, and $H_p$ is an $\eta$-valid sequence except that it may not be maximal. $\completesequence$ then runs the while-loop of the log-adaptive algorithm starting with $H_p$ and the existing copies induced by the last collection $\calDH$ to return a maximal sequence of $H_p$ (with respect to $f, u$).
\end{itemize}
We note that all facilities added by \completesol and \completesequence are $0$-openable and regular. 
We also note that, given $\calD_{p-1}$ and $\calH$, \completesol generates $\calD_{p}$ by \completeD$(\theta=(1+\epsilon^3)^{p-1}, \calD_{p-1}, (H_1,\dots, H_p))$.
Additionally, for intuition, observe that, with the above notation, the set of facilities opened by $\completesol_{f,u}(\emptyset, (\calD_0))$ is equivalent to that opened by the log-adaptive algorithm, where $\calD_0$ denotes the collection of copies in which each client has only one copy with $(ct=A,\mu=1,\alpha=1)$. (There is no free facility here.) 
To simplify notation, we only use $\completesequence$ when the partial solution $\calH$ and $\calDH$ are clear from the context, so we simply write $\completesequence_{(f,u)}(\langle h_1, \ldots, h_\ell \rangle)$. Furthermore, we omit $(f,u)$ when the parameters are clear from the context.

\subsection{\mergealg}
\label{subsec:merging_algorithm}

This subsection is almost identical to Subsection 5.3 of \cite{CCGGLW2026kmeans}. The main difference is the proof of \Cref{claim:1ststep_etavalid} in \Cref{sec:parameters_identical}.

We now present our procedure \mergealg that \emph{walks} between two solutions. We will maintain two solutions $(\calH, \calDH, f, u)$ and $(\calH', \calDH', f', u')$ whose numbers of open regular facilities \emph{sandwich $k$}. (Formally, let us say $a$ and $b$ sandwich $k$ if $a<k<b$ or $b<k<a$.) We will gradually make them closer until one of them opens exactly $k$ regular facilities. 
We start from the two initial solutions obtained using the standard method. Note that the minimum nonzero distance is $1$ and the maximum squared distance is $\Maxdist = \poly(n)$, which implies that $\sopt \leq n\Maxdist$. 

\paragraph{Initialization.} We binary search on $f$ in the interval $[1/n^2, 4n\Maxdist]$ and run the routine $\completesol_{f, u}(\emptyset, \langle\calD_0\rangle)$ (with zero $u$).
When $f = 1/n^2$, every client $j \in D$ satisfies $d(j, S) = 0$ if $d(j, \mathcal{F}) = 0$ and $d^2(j, S) \leq d^2(j, \mathcal{F}) + 2/n^2$ otherwise, as each client can open its closest facility by itself when $\alpha_j = d^2(j, \mathcal{F}) + 2/n^2$. If this solution opens at most $k$ centers, then it is already a $(1+2/n)$-approximation. On the other hand, when $f = 4n\Maxdist$, the algorithm will open exactly one facility, since the first facility is open when $\theta \geq 2\Maxdist$, so every client becomes immediately inactive when it is open. 
Therefore, the binary search yields two solutions $(\calH, \calDH, f, u)$ and $(\calH', \calDH', f', u)$ where $(\calH, \calDH, f, u)$ opens less than $k$ facilities and $(\calH', \calDH', f', u)$ opens more than $k$ facilities where $f \leq f' \leq f + \eta$.

\paragraph{General Step.}
\mergealg works on phase $p = 1, 2, \dots$ iteratively, and ensures that the two solutions are identical up to the $p$-th phase. Formally, \mergealg takes two solutions $(\calH, \calDH, f, u)$ and $(\calH', \calDH', f', u')$ as well as the current phase $p$ as input. The following promises on them will be satisfied whenever we call \mergealg. 
\begin{enumerate}[label=(\roman*)]
    \item Both consist of $\eta$-valid sequences.
    \item The numbers of open regular facilities by the two solutions sandwich $k$. 
    \item %
    At most one parameter, which we call the \emph{difference parameter}, differs by at most $\eta$ between the two solutions, i.e., if $f\neq f'$ then $|f-f'| \leq \eta$ and $u(\tilde i) = u'(\tilde i)$ for every $\tilde i \in \Sf$; otherwise if $f=f'$ there is at most one free facility $\tilde i\in \Sf$ so that $u(\tilde i) \neq u'(\tilde i)$ and $|u(\tilde i) - u'(\tilde i)| \leq \eta$. (We don't consider $\pi, \pi'$ defined in the next bullet as \emph{difference parameters}.)
    \item \label{condition-iv} The probability that each client is directly connected to any facility is almost the same in both solutions, i.e., we have that for every client $j$, $|\pi(j) - \pi'(j)| \leq \eta/\Maxdist$, where $\pi, \pi'$ are the probabilities of direct connections for the collection of copies $\calD_{p-1}$ and $\calD'_{p-1}$, respectively.
    \item $\calH = (H_1, H_2, \ldots, H_L)$ and $\calH' = (H'_1, H'_2, \ldots, H'_{L'})$ have a common prefix of length $p-1$, i.e., $H_i = H'_i$ for $i=1,2, \ldots, p-1$. Furthermore, the remaining sequences $H_p, \ldots, H_L$ and $H'_p,\ldots, H'_{L'}$ only contain regular facilities and are $0$-valid with respect to the parameters $f, u$ and $f', u'$, respectively. (This implies that $\calH$ and $\calH'$ open the same set of free facilities.)
\end{enumerate}
Given the promises, the goal of \mergealg is to 
\begin{enumerate}[label=(\arabic*)]
    \item either find a solution of $\eta$-valid sequences that opens exactly $k$ regular facilities and $O(\log n)$ free facilities, or 
    \item produce two new solutions that satisfy the above promises for phase $p + 1$. 
\end{enumerate}

Since the maximum squared distance between any two points is $\Maxdist \leq \poly(n)$ and $f \leq 4n\Maxdist$, 
before $\theta$ reaches $6\Maxdist$, at least one facility will open and every client will become inactive. 
Therefore, for some threshold $p^* = O(\log n / \epsilon^3)$, once $\calH$ and $\calH'$ agree on the first $p^*$ phases, they open exactly the same set of facilities which violates the promise that the numbers of open regular facilities sandwich $k$. Therefore, case (1) must happen for some $p \leq p^*$, which yields a solution of $\eta$-valid sequences that opens exactly $k$ regular facilities and $O(\log n)$ free facilities. 
Combined with \Cref{thm:robust}, this proves our main bicriteria guarantee. Indeed, we then have a solution consisting of open centers $S$ with $|S| = k  + O(\log n / \epsilon^3)$ such that, by \Cref{thm:robust},
\begin{align*}
    \sum_{j\in D} w(j) \cdot d^2(j, S) &\leq \frac{\Gamma}{1-\delta} \cdot \left( \optlpfl(f) - (f-\eta n) \cdot k \right) \leq (\Gamma+O(\epsilon)) \sopt\,,
\end{align*}
where, for the last inequality, we used that $\eta = 2^{-n}$ and $\sopt\geq 1$ (since any non-zero distance is at least $1$). The main theorem then follows by scaling $\epsilon$ appropriately.

In the remainder of this section, we present \mergealg for two solutions and a given phase $p$. It consists of four subroutines, presented in the subsequent subsections respectively. Each of the first three subroutines achieves (1) or (2) without condition \ref{condition-iv} above (which ends the whole procedure or lets us move on to the fourth subroutine), or sets up the stage for the next subroutine. The third subroutine always achieves (1) or (2) without condition \ref{condition-iv}.
The fourth subroutine, then achieves (2), for those not achieving (1), by ensuring condition \ref{condition-iv}.

\subsubsection{Making Parameters Identical}
\label{sec:parameters_identical}
By renaming, we assume that if $f\neq f'$ is the difference parameter then $f' < f$, and if the difference parameter is $u(\tilde i) \neq u'(\tilde i)$ then $u(\tilde i) < u'(\tilde i)$.

\begin{claim}
    $H'_1, H'_2, \ldots, H'_p$ are $\eta$-valid sequences with respect to parameters $f, u$ and $\calD_0, \calD_1,\dots, \calD_{p-1}$, respectively. On the other hand, $\calD_1, \ldots, \calD_{p-1}$ are valid collections of copies with respect to parameters $f, u$ and $(H'_1, \ldots, H'_{p-1})$.
    \label{claim:1ststep_etavalid}
\end{claim}
\begin{proof}
As the first $p-1$ sequences are identical, we have that $H'_1, \ldots, H'_{p-1}$ are $\eta$-valid with respect to parameters $f, u$ and $\calD_0,\dots, \calD_{p-2}$, respectively. For the same reason, $\calD_1, \ldots, \calD_{p-1}$ are valid with respect to parameters $f, u$ and $(H'_1,\dots, H'_{p-1})$.
It remains to show that $H'_p$ is $\eta$-valid with respect to parameters $f, u$ and $\calD_{p-1}$. With an abuse of notation, we will use $\calH$ to denote the sequence of sequences $(H'_1, \ldots, H'_{p-1})$ in the rest of the proof.

Now suppose that $|f-f'|\leq \eta$ and $u(\tilde i) \neq u'(\tilde i)$ (if exists). If such $\tilde{i}$ does not exist, we take any $\tilde{i}$. 
$H'_p$ was $0$-valid with respect to $f',u',\calD'_{p-1}$.
We have that $u'(\tilde i) - \eta \leq u(\tilde i) \leq u'(\tilde i)$. Further, $|\tilde{\pi}(j)-\tilde{\pi}'(j)|\leq \eta/\Maxdist$ for each client $j$, where $\tilde{\pi}$ and $\tilde{\pi}'$ are the probabilities that $j$ is directly connected in $\calD_{p-1}$ and $\calD'_{p-1}$, respectively.

Let $H'_p = \langle h_1, \dots, h_{\ell} \rangle$ and consider $h_t$ for some $t \in [\ell]$, which is $0$-openable with respect to $f', u', \theta = (1 + \epsilon^3)^{p - 1}, \calD'_{p-1}$ and some super-sequence $\calH' \supseteq \calH \oplus (\langle h_1, \dots, h_{t-1} \rangle$). Let $H_{\rm last}$ be the final sequence in $\calH'$. Let $S, S'$ be the set of facilities open in $\calH$ and $\calH'$, respectively. 

Let $d'(\cdot,\cdot)$ denote the pairwise distance defined according to $u'$.
Consider the collection of copies $\calD'_{\rm new} = \text{\retrieveD}(\calD'_{p-1}, H_{\rm last})$, where the distances in the procedure follow $d'$. 
Let $A'$ denote the set of clients whose copies are active in $\calD'_{\rm new}$.
Let $\alpha'_j$ be the $\alpha$-value of the active or indirectly connected copy of client $j$ in $\calD'_{\rm new}$.
If no such copy exists, we let $\alpha'_j = 0$, which will not be used later.
Let $\pi'(j)$ be the probability that $j$ is directly connected in $\calD'_{\rm new}$.
Similarly, we define $d(\cdot,\cdot)$ according to $u$, and define $(A, \{ \alpha_j \}, \{\pi(j)\}, d(\cdot,\cdot))$ according to $\calD_{\rm new} = \text{\retrieveD}(\calD_{p-1}, H_{\rm last})$ under distances $d$.

As the only difference parameter is $u(\tilde i) \in [u'(\tilde i) - \eta, u'(\tilde i)]$, we have $d \leq d'$, meaning it is strictly easier to cross the metric distance thresholds for $IC$ and $DC$ under $d$. Thus, we can observe that $A \subseteq A'$.

For simplicity, we will consider the viewpoint of \Cref{remark:robustopenable} for $\eta$-openability.
Based on the fact that $h_t$ was $0$-openable with respect to the former setting with $(\tau'_j)_{j \in A'}$, let 
$(\tau_j)_{j \in A}$ be defined as $\tau_j = \min(\tau'_j, (1-\delta)d^2(j, S'))$ for $j \in A$. 
Let us check if $h_t$ is $\eta$-openable with respect to the latter setting 
with $(\tau_j)_{j \in A}$ by checking the conditions in \Cref{remark:robustopenable}.
The first two bullets are straightforward:
\begin{itemize}
    \item $\tau_j = \alpha_j$ for every $j \in A \setminus B(h_t, \sqrt{\epsilon \theta})$: Since it was true for $A' \supseteq A$, it is clear that $\tau_j\leq \tau'_j=\alpha_j$. Since $j\in A \subseteq A'$, we have $\alpha_j=\alpha'_j=\tau'_j=\theta$ and $(1-\delta)d^2(j,S')>\alpha_j$, implying $\tau_j\geq \alpha_j$. 
    \item $\alpha_j \leq \tau_j \leq \min\{(1-\delta)d^2(j,S'),(1+\epsilon^3)\theta\}$ for every $j \in A \cap B(h_t, \sqrt{\epsilon \theta})$: Fix $j \in A \cap B(h_t, \sqrt{\epsilon \theta})$. The first inequality holds because $\tau'_j\geq \alpha'_j=\theta=\alpha_j$ and $(1-\delta)d^2(j,S')>\alpha_j$. 
    Since $A \subseteq A'$, $\tau'_j \leq \min\{(1-\delta)d'^2(j,S'),(1+\epsilon^3)\theta\}$. Since $\tau_j$ is defined as $\min(\tau'_j, (1-\delta)d^2(j,S'))$, the second inequality is satisfied.
\end{itemize}

Before discussing the third and fourth bullets, we establish a useful claim.

\begin{claim}
    \label{claim:closeness_alpha-values}
    Suppose that $\calH,\calDH$ and $\calH,\calDH'$ are two partial solutions opening the same set of facilities after phase $p-1$.
    We assume they are under $(f,u)$ and $(f',u')$ respectively.
    Suppose that $u, u'$ differ on at most one parameter by at most $\eta$.
    Fix any client $j$. 
    Let $\alpha, \alpha'$ be the $\alpha$-value of the copies of $j$ that is of type $IC$ or $A$ in $\calD_{p-1}$ and $\calD'_{p-1}$ (assuming both exist), respectively. 
    Then, $|\alpha-\alpha'| \leq \eta$.
\end{claim}
\begin{proof}
    Let $\theta = (1+\epsilon^3)^{p-1}$. 
    Suppose $\calH=(H_1,\dots, H_{p-1})$ and let $S = \cup_{r\in [p-1]} H_r$.
    Suppose $d,d'$ are the pairwise distances defined according to $u,u'$, respectively.
    All the following $\alpha$-values follow \Cref{claim:alpha_values_robust}.

    If $j$ has an active copy in both $\calD_{p-1}$ and $\calD'_{p-1}$, then $\alpha = \alpha' = \theta$.

    If $j$ has an indirectly connected copy in both $\calD_{p-1}$ and $\calD'_{p-1}$, then $\alpha = (1-\delta)d^2(j,S)$ and $\alpha' = (1-\delta)d'^2(j,S)$. Since $u,u'$ differ on at most one parameter by at most $\eta$, we have $|d^2(j,S) - d'^2(j,S)| \leq \eta$ and thus $|\alpha - \alpha'| \leq (1-\delta)\eta \leq \eta$.

    If $j$ has an active copy in one collection and an indirectly connected copy in the other collection, since $d \le d'$, it must be that $j$ has an active copy in $\calD'_{p-1}$ (under $d'$) and an indirectly connected copy in $\calD_{p-1}$ (under $d$). Then, we have $\alpha' = \theta$ and $\alpha = (1-\delta)d^2(j,S) \le \theta$. Since $j$ is active in $\calD'_{p-1}$, we have $(1-\delta)d'^2(j,S) > \theta$. Therefore, we have
    \begin{align*}
        |\alpha' - \alpha| &= \theta - (1-\delta)d^2(j,S) \\
        &\leq (1-\delta)d'^2(j,S) - (1-\delta)d^2(j,S) \leq (1-\delta)\eta \leq \eta. \qedhere
    \end{align*}
\end{proof}

For the third and fourth bullets, we construct an abstract bid function $bid_{A, \alpha, \tau, \pi, d, \gamma}(j)$ based on the metric openability terms, where $\gamma(j)$ is a client-specific target distance bound:
\begin{align*}
    bid_{A, \alpha, \tau, \pi, d, \gamma}(j) = \begin{cases}
    [\tau_j - \rho \gamma(j)]^+ & \text{if } j \in A, \\
    \pi(j)[(1-\delta)d^2(j,S') - \gamma(j)]^+ + (1-\pi(j)) [\alpha_j - \rho \gamma(j)]^+ & \text{if } j \notin A
    \end{cases}
\end{align*}
For the third bullet, we will set $\gamma(j) = (1-\delta)d^2(h_t, j)$. For the fourth bullet, we will set $\gamma(j) = d^2(i_0, j)$. In both cases, note how the $DC$ term subtracts $\gamma(j)$ directly, while the $IC$ and $A$ terms subtract $\rho \gamma(j)$.

Next, we evaluate $|bid_{A, \alpha, \tau, \pi, d, \gamma}(j) - bid_{A', \alpha', \tau', \pi', d', \gamma}(j)|$ for every client $j$. 
Recall that $|\tilde{\pi}(j)-\tilde{\pi}'(j)|\leq \eta/\Maxdist$.
We discuss the following cases:
\begin{itemize}
\item Suppose that $\tilde{\pi}(j), \tilde{\pi}'(j) > 1-\eta/\Maxdist$. Since the procedure \retrieveD doesn't turn directly connected copies into active or indirectly connected, we have $\pi(j), \pi'(j) > 1 - \eta/\Maxdist$ and thus $|\pi(j) - \pi'(j)| \leq \eta/\Maxdist$. In addition, we have $j\notin A$ and $j\notin A'$. Because \retrieveD doesn't change $\alpha$-values for indirectly connected or active copies, $\alpha_j=(1-\delta)d^2(j,S)$ and $\alpha'_j=(1-\delta)d'^2(j,S)$. Because $d^2(j,S'), d'^2(j,S')\leq \Maxdist$, and $\alpha'_j, \alpha_j\leq \theta \leq 6\Maxdist$, we have 
\begin{align*}
    &|bid_{A, \alpha, \tau, \pi, d, \gamma}(j) - bid_{A', \alpha', \tau', \pi', d', \gamma}(j)| \\
    &\qquad \leq 6\Maxdist \cdot |\pi(j)-\pi'(j)| + \max\{|(1-\delta)d^2(j,S') -  (1-\delta)d'^2(j, S')|, |\alpha_j - \alpha'_j|\}  \\
    &\qquad \leq 6\eta + \eta = 7\eta.
\end{align*}
\item Suppose that one of $\tilde{\pi}(j), \tilde{\pi}'(j)$ is strictly less than $1-\eta/\Maxdist$. Since $|\tilde{\pi}(j)-\tilde{\pi}'(j)|\leq \eta/\Maxdist$, both are strictly less than $1$. 
This implies $j$ has exactly one active or indirectly connected copy in both $\calD_{p-1}$ and $\calD'_{p-1}$. We will use $bid, bid'$ for $bid_{A,\alpha,\tau,\pi,d,\gamma}(j)$ and $bid_{A',\alpha',\tau',\pi',d',\gamma}(j)$ respectively for simplicity.
\begin{itemize}
\item If both copies become directly connected, then $\pi(j)=\pi'(j)=1$ and thus $|bid - bid'| \le |(1-\delta)d^2(j,S') - (1-\delta)d'^2(j,S')| \le \eta$.
\item If both copies remain active or indirectly connected, then $\pi(j)=\tilde{\pi}(j), \pi'(j)=\tilde{\pi}'(j)$ and thus $|\pi(j)-\pi'(j)|\leq \eta/\Maxdist$. By \Cref{claim:closeness_alpha-values} and the fact that \retrieveD doesn't modify the $\alpha$-values of active or indirectly connected copies, we have $|\alpha_j - \alpha'_{j}| \leq \eta$. 
\begin{enumerate}
\item If both copies are indirectly connected (i.e., $j\notin A'$), we use the same derivation as above to prove that $|bid - bid'| \leq 6\eta + \eta = 7\eta$.

\item If both copies are active (i.e., $j\in A$), we have $\tau_j\neq \tau'_j$ only if $\tau'_j>(1-\delta) d^2(j,S')$. In this case, we have $\tau_j = (1-\delta) d^2(j,S')$ and $\tau'_j \leq (1-\delta) d'^2(j,S')$. Therefore, $|\tau_j-\tau'_j| \leq (1-\delta) \cdot |d^2(j,S')-d'^2(j,S')|\leq \eta$ and we have $|bid-bid'|\leq \eta$. 

\item If $j\in A'\setminus A$ (this is the remaining case because $A\subseteq A'$), we have $\pi'(j) = 0$ and thus $\tilde\pi'(j) = 0$. 
This implies that $\tilde{\pi}(j) \leq \eta/M$ and thus $\pi(j) \leq \eta/M$. Therefore, $|\pi(j)-\pi'(j)|\leq \eta/M$. 
Furthermore, we have $\tau'_j \geq \theta \geq \alpha_j$, and $\tau'_j \leq (1-\delta) d'^2(j,S') \leq (1-\delta)(d^2(j,S') +\eta) \leq \alpha_j+\eta$ because $j\notin A$. 
Therefore, we have $|\tau'_j-\alpha_j|\leq \eta$.
Using the same derivation as above, we have $|bid-bid'|\leq 7\eta$.
\end{enumerate}

\item If one copy crosses the threshold to become directly connected while the other remains active or indirectly connected, since $d \le d'$, it must be that the copy under $d$ becomes directly connected ($\pi(j)=1$) while the copy under $d'$ remains active or indirectly connected ($\pi'(j)=\tilde{\pi}'(j)$).
We can characterize $bid_{A, \alpha, \tau, \pi, d, \gamma}(j)$ and $bid_{A', \alpha', \tau', \pi', d', \gamma}(j)$ as follows:
\begin{align}
    \label{eqn:bid-make-idential-parameters-case2.3}
    \begin{split}
    bid_{A, \alpha, \tau, \pi, d, \gamma}(j) &= [(1-\delta)d^2(j,S')-\gamma(j)]^+ , \\
    bid_{A', \alpha', \tau', \pi', d', \gamma}(j) &= \pi'(j)\cdot [(1-\delta)d'^2(j,S')-\gamma(j)]^+ + (1-\pi'(j))[\alpha'_{j}-\rho\gamma(j)]^+.
    \end{split}
\end{align}

Crucially, because $\pi(j)=1$ and $|\pi(j) - \pi'(j)| \leq \eta/\Maxdist$, we have $1 - \pi'(j) \leq \eta/\Maxdist$. 
To bound the difference between the bids, let $X = [(1-\delta)d^2(j,S')-\gamma(j)]^+$, $Y = [(1-\delta)d'^2(j,S')-\gamma(j)]^+$, and $Z = [\alpha'_{j}-\rho\gamma(j)]^+$.
We want to bound $|X - (\pi'(j) Y + (1-\pi'(j)) Z)|$. 
By the triangle inequality, this is at most:
\begin{align*}
    |X - (\pi'(j) Y + (1-\pi'(j)) Z)| &= |X - Y + (1-\pi'(j))(Y - Z)| \\
    &\leq |X - Y| + (1-\pi'(j))|Y - Z|~.
\end{align*}
First, we bound $|X - Y|$:
\[
    |X - Y| \leq (1-\delta)|d^2(j,S') - d'^2(j,S')| \leq \eta~.
\]
Second, we bound $|Y - Z|$. Since all pairwise squared distances are bounded by $\Maxdist$, we have $Y \leq \Maxdist$. Furthermore, because the log-adaptive algorithm terminates before $\theta$ exceeds $6\Maxdist$, we have $Z \leq \alpha'_j \leq \theta \leq 6\Maxdist$. Since both $Y$ and $Z$ are non-negative, their absolute difference is bounded by the maximum of their upper bounds, meaning $|Y - Z| \leq 6\Maxdist$.
Therefore, 
\[
    (1-\pi'(j))|Y - Z| \leq \frac{\eta}{\Maxdist} \cdot 6\Maxdist = 6\eta~.
\]
Putting this back into \cref{eqn:bid-make-idential-parameters-case2.3}, we conclude that:
\begin{align*}
    |bid_{A, \alpha, \tau, \pi, d, \gamma}(j) - bid_{A', \alpha', \tau', \pi', d', \gamma}(j)| \leq \eta + 6\eta = 7\eta~.
\end{align*}
\end{itemize}
\end{itemize}

Using this inequality (which gives a maximum difference of $7\eta$ across all cases), we can easily obtain for the sum over all clients:
\begin{align*}
\left|\sum_{j\in D} bid_{A, \alpha, \tau, \pi, d, \gamma}(j) - \sum_{j\in D} bid_{A', \alpha', \tau', \pi', d', \gamma}(j)\right| \leq \sum_{j\in D} 7\eta \leq 7n\eta~. 
\end{align*}
We check the last two bullets in \Cref{remark:robustopenable} as follows. For the third bullet, we define $\gamma(j) = (1-\delta)d^2(h_t, j)$. Because $h_t$ is a regular facility, $d^2(h_t, j) = d'^2(h_t, j)$. Because $h_t$ was $0$-openable with respect to $\theta, f', u', \calD'_{p-1}$, the left-hand side under $bid'$ evaluates to at least $\hat f$. The difference of at most $7n\eta$ ensures the third bullet is satisfied under $\theta, f, u, \calD_{p-1}$ with the required slack bound of $\hat{f}' - 7n\eta \geq \hat{f}-8n\eta$.

For the fourth bullet, we define $\gamma(j) = d^2(i_0, j)$. With the same abstract bid formulation and distance stability for the regular facility $i_0$, the sum over all clients under $bid'$ is at most $\hat f$. The difference of at most $7n\eta$ ensures the sum under $bid$ is at most $\hat f + 7n\eta \leq \hat f + 8n\eta$, thereby concluding the fourth bullet.
\end{proof}

\paragraph{Outcome of this section of \mergealg.}
Let
\[
    \calH'', \calDH'' = \completesol_{(f,u)}((H'_1, \ldots, H'_p), (\calD_0, \ldots, \calD_{p-1}))\,,
\]
which is a valid call by \Cref{claim:1ststep_etavalid}.
If $\calH''$ opens $k$ regular facilities, we return this solution. Otherwise, 
as $(\calH, \calDH, f,u)$ and $(\calH', \calDH', f', u')$ sandwich $k$, we must have that either $(\calH, \calDH, f, u)$ and $(\calH'', \calDH'', f, u)$, or $(\calH'', \calDH'', f, u)$ and $(\calH', \calDH', f', u')$ sandwich $k$.

On the one hand, if $(\calH'', \calDH'', f, u)$ and $(\calH', \calDH', f',u')$ sandwich $k$, then we made progress in that we now have two solutions with a common prefix of length $p$ (instead of $p-1$) and exactly one parameter differs by at most $\eta$. 
However, the new collection of copies $\calD''_p$ and $\calD'_p$ may not satisfy condition \ref{condition-iv} regarding the bounds on the probabilities of direct connections ($\pi$). 
In this case, we proceed to \Cref{sec:approx_equal_rho} of phase $p$.

On the other hand, if $(\calH, \calDH, f,u)$ and $(\calH'', \calDH'', f, u)$ sandwich $k$, then these two solutions have the exact same parameters $f$ and $u$, have $\calD_{r}=\calD''_{r}$ for every $r\leq p-1$, and their first $\eta$-valid sequences that differ are $H_p$ and $H'_p$. In this case, we rename $\calH''$ to $\calH'$ (and $\calDH''$ to $\calDH'$) and proceed to the next step.

\subsubsection{\texorpdfstring{Maximize Common Prefix of $H_p$ and $H'_p$}{Maximize Common Prefix of Hp and H'p}}
\label{sec:common_prefix}
We are given two solutions $(\calH, \calDH, f,u)$ and $(\calH', \calDH', f, u)$ on the same parameters $f,u$ that sandwich $k$. 
It is guaranteed that $\forall r\leq p-1, \calD_{r}=\calD'_{r}$ in the history of collections of copies.
Moreover, if we let $\calH = (H_1, \ldots, H_L)$ and $\calH' = (H'_1, \ldots, H'_{L'})$, then $H_i = H'_i$ for $i= 1, \ldots, p-1$ and $H_p \neq H'_p$. In addition, $H_p$ and $H'_p$ only contain regular facilities. 

In this step of \mergealg, we increase the common prefix of $H_p$ and $H'_p$ by ``slowly'' modifying $H'_p$. As $H_p$ only contains regular facilities, we let $H_p = \langle i_1, i_2, \ldots, i_\ell \rangle$. Moreover, as the parameters $(f,u)$ and $\calD_1,\dots, \calD_{p-1}$ are clear from the context, we omit them in the following when, e.g., using the definition of $\eta$-openability, and calling the procedures \completesol and \completesequence.\footnote{When we use \completesol$(H_1,\dots, H_{p-1}, H''_p)$, we are actually referring to \completesol$(\calH''=(H_1,\dots, H_{p-1}, H''_p), \calDH''=(\calD_0,\dots,\calD_{p-1}))$. When we use \completesequence$(H''_p)$, we are actually referring to \completesequence$(\calH''=(H_1,\dots, H_{p-1}), H''_{p}, \calDH''=(\calD_0,\dots,\calD_{p-1}))$.}

This subsection proceeds similarly to the approach in \cite{charikar2025kmeans}. 

\begin{mdframed}[hidealllines=true, backgroundcolor=gray!15]
\vspace{-5mm}
\paragraph{Transforming $H'_p$ step-by-step so that $H_p$ becomes a prefix.}\ \\

\noindent Repeat the following until all of $H_p$ becomes a prefix of $H'_p$.\\

\begin{itemize}
        \item Let $q$ be the largest index so that $H'_p = \langle i_1, i_2, \ldots, i_q, h'_{q+1}, \ldots, h'_{\ell_1}\rangle$, i.e., $q$ equals the length of the common prefix $i_1, i_2, \ldots, i_q$ of $H_p$ and $H'_p$.
            \item Update $H'_p$ by inserting a free copy of $i_{q+1}$, denoted $\tilde i_{q+1}$, at the end of $H'_p$ with $u(\tilde i_{q+1}) = 0$. That is, we get the sequence
            \begin{gather*}
                H'_p = \langle i_1, i_2, \ldots, i_q, h'_{q+1}, \ldots, h'_{\ell_1}, \tilde{i}_{q+1} \rangle.
            \end{gather*}

            \item For $t = q+1 , \ldots, \ell_1$,
            \begin{itemize}
                \item Update $H'_p$ by first removing $h'_t$ and then completing the sequence by a call to \completesequence. More explicitly, if 
\begin{align*}
                 H'_p = &\langle i_1, \ldots, i_q, h'_t, h'_{t+1} \ldots, h'_{\ell_1}, \tilde i_{q+1}, h'_{\ell_1+1}, \ldots, h'_{\ell_2}\rangle\\
\intertext{then we delete $h'_t$ to obtain}
                    H''_p = &\langle i_1, \ldots, i_q,  h'_{t+1} \ldots, h'_{\ell_1}, \tilde i_{q+1}, h'_{\ell_1+1}, \ldots, h'_{\ell_2}\rangle\\
\intertext{and finally update $H'_p = \completesequence(H''_p)$ which will be of the form}
              &\langle i_1, \ldots, i_q, h'_{t+1}, \ldots, h'_{\ell_1}, \tilde i_{q+1}, h'_{\ell_1+1}, \ldots, h'_{\ell_2}, \ldots, h'_{\ell_3}\rangle\,,
\end{align*}
                where $h'_{\ell_2+1},\ldots, h'_{\ell_3}$ are the facilities added by \completesequence (which could be empty).
            \end{itemize}

            \item After removing $h'_{q+1}, \ldots, h'_{\ell_1}$, we have
            \begin{align*}
                H'_p &= \langle i_1, i_2, \ldots, i_q, \tilde{i}_{q+1}, h'_{x}, \ldots, h'_{y}\rangle\,.\\
            \intertext{We update $H'_p$ by replacing $\tilde{i}_{q+1}$ with its regular copy $i_{q+1}$, i.e., we obtain}
                H'_p &= \langle i_1, i_2, \ldots, i_q, i_{q+1}, h'_{x}, \ldots, h'_{y}\rangle\,.
            \end{align*}
\end{itemize}
\end{mdframed}
We have the following claim.

\begin{claim}
    Throughout the updates to $H'_p$, $H'_1, \ldots, H'_p$ remain $\eta$-valid sequences, and $H'_p$ contains at most one free facility.
    \label{claim:2ndstep_valid_free}
\end{claim}
\begin{proof}
    As we never update $f$ or the value $u(\cdot)$ of a free facility in $H'_1, \ldots, H'_{p-1}$, these sequences clearly remain $\eta$-valid. 

    We continue to argue that $H'_p$ remains $\eta$-valid throughout its updates. By \Cref{claim:1ststep_etavalid}, this is true before any updates to $H'_p$. We now analyze the three types of updates to $H'_p$. 
    \begin{itemize}
        \item Updating $H'_p$ by inserting a free copy of $i_{q+1}$ maintains that it is $\eta$-valid as every regular facility stays $\eta$-openable in the sequence (they have the same prefix as before) and the sequence remains maximal.
        \item Updating $H'_p$ by removing $h'_t$ and calling \completesequence. Removing a $h'_t$ leaves the new prefix as a subsequence of the old prefix. Thus, each subsequent regular facility is still $\eta$-openable with respect to the original sequence padded with $h'_t$ (which acts as a valid super-sequence $\calH'$). Finally, the call to \completesequence adds facilities until no more can be opened, ensuring that the sequence is maximal.
        \item Updating $H'_p$ by replacing $\tilde i_{q+1}$ with its regular copy $i_{q+1}$. First note that, since $u(\tilde i_{q+1}) = 0$, this does not impact whether the following regular facilities are $\eta$-openable (with respect to a super-sequence), as the distances $d^2(\tilde{i}_{q+1}, \cdot)$ and $d^2(i_{q+1}, \cdot)$ are perfectly identical. By the same argument, the sequence stays maximal. Finally, to see that $i_{q+1}$ is $\eta$-openable (with respect to a potential super-sequence), notice that the prefix $i_1, \ldots, i_q$ is the same as that in $H_p$, which is already known to be an $\eta$-valid sequence. 
    \end{itemize}

    We complete the proof of the claim by arguing that $H'_p$ contains at most one free facility at any point in time. This follows from the facts that (1) $H'_p$ initially contains no free facilities, (2) in one repetition, the single free facility $\tilde i_{q+1}$ is inserted into $H'_p$, and (3) $\tilde i_{q+1}$ is strictly replaced by its regular copy $i_{q+1}$ before continuing to the next repetition.
\end{proof}

\newcommand{\Hbefore}{\ensuremath{H^{\text{before}}_p}\xspace}
\newcommand{\Hafter}{\ensuremath{H^{\text{after}}_p}\xspace}

\paragraph{Outcome of this section of \mergealg.} We distinguish three cases.

\paragraph{Case 1.} If at some point $H'_p$ is such that $\completesol(H'_1, H'_2, \ldots, H'_p)$ opens exactly $k$ facilities, then we return that solution.

\paragraph{Case 2.} Otherwise, suppose there is an update to $H'_p$ so that if we let \Hbefore and \Hafter denote $H'_p$ before and after this update, respectively, then $\completesol(H_1, H_2, \ldots, \Hbefore)$ and $\completesol(H_1, H_2, \ldots, \Hafter)$ sandwich $k$. There are three kinds of updates to $H'_p$. However, note that the update where $\tilde i_{q+1}$ is replaced by its regular copy $i_{q+1}$ cannot satisfy the conditions of this case. Indeed, as $u(\tilde i_{q+1}) = 0$, we have
\[
    \completesol(H_1, \ldots, \Hbefore) = \completesol(H_1, \ldots, \Hafter) 
\]
in this case, which contradicts that they sandwich $k$. 

We proceed to analyze the two other kinds of updates.
If the update to $H'_p$ is to insert a free facility $\tilde{i}_{q+1}$ with $u(\tilde{i}_{q+1}) = 0$, then
\begin{align*}
    \Hbefore &= \langle i_1, i_2, \ldots, i_q, h'_{q+1}, \ldots, h'_{\ell_1} \rangle\\
    \Hafter &= \langle i_1, i_2, \ldots, i_q, h'_{q+1}, \ldots, h'_{\ell_1}, \tilde{i}_{q+1} \rangle
\end{align*}
Note that the solution $\completesol(H_1, H_2, \ldots, \Hbefore)$ is equivalent to $\completesol(H_1, H_2, \ldots, \Hafter)$ if we set $u(\tilde{i}_{q+1}) = 10\Maxdist$ (except for the free facility $\tilde i_{q+1}$). We can thus do binary search on $u(\tilde{i}_{q+1})$ to find two values $u'(\tilde{i}_{q+1})$ and $u(\tilde{i}_{q+1})$ with $|u'(\tilde{i}_{q+1}) - u(\tilde{i}_{q+1})| \leq \eta$ so that the solutions $\completesol_{(f,u')}(H_1, \ldots, \Hafter)$ and $\completesol_{(f,u)}(H_1, \ldots, \Hafter)$ sandwich $k$ (where $u$ and $u'$ are identical except for $\tilde{i}_{q+1}$). 
We can thus proceed to \Cref{sec:approx_equal_rho} of phase $p$, since they have the first $p$, instead of $p-1$, $\eta$-valid sequences in common and exactly one difference parameter. 
Their maximality follows from that of \Hbefore. 

Finally, consider when the update to $H'_p$ is of the second type, i.e., $u(\tilde{i}_{q+1}) = 0$ and 
\begin{align*}
    \Hbefore  &= \langle i_1, \ldots, i_q, h'_t, h'_{t+1} \ldots, h'_{\ell_1}, \tilde i_{q+1}, h'_{\ell_1+1}, \ldots, h'_{\ell_2}\rangle\\
    \Hafter &= \langle i_1, \ldots, i_q, h'_{t+1}, \ldots, h'_{\ell_1}, \tilde i_{q+1}, h'_{\ell_1+1}, \ldots, h'_{\ell_2}, \ldots, h'_{\ell_3}\rangle
\end{align*}
In this case, $\completesol(H_1, \ldots, \Hbefore)$ and $\completesol(H_1, \ldots, \Hafter)$ are two solutions with the same parameters $f, u$, that sandwich $k$ and we proceed to the next step with the following properties of \Hbefore and \Hafter:
\begin{itemize}
    \item Both sequences contain one free facility, $\tilde{i}_{q+1}$.
    \item \Hbefore and \Hafter are identical except that (1) \Hbefore contains a regular facility $h'_t$ not present in \Hafter, and (2) \Hafter contains a (potentially empty) suffix of regular facilities not present in \Hbefore.
\end{itemize}

\paragraph{Case 3.} None of the previous cases apply. Then, completing the solution with the initial $H'_p$ and the final $H'_p$ cannot sandwich $k$. Since then (assuming the first case does not apply) there must be an update to $H'_p$ that results in two solutions that sandwich $k$, since the completion of $H_p$ and the completion of the initial $H'_p$ sandwich $k$. The above implies that if we let $\Hbefore = H_p$ and $\Hafter$ equal the final $H'_p$, then $\completesol(H_1, \ldots, \Hbefore)$ and $\completesol(H_1, \ldots, \Hafter)$ sandwich $k$. Further note that since $\Hafter$ equals $H'_p$ after the final updates, $\Hbefore$ appears as a prefix of $\Hafter$ and both sequences only contain regular facilities. To summarize, in this case, $\completesol(H_1, \ldots, \Hbefore)$ and $\completesol(H_1, \ldots, \Hafter)$ are two solutions with the same parameters $f, u$, that sandwich $k$ and we proceed to the next step with the following properties of \Hbefore and \Hafter:
\begin{itemize}
    \item Both sequences contain no free facilities. 
    \item \Hbefore and \Hafter are identical except that \Hafter contains a suffix of regular facilities not present in \Hbefore.
\end{itemize}

\subsubsection{Removing Extra Facilities}
\label{sec:extra_facilities}
This subsection proceeds almost identically to that of \cite{charikar2025kmeans} (except that we will go to the next subroutine instead of finishing phase $p$). As in \Cref{sec:common_prefix}, the parameters $(f,u)$ and $\calD_1,\dots, \calD_{p-1}$ are clear from the context, and we will omit them when using the definition of $\eta$-openability and calling the function \completesol.

The input to this step consists of two solutions $\completesol(H_1, \ldots, \Hbefore)$ and $\completesol(H_1,\ldots, \Hafter)$ that have the same parameters $f, u$ and sandwich $k$. Moreover, $\Hbefore$ and $\Hafter$ both contain at most one free facility and they are identical except for two potential differences: 
\begin{itemize}
\item $\Hbefore$ may contain a facility not present in $\Hafter$;
\item $\Hafter$ contains a suffix of regular facilities (which may be empty) that is not present in $\Hbefore$.
\end{itemize}

To simplify notation in this section, we let $i_*$ denote the regular facility in $\Hbefore$ that is not present in $\Hafter$ if it exists (so $i_*$ corresponds to $h'_t$ in the previous section). We further let $h_1, \ldots, h_\ell$ denote the suffix of $\Hafter$ not present in $\Hbefore$.

We now transform $\Hafter$ into $\Hbefore$ one change at a time:

\begin{mdframed}[hidealllines=true, backgroundcolor=gray!15]
\vspace{-5mm}
\paragraph{Generation of $\eta$-valid sequences from $\Hafter$ to $\Hbefore$.}\ \\
\begin{itemize}
    \item If $\Hbefore$ contains the regular facility $i_*$ not present in $\Hafter$, then let $\tilde{i}_*$ be a free copy of $i_*$ with $u(\tilde{i}_*) =0$ and obtain $H^{0}_p$ from $\Hafter$ by adding $\tilde{i}_*$ at the end. Otherwise, we let $H^0_p$ equal $\Hafter$. 
    \item For $r= 1, \ldots, \ell$, obtain $H^r_p$ from $H^{r-1}_p$ by removing $h_r$. 
\end{itemize}
\end{mdframed}

We remark that $H^\ell_p$ is identical to $\Hbefore$ since we removed the whole suffix $h_1, \ldots, h_\ell$ present in $\Hafter$, except for potentially the facility $i_*$. That is, the only difference in $\Hbefore$ and $H^\ell_p$ may be the placement of $\tilde{i}_*$ and that $\tilde{i}_*$ is a free copy of the regular facility $i_*$ present in $\Hbefore$. However, since $u(\tilde{i}_*) = 0$, we have that the two sequences $\Hbefore$ and $H^\ell_p$ have an identical impact on \completesol, which is a fact we use in the proof of \Cref{claim:3rdstep_outcome}.

\begin{claim}
    For $r= 0, 1 \ldots \ell$, $H^r_p$ is an $\eta$-valid sequence with at most two free facilities.  
\end{claim}
\begin{proof}
    The sequence $H^r_p$ has at most two free facilities because $\Hafter$ has at most one free facility, and we potentially add one new free facility $\tilde i_*$.

    We now verify that $H^r_p$ is $\eta$-valid. For any regular facility $h'$ in the sequence, the prefix of $H^r_p$ before $h'$ is a subsequence of the prefix before $h'$ in $\Hafter$. Thus, as $h'$ was openable with respect to a super-sequence $\calH(h')$ in \Hafter, $h'$ must be $\eta$-openable with respect to the same super-sequence $\calH(h')$ in $H^r_p$. This ensures that the first property of \Cref{def:valid_sequence} is satisfied. To verify maximality, notice that since we added $\tilde i_*$ with $u(\tilde i_*) =0$ at the end, we have that the set of active clients after opening up the facilities in $H^r_p$ is a subset of the active clients after opening up the facilities in $\Hbefore$. Hence, the maximality of $\Hbefore$ implies the maximality of $H^r_p$.
\end{proof}

\begin{claim}
    \label{claim:3rdstep_outcome}
    One of the following is true.
    \begin{itemize}
        \item For one of the sequences $H^r_p$, $\completesol(H_1, H_2, \ldots, H^r_p)$ opens $k$ regular facilities.
        \item $\completesol(H_1, H_2, \ldots, \Hafter)$ and $\completesol(H_1, H_2, \ldots, H^{0}_p)$ sandwich $k$.
        \item $\completesol(H_1, H_2, \ldots, H^r_p)$ and $\completesol(H_1, H_2, \ldots, H^{r+1}_p)$ sandwich $k$ for an $r\in \{0, 1, \ldots, \ell-1\}$. 
    \end{itemize}
\end{claim}
\begin{proof}
    Assume the first bullet point does not hold, i.e., the completion of no $H^r_p$ leads to the opening of $k$ regular facilities. Now notice that the second and third bullet points ask whether consecutive solutions sandwich $k$, starting with $\Hafter$ and ending with $H^{\ell}_p$.
    By definition, $H^{\ell}_p$ equals $\Hbefore$ except potentially for the placement of $\tilde{i}_*$ and that $\tilde{i}_*$ is the free copy of $i_*$ present in $\Hbefore$. However, as $u(\tilde{i}_*) = 0$, the two sequences have an identical impact when we complete the solution, which implies
    \[
    \completesol(H_1, \ldots, H^{\ell}_p) = \completesol(H_1, \ldots, \Hbefore).
    \]
    Thus, as $\completesol(H_1, \ldots, \Hafter)$ and $\completesol(H_1, \ldots, \Hbefore)$ sandwich $k$, either the second or third bullet point of the statement must be valid (since no solution opened exactly $k$ regular facilities).
\end{proof}

\paragraph{Outcome of this section of \mergealg.} We distinguish three cases based on the first bullet point that is true in the above claim.

\paragraph{Case 1.} If one of the sequences $H^r_p$ is such that $\completesol(H_1, H_2, \ldots, H^r_p)$ opens $k$ regular facilities, then we return that solution.

\paragraph{Case 2.} Otherwise, if the two solutions $\completesol(H_1, H_2, \ldots, \Hafter)$ and \\$\completesol (H_1, H_2, \ldots, H^{0}_p)$ sandwich $k$, then $H^0_p$ is equal to $\Hafter$ except that $\tilde{i}_*$ was added at the end of $\Hafter$ with $u(\tilde{i}_*) = 0$. 
Note that the solution $\completesol(H_1, H_2, \ldots, \Hafter)$ is equivalent to $\completesol(H_1, H_2, \ldots, H^0_p)$ if we set $u(\tilde{i}_*) = 10\Maxdist$ (except for the presence of $\tilde{i}_*$). We can thus do a binary search on $u(\tilde{i}_*)$ to find two values $u'(\tilde{i}_*)$ and $u(\tilde{i}_*)$ with $|u'(\tilde{i}_*) - u(\tilde{i}_*)| \leq \eta$ so that the two solutions $\completesol_{(f,u')}(H_1, \ldots, H^0_p)$ and $\completesol_{(f,u)}(H_1, \ldots, H^0_p)$ sandwich $k$ (where $u$ and $u'$ are identical except for $\tilde{i}_*$). 
We can thus proceed to \Cref{sec:approx_equal_rho} of phase $p$, since they have the first $p$, instead of $p-1$, $\eta$-valid sequences in common and exactly one difference parameter. 

\paragraph{Case 3.} Finally, consider the case when none of the above cases apply and (thus by the above claim) there is an $r\in \{0, 1, \ldots, \ell-1\}$ so that $\completesol(H_1, H_2, \ldots, H^r_p)$ and $\completesol(H_1, H_2, \ldots, H^{r+1}_p)$ sandwich $k$.
The only difference between $H^r_p$ and $H^{r+1}_p$ is that $H^r_p$ contains $h_{r+1}$ which is not present in $H^{r+1}_p$. Let $H_p$ be the sequence obtained by adding a free copy $\tilde i_{r+1}$ of $h_{r+1}$ at the end of $H^{r+1}_p$ (which clearly maintains that it is $\eta$-valid), then 
\begin{align*}
    \completesol(H_1, H_2, \ldots, H_p) &= \completesol(H_1, H_2, \ldots, H^r_p) \mbox{ if $u(\tilde i_{r+1}) = 0$}\\
\intertext{and} 
    \completesol(H_1, H_2, \ldots, H_p) &= \completesol(H_1, H_2, \ldots, H^{r+1}_p) \mbox{ if $u(\tilde i_{r+1}) = 10\Maxdist$.}
\end{align*}
Hence, doing a binary search on $u(\tilde{i}_{r+1})$ as in the previous case can proceed to \Cref{sec:approx_equal_rho} of phase $p$, since they have the first $p$, instead of $p-1$, $\eta$-valid sequences in common and exactly one difference parameter.

\subsubsection{Making the Probability of Direct Connects Almost Identical}
\label{sec:approx_equal_rho}

The input of this step consists of solutions $(\calH, \calDH, f, u)$ and $(\calH, \calDH', f', u')$ such that $(f,u)$ and $(f',u')$ have at most one difference parameter. Moreover, the two solutions sandwich $k$. 
However, $\calD_p$ and $\calD'_p$ are simply obtained from $\calD_{p-1}$ and $\calD'_{p-1}$ respectively by running \completeD.
In this step, we will walk between the collection of copies $\calD_{p}$ and $\calD'_p$ following \Cref{def:valid-collection} so that condition \ref{condition-iv} holds for phase $p$. 

We follow the procedure below twice. Suppose that $S=\cup_{r\in [p]} H_r$. Suppose that $d, d'$ are the pairwise distances induced by $u, u'$. One run is to transform $(\calD_p, \calD_{p-1}, d)$ towards $(\calD'_p,\calD'_{p-1}, d')$.
The other run is to transform $(\calD'_p, \calD'_{p-1}, d')$ towards $(\calD_p, \calD_{p-1}, d)$.

\begin{mdframed}[hidealllines=true, backgroundcolor=gray!15]
\vspace{-5mm}
\paragraph{Transforming connection types in $(\calD,\tilde{\calD}, d)$ towards $(\calD', \tilde{\calD}', d')$}\ \\

\noindent Repeat the following.\\
\begin{itemize}
        \item Find a copy $j$ in $\calD$ and $j'$ in $\calD'$ such that
         \begin{itemize}
            \item $j$ and $j'$ correspond to the same client,
            \item $j$ and $j'$ are both of type $IC$ or $A$ in $\tilde{\calD}$ and $\tilde{\calD}'$,
            \item $j$ is of type $DC$ in $\calD$ while $j'$ is of type $IC$ or $A$ in $\calD'$.
         \end{itemize}
        If no such client exists, then stop the process.
        \item Update $\calD$ by changing $j$ to $IC$ type and update the $\alpha$-value $\alpha_j$ to $(1-\delta)d^2(j,S)$.
\end{itemize}
\end{mdframed}

After running the two procedures, we can obtain a sequence of solutions $\tilde{\calD}_0=\calD_p, \tilde{\calD}_1, \ldots, \tilde{\calD}_m$ by the first run, while another sequence of solutions $\tilde{\calD}'_0=\calD'_p, \tilde{\calD}'_1, \ldots, \tilde{\calD}'_{m'}$ is obtained by the second run.
First, we show that all $\tilde{\calD}_r$ and $\tilde{\calD}'_{r}$ are valid collections of copies.

\begin{claim}
    \label{clm:validD-duringwalk}
    Consider time $\theta=(1+\epsilon^3)^{p-1}$.
    For any $r\in [m]$, $\tilde{\calD}_r$ is valid with respect to $\theta, (H_1,\dots, H_p), \calD_{p-1}$.
    For any $r\in [m']$, $\tilde{\calD}'_r$ is valid with respect to $\theta, (H'_1,\dots, H'_p), \calD'_{p-1}$.
\end{claim}
\begin{proof}
    We prove the claim for $\tilde{\calD}_r$; the proof for $\tilde{\calD}'_r$ is symmetric.
    We proceed by induction on $r$.

    For the base case when $r=0$, because $\tilde{\calD}_0 = \calD_p$, by the definition of \completeD, we have that $\tilde{\calD}_0$ is perfectly valid with respect to $\theta, (H_1,\dots, H_p), \calD_{p-1}$.

    Next, we consider the induction step when $r>0$.
    Suppose $\tilde{\calD}_{r-1}$ is valid with respect to $\theta, (H_1,\dots, H_p), \calD_{p-1}$.
    In the procedure, we update $\tilde{\calD}_{r-1}$ to $\tilde{\calD}_r$ by changing a copy $j$ from type $DC$ to type $IC$. To show that $\tilde{\calD}_r$ is valid with respect to $\theta, (H_1,\dots, H_p), \calD_{p-1}$, we need to show that its original $\alpha$-value in $\calD_{p-1}$ satisfies the slack condition: $\rho(1-\delta)d^2(j,S) \leq \alpha_j \leq \rho(1-\delta)d^2(j, S) + 4\eta$.

    First, because $j$ was $IC$ or $A$ in $\calD_{p-1}$ but became $DC$ in $\calD_p$, it must have triggered the $DC$ conversion during \retrieveD against the newly opened facilities $H_p$. Therefore, it strictly satisfied $\alpha_j \geq \rho(1-\delta)d^2(j, H_p)$. Since $S = S_{\text{old}} \cup H_p$, we trivially have $d^2(j, S) \leq d^2(j, H_p)$, satisfying the lower bound $\alpha_j \geq \rho(1-\delta)d^2(j, S)$.

    Second, suppose that $j'$ is the corresponding copy of $j$ in $\calD'_{p-1}$ (used in the above procedure), with dual value $\alpha'_{j'}$. Since $j, j'$ are of type $IC$ or $A$ in $\calD_{p-1}, \calD'_{p-1}$ respectively, and $u, u'$ differ in at most one parameter by at most $\eta$, \Cref{claim:closeness_alpha-values} gives $|\alpha_j-\alpha'_{j'}| \leq \eta$.
    
    To bound $\alpha'_{j'}$, we look at its behavior under $d'$. Because $j'$ was $IC$ or $A$ in the valid collection $\calD'_{p-1}$, its value is exactly $(1-\delta)d'^2(j, S_{\text{old}})$ or $\theta$ (with $\theta < (1-\delta)d'^2(j, S_{\text{old}})$). In either case, since $\rho > 1$, it strictly satisfies $\alpha'_{j'} \leq \rho(1-\delta)d'^2(j, S_{\text{old}})$. Furthermore, since it remains of type $IC$ or $A$ in $\calD'_{p}$, it did \emph{not} trigger the \retrieveD conversion under $d'$ against $H_p$, meaning $\alpha'_{j'} < \rho(1-\delta)d'^2(j, H_p)$. 
    Combining these bounds against $S_{\text{old}}$ and $H_p$, we establish its bound against the full set $S = S_{\text{old}} \cup H_p$:
    \[
        \alpha'_{j'} \leq \rho(1-\delta) \min(d'^2(j, S_{\text{old}}), d'^2(j, H_p)) = \rho(1-\delta)d'^2(j, S)~.
    \]

    Finally, because $u$ and $u'$ differ by at most $\eta$, we have $d'^2(j, S) \leq d^2(j, S) + \eta$. Therefore, we can tightly bound $\alpha_j$ against $d$:
    \begin{align*}
        \alpha_j \leq \alpha'_{j'} + \eta &\leq \rho(1-\delta)d'^2(j, S) + \eta \\
        &\leq \rho(1-\delta)(d^2(j, S) + \eta) + \eta \\
        &= \rho(1-\delta)d^2(j, S) + \rho(1-\delta)\eta + \eta \\
        &\leq \rho(1-\delta)d^2(j, S) + (1+\rho)\eta\\
        &\leq \rho(1-\delta)d^2(j,S) + 4\eta~.
    \end{align*}
    Therefore, the original $\alpha_j$ value securely satisfies the required $4\eta$ validity bounds. After updating $j$'s type to $IC$ and setting its new $\alpha$-value to $(1-\delta)d^2(j,S)$, all conditions in \Cref{def:valid-collection} hold for $\tilde{\calD}_r$. This completes the induction.
\end{proof}

Second, we show that $\tilde{\calD}_m$ and $\tilde{\calD}'_{m'}$ are close enough so that condition \ref{condition-iv} holds for them.

\begin{claim}
    \label{clm:closeD-afterwalk}
    Fix any client $j$.
    Suppose $\pi(j), \pi'(j)$ are the probabilities of direct connects for $j$ in $\tilde{\calD}_m$ and $\tilde{\calD}'_{m'}$ respectively.
    Then, $|\pi(j) - \pi'(j)| \leq \eta/\Maxdist$.
\end{claim}
\begin{proof}
    Because client $j$ has at most one copy of type $IC$ or $A$ in both $\tilde{\calD}_m$ and $\tilde{\calD}'_{m'}$, we can use the mass of this unique $IC$ or $A$ copy (if any) to determine the exact $DC$ probabilities $\pi(j)$ and $\pi'(j)$.
    Suppose $\tilde{\pi}(j), \tilde{\pi}'(j)$ are the probabilities of direct connects for $j$ in $\calD_{p-1}$ and $\calD'_{p-1}$ respectively. Let $\tilde{j}$ and $\tilde{j}'$ be the unique copies of type $IC$ or $A$ for client $j$ in $\calD_{p-1}$ and $\calD'_{p-1}$ respectively (if they exist).

    If one of them does not exist (meaning the client is fully $DC$ in that historical state), because we have the invariant $|\tilde{\pi}(j)-\tilde{\pi}'(j)|\leq \eta/\Maxdist$, we must have $\tilde{\pi}(j), \tilde{\pi}'(j) \geq 1-\eta/\Maxdist$. The walking procedure only acts to revert copies that were $IC/A$ in the previous phase; it does not convert historical $DC$ copies. Thus, the historical $DC$ mass is permanent, meaning $\pi(j) \geq \tilde{\pi}(j)$ and $\pi'(j) \geq \tilde{\pi}'(j)$. This mathematically restricts both final probabilities to the top of the interval $[1-\eta/\Maxdist, 1]$, strictly ensuring $|\pi(j) - \pi'(j)| \leq \eta/\Maxdist$.

    Next, we consider the case when both $\tilde{j}$ and $\tilde{j}'$ exist. Because we run the bidirectional walking procedures until no more possible updates can be made, it is impossible for $\tilde{j}$ and $\tilde{j}'$ to end up with divergent types (one $DC$ and one $IC/A$). Therefore, we have only two possible atomic outcomes for these specific copies:
    
    \begin{enumerate}
        \item They both evaluate to type $DC$ in $\tilde{\calD}_m$ and $\tilde{\calD}'_{m'}$. Because $\tilde{j}$ and $\tilde{j}'$ hold the entirety of the non-$DC$ mass (exactly $1-\tilde{\pi}(j)$ and $1-\tilde{\pi}'(j)$ respectively), converting them to $DC$ forces the total $DC$ probability to $\tilde{\pi}(j) + (1-\tilde{\pi}(j)) = 1$. Thus, $\pi(j) = \pi'(j) = 1$, yielding a difference of $0$.
        \item They both evaluate to type $IC/A$, meaning the procedure reverted any $DC$ conversions from this phase. In this case, the $DC$ probabilities remain exactly their historical values: $\pi(j)=\tilde{\pi}(j)$ and $\pi'(j)=\tilde{\pi}'(j)$. By our inductive promise, their difference is bounded by $\eta/\Maxdist$.
    \end{enumerate}
    In all cases, we strictly maintain the invariant $|\pi(j) - \pi'(j)| \leq \eta/\Maxdist$.
\end{proof}

\paragraph{The outcome of this section of \mergealg.}
Because the previous subroutines successfully aligned the opened facilities up to phase $p$, we have $H'_i = H_i$ for all $i \le p$. We can thus obtain a sequence of solutions as follows:
\begin{align*}
    \calH^{(0)}, \calDH^{(0)} &= \completesol_{f,u}((H_1,\dots, H_p), (\calD_0, \dots, \calD_{p-1}, \tilde{\calD}_0))\\
    \calH^{(1)}, \calDH^{(1)} &= \completesol_{f,u}((H_1,\dots, H_p), (\calD_0, \dots, \calD_{p-1}, \tilde{\calD}_1))\\
    \hfill
    \vdots
    \hfill
    \\
    \calH^{(m)}, \calDH^{(m)} &= \completesol_{f,u}((H_1,\dots, H_p), (\calD_0, \dots, \calD_{p-1}, \tilde{\calD}_m))\\
    \calH^{(m+1)}, \calDH^{(m+1)} &= \completesol_{f',u'}((H_1,\dots, H_p), (\calD'_0, \dots, \calD'_{p-1}, \tilde{\calD}'_{m'}))\\
    \calH^{(m+2)}, \calDH^{(m+2)} &= \completesol_{f',u'}((H_1,\dots, H_p), (\calD'_0, \dots, \calD'_{p-1}, \tilde{\calD}'_{m'-1}))\\
    \hfill
    \vdots
    \hfill
    \\
    \calH^{(m+m'+1)}, \calDH^{(m+m'+1)} &= \completesol_{f',u'}((H_1,\dots, H_p), (\calD'_0, \dots, \calD'_{p-1}, \tilde{\calD}'_0))
\end{align*}

It is clear that the first one and the last one sandwich $k$. Therefore, there must be two consecutive solutions in the above sequence that sandwich $k$. 
We can discuss two cases then. 

\paragraph{Case 1.} If $(\calH^{(m)},\calDH^{(m)})$ and $(\calH^{(m+1)}, \calDH^{(m+1)})$ sandwich $k$, according to \Cref{clm:validD-duringwalk} and \Cref{clm:closeD-afterwalk}, we can proceed to phase $p+1$ of \mergealg with these two solutions. Since the connection types in $\tilde{\calD}_m$ and $\tilde{\calD}'_{m'}$ are perfectly aligned, the only difference between the two solutions is the original difference parameter (either $f \neq f'$ or $u \neq u'$). They share the first $p$ sequences in common, satisfy condition \ref{condition-iv}, and differ by exactly one difference parameter.

\paragraph{Case 2.}
Suppose that $(\calH^{(r)},\calDH^{(r)})$ and $(\calH^{(r+1)},\calDH^{(r+1)})$ sandwich $k$ for some $r\neq m$.
Without loss of generality, we assume that $r < m$ (the other case is symmetric).
Suppose $j$ is the copy whose connection type differs in $\tilde{\calD}_r$ and $\tilde{\calD}_{r+1}$, where $j$ is of type $DC$ in $\tilde{\calD}_r$ while $j$ is of type $IC$ in $\tilde{\calD}_{r+1}$.
We observe that condition \ref{condition-iv} only fails on the client corresponding to copy $j$. 

We can further split this copy $j$ by choosing $\mu'\in [0,\mu(j)]$ such that one copy $j^{(1)}$ has $IC$ type with probability $\mu'$ and the other copy $j^{(2)}$ has $DC$ type with probability $\mu(j)-\mu'$.
We denote the resulting collection of copies by $\tilde{\calD}^{\rm sp}(\mu')$. 
According to \Cref{clm:validD-duringwalk}, because $\tilde{\calD}_r$ and $\tilde{\calD}_{r+1}$ are valid collections of copies, any interpolation $\tilde{\calD}^{\rm sp}(\mu')$ is a valid collection of copies.
Therefore, we can obtain a solution $(\calH^{\rm sp}(\mu'), \calDH^{\rm sp}(\mu'))$ by calling $\completesol_{f,u}((H_1,\dots, H_p), (\calD_0, \dots, \calD_{p-1}, \tilde{\calD}^{\rm sp}(\mu')))$.

By binary searching $\mu'$, we can find two values $\mu_1'$ and $\mu_2'$ with $|\mu_1'-\mu_2'|\leq \eta/\Maxdist$ in polynomial time such that the two solutions $(\calH^{\rm sp}(\mu'_1), \calDH^{\rm sp}(\mu'_1))$ and $(\calH^{\rm sp}(\mu'_2), \calDH^{\rm sp}(\mu'_2))$ sandwich $k$.

Because the overall probabilities of direct connections for client $j$ in the two resultant solutions $(\calDH^{\rm sp}(\mu'_1))_p$ and $(\calDH^{\rm sp}(\mu'_2))_p$ differ by exactly $|\mu'_1 - \mu'_2|$, and we bounded this difference by $\eta/\Maxdist$, these two solutions satisfy condition \ref{condition-iv}. 
We can then proceed to phase $p+1$ of \mergealg with these two solutions: they have the first $p$ $\eta$-valid sequences in common, utilize identical parameters $(f, u)$, and the newly established parameter between them is exactly the connection probability mass $\mu'$ (which is not viewed as a difference parameter).

\subsection{Analysis of the Robust Algorithm}
\label{sec:robust-analysis}

In this subsection, we prove \Cref{thm:robust}, restated as follows, showing that a solution $(\calH=(H_1,\dots, H_L), \calDH=(\calD_0,\calD_1,\dots, \calD_L))$ consisting of $\eta$-valid sequences and valid collections of copies yields an almost LMP $\Gamma$-approximation. This subsection follows the same logic as Subsection 5.4 from \cite{CCGGLW2026kmeans}, the main difference is in establishing the constraints for the Dual Payment, which are different for this new algorithm.

\thmrobust*

Our analysis closely follows that of \logadaptalg in \Cref{sec:log_adaptivity}. However, one key difference is that in the current setting, each client $j\in D$ may have multiple copies in its collection of copies $D_j$.
Older setting in \Cref{sec:log_adaptivity} can be seen as a special case where each client has only one weighted copy.
To close the gap between the two settings, for each copy $\tilde{j} \in \calD_L$ in the final collection of copies, it induces a corresponding weighted client with weight $\mu(\tilde{j})$, which is equal to the probability mass of the copy $\tilde{j}$ in $\calD_L$.
For simplicity, in the rest of this subsection, we will call these induced clients by the copies simply as ``clients'' and call the clients in $D$ as ``original clients''.
We will simply use $j$ to denote such a client, and use $oc(j)$ to denote its corresponding ``original client''.
Also, we will use the terminology clients and copies interchangeably later in the analysis.

We can keep track of how each copy splits in each $\calD_p$ for $p=1,\dots,L$. 
In this way, for each copy in $\calD_L$, we can know which copy it belongs to in any $\calD_p$ for each $p=1,\dots, L$.
For simplicity, we will split each copy $j$ in $\calD_p$ into multiple copies $j_1, \dots, j_r$ in $\calD_L$ which all belong to it.
These $j_1, \dots, j_r$ will keep their own probability mass in $\calD_p$, but their connection types ($A, DC, IC$) and $\tilde{\alpha}$-values (if not of type $DC$) will be inherited from $j$ in $\calD_p$.
After processing this splitting, we can view $\calD_1,\dots, \calD_{L}$ as the history of how each copy in $\calD_L$ evolves from phase $1$ to phase $L$.
Further, the split process in the definition of ``valid collection of copies'' (\cref{def:valid-collection}) can be regarded as turning some copies in each $\calD_p$ from $DC$ to $IC$.

One more issue to address is that in the collections of copies, we only maintain $\alpha$-values for active or indirectly connected copies. We may lose track of the $\alpha$-values of directly connected copies. To handle this issue, we define the $\alpha$-value for each copy using the following procedure. We remark that this procedure is only for analysis purposes. The sequence of sequences $\calH$ is enough for us to recover a solution. Hence, we do not need to pay attention to the time complexity of this procedure.

\begin{figure}[ht!]
\begin{center}
\begin{minipage}{1.0\textwidth}
\begin{mdframed}[hidealllines=true, backgroundcolor=gray!15]

\begin{algorithm}[\recoverA] \ \\[0.2cm]\label{alg:recover-alpha}
\textbf{Initialization:} Set $\theta=1$, $S=\emptyset$, $A=\calD_L$, $DC=\emptyset$, $IC=\emptyset$, and $\alpha_j=\theta,R_j=0$ for every $j\in \calD_L$.\\

\noindent For each $p\in \{1,\dots, L\}$: \\[-0.5cm]
\begin{itemize}
    \item Let $\theta = (1+\epsilon^3)^{p-1}$.
    \item \textbf{Stage 1.} For each facility $h_t$ in the sequence $H_p$, perform the following steps:
    \begin{enumerate}[label=\Roman*)]
        \item If $h_t$ is free, add $h_t$ to $S$ and jump to step IV). Otherwise, find a super-sequence $\calH' \supseteq \langle H_1,\dots, H_{p-1}\rangle \oplus(\{h_1,\dots,h_{t-1}\})$ such that $h_t$ is $\eta$-openable with respect to $\calH'$. Let $A'$ be the set of active clients induced by $\calD_{new}=\text{\retrieveD}(\calD_{p-1}, \calH'-\langle H_1,\dots, H_{p-1}\rangle)$.
        \item Compute $(\tau_j)_{j\in A'}$ so that $h_t$ is $\eta$-openable with $(\tau_j)_{j\in A'}$.
        \item Add $h_t$ to $S$. Set $\alpha_j\leftarrow \tau_j$ for every $j\in A'\cap B(h_t,\sqrt{\epsilon\theta})$.
        \item For every $j\in A\cup IC$:
        \begin{itemize}
            \item If $\alpha_j\geq \rho(1-\delta) d^2(j,h_t)$, move $j$ to $DC$, and set $R_j$ as follows:
            \begin{itemize}
                \item if $h_t$ is regular and $j$ is directly connected in $\calD_{new}$, set $R_j = \sqrt{\alpha_j/\rho}$;
                \item otherwise, set $R_j = \sqrt{1-\delta} \cdot d(j,h_t)$.
            \end{itemize}
            \item Otherwise, if $j\in A$ and $\alpha_j\geq (1-\delta)d^2(j,h_t)$, move $j$ to $IC$.
        \end{itemize}
    \end{enumerate}

    \item \textbf{Stage 2.} Perform the following step and move to the next phase:
    \begin{enumerate}[label=\roman*)]
        \item For every $j\in IC$, set $\alpha_j\leftarrow (1-\delta)d^2(j,S)$.
        \item For every $j\in A$, set $\alpha_j\leftarrow \min\{(1+\epsilon^3)\theta, (1-\delta)d^2(j, S)\}$, and move it to $IC$ when $\alpha_j=(1-\delta)d^2(j,S)$.
        \item For every $j\in DC$, if $j$'s type in $\calD_p$ is indirectly connected, move $j$ to $IC$ and set $\alpha_j\leftarrow (1-\delta)d^2(j,S)$, and $R_j \gets 0$.
        \item Update $\theta\leftarrow(1+\epsilon^3)\theta$.
    \end{enumerate}
\end{itemize}
\end{algorithm}
\end{mdframed}
\end{minipage}
\end{center}
\end{figure}

For \recoverA, we have the following remarks:
\begin{itemize}
    \item In the definition of the process, we will view free facilities and regular facilities equivalently (especially when we run \retrieveD), but in the later analysis of dual payment and \cref{thm:robust}, we still distinguish between them.
    \item The steps I) and II) in Stage 1 are well-defined since $H_p$ is $\eta$-valid with respect to $\theta, (H_1,\dots, H_{p-1}), \calD_{p-1}$ (\cref{def:valid_sequence}).
    \item In Step IV) in Stage 1, we have two different ways to define $R_j$ variables for newly directly connected clients. The reason we need different definitions is that $h_t$ is only openable with respect to a super-sequence. If a client $j$ is viewed as directly connected w.r.t. the super-sequence, we define $R_j$ maximally, by enforcing $R^2_j=\alpha_j/\rho$, because the $(\rho-1)\alpha_j/\rho$ part of $\alpha_j$ will no longer be used for opening other facilities (see, the proof \cref{lem:approx-guarantee-robust}). Otherwise, we define $R_j$ as usual, by $R_j = \sqrt{1-\delta} d(j,h_t)$.
    This is a new feature for our analysis, compared with \cite{CCGGLW2026kmeans}. 
    \item We can show that \recoverA produces the same solution as $(\calH, \calDH)$ by comparing \recoverA with \retrieveD, \completeD and the definition of valid collection of copies (\cref{def:valid-collection}):
    \begin{itemize}
        \item In Stage 1, step III) of \recoverA opens facilities in $H_p$ one by one; and then step IV) of \recoverA updates the $\alpha$-values and connection types of all clients in exactly the same way as \retrieveD.
        \item In Stage 2, steps i) and ii) of \recoverA update the $\alpha$-values and connection types of all clients in exactly the same way as the second step of \completeD. Step iii) of \recoverA moves some directly connected clients to indirectly connected, which is exactly the same as the split process in \cref{def:valid-collection} under our viewpoint. We shall remark that all copies moved from $DC$ to $IC$ are those in $IC\cup A$ in $\calD_{p-1}$ because of our algorithm in \cref{sec:approx_equal_rho}.
    \end{itemize}
\end{itemize}

Throughout the analysis, we let $\alpha_j^*$ be the final $\alpha$-value of client $j\in \calD_L$.
With this \recoverA procedure, we can prove two key lemmas, \cref{lem:approx-guarantee-robust,lem:dual-feasibility-robust}, that are analogs of {\Cref{lem:approx_guarantee_logadaptive,lem:dual-feasibility-logadaptive}}, respectively. %
The proofs of them are almost identical to that for \logadaptalg. There are only three minor changes:
\begin{enumerate}
    \item \Cref{def:robustopenable} of $\eta$-openability allows for a facility to be paid up to an $8n\eta$-additive difference and ensures every facility not to be overbid up to an $8n\eta$-additive difference. The first difference only appears in \Cref{lem:approx-guarantee-robust} below, while the second difference only appears in \cref{lem:dual-feasibility-robust}. These differences reflect in step I) and II) of stage 1. 

    \item \Cref{def:valid_sequence} of an $\eta$-valid sequence allows $h_t$ is $\eta$-openable with respect to a super sequence $\calH'\supseteq \calH\oplus(\{h_1,\dots,h_{t-1}\})$. 
    This difference  appears in the proof of \cref{lem:approx-guarantee-robust}, and (the last part of) the proof of \Cref{lem:dual-feasibility-robust}, an analog of \Cref{lem:dual-feasibility-logadaptive}. This difference reflects in step I), II) and IV) of stage 1. In particular, a significant difference is in the proof of $r_{jt}\leq R_j$ (\cref{lem:rjt_leq_Rj}). 

    \item \cref{def:valid-collection} of valid collections of copies allows for some directly connected copies to be moved to indirectly connected in Stage 2 of each phase. This difference only appears in (the last part of) the proof of \Cref{lem:approx-guarantee-robust} and the (first part of) the proof of \cref{lem:dual-feasibility-robust}. This difference reflects in step iii) of stage 2. 
\end{enumerate}
The combination of these two lemmas will imply \Cref{thm:robust}.

\begin{proof}[Proof of \Cref{thm:robust}]
For each original client $x\in D$, let
\[
    D_x=\{j\in \calD_L: oc(j)=x\}.
\]
Define
\[
    \alpha_x
    :=
    \sum_{j\in D_x} w(j)\cdot
    \bigl(\alpha^*_j-(\rho-1)R_j^2\bigr).
\]
Since all copies of $x$ are located at the same point as $x$, they have the same
distance to every facility as $x$ does. Also,
\[
    \sum_{j\in D_x}w(j)=1.
\]

Let
\[
    \Gamma' := \Gamma+201\sqrt{\eps},
    \qquad
    \Lambda := \Gamma' + 20n\eta,
\]
and run the robust algorithm with
\[
    \hat f := \Lambda f + 12n\eta.
\]
Equivalently,
\[
    \hat f-12n\eta=\Lambda f.
\]

By \Cref{lem:approx-guarantee-robust}, we have
\begin{align}
    \sum_{x\in D} d^2(x,S)
    &=
    \sum_{j\in \calD_L} w(j)d^2(j,S) \notag\\
    &\le
    \frac{
        \sum_{j\in \calD_L}w(j)
        \bigl(\alpha^*_j-(\rho-1)R_j^2\bigr)
        -|\Sr|(\hat f-12n\eta)
    }{1-\delta} \notag\\
    &=
    \frac{
        \sum_{x\in D}\alpha_x
        -|\Sr|\Lambda f
    }{1-\delta}.
    \label{eqn:approx-ratio-robust-corrected}
\end{align}

We next prove that $\alpha/\Lambda$ is feasible for the dual of the facility
location LP with opening cost $f$. Fix any regular facility $i$. By
\Cref{lem:dual-feasibility-robust}, convexity of $[z]^+=\max\{z,0\}$, and
$\sum_{j\in D_x}w(j)=1$, we get
\begin{align}
    \sum_{x\in D}[\alpha_x-\Gamma'd^2(i,x)]^+
    &=
    \sum_{x\in D}
    \left[
        \sum_{j\in D_x}w(j)
        \bigl(\alpha^*_j-(\rho-1)R_j^2\bigr)
        -\Gamma'd^2(i,x)
    \right]^+ \notag\\
    &=
    \sum_{x\in D}
    \left[
        \sum_{j\in D_x}w(j)
        \bigl(\alpha^*_j-(\rho-1)R_j^2-\Gamma'd^2(i,j)\bigr)
    \right]^+ \notag\\
    &\le
    \sum_{x\in D}\sum_{j\in D_x}w(j)
    \left[
        \alpha^*_j-(\rho-1)R_j^2-\Gamma'd^2(i,j)
    \right]^+ \notag\\
    &\le
    \hat f+8n\eta.
    \label{eqn:dual-feasibility-original-robust-corrected}
\end{align}

Let
\[
    D^*
    :=
    \{x\in D:\alpha_x>\Lambda d^2(i,x)\}.
\]
If $D^*=\emptyset$, then
\[
    \sum_{x\in D}[\alpha_x-\Lambda d^2(i,x)]^+=0
    \le \Lambda f.
\]
Otherwise,
\begin{align}
    \sum_{x\in D}[\alpha_x-\Lambda d^2(i,x)]^+
    &=
    \sum_{x\in D^*}
    \bigl(\alpha_x-\Lambda d^2(i,x)\bigr) \notag\\
    &=
    \sum_{x\in D^*}
    \bigl(\alpha_x-\Gamma'd^2(i,x)\bigr)
    -20n\eta\sum_{x\in D^*}d^2(i,x) \notag\\
    &\le
    \sum_{x\in D}[\alpha_x-\Gamma'd^2(i,x)]^+
    -20n\eta\sum_{x\in D^*}d^2(i,x) \notag\\
    &\le
    \hat f+8n\eta
    -20n\eta\sum_{x\in D^*}d^2(i,x).
    \label{eqn:dual-feasibility-shifted-intermediate}
\end{align}
Every $x\in D^*$ satisfies
$d^2(i,x)\ge 1$ (because of the assumptions we made in \cref{lem:aspectratio,,lem:aspectratio-euclid2} for the metric and Euclidean cases respectively). Since $D^*\neq\emptyset$, this implies
\[
    \sum_{x\in D^*}d^2(i,x)\ge 1.
\]
Therefore, from \cref{eqn:dual-feasibility-shifted-intermediate},
\[
    \sum_{x\in D}[\alpha_x-\Lambda d^2(i,x)]^+
    \le
    \hat f+8n\eta-20n\eta
    =
    \hat f-12n\eta
    =
    \Lambda f.
\]
Thus, for every regular facility $i$,
\[
    \sum_{x\in D}
    \left[
        \frac{\alpha_x}{\Lambda}-d^2(i,x)
    \right]^+
    \le f.
\]
Together with $\alpha_x\ge 0$, this shows that $\alpha/\Lambda$ is feasible for
the dual of the facility location LP with opening cost $f$. Hence,
\[
    \sum_{x\in D}\frac{\alpha_x}{\Lambda}
    \le
    opt_{\rm LP}(f).
\]

Plugging this into \cref{eqn:approx-ratio-robust-corrected}, we obtain
\begin{align*}
    \sum_{x\in D}d^2(x,S)
    &\le
    \frac{\Lambda}{1-\delta}
    \left(
        \sum_{x\in D}\frac{\alpha_x}{\Lambda}
        -|\Sr|f
    \right)\\
    &\le
    \frac{\Lambda}{1-\delta}
    \left(
        opt_{\rm LP}(f)-|\Sr|f
    \right)\\
    &=
    \frac{\Gamma+201\sqrt{\eps}+20n\eta}{1-\delta}
    \left(
        opt_{\rm LP}(f)-|\Sr|f
    \right).
\end{align*}
This completes the proof.
\end{proof}

Note the weights satisfy $\sum_{j\in \calD_L} w(j) = n$ because each original client has total weight $1$ over all its copies.
In the rest of this subsection, we focus on proving \cref{lem:approx-guarantee-robust,,lem:dual-feasibility-robust}. Because it is clear in the context we will only consider the copies as the clients, we will assume $D=\calD_L$ in the proof and omit the ``original clients'' for simplicity.
We start by showing the dual payment guarantee with respect to $\alpha^*$, and then we prove that $(\alpha^*_j-(\rho-1)R^2_j)_{j\in D}$ is a feasible solution to the dual after appropriate scaling.

\subsubsection{Dual Payment}

\begin{lemma}
    \label{lem:approx-guarantee-robust}
    We have $\sum_{j\in\calD_L} w(j) \cdot \left[ (1-\delta)d^2(j,S) + (\rho-1)R_j^2 \right] + \sum_{i\in\Sr}(\hat f - 12n\eta) \leq \sum_{j\in \calD_L} w(j) \cdot \alpha^*_j$.
\end{lemma}

\begin{proof}
    First, let us show that it suffices to prove that at the end of \recoverA, we have
    \begin{gather}
        \sum_{j\in DC} w(j) \cdot \alpha_j \geq \sum_{j\in DC}w(j) \cdot \left[ (1-\delta)d^2(j, S) + (\rho-1)R_j^2 \right] + \sum_{i\in \Sr}(\hat f - 12n\eta)\,.
        \label{ineq:approxgamma-robust}
    \end{gather}
    Indeed, at the end of \recoverA, $A=\emptyset$. For every $j\in IC$, we have $\alpha_j=(1-\delta)d^2(j, S)$ by the construction of the update steps in Stage 2. Because $R_j = 0$ for $j \in IC$, their contribution naturally satisfies the required bound. Therefore, \cref{ineq:approxgamma-robust} implies the lemma.

    To this end, we prove by induction on $p$ that after the $p$-th phase in \recoverA, \cref{ineq:approxgamma-robust} holds. The base case is $p=0$, i.e., at the beginning of the \recoverA, where $DC, S=\emptyset$. Next, we show that this invariant is preserved after each phase of the algorithm.

    The first case is that we only open free facilities in Stage 1 of the $p$-th phase. In this case, the term $\sum_{i\in \Sr} (\hat{f} - 12n\eta)$ does not change. Suppose $DC'$ are the clients that are moved from $IC\cup A$ to $DC$ after stage 2 of this phase. 
    In an earlier remark, we have argued that there is no copy moved from $DC$ to $IC\cup A$. 
    Consider any $j\in DC'$. Due to step IV) of Stage 1, we have $\alpha_j\geq \rho(1-\delta)d^2(j, h)$, where $h\in S$ is the first open facility satisfying the inequality. Furthermore, for this newly directly connected client, $R_j = \sqrt{1-\delta}d(j, h)$.
    As established in \Cref{sec:log_adaptivity} (see the proof of \cref{lem:approx_guarantee_logadaptive}), we have $(1-\delta)d^2(j,S) + (\rho-1)R_j^2 \le \rho(1-\delta)d^2(j,h) \leq \alpha_j$. 
    Thus, $j$'s contribution to the LHS of \cref{ineq:approxgamma-robust} is no less than that to the RHS. For other clients in $DC\setminus DC'$, their contributions to the LHS do not change while the RHS may only decrease because $S$ only grows. Therefore, the inequality still holds.

    The second case is that we open at least one regular facility $i$ in Stage 1 of the $p$-th phase. The rest of the proof will focus on this case.
    Let us consider each step of stage 1, where we open a facility $i$, i.e., adding it to $S$. 
    Suppose that $(\alpha, S, A, IC, DC, \theta)$ denotes the state right before opening $i$. 
    Suppose that $(S', A', IC', DC')$ denote the set of open facility in the super-sequence $\calH'$ considered in the step $i$ is opened, the set of active clients with respect to $\calD_{new}$ (as in Step I) of the process, defined in terms of $\calH'$), the set of indirectly connected clients with respect to $\calD_{new}$ and the set of directly connected clients with respect to $\calD_{new}$. Let $(\tau_j)_{j\in A'}$ be the $\tau$-variables when opening $i$.
    Because $\calH'$ is a super sequence of open facilities, we have $S'\supseteq S$, $DC'\supseteq DC$ and $A'\subseteq A$.
    
    According to the third bullet of $\eta$-openability (\cref{def:robustopenable}), we have 
    \begin{align*}
        \hat f - 8 n\eta \leq \sum_{j\in D} w(j) \cdot \bid(j)~,
    \end{align*}
    where
    \begin{align*}
        \bid(j) = \begin{cases}
            [\tau_j - \rho(1-\delta) d^2(j,i)]^+ & \text{if $j\in A'$;}\\
            [\alpha_j - \rho(1-\delta) d^2(j,i)]^+ & \text{if $j\in IC'$;}\\
            (1-\delta) [d^2(j,S')-d^2(j,i)]^+ & \text{if $j\in DC'$.}
        \end{cases}
    \end{align*}

    Next, we have a case-by-case discussion on the client's contribution to both sides of \cref{ineq:approxgamma-robust} for each $j\in \calD_L$.
    Our goal is to show that the contribution to the LHS minus the contribution to the RHS is at least $w(j) \cdot \bid(j)$. 
    Because of the third bullet of $\eta$-openability, we have the total contribution to the LHS is greater than that to the RHS by an additive factor of $4n\eta$. Throughout the discussion, for simplicity, we can ignore the $w(j)$ coefficient, because we have such a coefficient on their contributions to both sides.
    \begin{itemize}
        \item If $j\in A'$, this means $j\in A$. 
        \begin{itemize}
            \item If $j$ is not $DC$ after opening $i$, the contributions to both sides are $0$ and we have $\tau_j<\rho(1-\delta)d^2(j,i)$, and thus $\bid(j) = [\tau_j-\rho(1-\delta)d^2(j,i)]^+=0$.
            \item If $j$ is $DC$ after opening $i$, the contributions to both sides are $\tau_j$ and $(1-\delta)d^2(j,i) + (\rho-1)R_j^2$. Because $j\notin DC'$, we have $R_j = \sqrt{1-\delta} d(j,i)$ and thus the difference is $\tau_j-\rho(1-\delta)d^2(j,i)$. Because $j$ becomes $DC$ after opening $i$, this quantity is greater than $0$. We have the difference equal to $\bid(j)$.
        \end{itemize}
        \item If $j\in IC'$, this means $j\in A\cup IC$. 
        \begin{itemize}
            \item If $j$ is not $DC$ after opening $i$, the contributions to both sides are $0$ and we have $\alpha_j<\rho(1-\delta)d^2(j,i)$, and thus $\bid(j) = [\alpha_j-\rho(1-\delta)d^2(j,i)]^+=0$.
            \item If $j$ is $DC$ after opening $i$, the contributions to both sides are $\alpha_j$ and $(1-\delta)d^2(j,i) + (\rho-1)R_j^2$. Because $j\notin DC'$, we have $R_j = \sqrt{1-\delta} d(j,i)$ and thus the difference is $\alpha_j-\rho(1-\delta)d^2(j,i)$. Because $j$ becomes $DC$ after opening $i$, this quantity is greater than $0$. We have the difference equal to $\bid(j)$.
        \end{itemize}
        \item If $j\in DC$, this means $j\in DC'$. Then, 
        the contribution of it to the LHS is $0$. Its contribution to the RHS is $(1-\delta) \cdot(d^2(j,S\cup \{i\})-d^2(j,S))$ because $R_j$ is not changed. Further, we have $d^2(j,S\cup \{i\}) - d^2(j,S) = - [d^2(j,S)-d^2(j,i)]^+ \leq -\bid(j)/(1-\delta)$ because $S\subseteq S'$ and $d^2(j,S)\geq d^2(j,S')$.
        Therefore, the difference is no less than $\bid(j)$.
        \item If $j\in DC'\setminus DC$, we discuss two cases.
        \begin{itemize}
            \item If $j$ is not moved to $DC$ after opening $i$, the contributions to both sides are $0$. Note that $\alpha_j\geq \rho (1-\delta) d^2(j,S')$ because $j\in DC'$. However, as $\alpha_j<\rho (1-\delta) d^2(j,S\cup \{i\})\leq \rho(1-\delta)d^2(j,i)$, we have $d^2(j,S')<d^2(j,i)$. This implies that $\bid(j)=0$. Therefore, the difference equals to $\bid(j)$.
            \item If $j$ is moved to $DC$ after opening $i$, we set $R_j = \sqrt{\alpha_j/\rho}$ at this time. Its contribution to the LHS is $\alpha_j$. Its contribution to the RHS is $(1-\delta) d^2(j,S\cup \{i\}) + \frac{\rho-1}{\rho} \alpha_j$. The difference is then $\alpha_j/\rho - (1-\delta) d^2(j,S\cup \{i\})$. Note that because $j\notin DC$ but is directly connected after opening $i$, $d^2(j,S\cup \{i\})=d^2(j,i)$. 
            The difference further equals to $\alpha_j/\rho - (1-\delta)d^2(j,i)$, which is greater than $0$. Note that $\alpha/\rho \geq (1-\delta) d^2(j,S')$ because $j\in DC'$, we have the difference $\geq (1-\delta) (d^2(j,S')-d^2(j,i))$. Combining these two lower bounds on the difference, we conclude that the difference is lower bounded by $\bid(j)$.
        \end{itemize}
    \end{itemize}

    Then, in stage 2, we only move some clients from $DC$ to $IC$ (step iii)). For each such client $j$, its contribution to the LHS of \cref{ineq:approxgamma-robust} drops by $\alpha_j$ while its contribution to the RHS drops by $(1-\delta)d^2(j,S) + (\rho-1)R_j^2$. Note that our definition ensures that $R_j^2 \geq \alpha/\rho \geq (1-\delta)d^2(j,S)$ or $R_j^2=(1-\delta)d^2(j,i)\geq (1-\delta) d^2(j,S)$ for some $i\in S$. 
    According to \cref{def:valid-collection}, for each such client $j$ moved from DC to IC, we are guaranteed that $\alpha_j < \rho(1-\delta)d^2(j,S)+4\eta \leq (1-\delta)d^2(j,S) + (\rho-1)R_j^2 + 4\eta$. 
    Therefore, the decrease of the LHS is at most the decrease of the RHS plus an additive $4\eta$ per client. Over all such clients, the total extra drop is at most $4\eta \sum w(j) = 4n\eta$.
    Since we accumulated at least $4n\eta$ surplus in stage 1, the overall invariant \cref{ineq:approxgamma-robust} still safely holds.
    This completes the induction and thus the proof.
\end{proof}

\subsubsection{Dual Feasibility}

We prove \Cref{lem:dual-feasibility-robust} for dual feasibility. Throughout this part, we will assume that $\Gamma>\rho+100\sqrt{\epsilon}$.

\begin{lemma}
    \label{lem:dual-feasibility-robust}
    For each regular facility $i$, we have 
    \[\sum_{j\in \calD_L} w(j) \cdot [\alpha^*_j-(\rho-1)R^2_j-(\Gamma+201\sqrt{\eps})\cdot d^2(i,j)]^+ \leq \hat{f} + 8n\eta,\]
    where
    \begin{itemize}
        \item if $d$ is a (general) metric, we choose $\rho=2.5, \Gamma=4.9$; or
        \item if $d$ is a Euclidean metric, we choose $\rho=2, \Gamma = 3+\ln(2)$.
    \end{itemize}
\end{lemma}

The following claim is an analog of \cref{clm:no_overbidding}.

\begin{claim}[No Over-Bidding]
\label{clm:robustnooverbidding}
At any point in the algorithm, for every facility $i_0 \in \mathcal{F}$, we have
$$ \sum_{j \in DC} w(j) \cdot [(1-\delta)d^2(j,S) - d^2(j,i_0)]^+ + \sum_{j \in A \cup IC} w(j) \cdot [\alpha_j - \rho d^2(j,i_0)]^+ \le \hat{f}. $$
\end{claim}
\begin{proof}
\textbf{Stage 1:} Consider an arbitrary facility $i_0 \in \mathcal{F}$. Suppose that a facility $i'$ is opened with dual increases $(\tau_j)_{j \in A}$. The opening of $i'$ has three potential effects on the clients: (1) it may strictly increase the $\alpha$-values of active clients $j \in A \cap B(i', \sqrt{\epsilon\theta})$ to $\tau_j$ and subsequently make them directly connected ($DC$), (2) it may make some existing active or indirectly connected ($IC$) clients directly connected to $i'$, and (3) it may make some active clients indirectly connected. 

Let us analyze how the left-hand side of our constraint is impacted by these changes. First, observe that a client $j$ that was already directly connected before the opening of $i'$ can only lower its contribution, since the distance $d^2(j, S)$ can only decrease when $i'$ is added to $S$.

For a client $j$ that was active or indirectly connected and now becomes a direct connect, its contribution changes from%
$w(j) \cdot [\alpha_j - \rho d^2(j,i_0)]^+$ %
to $w(j) \cdot [(1-\delta)d^2(j,S \cup \{i'\}) - d^2(j,i_0)]^+$. 

If $j$ is active and $\tau_j>\alpha_j=\theta$, this means that $j$ is directly connected to facility $i'$, which has just been opened in the step, such that $d^2(j,i')\leq \eps \theta$.
This implies $j$ will be directly connected to $i'$ and $(1-\delta) \rho d^2(j,S \cup \{i'\}) = (1-\delta) \rho d^2(j,i') \leq  \alpha_j$ because $\eps$ is sufficiently small.
Hence, we have $$[(1-\delta)d^2(j,S \cup \{i'\}) - d^2(j,i_0)]^+ \leq [(1-\delta) \rho d^2(j,S \cup \{i'\}) - \rho d^2(j,i_0)]^+ \leq [\alpha_j - \rho d^2(j,i_0)]^+.$$

Next, we consider the case $j$ is active or indirectly connected but $\tau_j=\alpha_j$.
Because the client became directly connected, its dual value must have dominated the connection cost, meaning%
$\alpha_j \ge \rho(1-\delta)d^2(j,i')$. Since $\rho > 1$ and $[x]^+ \ge 0$, we have:
\begin{align*}
[(1-\delta)d^2(j,S \cup \{i'\}) - d^2(j,i_0)]^+ &\le \left[ \frac{\tau_j}{\rho} - d^2(j,i_0) \right]^+ \\
&\le \left[ \tau_j - \rho d^2(j,i_0) \right]^+.
\end{align*}
Thus, its contribution does not increase. 

With the above discussions, we prove that the invariant holds throughout Stage 1. 

\textbf{Stage 2:} In the second stage, we have three steps: decreasing the $\alpha$-values of clients in $IC$ to $(1-\delta)d^2(j,S)$; increasing the $\alpha$-values of clients in $A$ to $\min\{(1+\epsilon^3)\theta, (1-\delta)d^2(j,S)\}$; and moving some clients from $DC$ to $IC$. 

The first step can only decrease the left-hand side of the claim. The third step changes the contribution of a moved client from $[(1-\delta)d^2(j,S) - d^2(j,i_0)]^+$ to $[\alpha_j - \rho d^2(j,i_0)]^+ = [(1-\delta)d^2(j,S) - \rho d^2(j,i_0)]^+$. Since $\rho > 1$, this also strictly decreases its contribution. Hence, it suffices to show that the continuous increase in the second step does not violate the claim.

Assume towards contradiction that at the end of the phase, if we increase $\alpha_j$ of every $j \in A$ to $\min\{(1+\epsilon^3)\theta, (1-\delta)d^2(j,S)\}$, the claim is violated for a non-empty subset of facilities $F'$.

We select a ``minimal'' such counter-example: let $\tau' \le (1+\epsilon^3)\theta$ be the smallest value such that if we set $\alpha'_j := \min\{\tau', (1-\delta)d^2(j,S)\}$, it satisfies
\begin{align*}
\sum_{j \in A} w(j)[\alpha'_j - \rho d^2(j,i)]^+ &+ \sum_{j \in IC} w(j)[\alpha_j - \rho d^2(j,i)]^+ \\
&+ \sum_{j \in DC} w(j)[(1-\delta)d^2(j,S) - d^2(j,i)]^+ = \hat{f}
\end{align*}
for some $i \in F'$. We remark that $\tau' \ge \theta$ since the constraint was strictly satisfied at the end of Phase 1. Furthermore, $i \notin S$ is not yet open; if it were open, there would be no active client $j \in A$ for which $\alpha'_j - \rho d^2(j,i)$ is strictly positive (since $\alpha'_j \le (1-\delta)d^2(j,S) \le (1-\delta)d^2(j,i) < \rho d^2(j,i)$). Hence, the increase in $\alpha$-values cannot cause $i$ to violate the constraint if it were already opened, which contradicts $i \in F'$.

We now show that this facility $i$, combined with dual values $\tau_j = \alpha'_j$ for $j \in A \cap B(i, \sqrt{\epsilon\theta})$ and $\tau_j = \alpha_j$ for $j \in A \setminus B(i, \sqrt{\epsilon\theta})$, satisfies the conditions of $\eta$-openability. In other words, $i$ is an openable facility that was not yet opened, which contradicts the proper completion of the first stage's while-loop. 

We verify the conditions of $\eta$-openability one-by-one. The first two bullets are directly satisfied by our choice of $\tau_j \le \tau'$ and the definition of $\alpha'_j$. For the third bullet (facility $i$ is paid for):
\begin{align*}
&\sum_{j \in A} w(j) \cdot [\tau_j - \rho(1-\delta)d^2(j,i)]^+ + \sum_{j \in IC} w(j) \cdot [\alpha_j - \rho(1-\delta)d^2(j,i)]^+ \\
&\quad + \sum_{j \in DC} w(j) \cdot [(1-\delta)d^2(j,S) - (1-\delta)d^2(j,i)]^+ \\
&= \sum_{j \in A \cap B(i, \sqrt{\epsilon\theta})} w(j) \cdot [\alpha'_j - \rho(1-\delta) d^2(j,i)]^+ + \sum_{j \in A \setminus B(i, \sqrt{\epsilon\theta})} w(j) \cdot [\alpha_j - \rho(1-\delta)d^2(j,i)]^+ \\
&\quad + \sum_{j \in IC} w(j) \cdot [\alpha_j - \rho(1-\delta) d^2(j,i)]^+ + \sum_{j \in DC} w(j) \cdot [(1-\delta)d^2(j,S) - (1-\delta)d^2(j,i)]^+ \\
&\ge \sum_{j \in A \cap B(i, \sqrt{\epsilon\theta})} w(j) \cdot [\alpha'_j - \rho d^2(j,i)]^+ + \sum_{j \in A \setminus B(i, \sqrt{\epsilon\theta})} w(j) \cdot [\alpha'_j - \rho d^2(j,i)]^+ \\
&\quad + \sum_{j \in IC} w(j) \cdot [\alpha_j - \rho d^2(j,i)]^+ + \sum_{j \in DC} w(j) \cdot [(1-\delta)d^2(j,S) - d^2(j,i)]^+ \\
&= \hat{f}. %
\end{align*}
Here, for the subset outside the ball $A \setminus B(i, \sqrt{\epsilon\theta})$, we used the fact that $\alpha_j - \rho(1-\delta)d^2(j,i) \ge (1+\epsilon^2)\alpha_j - \rho d^2(j,i) \ge \alpha'_j - \rho d^2(j,i)$ when $d^2(j,i) \ge \epsilon\theta$ (recalling that $\alpha_j = \theta$ for all active clients). 

It remains to verify the fourth bullet of $\eta$-openability (dual feasibility). Suppose toward contradiction that there is a facility $i_0$ such that
$$ \sum_{j \in DC} w(j)[(1-\delta)d^2(j,S) - d^2(j,i_0)]^+ + \sum_{j \in IC} w(j)[\alpha_j - \rho d^2(j,i_0)]^+ + \sum_{j \in A} w(j)[\tau_j - \rho d^2(j,i_0)]^+ > \hat{f}. %
$$
Because $\tau_j \le \alpha'_j$ for all $j \in A$, this implies:
\begin{align*}
&\sum_{j \in DC} w(j)[(1-\delta)d^2(j,S) - d^2(j,i_0)]^+ + \sum_{j \in IC} w(j)[\alpha_j - \rho d^2(j,i_0)]^+ + \sum_{j \in A} w(j)[\alpha'_j - \rho d^2(j,i_0)]^+ \\
&\ge \sum_{j \in DC} w(j)[(1-\delta)d^2(j,S) - d^2(j,i_0)]^+ + \sum_{j \in IC} w(j)[\alpha_j - \rho d^2(j,i_0)]^+ + \sum_{j \in A} w(j)[\tau_j - \rho d^2(j,i_0)]^+ \\
&> \hat{f}.  %
\end{align*}
which directly contradicts the minimality of $\tau'$. Thus, $i$ is openable, representing a contradiction.
\end{proof}

Next, we fix an arbitrary regular facility $i \in \mathcal{F}$ and show that the dual-feasibility for $i$ is satisfied at the end of the robust algorithm.
We say that facility $i$ is frozen if there exists some open facility $i_0$ such that $$d(i,i_0)\leq 2\cdot \sqrt{\eps\theta}~.$$ 

\paragraph{Modified $\alpha$-values.} For analysis purposes, we introduce a modification of the $\alpha$-values for each client (copy) $j \in \calD_L$ in terms of the fixed facility $i$. We denote this modified $\alpha$-value as $\alpha^c_j$.
This modified $\alpha^c$-value will be used to write the analog of the dual constraints. 

The modifications happen only when we open a facility $i_0$ that does not freeze $i$, i.e., when $d(i,i_0)>2\sqrt{\eps\theta}$.
More specifically, at the stage with time $\theta$, if we open such a facility $i_0$ that does not freeze $i$, we keep $\alpha^c_j$ as the value before opening $i_0$. That is, we do not increase $\alpha^c_j$ to $\tau_j$ if $j\in B(i_0,\sqrt{\epsilon\theta})$ (step 1b) for such $i_0$. In other steps (step 1b for a facility $i_0$ that freezes $i$ or step 2), we still modify $\alpha^c_j$ in the same way we modify $\alpha_j$. 

In the rest of this subsection, we will consider $\alpha^c$ instead of $\alpha^*$ in writing the constraints. 
To ensure that the $\alpha^*$-value can ensure the dual feasibility, we first show that $\alpha_j^c, \alpha_j^*$ are close to each other for every client $j \in \calD_L$.

The following lemma and its proof are identical to \Cref{lem:alphaclose}.

\begin{lemma}
    For any client $j\in \calD_L$, we have $\alpha^*_j-\alpha^c_j \leq \epsilon^2\cdot d^2(j,i)$.
\end{lemma}

Now, we assume that 
\begin{align}
    \label{eqn:Dstar-robust}
    D^*= \{j\in \calD_L: \alpha^c_j - (\rho-1)R_j^2 > \Gamma d^2(j,i)\}~.
\end{align} 
Then, it suffices to show that 
\begin{align*}
\sum_{j \in D^*} w(j) \cdot (\alpha_j^c - (\rho-1)R_j^2)  \le \hat{f} + 8n\eta + (\Gamma+200\sqrt{\eps})  \cdot \sum_{j \in D^*} w(j) \cdot d^2(j,i)~.
\end{align*}

\paragraph{Establishing the constraints.} 
Up to this point, our arguments closely mirror the dual feasibility proof for the log-adaptive algorithm in \Cref{sec:log_adaptivity}. However, the remaining part of the proof needs some changes from \Cref{sec:log_adaptivity}, due to the fact that a facility may be $\eta$-openable with respect to a super-sequence instead of the sequence that the algorithm actually executes.

Let $i_0$ be the facility that freezes $i$, which opens in (stage 1 of) phase $p$. Let $\theta = (1+\epsilon^3)^{p-1}$ be the active dual parameter at the beginning of phase $p$. Let $\calH$ be the sequence of opened facilities up to $i_0$, and let $\mathcal{H'} \supseteq \mathcal{H}$ be the super-sequence with respect to which $i_0$ is $\eta$-openable. 

As in \Cref{sec:log_adaptivity}, we order the clients in $D^*$ according to the order they are removed from $A$ (in the algorithm), breaking ties in the increasing order of their $\alpha^c$-values. The only additional rule is for the clients who are in $A$ right before $i_0$ is opened in $\calH$; let us call this set of active clients $A$. By \Cref{clm:robustfrozenstay}, they are the last several clients in the above order of $D^*$. Let $A^F \subseteq A$ be the clients in $D^*$ who are still active right before $i_0$ is opened according to the collection of copies induced by \text{\retrieveD}$(\calD_{p-1}, H'_p)$. Let $A'=(D^* \cap A)\setminus A^F$ be the clients that become inactive during the extra prefix in $\calH'$. In our ordering for $D^*\cap A$, we place $A'$ first and $A^F$ after, sorting within each group in the increasing order of their $\alpha^c_j$ values. 

Let $a$ be the index of the last client in $D^* \setminus A^F$, meaning the clients in $[a+1:s]$ are exactly $A^F$. For each client $t \in [a]$, we define $S^t, DC^t,$ and $IC^t$ to be the sets of opened facilities, directly connected clients in $[t-1]$, and indirectly connected (or active) clients in $[t-1]$, evaluated at the beginning of the exact stage $t$ becomes inactive. 
For each client $t\in [a+1:s]$, we define $S^t, DC^t,$ and $IC^t$ to be the sets of opened facilities, directly connected clients in $[t-1]$, and indirectly connected (or active) clients in $[t-1]$, evaluated right before $t$ becomes inactive in $\calH'$. 
Similarly, let $S^F, IC^F,$ and $DC^F$ be the set of opened facilities, indirectly connected, and directly connected clients right before $i_0$ is opened evaluated with respect to the super-sequence $\calH'$. 

We remark that the definitions of $S^F, IC^F, DC^F,$ and $A^F$ here depend on the super-sequence $\calH'$, while in \Cref{sec:log_adaptivity}, they only depend on the sequence $\calH$ executed by the algorithm. To complete the proof, we will now apply the unified constraint lemmas (\Cref{lem:dual-feasibility-simpler-metric,lem:dual-feasibility-simpler-euclidean,lem:dual-feasibility-simpler-euclidean-special}) evaluated over this super-sequence state.

The two following claims are analogs of Claims \ref{clm:notfrozen} and \ref{clm:frozenstay}. The proof of \cref{clm:robustnotfrozen} is identical to that of \ref{clm:notfrozen}. For \cref{clm:robustfrozenstay}, we strengthen the argument to ``client $j$ becomes directly connected'' upon ``client $j$ becomes active''. 

\begin{claim}
\label{clm:robustnotfrozen}
For any $t \in D^*$ that becomes inactive strictly before $i$ becomes frozen,
\[
\sum_{j \in DC^t} w(j) [(1-\delta)d^2(j,S^t) - d^2(j,i)]^+ 
+ \sum_{j \in IC^t} w(j) [\alpha_j^c - \rho d^2(j,i)]^+ 
+ \sum_{j > t} w(j) [\alpha_t^c - \rho d^2(j,i)]^+ \le \hat{f} + 8n\eta.
\]
\end{claim}

\begin{claim}
\label{clm:robustfrozenstay}
    If $i$ is frozen by $i_0$ with $(\tau_j)_{j\in A}$, then for every $j\in D^*\cap A$, $j$ becomes directly connected when opening $i_0$ and $\alpha_j^c\leq\tau_j$.
\end{claim}
\begin{proof}
    Applying the same proof of \cref{clm:frozenstay}, we can show that $j$ is inactive. Next, we present a short argument for why it is directly connected for our choices of $\Gamma$.

    Note that $\theta \leq \alpha^c_j$ and $\alpha^c_j\geq \Gamma \cdot d^2(j,i)$ because $j\in D^*\cap A$. 
    Note that $d(j,i_0) \leq  d(j,i)+d(i,i_0) \leq d(j,i) + 2\sqrt{\eps\theta}$. 
    This implies that 
    \begin{align*}
        d^2(j,i_0) \leq d^2(j,i) + 4\eps\theta + 4\sqrt{\eps\theta} d(j,i) &\leq \alpha^c_j/\Gamma + 4\eps\theta + 4\sqrt{\eps\theta} \sqrt{\alpha^c_j/\Gamma}
        \\
        &\leq \alpha^c_j /\Gamma + 4\eps \alpha_j^c + 4\sqrt{\epsilon/\Gamma} \alpha^c_j 
        \\
        &\leq (1/\Gamma + 8\sqrt{\eps})\alpha^c_j.
    \end{align*}
    Because $\Gamma>\rho + 100\sqrt{\eps}$, $\rho\in [2,3]$ and $\eps$ is sufficiently small, we have $\alpha^c_j\geq \rho d^2(j,i_0)$.
\end{proof}

The following claim is the analog of \Cref{clm:3rdconstraint}.

\begin{claim}
\label{clm:robust3rdconstraint}
For any $t\in [s]$,
\[
\sum_{j\in DC^t}w(j)((1-\delta)d^2(j, S^t) - d^2(j, i)) + \sum_{j\in IC^t}w(j)(\alpha_j^c - \rho d^2(j, i)) + \sum_{j \geq t}w(j)(\alpha_t^c-\rho d^2(j,i)) \le \hat{f} + 8n\eta.
\]
\end{claim}
\begin{proof}
    Consider the point in the execution evaluated against the super-sequence $\calH'$ at which $t$ becomes inactive. As established by \Cref{clm:robustfrozenstay}, if a facility $i_0$ freezes $i$, every client $j \in D^*$ that is active right before $i_0$ is opened becomes inactive during the exact step $i_0$ is opened. Therefore, no client in $D^*$ can become inactive \emph{after} $i_0$ is opened in $\calH'$. This partitions our ordered clients $t \in D^*$ into exactly two cases:

    \paragraph{Case 1: $t \in [a]$.}
    Client $t$ becomes inactive strictly before $i_0$ is opened in $\calH'$. Because every facility in the valid super-sequence $\calH'$ is $\eta$-openable, the No Over-Bidding invariant (\Cref{clm:robustnooverbidding}) rigorously holds at every sequential step of its execution. We evaluate this depending on the stage in which $t$ becomes inactive:
    
    \textbf{Stage 1:} Suppose $t$ becomes inactive due to the discrete opening of a facility $h_k \in \calH'$. Because $h_k$ precedes $i_0$ (the facility that freezes $i$), this event happens strictly before $i$ is frozen. Therefore, $h_k$ does not freeze $i$. By our definition of $\alpha^c$, no client in $D^*$ 
    increases its $\alpha^c$-value, 
    meaning the modified dual $\alpha^c$ does not incorporate any Step 1b increases from $h_k$. Thus, $\alpha_t^c$ is exactly the active dual parameter $\theta$ right before $h_k$ is opened. We evaluate the No Over-Bidding constraint at the state right before $h_k$ opens:
    \[
    \sum_{j \in DC^t} w(j)[(1-\delta)d^2(j,S^t) - d^2(j,i)]^+ + \sum_{j \in IC^t \cup A^t} w(j)[\alpha_j - \rho d^2(j,i)]^+ \le \hat{f} + 8n\eta.
    \]
    For $j \in DC^t$, the term trivially bounds $(1-\delta)d^2(j,S^t) - d^2(j,i)$. For $j \in IC^t$, the modified dual $\alpha^c$ never exceeds the true dual prior to a freezing step, so $\alpha_j \ge \alpha_j^c$. For $j \ge t$, because our ordering places clients by the time they become inactive, they are still active in $A^t$ at this moment, meaning their actual dual is $\alpha_j = \theta = \alpha_t^c$. Thus, $[\alpha_j - \rho d^2(j,i)]^+ \ge \alpha_t^c - \rho d^2(j,i)$.

    \textbf{Stage 2:} Suppose $t$ becomes inactive due to the continuous update of $\alpha$-values in Stage 2. We evaluate the No Over-Bidding constraint at the exact moment $t$ hits its threshold. At this moment, the modified dual exactly tracks the true dual, so $\alpha_t^c = \alpha_t$. For $j \in IC^t$, we again have $\alpha_j \ge \alpha_j^c$. For $j \ge t$, these clients either remain active or become inactive at the same moment but are sorted after $t$ (meaning $\alpha_j^c \ge \alpha_t^c$). In both cases, their true dual at this moment satisfies $\alpha_j \ge \alpha_t = \alpha_t^c$. Thus, $[\alpha_j - \rho d^2(j,i)]^+ \ge \alpha_t^c - \rho d^2(j,i)$.

    In both Stage 1 and Stage 2, the subsets $DC^t, IC^t,$ and $\{j \in D^* : j \ge t\}$ are mutually disjoint. Restricting the summations to these subsets and dropping the non-negative $[\cdot]^+$ operators rigorously establishes the algebraic lower bound:
    \[
    \sum_{j \in DC^t} w(j) ((1-\delta)d^2(j,S^t) - d^2(j,i)) + \sum_{j \in IC^t} w(j) (\alpha_j^c - \rho d^2(j,i)) + \sum_{j \ge t} w(j) (\alpha_t^c - \rho d^2(j,i)) \le \hat{f} + 8n\eta.
    \]

    \paragraph{Case 2: $t \in [a+1 : s]$.}
    Client $t$ remains active right before $i_0$ is opened in $\calH'$ (meaning $t \in A^F$) and becomes inactive exactly when $i_0$ is opened. Because all clients in $A^F$ become inactive simultaneously at this step, and ties in our ordering are broken by increasing $\alpha^c$-value, we are guaranteed that $\alpha_t^c \le \alpha_j^c$ for all $j \ge t$. Furthermore, evaluated right before $i_0$ is opened in $\calH'$, we identically have $S^t = S^F$, $DC^t \subseteq DC^F$, and $IC^t = IC^F \cup \{j \in A^F : j < t\}$.

    Consider the fourth bullet of $\eta$-openability (dual feasibility) for $i_0$ evaluated at the fixed facility $i$, just before $i_0$ is opened in $\calH'$:
    \[
    \sum_{j \in DC^F} w(j) [(1-\delta)d^2(j,S^F) - d^2(j,i)]^+ 
    + \sum_{j \in IC^F} w(j) [\alpha_j - \rho d^2(j,i)]^+ 
    + \sum_{j \in A^F} w(j) [\tau_j - \rho d^2(j,i)]^+ \le \hat{f} + 8n\eta.
    \]
    
    Since the $[\cdot]^+$ terms are strictly non-negative, restricting the summations to disjoint subsets of interest maintains the upper bound. We then drop the non-negative operators entirely to establish a strict algebraic lower bound:
    \begin{itemize}
        \item For $j \in DC^t \subseteq DC^F$: Because $S^t = S^F$, we have: 
        $[(1-\delta)d^2(j,S^F) - d^2(j,i)]^+ \ge (1-\delta)d^2(j,S^t) - d^2(j,i)$.
        
        \item For $j \in IC^t \cap IC^F$: The modified dual never exceeds the true dual prior to freezing, so $\alpha_j \ge \alpha_j^c$. Thus:
        $[\alpha_j - \rho d^2(j,i)]^+ \ge \alpha_j - \rho d^2(j,i) \ge \alpha_j^c - \rho d^2(j,i)$.
        
        \item For $j \in IC^t \cap A^F$: These clients are in $A^F$ but ordered before $t$. By \Cref{clm:robustfrozenstay}, we have $\tau_j \ge \alpha_j^c$. Evaluated in the summation over $A^F$, we have:
        $[\tau_j - \rho d^2(j,i)]^+ \ge \tau_j - \rho d^2(j,i) \ge \alpha_j^c - \rho d^2(j,i)$.
        (Combining this with the previous bullet perfectly reconstructs the $\alpha_j^c - \rho d^2(j,i)$ term for all $j \in IC^t$).
        
        \item For $j \ge t$: These clients are exactly $\{j \in A^F : j \ge t\}$. By \Cref{clm:robustfrozenstay}, $\tau_j \ge \alpha_j^c$. Applying our tie-breaking rule, $\alpha_j^c \ge \alpha_t^c$, meaning $\tau_j \ge \alpha_t^c$. Evaluated in the $A^F$ sum, we have:
        $[\tau_j - \rho d^2(j,i)]^+ \ge \tau_j - \rho d^2(j,i) \ge \alpha_t^c - \rho d^2(j,i)$.
    \end{itemize}
    
    Because all evaluated subsets are disjoint and strictly contained within the original $\eta$-openability sums, summing these lower bounds yields:
    \[
    \sum_{j \in DC^t} w(j) ((1-\delta)d^2(j,S^t) - d^2(j,i)) 
    + \sum_{j \in IC^t} w(j) (\alpha_j^c - \rho d^2(j,i)) 
    + \sum_{j \ge t} w(j) (\alpha_t^c - \rho d^2(j,i)) \le \hat{f} + 8n\eta.
    \]
    This matches the required inequality, completely resolving the case.
\end{proof}

For any client $t \in [s]$ and $j \in DC^t$, define
\[
r_{jt} := \sqrt{1-\delta}d(j,S^t).
\]
Next, we prove that $r_{jt}\leq R_j$ for every $t\in [s]$ and $j\in DC^t$.
\begin{lemma}
\label{lem:rjt_leq_Rj}
    For every $t\in [s]$ and $j\in DC^t$, we have $r_{jt}\leq R_j$. 
\end{lemma}
\begin{proof}
    We discuss two cases. 

    The first case is if $t\in [a]$. 
    In this case, if $j\in DC^t$, it means $j\in DC$ and $(r_{jt})^2 = (1-\delta)d^2(j,S)$ at the beginning of the stage $t$ becomes inactive. 
    Therefore, at the beginning of the stage, we have defined $R_j$ as $R_j=\sqrt{1-\delta} d(j,h)$ for some $h\in S$ or $R_j=\sqrt{\alpha_j/\rho}$. In both case, we can derive $r_{jt}\leq R_j$ because $d^2(j,S)\leq d^2(j,h)$ and $(1-\delta)d^2(j,S)\leq \alpha_j/\rho$.

    The second case is if $t>a$. 
    Let $S$ be the set of open facilities right before $i$ is frozen. We have $S=S^t$ by our earlier definition for this case. If $j\in DC^t$, there are two subcases:
    \begin{itemize}
        \item Suppose that $j\in DC$ right before $i$ is frozen. This means $(r_{jt})^2 = (1-\delta)d^2(j,S)$, and $R_j=\sqrt{1-\delta} d(j,h)$ or $R_j=\sqrt{\alpha_j/\rho}$. As discussed earlier in this proof, we have $r_{jt}\leq R_j$.
        \item Otherwise (corresponding to the case $j$ is directly connected only when it is evaluated in the super sequence $\calH'$), we have $R_j=\sqrt{\alpha_j/\rho}$ because $j$ becomes directly connected immediately after $i$ is frozen (\cref{clm:robustfrozenstay}). We also have $(1-\delta) d^2(j,S^t)\leq \alpha_j/\rho$ because $j$ is directly connected with respect to $S^t$. Therefore, we have $r_{jt}\leq R_j$ by the definition of $r_{jt}$.
    \end{itemize}
    A final remark is that in the proof, we are considering the final real $\alpha$-values instead of the $\alpha^c$-values.
\end{proof}

\subsubsection*{General metric}

Now, we will show \Cref{lem:dual-feasibility-robust} for $\rho = 2.5$ and $\Gamma = 4.9$. To this end, we show that $(IC^t), (DC^t), (\alpha_j^c), (R_j), (r_{jt})$ satisfy all the constraints of \Cref{lem:dual-feasibility-simpler-metric} (with the right-hand side of the capacity constraint replaced by $\hat{f} + 8n\eta$). 

The proof of this is strictly identical to the one in \Cref{sec:log_adaptivity}, with only two minor adaptations due to the super-sequence:
First, the last constraint (the capacity bound) is now directly provided by \Cref{clm:robust3rdconstraint} with the updated right-hand side $\hat{f} + 8n\eta$.
Second, we will use \cref{lem:rjt_leq_Rj} to argue that $r_{jt}\leq R_j$ still holds. 

The remaining bounds, including $\alpha_t^c \le (r_{jt} + d(j,t))^2$, follow exactly the same triangle inequality logic as in the log-adaptive part because $\alpha_t^c \le (1-\delta)d^2(t, S^t)$ remains valid for all $t$ (for $t \in [a+1:s]$, we have $\alpha_t^c \le \tau_t \le (1-\delta)d^2(t, S^F) = (1-\delta)d^2(t, S^t)$ by the second bullet of $\eta$-openability). 

Hence, applying \Cref{lem:dual-feasibility-simpler-metric} with the adjusted right-hand side, we get:
\[
\sum_{j\in [s]}w(j)\cdot (\alpha_j^c - (\rho-1)R_j^2)\le \hat{f} + 8n\eta + \Gamma\cdot \sum_{j\in [s]}w(j)\cdot d^2(j,i).
\]
This completes the proof of \Cref{lem:dual-feasibility-robust} for $\rho=2.5$ and $\Gamma = 4.9$.

\subsubsection*{Euclidean metric}

Now, we will show \Cref{lem:dual-feasibility-robust} for $\rho=2$, $\Gamma=3+\ln 2$, and $d$ a Euclidean metric. Let $a$ be the index of the last client that becomes inactive strictly before $i_0$ is opened in the super-sequence $\calH'$.

The following claim is the analog of \Cref{clm:2nd-constraint-euclidean}.

\begin{claim}
\label{clm:robust2nd-constraint-euclidean}
For each $t\in [a]$, $j\in DC^t$ and $\ell\in IC^t \cup \{t\}$, we have:
\[
    \alpha_{\ell}^c \le (r_{jt} + d(j, \ell))^2.
\]
\end{claim}
\begin{proof}
    We first prove the bound for $t \in [a]$. Consider the exact state in the execution evaluated at the beginning of the stage where $t$ becomes inactive. By definition, $S^t$ is the set of open facilities at this state. 
    
    Consider any client $\ell \in IC^t \cup \{t\}$. As $\ell \le t$, $\ell$ becomes inactive at this state or earlier. We prove that $\alpha^c_\ell \le (1-\delta)d^2(\ell, S^t)$ by discussing two cases:
    \begin{itemize}
        \item If $\ell$ became inactive strictly before the phase containing $t$'s inactivation, then $\ell$ is indirectly connected at the beginning of this phase. Because of the Stage 2 continuous updates at the end of the previous phase, we have $\alpha^c_{\ell} \le \alpha_{\ell} = (1-\delta)d^2(\ell,S_{prev})$. Because the set of open facilities only grows, $S_{prev} \subseteq S^t$, meaning the distance can only decrease. Thus, $\alpha^c_{\ell} \le (1-\delta)d^2(\ell,S^t)$.
        
        \item Otherwise, $\ell$ becomes inactive during the same phase as $t$ (this includes the case $\ell=t$). Let $\theta_t$ be the active dual parameter at the beginning of this phase. Since $\ell$ is active at this time, $\alpha^c_\ell \le \alpha_\ell = \theta_t$. We must strictly have $\theta_t \le (1-\delta)d^2(\ell, S^t)$, because if $\theta_t > (1-\delta)d^2(\ell, S^t)$, $\ell$ would have been moved to $IC$ during the Stage 2 continuous update of the previous phase (when the threshold exceeded the connection cost), contradicting that $\ell$ is still active at the beginning of the current phase. Therefore, $\alpha^c_{\ell} \le \theta_t \le (1-\delta)d^2(\ell,S^t)$.
    \end{itemize}

    We have established that $\alpha_\ell^c \le (1-\delta)d^2(\ell, S^t)$.
    Because the triangle inequality holds for the Euclidean metric $d$, we strictly have $d(\ell, S^t) \le d(j, S^t) + d(j, \ell)$. Taking the square root of our established bound, we get:
    \begin{align*}
        \sqrt{\alpha^c_\ell} &\le \sqrt{1-\delta} d(\ell, S^t) \\
        &\le \sqrt{1-\delta} (d(j, S^t) + d(j, \ell)) \\
        &= r_{j,t} + \sqrt{1-\delta}d(j, \ell).
    \end{align*}
    
    Since $\delta > 0$, we have $\sqrt{1-\delta} < 1$. Because metric distances are strictly non-negative, replacing the scaled distance with the full distance preserves the inequality:
    \[ \sqrt{\alpha^c_\ell} \le r_{j,t} + d(j, \ell). \]
    
    Squaring both sides yields $\alpha^c_{\ell}\leq (r_{j,t}+d(j,\ell))^2$,
    completing the proof.
\end{proof}

We will make several cases according to how $i$ gets frozen.

\paragraph{Case 1: All clients in $D^*$ become inactive strictly before $i$ is frozen.} 
In this case, we have $a=s$. Let us show that $(IC^t), (DC^t), (\alpha_j^c), (R_j),$ and $(r_{jt})$ satisfy all the constraints of \Cref{lem:dual-feasibility-simpler-euclidean}, where the right-hand side of the capacity bound is adjusted to $\hat{f} + 8n\eta$.

First, constraint (1) holds because of \cref{lem:rjt_leq_Rj}.
Second, since $a=s$, \Cref{clm:robust2nd-constraint-euclidean} implies that constraint (2) is strictly satisfied for all $t \in [s]$. 
Finally, \Cref{clm:robust3rdconstraint} directly establishes constraint (3) with the relaxed right-hand side $\hat{f} + 8n\eta$. 

Hence, we can seamlessly apply \Cref{lem:dual-feasibility-simpler-euclidean} (replacing $\hat{f}$ with $\hat{f} + 8n\eta$) to obtain:
\[
    \sum_{j\in [s]}w(j)\cdot(\alpha_j^c-(\rho-1)R_j^2)\le \hat{f} + 8n\eta + \Gamma\cdot \sum_{j\in [s]}w(j)\cdot d^2(j,i).
\]
This perfectly matches our target inequality, completing the proof for this case.

\paragraph{Case 2: Some clients in $D^*$ become inactive when $i$ is frozen.}
In this case, we assume that $i_0$ freezes $i$ in the super-sequence $\calH'$. Let $IC^F$ be the set of indirectly connected clients right before $i_0$ is opened in $\calH'$, and let $A^F$ be the set of active clients in $D^*$ right before $i_0$ is opened (meaning $a < s$). We define:
\[
    L := \{j \in IC^F : d(j,i) > \sqrt{\epsilon\theta}\}.
\]
Note that $IC^F$ is disjoint from $A^F$, meaning $j \notin IC^F$ for any $j \in [a+1 : s]$.

We will further discuss two subcases to prove \Cref{lem:dual-feasibility-robust}.

\paragraph{Subcase 2(a): $[a+1:s]\cap B(i_0,\sqrt{\eps\theta})=\emptyset$ or $\sum_{j\in L}w(j)>\eps\cdot\sum_{j\in [a+1:s]}w(j)$.} 
We redefine $DC^s$ and $IC^s$ as the respectively directly connected and active or indirectly connected client in $[s-1]$ at the beginning of the stage $s$ becomes inactive. First, let us bound the total increment of $\alpha_j^c$ when $i_0$ is opened.

The following claim is the analog of \Cref{clm:increment-bound}.

\begin{claim}
\label{clm:robustincrement-bound}
The total increment of $\alpha^c_j$ for $j \in D^*$ when $i_0$ is opened is at most $\epsilon \sum_{j \in D^*} w(j) d^2(j,i)$.
\end{claim}
\begin{proof}
    If $[a+1:s] \cap B(i_0, \sqrt{\eps\theta}) = \emptyset$, or if no active client in $[a+1:s]$ strictly increases its $\alpha^c$-value when $i_0$ is opened, the total increment for all $j \in D^*$ is exactly $0$. Because weights and squared distances are non-negative, the bound $0 \le \epsilon \sum_{j\in D^*} w(j) d^2(j,i)$ holds trivially.
    
    Next, we consider the case where $\sum_{j\in L}w(j) > \epsilon \cdot \sum_{j\in [a+1:s]}w(j)$.

    Let $U = [a+1:s] = A^F$ denote the set of clients in $D^*$ that become inactive exactly when $i$ is frozen (i.e., when $i_0$ is opened in $\calH'$). For any client $j \in [1:a]$, $j$ became inactive strictly before this step. By the algorithm's rules, the dual values of inactive clients are never increased during Stage 1. Therefore, the total increment of $\alpha^c_j$ across all $j \in D^*$ is exactly the sum of the increments for the active clients in $U$.

    Right before $i_0$ is opened, every $j \in U$ is active, meaning its dual value is exactly $\alpha_j = \theta$. When $i_0$ is opened, its modified dual value $\alpha^c_j$ is bounded by $\tau_j$. By the second bullet of $\eta$-openability (\Cref{def:robustopenable}), the maximum increased value is bounded by $\tau_j \le (1+\epsilon^3)\theta$. Thus, for every $j \in U$, the increment is bounded by $\epsilon^3 \theta$.
    
    Summing this over all $j \in U$, the total increment for $D^*$ is bounded by:
    $$ \sum_{j \in U} w(j)(\alpha^c_j - \alpha_j) \le \epsilon^3 \theta \sum_{j \in U} w(j). $$

    By our Subcase 2(a) assumption, we have $\sum_{j\in L}w(j) > \epsilon \sum_{j\in U}w(j)$, which algebraically rearranges to:
    $$ \sum_{j \in U} w(j) < \frac{1}{\epsilon} \sum_{j \in L} w(j). $$

    Substituting this weight bound into our total increment strictly yields:
    $$ \sum_{j \in U} w(j)(\alpha^c_j - \alpha_j) < \epsilon^3 \theta \left( \frac{1}{\epsilon} \sum_{j \in L} w(j) \right) = \epsilon^2 \theta \sum_{j \in L} w(j). $$

    Recall the definition of the subset $L = \{j \in IC^F : d(j,i) > \sqrt{\epsilon\theta}\}$. For every client $j \in L$, we strictly have $d^2(j,i) > \epsilon\theta$, which rearranges to $\theta < \frac{1}{\epsilon} d^2(j,i)$. Substituting this geometric upper bound for $\theta$ into our increment expression, we obtain:
    $$ \sum_{j \in U} w(j)(\alpha^c_j - \alpha_j) < \epsilon^2 \sum_{j \in L} w(j) \left( \frac{d^2(j,i)}{\epsilon} \right) = \epsilon \sum_{j \in L} w(j) d^2(j,i). $$
    
    Finally, note that $L \subseteq IC^F \subseteq [1:a] \subset D^*$. Because client weights $w(j)$ are strictly positive and squared metric distances $d^2(j,i)$ are non-negative, the sum over the subset $L$ is trivially bounded by the sum over the entire set $D^*$:
    $$ \epsilon \sum_{j \in L} w(j) d^2(j,i) \le \epsilon \sum_{j \in D^*} w(j) d^2(j,i). $$
    
    This matches the target inequality, completely resolving the claim.
\end{proof}

Applying \cref{clm:robustnooverbidding} at the beginning of the stage that $s$ becomes inactive, this lemma implies the following bound:
\begin{align*}
    w(s) (\alpha^c_s-\rho d^2(s,i)) + \sum_{\ell\in IC^s} w(\ell)(\alpha^c_\ell-\rho d^2(\ell,i)) + \sum_{j\in DC^s} w(j) (r^2_{js}-d^2(j,i)) \\
    \leq (\hat{f} + 8n\eta) + \epsilon \sum_{j\in D^*} w(j)d^2(j,i).
\end{align*}

Because we strictly redefined $IC^s, DC^s,$ and $S^s$ according to the sets $IC, DC, S$ at the beginning of the stage that $s$ becomes inactive, \Cref{clm:robust2nd-constraint-euclidean} perfectly generalizes for $t=s$.
Thus, we can cleanly apply \Cref{lem:dual-feasibility-simpler-euclidean-special}, replacing $\hat{f}$ with $(\hat{f} + 8n\eta) + \epsilon\sum_{j\in D^*}w(j)d^2(j,i)$ and using $L=\emptyset$ and $\epsilon'=0$ (the internal $\epsilon$-parameter in the lemma). This directly gives:
\[
    \sum_{j\in [s]}w(j)\cdot (\alpha_j^c - (\rho-1)R_j^2)\le (\hat{f} + 8n\eta) + (\Gamma + \epsilon)\cdot\sum_{j\in [s]}w(j)\cdot d^2(j,i).
\]
This completes the proof for this subcase.

\paragraph{Subcase 2(b): $[a+1:s]\cap B(i_0,\sqrt{\eps\theta})\neq \emptyset$ and $\sum_{j\in L}w(j)\le \eps\cdot\sum_{j\in [a+1:s]}w(j)$.}
In this case, we again define $DC^s = DC^F$ and $IC^s = IC^F \cup A^F$ as the sets of directly connected and active/indirectly connected clients respectively, right before $s$ becomes inactive (i.e., right before $i_0$ is opened evaluated in the super-sequence $\calH'$). Let $S^s = S^F$.

We show that $a$, $L$, $(IC^t), (DC^t), (\alpha_j^c), (R_j),$ and $(r_{jt})$ strictly satisfy all the constraints of \Cref{lem:dual-feasibility-simpler-euclidean-special}, with the capacity constraint relaxed by $8n\eta$. Conditions (a), (b), (c), (1), (2), and (3) of the lemma hold by the same arguments established for Case 1 and the structural definitions of the sets. It remains to rigorously prove that condition (4) is satisfied. Because $L \subseteq IC^F$, $IC^F \subseteq [a]$, and $IC^F \subseteq IC^s$, we have $L \subseteq [a] \cap IC^s$. The following claim establishes the required algebraic bounds for condition (4).

The following claim is the analog of \Cref{clm:4th-constraint-euclidean}.

\begin{claim}
\label{clm:robust4th-constraint-euclidean}
The following hold:
\begin{itemize}
    \item $\sum_{\ell\in L} w(\ell) \leq \epsilon \cdot \sum_{\ell\in [a+1:s]} w(\ell)$;
    \item for any $j\in DC^s$ and $\ell\in IC^F\setminus L$, we have $\alpha^c_{\ell} \leq (1+20\sqrt{\epsilon}) \cdot (r_{js}+d(j,\ell))^2$;
    \item for any $j\in DC^s$ and $\ell \in [a+1:s]$, we have $\alpha^c_{\ell} \leq (r_{js}+d(j,\ell))^2$;
    \item for any $j\in DC^s$, $\ell\in L$, and $\ell'\in [a+1:s]$, we have $\alpha^c_{\ell} \leq (r_{js}+d(j,\ell'))^2$.
\end{itemize}
\end{claim}
\begin{proof}
    We prove the four bullets of the claim one by one.
    
    \paragraph{First bullet.} By the definition of Subcase 2(b), this weight bound is our direct assumption. Substituting the dummy variable $\ell$ directly yields the first bullet.

    \paragraph{Second bullet.} Consider any $j\in DC^s$ and $\ell\in IC^F\setminus L$. Let $\theta$ be the active dual parameter right before $i_0$ is opened in $\calH'$. Because $\ell \notin L$, its distance to $i$ satisfies $d(\ell, i) \le \sqrt{\epsilon\theta}$. Because $i_0$ freezes $i$, we know $d(i, i_0) \le 2\sqrt{\epsilon\theta}$. By the metric triangle inequality, $d(\ell, i_0) \le d(\ell, i) + d(i, i_0) \le 3\sqrt{\epsilon\theta}$. 
    
    To bridge the distance to the open facilities $S^s$, we instantiate an active client $\ell' \in [a+1:s] \cap B(i_0, \sqrt{\epsilon\theta})$, which is guaranteed to exist by the Subcase 2(b) definition. Because $\ell'$ is active, we strictly know $\theta \le (1-\delta)d^2(\ell', S^s)$, which geometrically implies $d(\ell', S^s) \ge \sqrt{\theta}$.
    
    Applying the triangle inequality across $\ell, \ell',$ and $S^s$ yields the core distance relationship:
    $$d(\ell, S^s) \ge d(\ell', S^s) - d(\ell, \ell') \ge d(\ell', S^s) - d(\ell, i_0) - d(i_0, \ell') \ge \sqrt{\theta} - 4\sqrt{\epsilon\theta}~.$$
    
    Applying the triangle inequality to $j, \ell,$ and $S^s$, and scaling by $\sqrt{1-\delta}$, we directly obtain (because $\delta = \epsilon/2$):
    $$(r_{js} + d(j, \ell))^2 \ge (1-\delta)d^2(\ell, S^s) \ge \theta (1-\delta) \left( 1 - 4\sqrt{\epsilon} \right)^2 \ge \theta\left(1-\epsilon/2\right)\left(1 - 8\sqrt{\epsilon}\right) > \theta(1-9\sqrt{\epsilon})~.$$
    
    Squaring both sides and applying the Taylor expansion $(1 - 9\sqrt{\epsilon})^{-1} \le 1 + 10\sqrt{\epsilon}$ (valid for our sufficiently small $\epsilon \in (0, 10^{-100})$) provides the exact upper bound for $\theta$:
    $$\theta \le (1 + 10\sqrt{\epsilon}) (r_{js} + d(j, \ell))^2~.$$
    
    Because $\ell \in IC^F$, it is indirectly connected right before $i_0$ opens, meaning its modified dual value is strictly bounded by the active parameter: $\alpha^c_{\ell} \le \alpha_\ell \le \theta$. Substituting our geometric bound yields the final inequality:
    $$\alpha^c_{\ell} \le (1+20\sqrt{\epsilon}) (r_{js} + d(j, \ell))^2~.$$

    \paragraph{Third bullet.} Consider any $j\in DC^s$ and $\ell\in [a+1:s] = A^F$. Because $\ell$ is active right before $i_0$ is opened, its maximum possible modified dual value is dictated by the second bullet of the $\eta$-openability condition (\Cref{def:robustopenable}), which strictly enforces $\tau_\ell \le (1-\delta)d^2(\ell, S^s)$. By \Cref{clm:robustfrozenstay}, we bypass the parameter $\theta$ entirely:
    $$\alpha^c_\ell \le \tau_\ell \le (1-\delta)d^2(\ell, S^s)~.$$
    
    Applying the triangle inequality to $j, \ell,$ and $S^s$, we have $d(\ell, S^s) \le d(j, S^s) + d(j, \ell)$. Multiplying by $\sqrt{1-\delta}$ and substituting $r_{js} = \sqrt{1-\delta}d(j, S^s)$ yields:
    $$\sqrt{1-\delta}d(\ell, S^s) \le r_{js} + \sqrt{1-\delta}d(j, \ell) \le r_{js} + d(j, \ell)~.$$
    
    Squaring both sides and substituting into our openability bound gives exactly:
    $$\alpha^c_\ell \le (1-\delta)d^2(\ell, S^s) \le (r_{js} + d(j, \ell))^2~,$$
    completing the proof of the third bullet.

    \paragraph{Fourth bullet.} Consider any $j\in DC^s$, $\ell\in L$, and $\ell'\in [a+1:s]$. Because $\ell'$ is active right before $i_0$ is opened, we strictly have $\theta \le (1-\delta)d^2(\ell', S^s)$.
    
    Applying the triangle inequality to $j, \ell',$ and $S^s$, we have $d(\ell', S^s) \le d(j, S^s) + d(j, \ell')$. Multiplying by $\sqrt{1-\delta}$ yields:
    $$\sqrt{1-\delta}d(\ell', S^s) \le r_{js} + \sqrt{1-\delta}d(j, \ell') \le r_{js} + d(j, \ell')~.$$
    
    Squaring both sides provides the geometric upper bound for $\theta$:
    $$\theta \le (1-\delta)d^2(\ell', S^s) \le (r_{js} + d(j, \ell'))^2~.$$
    
    Finally, because $\ell \in L \subseteq IC^F$, it is strictly indirectly connected before $i_0$ opens. According to the algorithm, its dual value is never increased during Stage 1 updates, ensuring $\alpha^c_{\ell} \le \alpha_\ell \le \theta$. Combining this with our geometric upper bound gives the exact final relation:
    $$\alpha^c_{\ell} \le \theta \le (r_{js} + d(j, \ell'))^2~,$$
    completing the proof of the fourth bullet.
\end{proof}

Because \Cref{clm:robust4th-constraint-euclidean} perfectly satisfies condition (4) of \Cref{lem:dual-feasibility-simpler-euclidean-special} (since $IC^F \setminus L \cup [a+1:s] = IC^s \setminus L$, the second and third bullets map to the second condition of the lemma, while the fourth bullet maps directly to the third condition), we can apply \Cref{lem:dual-feasibility-simpler-euclidean-special}. 

Replacing the capacity bound $\hat{f}$ with $\hat{f} + 8n\eta$ throughout the linear combination evaluated in the lemma, we obtain:
\[
    \sum_{j\in [s]}w(j)\cdot(\alpha_j^c-(\rho-1)R_j^2)\le \hat{f} + 8n\eta + (\Gamma + 200\sqrt{\eps})\cdot\sum_{j\in [s]}w(j)\cdot d^2(j,i).
\]
This identically satisfies our targeted dual feasibility bound required for the robust algorithm. This completes the proof of \Cref{lem:dual-feasibility-robust} for the Euclidean metric, and therefore the proof of \Cref{thm:robust}.

\section{\texorpdfstring{$({4}+O(\sqrt{\eps}))$-Approximation for  $(\frac{\zeta}{\log n})$-{Stable} Instances}{error rendering title}}
\label{sec:centerremoval}
In this section, we give a $({4}+O(\sqrt{\eps}))$-approximation algorithm for 
$(\frac{\zeta}{\log n})$-stable metric $k$-means instances, hence proving \Cref{thr:mainMetricStable}. We will reuse most of the notation and proofs in \cite{charikar2025kmeans}, and only mention the main differences.

To simplify notation, we let $\eps$ be the minimum of $\eps$ and $\zeta$ in the above theorem and further assume that $\eps<1/12$ is sufficiently small. With this notation, we present a $({4}+O(\sqrt{\eps})$-approximation algorithm for $\eps/\log(n)$-stable instances. This implies the above theorem as any $\zeta/\log(n)$-stable instance is also $\zeta'/\log(n)$-stable with $\zeta' \leq \zeta$.

The algorithm and the proof of the above result follow several steps where we iteratively simplify the instance by adding more structure. Finally, we obtain enough structure to solve the problem by maximizing a submodular function subject to a partition matroid constraint. We have not optimized constants in favor of simplifying the description, and throughout this section we let $O_\eps(\log n)$ denote $O(\log(n)/\eps^{O(1)})$.

\paragraph{Definitions of leaders.} 
We use the following definitions throughout this section. Given a set of centers $A\subseteq \facilities$, we let $\clcost(A):=\sum_{p\in \clients}\dist^2(p,A)$ denote the cost of the corresponding clustering (where each client $p$ is assigned to the closest center in $A$ and pays the corresponding distance). Sometimes we will consider a (possibly suboptimal) assignment $\mu:\clients \rightarrow A$ of clients to centers in $A$, and let $\clcost(A,\mu):=\sum_{p\in \clients}\dist^2(p,\mu(p))$ be the corresponding cost. Let $\opt$ be a fixed optimum solution for the $k$-means instance, and let $\sopt = \clcost(\opt)$ be its cost. 
For a client $p\in \clients$, let $\opt_p:=\dist^2(p,\opt)$ 
be the cost paid by $p$ in $\opt$.

Let $C^*$ be an $\opt$ cluster with center $c^*\in \opt$. We let $\avg_{C^*, \opt}$ be the squared distance from $c^*$ to the $\eps |C^*|$th closest client of $C^*$ to $c^*$.
We further let $C^*_\avg$ be the set of clients of $C^*$ at a squared distance at most
$\avg_{C^*, \opt}$ from $c^*$, we refer to these clients
as \emph{leaders} for $C^*$. So $\avg_{C^*, \opt}$ is the \emph{maximum squared distance} from a leader to its center $c^*$. Also note that we have, $|C^*_{avg}|\geq {\eps}|C^*|$. 
We thus have that $\avg_{C^*, \opt}$ is upper bounded by $\frac{1}{1-\eps}$ times the average cost $\frac{1}{|C^*|} \sum_{p\in C^*} \dist^2(p, c^*)$, and at the same time $C^*_{avg}$ contains a constant fraction of the clients. 

\paragraph{Partition matroids.}
In \Cref{sec:submodularopt}, we formulate a submodular function that we then maximize over a partition matroid using known results. We define those concepts and also state the result that we will use. First, recall the definition of a matroid and partition matroid. 
A \emph{matroid} is a tuple $(E, \mathcal{I})$ defined on a ground set $E$ with a family of independent sets $\mathcal{I}$ that satisfy: (1) if $A \in \mathcal{I}$ and $A' \subseteq A$, then $A' \in \mathcal{I}$; and (2) if $A, B \in \mathcal{I}$ and $|A| < |B|$, then there exists an element $e \in B - A$ such that $A \cup \{e\} \in \mathcal{I}$.
  In the special case of  a \emph{partition matroid}, the ground set $E$ is partitioned into disjoint subsets $E_1, E_2, \dots, E_k$, and there are non-negative integers $r_1, r_2, \dots, r_k$ such that:
\begin{itemize}
    \item A subset $I \subseteq E$ is independent if and only if $|I \cap E_i| \leq r_i$ for each $i = 1, 2, \dots, k$.
\end{itemize}
In other words, each subset $E_i$ has a capacity limit $r_i$, and an independent set $I$ cannot contain more than $r_i$ elements from $E_i$.
To work with matroids (to get a running time that is polynomial in the size of the ground set $E$), we often use \emph{independence queries}, which allow us to determine if a given subset $A \subseteq E$ is independent (i.e., whether $A \in \mathcal{I}$). Specifically, an independence query on a subset $A$ returns {true} if $A$ is independent and {false} otherwise.

\paragraph{Submodular functions.} Having defined matroids, we proceed to define non-negative monotone submodular functions. For a finite ground set $E$, a set function $f: 2^E \to \mathbb{R}$ is \emph{submodular} if it satisfies the diminishing returns property: for every pair of sets $A \subseteq B \subseteq E$ and any element $e \in E - B$, it holds that
\[
f(A \cup \{e\}) - f(A) \geq f(B \cup \{e\}) - f(B).
\]
Intuitively, this means that adding an element $e$ to a smaller set $A$ provides at least as much additional value as adding $e$ to a larger set $B$.
In addition, $f$ is called \emph{monotone} if for any pair of sets $A \subseteq B \subseteq E$, we have $f(A) \leq f(B)$. In other words, adding elements to a set does not decrease the function value. Finally, $f$ is \emph{non-negative} if $f(A) \geq 0$ for all $A \subseteq E$. 

A celebrated result~\cite{CalinescuCPV11} gives a polynomial-time $(1-1/e)$-approximation algorithm for the problem of maximizing a non-negative monotone submodular function over a matroid constraint. While the original result was a randomized algorithm, a recent striking result~\cite{BuchbinderFeldman24}, building upon the work of~\cite{FilmusW14}, gives a deterministic algorithm with the same guarantee. We remark that polynomial time here refers to a running time that is polynomial in the size of the ground set $E$.   To summarize, they show 
the following theorem.
\begin{theorem}%
    Let $f: 2^E \to \mathbb{R}$ be a non-negative monotone submodular function that we can evaluate in polynomial time, and let $(E, \mathcal{I})$ be a matroid on the same groundset, for which we can answer independence queries in polynomial time. Then, for every $\zeta>0$, there is a polynomial-time algorithm that outputs a set $X\subseteq E$ with $X \in \calI$ so that
    \[
       f(X) \geq (1-1/e - \zeta) \cdot \max_{X^*\in \calI} f(X^*)\,.  
    \]
    \label{thm:submodularmatroidoptimization}
\end{theorem}

\subsection{Properties of a Locally Optimal Solution}
\label{sec:localSearchAnalysis}
Let $S$ be a {locally optimal} solution output by 
the following standard local search algorithm. We remark that its running time is polynomial since by \Cref{lem:aspectratio} the cost of a solution is a polynomially bounded integer and, at each improving step, the cost decreases by at least one.

\begin{mdframed}[hidealllines=true, backgroundcolor=gray!15]
\vspace{-5mm}
\paragraph{Classic Local Search Procedure for $k$-Means.}\ \\
\begin{enumerate}
    \item $S \gets $ Arbitrary solution for $k$-Means.
    \item While there exists a solution $S'$ such that $|S \Delta S'|\le 2$ and $\clcost(S') <\clcost(S)$:~~
    \begin{enumerate}
        \item $S \gets S'$.
    \end{enumerate}
    \item Output $S$.
\end{enumerate}
\end{mdframed}
Local search achieves a $\apxLSkmeans$ approximation for
$k$-Means, see \cite{GT08}.
\begin{lemma}
    $\clcost(S) \leq \apxLSkmeans\cdot \sopt$.
    \label{lemma:localsearchis5approximation}
\end{lemma}

In the remaining part of the algorithm and its analysis we fix $S$, and 
 recall that we fixed the reference optimal solution $\opt$. We distinguish different types of clusters of $\opt$ and analyze their properties in the next two sections. We let $S_p:=\dist^2(p,S)$ be the cost paid
by $p$ in $S$ and recall that $\opt_p$ denotes the cost $p\in \clients$ pays in $\opt$.

\subsection{Pure Clusters}
\label{sec:pure}
 We say that a cluster $C^*$ of $\opt$ is \emph{pure} 
(w.r.t. $S$) if there exists a cluster $C'$ of $S$ such that $|C' \Delta C^*| \le {3\eps} \min(|C^*|,|C'|)$. In which
case we also say that $C'$ is pure w.r.t. $\opt$. We say that $C'$ and $C^*$ are associated.
The following lemma follows from the fact that the input is $\beta$-stable
for $\beta := \eps/\log n$ and that the solution $S$  is a $\apxLSkmeans$-approximation by \Cref{lemma:localsearchis5approximation}. Given a client or center $a$ and a distance $r$, we let $B(a,r)$ be the set of clients and centers at distance at most $r$ from $a$. 
\begin{lemma}
\label{lem:numnonpure}
\cite{charikar2025kmeans} The number $k_{imp}$ of clusters of $\opt$  that are not pure w.r.t. $S$ (and thus the number of clusters of $S$ which are not pure w.r.t $\opt$)
    is at most {$\frac{\log n}{\eps^3}$}. %
\end{lemma}

Let $C^*$ be a pure cluster of $\opt$ and $C'$ be the associated cluster of $S$. Let
$c^*$ be the center of $C^*$ and ${c'}$ be the center of ${C'}$. We define $t(c^*) := {c'}$.
For any client
$p \in C^*$, let $r(p) := \dist^2(p, {c'}) = \dist^2(p, t(c^*))$.
We have the following lemma which is the only lemma
that requires that the solution $S$ is obtained via the local search (the proof of the previous lemma only used that it was a constant-factor approximation). In words, it says that $S$ approximates the connection cost of clients belonging to pure clusters of the optimal solution almost perfectly.

\begin{lemma}
\label{lem:purecost}\cite{charikar2025kmeans}
    Let $\clientspure \subseteq \clients$ be the subset of clients that belong to pure clusters  of the optimal solution $\opt$. We have
    \begin{gather*}
        \sum_{p\in \clientspure} r(p) \leq \sum_{p\in \clientspure} \opt_p + O(\sqrt{\eps} \cdot \sopt)\,.
    \end{gather*}
\end{lemma}

\subsection{\texorpdfstring{Cheap Clusters and $D$-Sample Process to Hit Expensive Clusters}{Cheap Clusters and D-Sample Process to Hit Expensive Clusters}}
\label{sec:dsampleproc}
In the previous section, we showed that the cost of pure clusters in $\opt$ is close to their optimal cost in our local search solution $S$. Our goal in this section is to ``hit'' those non-pure clusters of $\opt$
 that have a high cost in $S$.
We say that a non-pure cluster $C^*$ of $\opt$ with center $c^*$ is \emph{basic-cheap} if the total cost in 
$S$ of the clients in $C^*_\avg$ is less than $\eps^{5}\sopt/\log n$ or 
if there exists a center of $S$ at squared distance at most
{$(1+\eps)\avg_{C^*,\opt} + \eps \gamma_{C^*}$} from $c^*$, where $\gamma_{C^*} := \frac{1}{|C^*|} \sum_{p\in C^*} (\opt_p + S_p)$; in the latter case we also say that $C^*$ is \emph{covered} by $S$.

The following procedure,  that aims to ``hit'' all clusters of $\opt$ that are non-pure and non-basic-cheap,  is inspired by the classic $k$-means++ algorithm and its
    so called $D^2$-sampling procedure to
    sample clients proportional to their cost in solution $S$. 
\begin{mdframed}[hidealllines=true, backgroundcolor=gray!15]
\vspace{-5mm}
\paragraph{$s$-$D^2$-Sample Process. }\ \\
\begin{enumerate}
    \item Let $W$ be the set obtained by taking $s$ independent samples of clients where, in each sample, client  $p$ is sampled with probability $\frac{\dist^2(p, S)}{\clcost(S)}$.
    \item Output $W$. 
\end{enumerate}
\end{mdframed}
Our algorithm obtains a set $W$ of clients by sampling $s^*$ using the $s^*$-$D^2$-sample process, where
\[
    s^* = \apxLSkmeans \cdot \frac{\log n}{\eps^{5}}\ln\left( \frac{81}{\eps^{2}}\right)\,.
\]
We have the following lemma that says that we hit all non-pure and non-basic-cheap clusters except for a set $V$ of clusters that have an insignificant cost in $S$.

For a non-pure cluster $C^*$ of $\opt$ with center $c^*\in \opt$, we let $t(c^*)$ be the closest center in $S$ to $c^*$, and for any $p\in C^*$, we let the replacement cost of $p$ be $r(p):=\dist^2(p,t(c^*))$. 
\begin{lemma}
\label{lem:probcost}\cite{charikar2025kmeans}
Consider the random set $W$ of clients obtained from the $s^*$-$D^2$-sample process.  Let $V$ be the set of clusters $C^*$ of $\opt$ that are non-pure and non-basic-cheap  such that $C^*_\avg  \cap W = \emptyset$. 
With probability at least $1-\eps$, 
\begin{align}\sum_{C^* \in V} \sum_{p \in C^*} r(p) \le \eps \cdot \sopt\,.
\label{eq:probcost}
\end{align}
\end{lemma}

\paragraph{Successful $W$.}
We say that the set of clients $W$ obtained from the $s^*$-$D^2$-sample process is a 
\emph{successful} sample  if~\eqref{eq:probcost} holds.
We define the following two types of $\opt$ clusters. We say that 
a non-pure cluster $C^*$ of $\opt$ with center $c^*$ is \emph{cheap} if it is basic-cheap 
or is in $V$; a  non-pure cluster is 
\emph{expensive} otherwise.  In other words, a cluster $C^*$ of the optimal solution is expensive if it is a non-pure and non-basic-cheap cluster such that $C^*_{avg} \cap W \neq \emptyset$. 

Assuming that $W$ is successful, we have a good bound on the cost of the clients assigned to cheap centers in the optimal solution. 
\begin{lemma}
\label{lem:cheapcost}\cite{charikar2025kmeans}
    Let $\clientscheap \subseteq \clients$ be the subset of clients that belong to cheap clusters of the optimal solution $\opt$. If $W$ is a successful sample, 
    \begin{gather*}
        \sum_{p\in \clientscheap} r(p) \leq \sum_{p\in \clientscheap} 4\,\opt_p + O(\sqrt{\eps} \cdot \sopt)\,.
    \end{gather*}
\end{lemma}

Given $S$ and $W$, the set of clusters of the optimal solution $\opt$ is thus partitioned into pure clusters, cheap clusters, and expensive clusters. We let $\ho$ be the centers of $\opt$ that are expensive. We further define the set $\clientsexpensive \subseteq \clients$  to be the subset of clients that belong to the expensive clusters in $\opt$. We have thus partitioned $D$ into sets $\clientspure, \clientscheap$, and $\clientsexpensive$ depending on the type of optimal cluster they belong to. \Cref{lem:purecost}
bounds the cost of the clients in $\clientspure$ in $S$ and \Cref{lem:cheapcost} bounds the cost of the clients in $\clientscheap$. For future reference, we summarize these two lemmas  (by weakening the upper bound for clients in $\clientspure$).
\begin{lemma}
Assuming  $W$ is a successful sample, 
    \[
    \sum_{p\in D - \clientsexpensive} r(p) \leq \sum_{p \in D - \clientsexpensive} 4 \opt_p + O(\sqrt{\eps}) \cdot \sopt\,.
    \]
    \label{lemma:convenience_upper_bound}
\end{lemma}

It follows that the connection cost in solution $S$ of clients in $\clients - \clientsexpensive$ is roughly within a factor $4$ of their cost in the optimal solution. The remaining part of the algorithm is thus devoted to modifying $S$ to obtain a small connection cost of the clients in $\clientsexpensive$ without significantly increasing the cost of the other clients.

\subsection{Identifying Balls of Expensive Clusters}
\label{sec:ballguesses}

By the definition of expensive clusters, each expensive cluster $C^*$ of $\opt$ satisfies $C^*\cap W \neq \emptyset$ where $W$ is the sample obtained by running the sampling procedure of the last subsection. Here, our goal is to guess a subset $\calB$ of balls so that for each expensive cluster $C^*$ there is a ball $B(\ell, \sqrt{\rho}) \in \calB$ so that $\ell$, which we refer to as a leader, is in  $C^*_{avg}$ and $\rho$ is a very close approximation to $\avg_{C^*, \opt}$. This guarantees that the center $c^* \in \ho$ of cluster $C^*$ is in $B(\ell, \sqrt{\rho})$ and that no other center in $B(\ell, \sqrt{\rho})$ is "too" far from $c^*$ since the radius of the ball is $\approx \sqrt{\avg_{C^*, \opt}}$. The balls $\calB$ will then be used when we approximate the centers in $\ho$ via submodular function maximization subject to a partition matroid constraint (the balls will correspond to the partitions of the matroid). 
As the algorithm does not know the set of expensive clusters, it enumerates all subsets of $W$, and for each guessed point, it enumerates a constant number of radii.

We say that a set of balls $\calB$ is a \emph{valid} set of balls if it satisfies the following: For each expensive cluster $C^*$ of $\opt$,  we have a ball $B(\ell, \sqrt{\rho}) \in \calB$ such that $\ell\in C^*_{avg}$ is a leader and $$\avg_{C^*,\opt} \leq \rho \leq  \avg_{C^*,\opt} + \eps \gamma_{C^*}, $$
where we recall that $\gamma_{C^*} = \frac{1}{|C^*|} \sum_{p \in C^*} (\opt_p + S_p)$.

\begin{lemma}[\cite{charikar2025kmeans}]
\label{lem:ballguesses} 
In polynomial time $n^{\eps^{-O(1)}}$ one can compute a collection $\calL_{\texttt{bal}}$ of sets of balls such that at least one set $\calB\in \calL_{\texttt{bal}}$ is a valid set of balls.
\end{lemma}

While our algorithm will try all possible sets in $\calL_{\texttt{bal}}$ (since it does not know which one is valid), we do our analysis  by considering the run of the algorithm when it selects this set  $\calB$ of valid balls.

\subsection{\texorpdfstring{\emph{Dummy} Centers and Mixed Solutions $M_O$ and $M_D$}{\emph{Dummy} Centers and Mixed Solutions MO and MD}}
\label{sec:dummycenters}

    Next, the algorithm creates, for each ball  $B(\ell, \sqrt{\rho}) \in \calB$ with center $\ell$ and
    radius $\sqrt{\rho}$, 
    a \emph{dummy} center $\delta$. The center $\delta$ is at distance $\sqrt{\rho}$ from $\ell$ and at distance 
   $\sqrt{\rho} + d(\ell, p)$ from any other input point $p$.

    Let $\Lambda$ be the set of dummy centers, so $|\Lambda| = |\calB|$. Recall that  $\clientsexpensive$ is the set of clients that are in an expensive cluster of $\opt$, i.e., served by a center
in $\ho$ in the optimal solution.

\begin{lemma}[\cite{charikar2025kmeans}]
\label{lem:structSminusS0}
Suppose the sample $W$ is successful, and that the set of balls $\calB$ is valid, then there exists a set of centers $S_0$ of $S$ that satisfies the following properties:
  \begin{enumerate}
  \item $|S_0| = |\ho| \le \frac{\log n}{\eps^3}$, and;
  \item $\forall c \in S_0$, there is no $c^* \in \opt-\ho$ such that $t(c^*) = c$, and;
  \item $\clcost(S - S_0 \cup \dummyset) \le
  9 \sum_{p \in \clientsexpensive}\opt_p + \sum_{p \in \clients {-} \clientsexpensive}
  r(p) +
  O(\eps \cdot{\sopt})) \le (9+O(\sqrt{\eps}))\sopt$.
  \end{enumerate}
  \label{lemma:S0properties}
\end{lemma}

\paragraph{Definition of the mixed solutions $M_O$ and $M_D$.}
Next, let us analyze
the cost of the swap $(\ho, S_0)$. Namely, of the solution $M_O = S - S_0 \cup 
\hat{\opt}$, where $M$ in $M_O$ stands for  ``mixed solution''  and the subscript $O$ stands for that the mixed solution  is obtained by swapping in some elements from $\opt$  (and removing $S_0$). We will also analyze the ``dummy'' version of $M_O$ where instead of swapping in $\ho$, we swap in the dummy centers ${\dummyset}$. We denote that solution by $M_D = S - S_0 \cup \dummyset$, where subscript $D$ stands for ``dummy''.  
\begin{lemma}[\cite{charikar2025kmeans}]
    We have  $d^2(p, \ho) \leq \opt_p$ if $p\in \clientsexpensive$ and $d^2(p, S- S_0)\leq r(p)$ if $p\in\clients - \clientsexpensive$.
    \label{lemma:costsMO}
\end{lemma}

Notice that $M_O$ contains $S- S_0$ and $\ho$. Therefore, if we sum up the above bounds for all clients we get
\begin{align*}
    \clcost(M_O) & \leq \sum_{p\in \clientsexpensive} \opt_p  + \sum_{p\in \clients {-} \clientsexpensive} r(p)\,, 
\end{align*}
which by \Cref{lemma:convenience_upper_bound} is at most $\sum_{p\in \clientsexpensive} \opt_p  + \sum_{p\in \clients {-}\clientsexpensive} 4 \opt_p + O(\sqrt{\eps}) \cdot \sopt$ (assuming that that the sample $W$ was successful). By swapping $S_0$ with $\ho$ we thus get a $1$-approximation on the clients  $\clientsexpensive$ and a $4$-approximation of the remaining clients (plus a small error term). Our  goal in the next sections will thus be to approximate this swap. The next few steps will be aimed at simplifying (or guessing parts of) $S_0$. The last step will then approximate $\ho$ and the remaining part of $S_0$ by submodular function maximization subject to a partition matroid constraint. For that part, it will be helpful to have a bound on the dummy solution as well; specifically, a modified version of it (see \Cref{lemma:costboundofMOandMDwithassignments} and \Cref{claim:Mprimebounds}). For intuition of those statements, let us here say that one can show 
\begin{align*}
        \clcost(M_D) & \leq \sum_{p\in \clientsexpensive} 9\,\opt_p  + \sum_{p\in \clients - \clientsexpensive} r(p)  + O({\sqrt{\eps}}) \cdot \sopt\,\,
\end{align*}
by observing that only the clients $\clientsexpensive$ have a different connection cost in $M_D$ than in $M_O$. 
Moreover, each such client $p\in \clientsexpensive$ that was previously assigned to a center $c^* \in \ho$, and belongs to cluster $C^*$ of $\opt$ with leader $\ell$, has an associated dummy center $\delta$ at squared distance 
$$
d^2(p,\delta)\leq 3d^2(p, c^*)+3d^2(c^*,\ell)+3d^2(\ell,\delta)\leq 3d^2(p, c^*) +  6\avg_{C^*, \opt} + 3\eps \gamma_{C^*}
$$ 
(assuming the set $\calB$ of balls is valid and using \Cref{lem:apxTriangleInequality3} with $\gamma=3$). Simplifications then give the stated inequality as $|C^*| \avg_{C^*, \opt} \leq (1+2\eps) \sum_{p\in C^*} \opt_p$ and $|C^*| \eps \gamma_{C^*} = \eps \sum_{p\in C^*} ( \opt_p + S_p)$.

\subsection{\texorpdfstring{Removal of Expensive Centers of $S_0$}{Removal of Expensive Centers of S0}}
\label{sec:removalofExpensive}

Thus, it remains to show that our algorithm can yield a good approximation to
the gain that the swap $(\ho, S_0)$ would provide. Of course, it will not 
necessarily be optimum, but we can show it
is enough to conclude 
the proof of our theorem.
We focus on the iteration of the algorithm
that considers a valid ball set of balls $\calB$. In particular, at this point given that $W$ is successful and $\calB$ valid, the number of balls in our ball guess is equal to $|S_0|=|\ho|=|\Lambda|\leq \frac{\log n}{\eps^3}$. Our algorithm then
makes use of the following Lemma in \cite{charikar2025kmeans} that takes as input the local search solution $S$ and the set $\dummyset$ of dummy centers.

    \begin{lemma}[\cite{charikar2025kmeans}]
    \label{lem:successguessprocess}
There is a procedure that runs in $n^{1/\eps^{O(1)}}$-time and produces a collection $\calLexp$ of at most $n^{1/\eps^{O(1)}}$ subsets of $S$ such that with probability at least $1-1/n$ there exists $\calQ\in \calLexp$ satisfying
    \begin{enumerate}
    \item $\calQ \subseteq S_0$, and
    \item The total cost in the solution $S - Q \cup \dummyset$ of the clusters with centers in $S_0 - \calQ$ is at most $\eps \cdot \sopt$.
    \end{enumerate}
\end{lemma}

\subsection{\texorpdfstring{Consistently Assigning Clients in Mixed Solutions and the sets  $\calU$ and $\calR$}{Consistently Assigning Clients in Mixed Solutions and the sets  U and R}}
\label{sec:consistentlyassigning}
While our algorithm tries all possible $\calQ \in \calLexp$, we focus  on the execution path of a set $\calQ$ of centers satisfying the properties of \Cref{lem:successguessprocess}, i.e., 
\begin{enumerate}
    \item 
      $\calQ \subseteq S_0$, and
      \item  the cost in the solution $S - \calQ \cup \dummyset$ of the clusters with centers in $S_0 - \calQ$ is at most $\eps \cdot \sopt$.
\end{enumerate}
    We define the solution $S_{\calQ} := S- \calQ \cup \dummyset$ and for a center $c\in S_{\calQ}$ we let $S_{\calQ}(c)$ be the subset of clients closest to $c$ (i.e., assigned to $c$) in the solution $S_{\calQ}$. The second property above allows us to consider each cluster $S_{\calQ}(c)$, $c\in S_0 - \calQ$, as contracted by paying a small extra cost of $\eps \cdot \sopt$. In particular, this allows us to give a ``consistent''  assignment $\mu_O$ of clients in the mixed solutions. Here, consistent means that all clients in $S_Q(c)$, for $c\in S_0 - \calQ$, are assigned to the same center in the mixed solutions where all of $S_0$ is removed.  
    
    \paragraph{Definition of $\mu_O$ and $\mu_D$.} {In the following we slightly deviate from the approach in \cite{charikar2025kmeans}. More specifically, the constant $4$ below is a $5$ in \cite{charikar2025kmeans}. The rest of the analysis remains however very similar.}
    We modify how he clients are assigned in the solution $M_O$ to obtain the assignment $\mu_O$.
    For every $c \in S_0 - \calQ$, all the clients in $S_{\calQ}(c)$ are reassigned to the same 
    center of $M_O$ as follows. Let $c_1$ be the 
    cluster of $\ho$ that is the closest to 
    $c$ and let $c_2$ be the center of $M_O-\ho = S - S_0$ that
    is the closest to $c$. 
    The clients of $S_{\calQ}(c)$ are all assigned to either $c_1$ or $c_2$:
    \begin{itemize}
        \item If $d^2(c,c_1) \le d^2(c,c_2)/{4}$, assign
        $S_{\calQ}(c)$ to $c_1$; $S_{\calQ}(c)$ is 
        called a \emph{type-1} cluster.
        \item Otherwise ($d^2(c,c_1) > d^2(c,c_2)/{4}$), assign
        $S_{\calQ}(c)$ to $c_2$; $S_{\calQ}(c)$ is 
        called a \emph{type-2} cluster.
    \end{itemize}
    Let  $T_1$ be the set of clients in a type-1 cluster and $T_2$ the set of clients in a type-2
    cluster. The remaining clients in $\clientsexpensive - (T_1 \cup T_2)$ are assigned to their closest centers in $\ho$, and the remaining clients in $(\clients - \clientsexpensive)- (T_1 \cup T_2)$ are assigned to their closest centers in $S- S_0$. Notice that, even neglecting the extra cost due to the consistent assignment, the above assignment of clients in $T_2$ is suboptimal when $d(c,c_2)>d(c,c_1)>d(c,c_2)/{4}$. The motivation for this reassignment is technical. More specifically, it will be used in Lemma \ref{lemma:costboundofMOandMDwithassignments} to have a better upper bound on the cost of clients in $C^*\cap T_1$ for a non-expensive cluster $C^*$ (case b2). This leads to a larger upper bound on the cost of clients in $C^*\cap T_2$ for an expensive cluster $C^*$ (case a3), which is however tolerable.

    We further obtain an assignment $\mu_D$ of clients in $M_D$ by modifying $\mu_O$ as follows:  each client $p$ with $\mu_O(p) \in \ho$ is assigned to the associated dummy center by $\mu_D$. So for each client $p$ we have $\mu_D(p) = \mu_O(p) $ unless $\mu_O(p)\in \ho$ in which case $\mu_D(p)$ equals the dummy center associated with $\mu_O(p)$.

    \paragraph{Sets $\calR$ and $\calU$.} By the definition of type-1 and type-2 cluster, we can split the
    set of centers of $S_0 - \calQ$ into two groups $\calR$ and $\calU$. Let $\calR$
    be the set of centers of $S_0- \calQ$ whose set of clients is completely assigned
    to a center of $\ho$ in the assignment $\mu_O$, and let $\calU = S_0 - \calQ- \calR$ be the remaining ones that are assigned to centers in $S- S_0$.

    We end this section by analyzing the costs of the assignments $\mu_O$ and $\mu_D$. {Recall that} we use the notation $\clcost(M_O,\mu_O)$ to denote the cost of $M_O$ equipped with assignment $\mu_O$ and, similarly $\clcost(M_D, \mu_D)$ for the cost of $M_D$ equipped with assignment $\mu_D$. We say that the selection of $(W, \calB, \calQ)$ is successful, if the sample $W$ selected in \Cref{sec:dsampleproc} is successful, the set of balls $\calB$ from \Cref{sec:ballguesses} is valid, and $\calQ$ selected in this section satisfies the properties of \Cref{lem:successguessprocess}. {In \cite{charikar2025kmeans} there is an analogue version of the next lemma, where (besides the mentioned difference in the constant $4$ versus $5$) only the sum $\clcost(M_O, \mu_O)+\clcost(M_D, \mu_D)$ is upper bounded. For reasons that will become clearer later, we need to bound the two quantities separately.}
\begin{lemma}\label{lemma:costboundofMOandMDwithassignments}
    If the selection
    of $(W, \calB, \calQ)$ is successful, one has
\begin{align*}
\clcost(M_O, \mu_O) & \leq \sum_{p\in \clientsexpensive- T_2} \opt_p + \sum_{p\in \clientsexpensive \cap T_2}{4}\OPT_p\\
& + \sum_{p\in (D-\clientsexpensive)- T_1}r(p)+\sum_{p\in (D-\clientsexpensive)\cap T_1}\frac{1}{{4}}r(p)+O(\sqrt{\eps})\sopt{\leq O(1)\sopt}\, ,
\end{align*}
and
\begin{align*}
\clcost(M_D, \mu_D) & \leq \sum_{p\in \clientsexpensive-T_2} 9\opt_p + \sum_{p\in \clientsexpensive \cap T_2}{4}\OPT_p\\
& + \sum_{p\in (D-\clientsexpensive)- T_1}r(p)+\sum_{p\in (D-\clientsexpensive)\cap T_1}\frac{9}{{4}}r(p)+O(\sqrt{\eps})\sopt{\leq O(1)\sopt}\,.
\end{align*}%
\end{lemma}
\begin{proof}
Throughout the proof of the lemma, we will repeatedly upper bound $d^2(p,\mu_D(p))$ in terms of $d^2(p,\mu_O(p))$ using the following bounds:
\begin{claim}\label{clm:costboundofMOandMDwithassignments}\cite{charikar2025kmeans}
Consider a client $p\in S_{\calQ}(c)$ with $\mu_O(p) \neq \mu_D(p)$ and so $\mu_O(p)\in \ho$ is the center of an expensive cluster $C^*$ of $\opt$. Then
\begin{enumerate}
\item $d^2(p, \mu_D(p))  \leq 3d^2(p, \mu_O(p)) + 6  ( \avg_{C^*, \opt} + \eps \gamma_{C^*})$, and
\item $d^2(p,\mu_D(p)) \leq \frac{2}{\sqrt{\eps}}d^2(p, S_\calQ) + 9(1+\sqrt{\eps})d^2(c, \mu_O(p))$.
\end{enumerate}
\end{claim}
We prove the lemma by distinguishing between expensive and non-expensive clusters $C^*$ of $\opt$. 

\paragraph{Case a: $C^*$ is an expensive cluster of $\opt$.} Let $c^*\in \ho$ be the center of $C^*$. Let $p\in C^*$ be a client in $C^*$ and let $c$ be its closest center in $S_\calQ$, i.e.,   $p\in S_\calQ(c)$.  We bound $d^2(p, \mu_O(p))$ {and} $d^2(p, \mu_D(p))$  by considering three cases:
\begin{description}
    \item[Case a1: $p\in C^* - (T_1 \cup T_2)$.]  By definition, we have  $\mu_O(p)$ is the closest center in $\ho$. Moreover, $\mu_O(p)$ equals the cluster center $c^*$ of $C^*$ since $p$ is closest to $c^*$ among all centers in $\opt \supseteq \ho$. {\Cref{lemma:costsMO} gives us $d^2(p,\mu_O(p))=\opt_p$.}
    Hence claim \ref{clm:costboundofMOandMDwithassignments}.1 implies,
    \[
        d^2(p, \mu_D(p)) \leq {3}\opt_p +  6  ( \avg_{C^*, \opt} + \eps \gamma_{C^*})\,.
    \]
\item[Case a2: $p\in C^* \cap T_1$.]  By definition of type-1 clusters,  $\mu_{O}(p) = c_1$ is the closest center in $\ho$ to $c$. Using \Cref{lem:apxTriangleInequality2} with $\gamma=1+\sqrt{\eps}$ twice,%
\begin{align*}
d^2(p, \mu_O(p)) & \leq \frac{2}{\sqrt{\eps}}d^2(p, S_{\cal Q})+(1+\sqrt{\eps})d^2(c,c_1)\\
& \leq \frac{2}{\sqrt{\eps}}d^2(p, S_{\cal Q})+(1+\sqrt{\eps})\left(\frac{2}{\sqrt{\eps}}d^2(p,S_{\cal Q})+(1+\sqrt{\eps})d^2(p,c_1)\right)\\
& \leq \frac{{5}}{\sqrt{\eps}}d^2(p, S_{\cal Q})+(1+O(\sqrt{\eps}))\opt_p.
\end{align*}
Furthermore, by Claim 
\ref{clm:costboundofMOandMDwithassignments}.2 and using similarly as above \Cref{lem:apxTriangleInequality2} with $\gamma=1+\sqrt{\eps}$,
\begin{align*}
d^2(p, \mu_D(p)) & \leq \frac{2}{\sqrt{\eps}} d^2(p, S_{\calQ}) + 9(1+\sqrt{\eps}) d^2(c, c_1) \\
& \leq \frac{2}{\sqrt{\eps}}d^2(p, S_\calQ)+9(1+\sqrt{\eps})(\frac{2}{\sqrt{\eps}}d^2(p, S_\calQ)+(1+\sqrt{\eps})d^2(p,c_1))\\
& \leq \frac{{21}}{\sqrt{\eps}}d^2(p, S_\calQ)+(9+O(\sqrt{\eps}))\opt_p\,.
\end{align*}

\item[Case a3: $p\in C^*\cap T_2$.] By the definition of type-2 clusters, $\mu_O(p) = c_2$ is the closest center to $c$ in $S- S_0$, and $d^2(c, c_2) \leq {4} d^2(c, \ho)$. So, by \Cref{lem:apxTriangleInequality2} with $\gamma=1+\sqrt{\eps}$, $d^2(c, c_2) \leq {4}(\frac{2}{\sqrt{\eps}} d^2(c,p) + (1+\sqrt{\eps})\opt_p)$ where we additionally used that $d^2(p, \ho) \leq \opt_p$ by \Cref{lemma:costsMO}.
As $\mu_O(p) = \mu_D(p)$ in this case, and using again \Cref{lem:apxTriangleInequality2} with $\gamma=1+\sqrt{\eps}$,
\begin{align*}
d^2(p, \mu_O(p)) {=} d^2(p, \mu_D(p)) & \leq \frac{2}{\sqrt{\eps}}d^2(p, c) + (1+\sqrt{\eps})d^2(c,c_2) \\
& \leq \frac{2}{\sqrt{\eps}}d^2(p, c) + (1+\sqrt{\eps}){4}(\frac{2}{\sqrt{\eps}}d^2(c,p) + (1+\sqrt{\eps})\opt_p) \\
& \leq \frac{{11}}{\sqrt{\eps}}d^2(p, S_\calQ) + ({4}+O(\sqrt{\eps})) \opt_p\,.
\end{align*}
\end{description}
By the above bounds, we have that 
\begin{align*}
& \sum_{p\in C^*} d^2(p, \mu_O(p)) \leq  \sum_{p\in C^* - (T_1 \cup T_2)} \opt_p +O(\sqrt{\eps})\sum_{p\in C^*}\OPT_p\\
& + \sum_{p\in C^*\cap T_1}\left(\opt_p + \frac{5}{\sqrt{\eps}} d^2(p, S_\calQ)\right)+\sum_{p\in C^*\cap T_2}\left({4}\opt_p + \frac{{11}}{\sqrt{\eps}} d^2(p, S_\calQ)\right).  
\end{align*}
Furthermore
\begin{align*}
& \sum_{p\in C^*} d^2(p, \mu_D(p)) \leq  \sum_{p\in C^* - (T_1 \cup T_2)} \left({3}\opt_p +  6(\avg_{C^*, \opt} + \eps \gamma_{C^*})\right) +O(\sqrt{\eps})\sum_{p\in C^*}\OPT_p \\
& + \sum_{p\in C^*\cap T_1}\left({9}\opt_p + \frac{{21}}{\sqrt{\eps}} d^2(p, S_\calQ)\right)+ \sum_{p\in C^*\cap T_2}\left({4}\opt_p + \frac{{11}}{\sqrt{\eps}} d^2(p, S_\calQ)\right)\,.  
\end{align*}
To complete the analysis for an expensive cluster $C^*$, it remains to upper bound the terms $6\eps\gamma_{C^*}$ and $6\avg_{C^*, \opt}$. By definition we have $|C^* - (T_1 \cup T_2)| \cdot 6\eps \gamma_{C^*}\leq  6\eps \sum_{p\in C^*} (\opt_p + S_p)$. For the other term, we claim that 
\[
    6 |C^* - (T_1 \cup T_2)|\cdot \avg_{C^*, \opt} \leq \sum_{p\in C^* - (T_1 \cup T_2)} 6\opt_p + 7\eps \cdot \sum_{p\in C^*} \opt_p\,. 
\]
This holds because there are at least $(1-\eps)|C^*|$ clients  in $C^*$ such that $\opt_p \geq \avg_{C^*, \opt}$, which implies 
\begin{align*}
    \sum_{p\in C^* - (T_1 \cup T_2)} 6\opt_p + 7\eps \cdot \sum_{p\in C^*} \opt_p &\geq 6(|C^* - (T_1 \cup T_2)|-\eps|C^*|)\cdot \avg_{C^*, \opt} + 7\eps(1-\eps)|C^*|\cdot \avg_{C^*, \opt} \\
    & \geq 6 |C^*- (T_1 \cup T_2)|\cdot \avg_{C^*, \opt}\,. 
\end{align*}
Combining this with a previous inequality, we obtain
\begin{align*}
& \sum_{p\in C^*} d^2(p, \mu_D(p)) \leq  \sum_{p\in C^* - (T_1 \cup T_2)} {9}\opt_p +O(\sqrt{\eps})\sum_{p\in C^*}(\OPT_p+S_p)\\
& + \sum_{p\in C^*\cap T_1}\left({9}\opt_p + \frac{{21}}{\sqrt{\eps}} d^2(p, S_\calQ)\right)+ \sum_{p\in C^*\cap T_2}\left({4}\opt_p + \frac{{11}}{\sqrt{\eps}} d^2(p, S_\calQ)\right)\,.  
\end{align*}
.

\paragraph{Case b: $C^*$ is a non-expensive cluster of $\opt$.} We proceed with a similar analysis. Indeed, consider a client $p\in C^*$  and let $c$ be its closest center in $S_\calQ$, i.e.,   $p\in S_\calQ(c)$.  We again upper bound $d^2(p, \mu_O(p))$ {and} $d^2(p, \mu_D(p))$  by considering three cases:

\begin{description}
    \item[Case b1: $p\in C^* - (T_1 \cup T_2)$.] By the definition of $\mu_O$, we have that $\mu_O(p)$ is the closest center in $S - S_0$ to $p$. So $\mu_D(p) = \mu_O(p)$ and one has by \Cref{lemma:costsMO} 
    $$
    d^2(p,\mu_O(p))=d^2(p,\mu_D(p))\leq r(p)\,.
    $$ 
\item[Case b2: $p\in C^* \cap T_1$.]  By the definition of type-1 clusters,  $\mu_{O}(p) = c_1$ is the closest center in $\ho$ to $c$. By \Cref{lem:apxTriangleInequality2} with $\gamma=1+\sqrt{\eps}$,
\begin{align*}
d^2(p, \mu_O(p))  &  \leq \frac{2}{\sqrt{\eps}}d^2(p, S_{\calQ})+(1+\sqrt{\eps})d^2(c, c_1)\,.
\end{align*}
Furthermore, by Claim \ref{clm:costboundofMOandMDwithassignments}.2,
\begin{align*}
d^2(p, \mu_D(p)) & \leq  \frac{2}{\sqrt{\eps}}d^2(p, S_{\calQ}) + 9(1+\sqrt{\eps}) d^2(c, c_1)\,.
\end{align*}
We proceed by upper bounding $d^2(c, c_1)$. 
Here we critically use the fact that, by the definition of $T_1$, one has $d^2(c, c_1) \leq d^2(c, c_2)/{4}$ where $c_2$ is the closest center to $c$ in $S- S_0$. Let $c_3$ be the closest center to $p$ in $S - S_0$. Then, using \Cref{lem:apxTriangleInequality2} with $\gamma=1+\sqrt{\eps}$, we obtain that 
$$
d^2(c, c_1) \leq \frac{1}{{4}}d^2(c, c_2)\leq \frac{1}{{4}}d^2(c,c_3) \leq \frac{2}{{4}\sqrt{\eps}}d^2(p,S_\calQ) + \frac{1+\sqrt{\eps}}{{4}}d^2(p, c_3).
$$ 
Moreover, we have $d^2(p, c_3) = d^2(p, S- S_0)\leq r(p)$ by \Cref{lemma:costsMO} and so
$d^2(c,c_1)  \le \frac{2}{{4}\sqrt{\eps}}d^2(p, S_\calQ) + \frac{1+\sqrt{\eps}}{{4}}r(p)$, which gives us the bounds
    \begin{align*}
    d^2(p, \mu_O(p))  \leq \frac{{5}}{\sqrt{\eps}}d^2(p, S_{\calQ})+ \frac{1+O(\sqrt{\eps})}{{4}}r(p)\,.
    \end{align*}
and
    \begin{align*}
    d^2(p, \mu_D(p)) \leq \frac{{21}}{\sqrt{\eps}}d^2(p, S_{\calQ}) + \frac{9+O(\sqrt{\eps})}{{4}}r(p)\,.
    \end{align*}
\item[Case b3: $p\in C^*\cap T_2$.] By the definition of type-2 clusters, $\mu_O(p) = c_2$ is the closest center to $c$ in $S- S_0$, and so $\mu_O(p) = \mu_D(p)$. Now by \Cref{lemma:costsMO} and applying twice \Cref{lem:apxTriangleInequality2} with $\gamma=1+\sqrt{\eps}$,
\begin{align*}
d^2(p, \mu_O(p)) {=d^2(p, \mu_D(p))}& \leq \frac{2}{\sqrt{\eps}}d^2(p,c) + (1+\sqrt{\eps})d^2(c, c_2)\\
& \leq \frac{2}{\sqrt{\eps}}d^2(p,c ) + (1+\sqrt{\eps})\left(\frac{2}{\sqrt{\eps}}d^2(p,c )+(1+\sqrt{\eps})d^2(p,c_2)\right)\\
& \leq \frac{5}{\sqrt{\eps}}d^2(p,c ) + (1+\sqrt{\eps})^2 d^2(p, S- S_0)\\
& \leq \frac{5}{\sqrt{\eps}}d^2(p, S_\calQ) + (1+O(\sqrt{\eps}))r(p).
\end{align*} 
\end{description}
Summing up the above bounds, for a non-expensive cluster $C^*$ of $\OPT$ we get
\begin{align*}
\sum_{p\in C^*} d^2(p, \mu_O(p)) & \leq \sum_{p\in C^*-(T_1\cup T_2)}r(p)+\sum_{p\in C^*\cap T_1}\left(\frac{{5}}{\sqrt{\eps}}d^2(p, S_{\calQ})+ \frac{1+O(\sqrt{\eps})}{{4}}r(p)\right)\\
& + \sum_{p\in C^*\cap T_2}\left(\frac{5}{\sqrt{\eps}}d^2(p, S_\calQ) + (1+O(\sqrt{\eps}))r(p) \right)
\end{align*}
and
\begin{align*}
\sum_{p\in C^*} d^2(p, \mu_D(p)) & \leq \sum_{p\in C^*-(T_1\cup T_2)}r(p)+\sum_{p\in C^*\cap T_1}\left(\frac{{21}}{\sqrt{\eps}}d^2(p, S_{\calQ})+ \frac{9+O(\sqrt{\eps})}{{4}}r(p)\right)\\
& + \sum_{p\in C^*\cap T_2}\left(\frac{5}{\sqrt{\eps}}d^2(p, S_\calQ) + (1+O(\sqrt{\eps}))r(p) \right)
\end{align*}

If we sum  up the above inequality for all non-expensive clusters and the bound obtained before  for the expensive clusters of $\opt$ we thus get
\begin{align*}
\clcost(M_O, \mu_O) & \leq \sum_{p\in \clientsexpensive- T_2} \opt_p + \sum_{p\in \clientsexpensive \cap T_2}{4}\OPT_p\\
& + \sum_{p\in (D-\clientsexpensive)- T_1}r(p)+\sum_{p\in (D-\clientsexpensive)\cap T_1}\frac{1}{{4}}r(p)+O(\sqrt{\eps})\sopt\, ,
\end{align*}
and
\begin{align*}
\clcost(M_D, \mu_D) & \leq \sum_{p\in \clientsexpensive-T_2} 9\opt_p + \sum_{p\in \clientsexpensive \cap T_2}{4}\OPT_p\\
& + \sum_{p\in (D-\clientsexpensive)- T_1}r(p)+\sum_{p\in (D-\clientsexpensive)\cap T_1}\frac{9}{{4}}r(p)+O(\sqrt{\eps})\sopt\, .
\end{align*}
Above we used that $\sum_{c \in S_0 - \calQ} \sum_{p \in S_{\calQ}(c)} d^2(p, S_\calQ) = \sum_{p\in T_1 \cup T_2} d^2(p, S_\calQ) \leq \eps \cdot \sopt$ (\Cref{lem:successguessprocess}) and $\clcost(S) \leq \apxLSkmeans \sopt$ (\Cref{lemma:localsearchis5approximation}) to bound the error term $O(\sqrt{\eps} \cdot \sopt)$.

The global upper bound of $O(1)\sopt$ simply follows in both cases from \Cref{lemma:convenience_upper_bound}, which says that $\sum_{p\in \clients - \clientsexpensive} r(p) \leq \sum_{p\in \clients - \clientsexpensive} 4\opt_p + O(\sqrt{\eps} \cdot \sopt)$ if the sample $W$ is successful.

\end{proof}

\subsection{\texorpdfstring{Removal of Cheap Centers in $S_0$}{Removal of Cheap Centers in S0}}
\label{sec:removalofCheap}

At this stage we have defined a clustering $S_{\calQ} = S - \calQ \cup \dummyset$, where $\dummyset$ is the set of dummy centers. Recall that the mixed solution $M_O$ equals $S-S_0 \cup \ho$. So $S_\calQ - \Lambda$ represents progress compared to $S$ in that we have already removed a subset $\calQ$ of $S_0$. Furthermore, with respect to the assignment of clients $\mu_O$, we have that the remaining centers $S_0 - Q$, which we want to remove, are divided into two sets $\calR$ and $\calU$. The clusters with centers in $\calR$ are completely assigned (by $\mu_O$) to centers in $\ho$, and every cluster with center in $\calU$ is completely assigned to a center in $S - S_0$. 
The task of this section is to make further progress by guessing the centers in $\calU$ and their reassignment to centers in $S- S_0$. This turns out to be a difficult task, and we will instead ``approximately'' guess a set $\bcalU$ and reassignment $\tmu$ of the clients of those clusters. Specifically, we will output a solution $S_{\calQ \cup \bcalU} = S_{\calQ} - \bcalU$ (obtained by removing $\bcalU$ from $S_{\calQ}$) and an assignment of clients $\tmu$ to centers in $S_\calQ-\bcalU$. This is done through the following lemma.   {Recall that for $c\in S_{\calQ}$, $S_{\calQ}(c)$ is the set of clients assigned to $c$ in $S_\calQ$, i.e., closest to $c$ among all centers in $S_\calQ$.}    
\begin{lemma}[\cite{charikar2025kmeans}]\label{lem:successcheapremove}
Given $S$ and $\calQ$ as described above, there is a polynomial-time algorithm that produces a collection $\calL_{cheap}$ of subsets $\bcalU$ of $S-\calQ$, and for each such $\bcalU$ and assignment $\tmu$ of clients to centers in $S_{\calQ \cup \bcalU} = S_\calQ-\bcalU$, such that at least one such pair $(\bcalU,\tmu)$ satisfies the following properties:
\begin{enumerate}
    \item $|\bcalU| = |\calU|$.
    \item $\bcalU \cap \calR = \emptyset$, i.e., $\bcalU$ does not contain any center of $S_0 - \calQ$ whose set of clients is completely assigned
    to a center of $\ho$ in the assignment $\mu_O$.
    \item $\tmu$ satisfies: \begin{enumerate}
        \item For every center $c\in S_{\calQ} - \bcalU$, we have $\tmu(p)= c$ for every $p\in S_{\calQ}(c)$.
        \item For a center $c\in \bcalU$, the clients in $S_\calQ(c)$ are reassigned to a center $c' \in S- \calQ - \bcalU - \calR$, i.e., $\tmu(p) = c'$ for every $p\in S_{\calQ}(c)$.
       \item The cost increase of the reassignment $\tmu$ of clients previously assigned to $\bcalU$ compared to that  of the reassignment $\mu_O$ of clients previously assigned to $\calU$ is bounded by $O({\sqrt{\eps}}) \cdot \sopt$:
       \begin{gather*}
        \sum_{c \in \calU \cup \bcalU}  \sum_{p\in S_\calQ(c)} \left( d^2(p, \tmu(p)) - d^2(p, \mu_O(p)) \right) = O (\sqrt{\eps}) \cdot \sopt\,. 
        \end{gather*}
    \end{enumerate}
\end{enumerate}
\end{lemma}

In words, the above properties of $\tmu$ say that we maintain the assignment of clients whose closest center remains the same, and the clients associated to the removed centers $\bcalU$ are reassigned to centers not in $\calR$ at no higher cost (up to $O(\sqrt{\eps}) \cdot \sopt$) than the cost of the reassignment of $\calU$ in the modified solution $M_O$.

\subsection{\texorpdfstring{Mixed Solutions after the Removal of $\bcalU$}{Mixed Solutions after the Removal of U}}    

We next focus on a pair $(\bcalU,\tmu)$ that satisfies the properties in the claim of Lemma \ref{lem:successcheapremove}. Recall that $S_{\calQ \cup \bcalU}$ consists of the centers $S' = S - \calQ - \bcalU$ and the dummy centers $\dummyset$.
Further recall that $M_O$ consists of centers $S- S_0 = S - \calQ - \calU - \calR$ and $\ho$.  $M_D$ is the same set of centers except that $\ho$ is replaced by $\dummyset$. Notice that some centers of $M_O$ and $M_D$ are now removed if $\bcalU - \calU \neq \emptyset$. To take care of this, we modify these mixed solutions to obtain $M'_O$ and $M'_D$. 
 
 We first define $M'_O$; the definition of $M'_D$ is then very similar. The centers of $M'_O$ are $S- \calQ - \bcalU
 - \calR$ and $\ho$. Hence, the difference between $M_O$  and $M'_O$  is that in $M_O$ we remove $\calU$ from $S$ and in $M'_O$ we remove $\bcalU$. In other words,  $M'_O = M_O \cup \calU - \bcalU$. Similarly, we let $M'_D = M_D \cup \calU - \bcalU$. We update the assignment $\mu_O$ of $M_O$ to an assignment $\mu'_O$ of $M'_O$ (and the assignment $\mu_D$ to $\mu'_D$). 
 Recall that $\mu_O$ assigns each client $p$ to its closest center in $\ho$ (if $p\in \clientsexpensive$) or its closest center in $S- S_0$ (if $p\in \clients- \clientsexpensive$) except for those clients that belong to a cluster $S_{\calQ}(c)$ with $c\in \calU\cup \calR$. Indeed, the clients in $S_\calQ(c)$ are all assigned to the center of $\ho$ that is closest to $c$  if $c\in \calR$, and if $c\in \calU$ they are all assigned to the center in $S - S_0$ that is closest to $c$. We also recall that $\mu_D$ is the same as $\mu_O$ except when $\mu_O(p) \in \ho$ in which case $\mu_O(p)$ is replaced by its corresponding dummy center.

 \paragraph{Definitions of $\mu_O'$ and $\mu_D'$.}
 For clients $p\in S_{\calQ}(c)$ with $c \not\in \calU \cup \bcalU \cup \dummyset$, we define $\mu_O'(p) = \mu_O(p)$. For $p\in S_{\calQ}(c)$  with $c\in \dummyset$, we let $\mu'_O(p)$ be the closest center in $\ho$.
 Finally, for $p\in S_{\calQ}(c)$ with $c\in \calU \cup \bcalU$, we define $\mu'_O(p) =  \tmu(p)$. 

 The assignment $\mu'_D$ is obtained in the same way from $\mu_D$ with the difference that we use $\dummyset$ instead of $\ho$: 
 For clients $p\in S_{\calQ}(c)$ with $c \not\in \calU \cup \bcalU \cup \dummyset$, we define $\mu_D'(p) = \mu_D(p)$. For $p\in S_{\calQ}(c)$  with $c\in \dummyset$, we let $\mu'_D(p)$ be the closest center in $\dummyset$.
 Finally, for $p\in S_{\calQ}(c)$ with $c\in \calU \cup \bcalU$, we define $\mu'_D(p) =  \tmu(p)$. 
 
 This completes the definition of $\mu'_O$ and $\mu'_D$.
  We remark that we have $\mu'_O(p) = \mu_O(p)$ and $\mu'_D(p) = \mu_D(p)$  for all $p\in S_\calQ(c)$ with $c\not \in \calU \cup \bcalU \cup \dummyset$.
    
    We continue by arguing that $\mu'_O$ is well-defined, i.e., that $\mu'_O(p) \in M_O'$ for every $p\in \clients$ (the proof for $\mu'_D$ is the same).
 This is immediate for a client $p\in S_\calQ(c)$ with $c \in \dummyset$. For a client $p\in S_{{\calQ}}(c)$ with $c\in \cal U \cup \bcalU$, it holds because, by \Cref{lem:successcheapremove}, $\tmu(p)$ equals $c\in M'_O$ if  $c \in \calU -  \bcalU$ and otherwise $\tmu(p) \in  S- \calQ - \bcalU - \calR\subseteq M'_O$.
 It remains to verify that $\mu_O(p) \in M_O'$ for a client $p\in S_\calQ(c)$ with $c\not \in \calU \cup \bcalU \cup \dummyset$. If $c\in \calR$, we have $\mu_O(p) \in \ho \subseteq M'_O$.  Similarly, if $c \not \in \calU \cup \bcalU \cup \Lambda \cup \calR $ and $p\in \clientsexpensive$ we have $\mu_O(p) \in \ho \subseteq M'_O$.  In the remaining case when $c \not \in \calU \cup \bcalU \cup \Lambda \cup \calR $ and $p\in \clients- \clientsexpensive$, $\mu_O(p)$ equals $p$'s closest center in $S- S_0$. As   $p\in S_\calQ(c)$, we thus have   $\mu_O(p) = c \in S - (S_0 \cup \bcalU) \subseteq M'_O$. 

 The following upper bound on $\clcost(M'_O, \mu'_O) + \clcost(M'_D, \mu'_D)$ is a fairly immediate consequence of the assumption that the cheap-removal process is successful (Property 3c of Lemma \ref{lem:successcheapremove}). 
 Similarly to before, we say that the selection of $(W, \calB, \calQ, \bcalU,\tmu)$ is successful, if the sample $W$ selected in~\ref{sec:dsampleproc} is successful, the set of balls $\calB$ from \Cref{sec:ballguesses} is valid,  $\calQ$ selected in \Cref{sec:removalofExpensive} satisfies the properties of \Cref{lem:successguessprocess}, and $\bcalU,\tmu$  selected in this section satisfies the properties of \Cref{lem:successcheapremove}. {Again, there is an analogue of the next claim in \cite{charikar2025kmeans}, where however only the sum $\clcost(M'_O, \mu'_O)+\clcost(M'_D, \mu'_D)$ is upper bounded. We need to upper bound the two quantities separately.}
 \begin{claim}\label{claim:Mprimebounds}
     If $(W, \calB, \calQ, \bcalU,\tmu)$ is successful,
$$
\clcost(M'_O, \mu'_O)\leq \clcost(M_O, \mu_O)+O(\sqrt{\eps}) \cdot \sopt\,,
$$
and
$$
\clcost(M'_D, \mu'_D)\leq \clcost(M_D, \mu_D)+O(\sqrt{\eps}) \cdot \sopt\,.
$$
 \end{claim}
 \begin{proof}
     For $x\in \{O, D\}$ the only differences between $\mu'_x$ and $\mu_x$ are clients in $S_{\calQ}(c)$ with $c\in \dummyset \cup \calU \cup \bcalU$.  Consider first a client $p \in S_{\calQ}(c)$  with $c \in \dummyset$. Then, as $c$ is the closest center among $(S - \calQ) \cup \dummyset$, which both contains $M_O - \ho$ and $M'_O - \ho$, we have that the closest center to $p$ in both $M'_O$ and $M_O$ is in $\ho$ (because by the definition of dummy centers, $d^2(p, \ho) \leq d^2(p, \dummyset)$). Similarly, the closest center to $p$ in both $M'_D$ and $M_D$ is in $\Lambda$. Hence, the definitions of $\mu'_O$ and $\mu'_D$ to assign $p$ to its closest center in $\ho$ and $\dummyset$, respectively, cannot increase the cost, i.e., $d^2(p, \mu'_O(p)) \leq d^2(p, \mu_O(p))$ and $d^2(p, \mu'_D(p)) \leq d^2(p, \mu_D(p))$ for such a client $p$. 
     
     Finally, for those clients $p\in S_\calQ(c)$ with $c\in\calU \cup \bcalU$ we have $\mu'_x = \tmu$ by definition and $d^2(p, \mu_O(p)) \leq d^2(p, \mu_D(p))$ since $\mu_O(p) = \mu_D(p)$ unless $\mu_O(p)\in \ho$ in which case $\mu_D(p)$ is the dummy center associated with $\mu_O(p)$ that can only be farther away from $p$ than $\mu_O(p)$.
     Hence
       \begin{gather*}
       \sum_{c \in \calU \cup \bcalU}  \sum_{p\in S_\calQ(c)} \left( d^2(p, \tmu(p)) - d^2(p, \mu_D(p)) \right) \leq \sum_{c \in \calU \cup \bcalU}  \sum_{p\in S_\calQ(c)} \left( d^2(p, \tmu(p)) - d^2(p, \mu_O(p)) \right) = O (\sqrt{\eps} \cdot {\sopt})\,,
        \end{gather*}
where in the equality we used Property (c) of \Cref{lem:successcheapremove}. 
\end{proof}

To better understand $\clcost(M'_O, \mu'_O)$, let us consider the cost $d(p, \mu'_O(p))$ of a single client $p$:
\begin{itemize}
    \item If $p \in S_{\calQ}(c)$ with $c\not\in \bcalU \cup \calR$ then $\tmu(p) = c$ by Property 3a of \Cref{lem:successcheapremove}.  Moreover,  $\mu'_O(p)$ assigns $p$ either to a center in $\ho$ or to one in $S - \calQ - \bcalU - \cal R$ and we have $d^2(p, c) \leq d^2(p, S - \calQ - \bcalU - \cal R)$.  Hence, in either case, we have
    \[
        d^2(p, \mu'_O(p)) \geq  d^2(p, \{\tmu(p)\} \cup \ho) = d^2(p, \{\tmu(p)\} \cup \ho \cup \dummyset) \,.
    \]
    \item If $p\in S_{\calQ}(c)$ with $c\in \calR$ then $\mu'_O(p) = \mu_O(p) =  c^*\in \ho$ where $c^*$ is the center of $\ho$ that is closest to $c$. So 
    \[
        d^2(p, \mu_O'(p)) = d^2(p, c^*)\,. 
    \]
    \item If $p \in S_{\calQ}(c)$ with $c\in \bcalU$ then $\mu'_O(p) = \tmu(p)$ and so
    \[
        d^2(p, \mu_O'(p)) \geq d^2(p, \{\tmu(p)\} \cup \ho \cup \dummyset).
    \]
\end{itemize}
Similarly, we can analyze $d^2(p, \mu_D(p))$ by replacing $\ho$ with the set $\dummyset$ of dummy centers (and $c^*$ by its associated dummy center). 
Summarizing, we have 
\begin{align*}
     \clcost(M'_O, \mu'_O) &\geq \sum_{c\in S_{\calQ} - \calR}\sum_{p\in S_{\calQ}(c)}  d^2\left(p, \{c\} \cup \ho \cup \dummyset\right) + \sum_{c\in \calR} \min_{c' \in \ho} \sum_{p\in S_{\calQ}(c)}  d^2\left(p, c' \right)\\
     \intertext{and}
     \clcost(M'_D, \mu'_D) &\geq \sum_{c\in S_{\calQ} - \calR}\sum_{p\in S_{\calQ}(c)}  d^2\left(p, \{c\}  \cup \dummyset\right) + \sum_{c\in \calR} \min_{c'\in \dummyset}\sum_{p\in S_{\calQ}(c)}  d^2\left(p,   c'\right)
\end{align*}

 Furthermore, by Properties 3a and 3b of \Cref{lem:successcheapremove}, we have that, for every $c\in \calR$, the set $S_{\calQ}(c)$ of clients assigned to $c$ in $S_{\calQ}$ equals the set $\tmu^{-1}(c)$ of clients assigned by $\tmu$.
 So, if we let $\{C_c\}_{c\in S_{\calQ \cup \bcalU}}$ be the partitioning of the set $\clients$ of clients according to $\tmu$, i.e., $C_c = \{p\in \clients \mid \tmu(p) =c\}$,  we can thus rewrite the above  bounds on the cost as 
\begin{align}
     \clcost(M'_O, \mu'_O) &\geq \sum_{c\not \in \calR}\sum_{p\in C_c}  d^2\left(p, \{c\} \cup \ho \cup \dummyset\right) + \sum_{c\in \calR}\min_{c' \in \ho}\sum_{p\in C_c}  d^2\left(p,  c'\right) \label{eq:MprimeObound}\\
     \intertext{and}
     \clcost(M'_D, \mu'_D) &\geq \sum_{c \not\in  \calR}\sum_{p\in C_c}  d^2\left(p, \{c\}  \cup \dummyset\right) + \sum_{c\in \calR}\min_{c' \in \dummyset}\sum_{p\in C_c}  d^2\left(p,  c'\right)\,.\label{eq:MprimeDbound}
\end{align}
In the next section, we use these bounds to give a polynomial-time algorithm that outputs a solution $S^*$ with $k$ centers so that (see \Cref{lemma:findingSstar})
\begin{align*}
    \clcost(S^*)  & \leq
    (1+6\sqrt{\eps}) \left({(1-\frac{1}{e})}\clcost(M'_O, \mu'_O) +{\frac{1}{e}}  \clcost(M'_D, \mu'_D) \right) + O(\sqrt{\eps})\cdot \sopt \\
     & \leq \sum_{p\in \clients} {4} \opt_p  + O(\sqrt{\eps}) \cdot \sopt
\end{align*}
where the second inequality is by \Cref{lemma:costboundofMOandMDwithassignments} {and} \Cref{claim:Mprimebounds}.
So the proof of \Cref{lemma:findingSstar}  in the next subsection is the final step in the proof of \Cref{thr:mainMetricStable} (see also \Cref{sec:stable:everythingtogether} where we put everything together to prove \Cref{thr:mainMetricStable}).

\subsection{\texorpdfstring{Finding $S^*$ via Submodular Optimization}{Finding S* via Submodular Optimization}}
\label{sec:submodularopt}

In this section, we give a polynomial-time algorithm for finding the solution $S^*$ by reducing the problem to maximizing a submodular function subject to a partition matroid constraint. Specifically, we prove the following lemma:
\begin{lemma}\label{lemma:findingSstar}
    We can in polynomial-time find a clustering $S^*$ with $k$ centers of cost at most {$(4+O(\sqrt{\eps}))\cdot \sopt$}.
\end{lemma}
Recall that at this point the algorithm calculated the local search solution $S$, sampled $W$, guessed $\calB$, $\calQ$, and $\bcalU$ with the assignment $\tmu$ of the clustering $S_{\calQ \cup \bcalU}$. We assume that all these choices were successful guesses, i.e., that $(W, \calB, \calQ, \bcalU, \tmu)$ is successful.  Further recall the notation that we partition the clients into the clusters $\{C_c\}_{c\in S_{\calQ \cup \bcalU}}$ where $C_c = \{ p\in \clients : \tmu(p) =c\}$.  

Our goal is to give a polynomial-time algorithm that finds approximations of the sets $\ho$ and $\calR$ that only have slightly worse cost. We start by defining the feasible set of candidates for $\ho$ as a partition matroid. Recall that $\ho$ contains one center from each ball in $\calB$.

\paragraph{Definition of partition matroid $\calM_\calB$.} 
We define the partition matroid that captures the constraint that we wish to open a center in each ball in $\calB$. 
Let $\facilities_\calB$ be the  (multi) subset of facility/center locations containing  $B\cap \facilities$ for each ball  $B\in \calB$. If a center $c$ is in multiple balls in $\calB$, then $\facilities_\calB$ contains one distinct copy of $c$ for each ball. We let $\facilities_{\calB}(B) \subseteq F_\calB$ denote the centers associated with $B\in \calB$. These sets satisfy the following two properties:
\begin{itemize}
    \item The sets $\facilities_{\calB}(B)$ partition $\facilities_\calB$.
    \item For $B \in \calB$, $\facilities_{\calB}(B)$ contains (a copy of) every center in $\facilities\cap B$.
\end{itemize}
The first property holds since we took a unique copy of each center for each ball, and they are thus disjoint: $\facilities_\calB(B) \cap \facilities_\calB(B') = \emptyset$ for distinct $B,B' \in \calB$. Indeed, while it is not better for the cost to open multiple copies of a center, we make the copies to ensure the above two properties. This allows us to define the partition matroid $\calM_\calB = (\facilities_\calB, \calI)$ where
\[
    \calI = \{X \subseteq \facilities_\calB : |X\cap F_\calB(B)| \leq 1 \mbox{ for every } B\in \calB\}\,. 
\]
Moreover, since there is exactly one ball in $\calB$ for each center in $\ho$, we have $\ho \in \calI$ (where we slightly abuse notation as we should take the copy of center $c^*\in \ho$ that belongs to its associated ball). 

\paragraph{Core, concentrated and hit clusters.} Our goal is now to define a submodular function $f$ so that we obtain a good approximation to $\ho$ by maximizing $f$ over the matroid constraint $\calM_\calB$.  In particular, the domain of $f$ is every subset of $\facilities_\calB$. However, we need some additional steps before defining $f$. In particular, we introduce the concept of the core of a cluster $C_c$ and the notions of concentrated and hit clusters, which allow us to simplify the structure of centers in $\calR$. For a cluster $C_c$, we define the \emph{core} of $C_c$ as
\begin{gather*}
    \core_c = \left\{p \in C_c: d^2(p, c) \leq \eps \cdot \frac{\clcost(S)}{|\calR|\cdot |C_c|}\right\}\,.
\end{gather*}
We further say that cluster $C_c$ is \emph{concentrated} if 
\begin{gather*}
    |\core_c| \geq (1-\eps) |C_c|
\end{gather*}
and it is \emph{hit} by a set $X$ of centers if there is a point $p \in \core_c$ such that
\begin{gather*}
    d^2(p, X) < d^2(p,c)\,.
\end{gather*}
For shortness we will sometimes say that a center is concentrated (resp., hit), if the corresponding cluster is so.

\paragraph{Guessing the centers of $\calR$ that are not concentrated.} We further simplify the task of finding the set of centers $\calR$  by guessing the centers of $\calR$ that are not concentrated. Specifically, partition the set $\calR$ into $\calR_0$ and $\calR_1$, where $\calR_1$ contains those centers of $\calR$ that are concentrated and $\calR_0$ contains those that are not. We can  correctly guess $\calR_0$ in polynomial time since
the following simple claim shows that it is a subset of the centers whose cluster costs at least $\eps^2 \cdot \sopt/|\calR|$, of which there are only $O(|\calR|/\eps^2)$ many, 
and so guessing $\calR_0$ can thus be done in time
$2^{O(|\calR|/\eps^2)}$.
\begin{claim}[\cite{charikar2025kmeans}]
    Suppose that $C_c$ is not concentrated. Then the cost of $C_c$ is at least $\eps^2 \cdot \sopt/|\calR|$.
\end{claim}

Now let $\calP$ be the potential centers that can be in $\calR$. Specifically, we let $\calP$ be the set that contains a center $c$ in $S -\calQ - \bcalU$ if $C_c$ equals the set of clients in $S_{\calQ}(c)$, where we recall that $S_{\calQ}(c)$ is the set of clients that are closest to $c$ in the clustering $S_{\calQ}$. Notice that the algorithm has all the information $S, \calB, \calQ, \bcalU$ and $\tmu$ to calculate $\calP$. Furthermore, by \Cref{lem:successcheapremove}, we have that no clients from $\bcalU$ were reassigned  by $\tmu$ to a center in $\calR$. So for $c\in \mathcal{R}$ we have $C_c = S_{\calQ}(c)$.  Hence, $\cal R \subseteq \calP$ and we can guess $\calR_0$ as follows:
\begin{mdframed}[hidealllines=true, backgroundcolor=gray!15]
\vspace{-5mm}
\paragraph{Guessing $\calR_0$}\ \\
\begin{enumerate}
    \item Let $\calC$ be the clusters of $\calP$ whose cost is at least $\eps^2 \cdot \sopt/|\calR|$.
    \item Output each subset of $\calC$.
\end{enumerate}
\end{mdframed}
By the above claim and the definition of $\calP$, one of the outputs is $\calR_0$.  Furthermore, the above guessing procedure outputs a family of polynomial many subsets. Indeed, we have $|\calR| \leq \log(n)/\eps^3$ by \Cref{lem:numnonpure}. Moreover, the total cost of the clusters in $\calP$ is $O(\sopt)$. To see this notice that the clusters corresponding to $\calP$ is a subset of the clusters of $S_{\calQ}$ and the cost of $S_{\calQ}$ is at most $\clcost(S - S_0  \cup \Lambda)$ (since $S_{\calQ}\supseteq S- S_0 \cup \Lambda$), which has cost at most $O(\sopt)$ by \Cref{lemma:S0properties}.  This implies that $|\calC| = O(\log(n)/\eps^5)$, so the total number of subsets is $n^{\eps^{-O(1)}}$. The algorithm proceeds by trying all possible subsets in the output, and we analyze the algorithm when it takes the correct guess of $\calR_0$.

\paragraph{Definition of the submodular function $f$. }
We first define another function $g$ on the same domain as $f$, i.e., on all subsets $X$ of $F_\calB$. We will then define $f$  by $f(X) = g(\emptyset) - g(X)$. Let the \emph{closed cost} of a cluster $C_c$ be defined as
\[
    \closedclcost_c(X) := 
    \begin{cases}
        \sum_{p\in C_c} d^2(p, \{c\} \cup X \cup \dummyset) & \mbox{if $C_c$ is hit by $X$,} \\
        \min_{c'\in X \cup \dummyset} \sum_{p\in \core_c} d^2(p, c') + \sum_{p\in C_c - \core_c} d^2(p, \{c\} \cup X \cup \dummyset) & \mbox{otherwise.} 
    \end{cases}
\]
Further, let $\calP_1$ be the potential centers for $\calR_1$: it contains each center $c\in \calP - \calR_0$ so that $c$ is concentrated, i.e., $|\core_c| \geq (1-\eps) |C_c|$. We remark that the algorithm can calculate this set $\calP_1$ as it only depends on $\calP$, the guessed set $\calR_0$, the value $\clcost(S)$, and the clusters $C_c$ defined by $\tmu$. Moreover, by definition, we have $\calR_1 \subseteq \calP_1$.
For a subset $X \subseteq \facilities_\calB$, we then define $g(X)$ to be the minimum value of 
\begin{gather*}
     \sum_{c \not \in \calR_0 \cup \calR'_1}\sum_{p\in C_c}  d^2\left(p, \{c\} \cup X \cup \dummyset\right) + \sum_{c\in \calR_0} \sum_{p\in C_c} d^2(p, X \cup \dummyset) + \sum_{c\in \calR'_1} \closedclcost_c(X) 
\end{gather*}
over all subsets $\calR'_1 \subseteq \calP_1$ with $|\calR_1'| = |\calR_1| = |\calR| - |\calR_0|$. 
In words, over the best $\calR_1'$, $g(X)$ is the cost of the solution obtained by removing the centers $\calR_0 \cup \calR_1'$  and assigning clients as follows:
\begin{itemize}
    \item If $p\in C_c$ for a remaining center $c \not\in \calR_0 \cup  \calR_1'$ or $p \not \in \core_c$ with $c\in \calR'_1$, $p$ is assigned to its closest center in $\{c\} \cup X \cup \dummyset$.
    \item If $p \in C_c$ for a removed center $c\in \calR_0$, $p$ is assigned to its closest center in $X \cup \dummyset$.
    \item If $p \in \core_c$ for a removed center $c\in \calR'_1$, $p$ is assigned to its closest center in $\{c\} \cup X \cup \dummyset$ if  $X$ hits $C_c$, and otherwise all clients in $\core_c$ are assigned to the same center $c'\in X \cup \dummyset$ that minimizes the cost.
\end{itemize}

We remark that this assignment is infeasible in the sense that it may assign clients to removed centers in $\calR_1$. Nevertheless, we relate the values of $g(\emptyset)$ and $g(\ho)$ to $\clcost(M'_D, \mu'_D)$ and $\clcost(M'_O, \mu'_O)$, respectively; and, we show that given an $X \subseteq \facilities_\calB$ that is independent in $M_\calB$, i.e., $X\in \calI$, we can in polynomial-time output $k$ centers $S^*$ whose cost is at most $(1+6\sqrt{\eps})g(X) + O(\sqrt{\eps}) \sopt$.
Finally, the definition of $g$ allows us to prove that $f$ (defined by $f(X) = g(\emptyset) - g(X)$) is a non-negative monotone submodular function.
\begin{lemma}[\cite{charikar2025kmeans}]
We have that $g$ and $f$  satisfy the following properties:
\begin{enumerate}
\item We can evaluate $g(X)$ in polynomial time for every given $X \subseteq \facilities_\calB$, and we can thus evaluate $f(X)$ in polynomial time. 
\item $f$ is a non-negative monotone submodular function.
\item The value $g(\emptyset)$ is at most $\clcost(M'_D, \mu'_D)$.
\item The value $g(\ho)$ is at most $\clcost(M'_O, \mu'_O)$.
\item  Given an $X \subseteq \facilities_\calB$ that is independent in $M_\calB$, i.e., $X\in \calI$, we can in polynomial-time output $k$ centers $S^*$ whose associated cost is at most $(1+6\sqrt{\eps}) g(X) + O(\sqrt{\eps}) \sopt$.
\end{enumerate}
\label{lemma:propertiesfandg}
\end{lemma}
This directly implies \Cref{lemma:findingSstar}.

\begin{proof}[Proof of \Cref{lemma:findingSstar}]

By the first property of \Cref{lemma:propertiesfandg}, we can evaluate $f$ in polynomial time. Moreover, it is easy to see that we can answer independence queries in polynomial time for any partition matroid, particularly for $\calM_\calB$. We can thus apply \Cref{thm:submodularmatroidoptimization} on $f$ and $\calM_\calB$  with $\zeta =  {\eps}$
to find a solution $X \in \calI$ such that
\begin{gather*}
    g(\emptyset) - g(X) =  f(X) \geq (1-1/e - {\eps})f(\ho) = (1-1/e - {\eps})( g(\emptyset) - g(\ho))\,,
\end{gather*}
where we used that $\ho$ is one feasible solution, i.e., $\ho \in \mathcal{I}$.
This in turn implies that{ 
$$
g(X)\leq (\frac{1}{e}+\eps)g(\emptyset)+(1-\frac{1}{e}-\eps)g(\ho).
$$}

Using the upper bounds on $g(\emptyset)$ and $g(\ho)$ of the above lemma we thus have found a set $X$ such that 
{ 
\begin{align*}
g(X) & \leq (\frac{1}{e}+\eps)\clcost(M'_D, \mu'_D)+(1-\frac{1}{e}-\eps)\clcost(M'_O, \mu'_O)\\
& \overset{\text{Claim \ref{claim:Mprimebounds}}}{\leq} \frac{1}{e}\clcost(M_D, \mu_D)+(1-\frac{1}{e})\clcost(M_O, \mu_O)+O(\sqrt{\eps})\sopt\,,
\end{align*}
where in the last inequality we also used that $\clcost(M_D, \mu_D)$ and $\clcost(M_O, \mu_O)$ are $O(1)\sopt$ by Lemma \ref{lemma:costboundofMOandMDwithassignments}.}
Now using the last property of the above lemma we can output a solution $S^*$ whose cost is at
most{
\begin{align*}
& (1+6\sqrt{\eps})\left(\frac{1}{e}\clcost(M_D, \mu_D)+(1-\frac{1}{e})\clcost(M_O, \mu_O)+O(\sqrt{\eps})\sopt\right)+O(\sqrt{\eps})\cdot \sopt\\
\leq & \frac{1}{e}\clcost(M_D, \mu_D)+(1-\frac{1}{e})\clcost(M_O, \mu_O)+O(\sqrt{\eps})\cdot \sopt\\
\overset{\text{Lem. \ref{lemma:costboundofMOandMDwithassignments}}}{\leq} &  
\sum_{p\in \clientsexpensive- T_2}(1+\frac{8}{e})\opt_p+\sum_{p\in \clientsexpensive \cap T_2}{4}\OPT_p\\
& +\sum_{p\in (D-\clientsexpensive)- T_1}r(p)+\sum_{p\in (D-\clientsexpensive)\cap T_1}\frac{1}{{4}}(1+\frac{8}{e})r(p)+O(\sqrt{\eps})\cdot \sopt\\
\leq & \sum_{p\in \clientsexpensive}4\opt_p+\sum_{p\in D-\clientsexpensive}r(p)+O(\sqrt{\eps})\cdot \sopt
\overset{\text{Lem. \ref{lemma:convenience_upper_bound}}}{\leq} \sum_{p\in D}4\opt_p +O(\sqrt{\eps})\cdot \sopt
\end{align*}} 
as required.

Finally, each of the above steps runs in polynomial time: the algorithm of \Cref{thm:submodularmatroidoptimization} is polynomial time, and the last property of \Cref{lemma:propertiesfandg} used to obtain $S^*$ is polynomial-time. 
Moreover, the number of guesses of $\calR_0$ is at most $n^{\eps^{-O(1)}}$, as argued after the description of that procedure. So we can, in polynomial time, try all possibilities and, among all solutions found (one for each guess of $\calR_0$), return one that minimizes the cost and, in particular, has cost at most that of $S^*$ (which was analyzed assuming the guess of $\calR_0$ was correct). We thus have  a polynomial time algorithm that returns a solution that satisfies the guarantee of the lemma.
\end{proof}

\subsection{\texorpdfstring{Putting Everything Together 
-- Proof of \Cref{thr:mainMetricStable}}{Putting Everything Together 
-- Proof of thr:mainMetricStable}}
\label{sec:stable:everythingtogether}
We now turn to proving \Cref{thr:mainMetricStable}. We first describe the algorithm we analyse.
\begin{mdframed}[hidealllines=true, backgroundcolor=gray!15]
\vspace{-5mm}
\paragraph{A $({4}+O(\sqrt{\eps}))$-Approximation for $k$-Means on $\eps/\log n$-
Stable Instances $k, \clients, \facilities, \dist$.}\ \\
\begin{enumerate}
    \item $S \gets $ Local Search on $(k, \clients, \facilities, \dist)$, see \Cref{sec:localSearchAnalysis}
    \item Let $W$ be the set produced by the
    $s^*$-$D^2$-Sample
    process (\Cref{sec:dsampleproc}) on $S$.
    \item For each $\calB{\in \calL_{\texttt{bal}}}$ {produced as in \Cref{lem:ballguesses}} on $W$ (\Cref{sec:ballguesses})   
    \begin{enumerate}
    \item Compute the set of dummy centers $\dummyset$.
        \item For each {$Q\in \calLexp$ produced as in \Cref{lem:successguessprocess} 
        (\Cref{sec:removalofExpensive})}:
        \begin{enumerate}
            \item \label{step:finalalg:submodular} For each $\bcalU{\in \calL_{cheap}}$, with the associated $\tmu$, {produced as in \Cref{lem:successcheapremove} }           (\Cref{sec:removalofCheap}):\newline 
\quad            solve the Submodular Instance associated with $(W,\calB,\calQ,\bcalU,\tmu)$ and compute the resulting solution $S_{(W,\calB,\calQ,\bcalU,\tmu)}$ for the $k$-Means problem (\Cref{sec:submodularopt})
        \end{enumerate}
    \end{enumerate}
    \item Output the solution $S_{(W,\calB,\calQ,\bcalU,\tmu)}$ of smallest cost. 
\end{enumerate}
\end{mdframed}

Our approach consists of running the above algorithm
$\log n$ times and take the minimum cost solution output.
In the remaining, we argue that the above algorithm yields a $({4}+O(\sqrt{\eps}))$-approximation with probability at least $(1-\eps)(1-1/n)$. \Cref{thr:mainMetricStable} then follows as the probability that all $\log n$ executions of the above algorithm would fail is at most $\left( 1- (1-\eps)(1-1/n)\right)^{\log n} < 1/n$, and so with high probability we output a $({4}+O(\sqrt{\eps}))$-approximation. 

We first discuss the algorithm's running time.
The local search algorithm runs in polynomial time. 
The submodular optimization step also runs in polynomial time by \Cref{lemma:findingSstar}.
The output of the $s^*$-$D^2$-Sample process (which is clearly a polynomial time procedure) is 
a subset of size $s^*$. By \Cref{lem:ballguesses}, we construct $\calL_{\texttt{bal}}$ in polynomial time, and $|\calL_{\texttt{bal}}| \leq n^{\eps^{-O(1)}}$. Similarly, by \Cref{lem:successguessprocess}, we construct $\calLexp$ in polynomial time and  $|\calLexp| \leq n^{\eps^{-O(1)}}$. %
{Finally, by \Cref{lem:successcheapremove}, we construct in polynomial-time 
 $n^{\eps^{-O(1)}}$ candidate pairs $\bcalU,\tmu$.} It follows that we solve $n^{\eps^{-O(1)}}$ submodular function optimization problems, each in polynomial time, and thus the total running time is polynomial.

We turn to proving the approximation guarantee.
We aim to show that there exists a 
$(W,\calB, \calQ, \bcalU, \tmu)$ that is successful with probability at least $(1-\eps/2)$.
Assuming this, the $({4}+O(\sqrt{\eps}))$-approximation follows by  \Cref{lemma:findingSstar}.

Our algorithm runs the local search algorithm and
thus provides us with a $\apxLSkmeans$ approximation $S$ that satisfies \Cref{lem:purecost}. Our algorithm next applies
the $s^*$-$D^2$-Sample process and by \Cref{lem:probcost} 
finds a successful $W$ with probability at least
$1-\eps$. Condition on having a successful
$W$, we then have that \Cref{lem:cheapcost} holds, and so does \Cref{lemma:convenience_upper_bound} by combining \Cref{lem:purecost} and \Cref{lem:cheapcost}.

Next, \Cref{lem:ballguesses} implies that one 
set of balls $\calB$ is a valid set of balls. This
set of balls induces a set of dummy centers 
$\dummyset$. From there, by \Cref{lem:successguessprocess} we have that 
with probability at least $1-1/n$, the 
procedure from \Cref{lem:successguessprocess} produces a set
of centers $\mathcal{Q}$ such that     \begin{enumerate}
    \item $\calQ \subseteq S_0$, and
    \item The total cost in solution $S {-} Q \cup \dummyset$ of the clusters in $S_0 {-} \calQ$ is at most $\eps \cdot \sopt$.
    \end{enumerate}
and so we have obtained a successful 
$(W,\calB,\calQ)$ and \Cref{lemma:costboundofMOandMDwithassignments}
applies. Condition on the event that we obtain
a successful $(W, \calB, \calQ)$, then, the procedure from \Cref{lem:successcheapremove} outputs a pair
$(\bcalU, \tmu)$ satisfying the properties in the claim of the lemma.
We thus have a $(W, \calB, \calQ, \bcalU,\tmu)$ such that $W$ is successful, the set of balls $\calB$ from \Cref{sec:ballguesses} is valid,  $\calQ$ selected in \Cref{sec:removalofExpensive} satisfies the properties of \Cref{lem:successguessprocess}, and $\bcalU,\tmu$  selected in this section satisfies the properties of \Cref{lem:successcheapremove} and
so $(W, \calB, \calQ, \bcalU,\tmu)$ is successful. Finally, note that the only probabilistic steps were the sampling of $W$ and of $\calQ$, and both steps are successful with probability at least $(1-\eps) \cdot (1-1/n)$, as required.

\section{\texorpdfstring{$(2+O(\sqrt{\eps}))$-Approximation for  $(\frac{\zeta}{\log n})$-{Stable} Euclidean Instances}{errorrendering-Approximation for  errorrendering-{Stable} {Euclidean} Instances}}
\label{sec:centerremoval_eucl}

In this section, we prove Theorem \ref{thr:mainEuclideanStable}. Recall that our goal is to compute a $2+O(\sqrt{\eps})$ approximate solution for $O(\zeta/\log n)$ stable instances. This improves on the $4+O(\sqrt{\eps})$ approximation in \cite{CCGGLW2026kmeans}, that however works in the more general metric setting, under the only restriction that the client locations are valid facility locations (while our improvement only works for the Euclidean case). Our algorithm is the same as in \cite{CCGGLW2026kmeans}, and our analysis is very similar too: in the following we will only focus on the differences in the analysis. 

\paragraph{Notation and Preliminaries.} To simplify the formulas, we let $\eps$ be the minimum of $\eps$ and $\zeta$ in the claim of Theorem \ref{thr:mainEuclideanStable} and further assume that $\eps<1/12$ is sufficiently small. With this notation, we present a $(2+O(\sqrt{\eps}))$-approximation algorithm for $\eps/\log(n)$-stable instances. This implies the above theorem as any $\zeta/\log(n)$-stable instance is also $\zeta'/\log(n)$-stable with $\zeta' \leq \zeta$.

Recall that, by known reductions, we can assume that we are given a set $\facilities$
of tentative centers, {with $D\subseteq \facilities$}, and we have to choose centers from that set. {Recall also that non-zero distances are between $1$ and $n^2/\eps^2$ according to Lemma \ref{lem:aspectratio-euclid}. We also assume w.l.o.g. that the optimal cost is positive (hence at least $1$).}

Given a cluster $C^*$ with center $c^*\in \facilities$, in the analysis it is convenient to consider the actual center of mass $\mu(C^*)$ of $C^*$. Obviously $\sum_{p\in C^*}d^2(\mu(C^*),p)\leq \sum_{p\in C^*}d^2(c^*,p)$. We will also need the following standard fact. 
\begin{lemma}\label{lem:centerMass}
For any $C^*\subseteq D$ and for any $c{,c^*}\in \Re^t$, 
$$
\sum_{p\in C^*}d^2(c,p)\leq |C^*|d^2(c,\mu(C))+\sum_{p\in C^*}d^2(\mu(C^*),p){\leq |C^*|d^2(c,\mu(C^*))+\sum_{p\in C^*}d^2(c^*,p)}.
$$
\end{lemma}

Given a set of centers $A\subseteq \facilities$ (not necessarily of cardinality $k$), we let $A_p=\dist^2(p,A)$ denote the cost associated to $p$ in the clustering induced by $A$, and $\clcost(A):=\sum_{p\in \clients}A_p$ be the total cost of $A$. The cluster in $A$ associated with $c\in A$ is denoted by $A(c)$. Sometimes we will consider a (possibly suboptimal) assignment $\mu:\clients \rightarrow A$ of clients to centers in $A$, and let $\clcost(A,\mu):=\sum_{p\in \clients}d^2(p,\mu(p))$ be the corresponding cost. Let $\opt$ be a fixed optimum solution under the restriction $\opt\subseteq \facilities$. Let also $\sopt = \clcost(\opt)$ be its cost.

\paragraph{Overview of our approach.} Our algorithm consists of a few steps which are discussed in the following subsections. The first step (see Section \ref{sec:localSearchAnalysis_eucl}) is to compute a candidate solution $S$ via standard local search. We will show that the points in almost all the clusters in $\opt$ (excluding a logarithmic size set) pay in $S$ roughly the same cost as in $\opt$. We call the centers $OPT_{pure}$ of such clusters \emph{pure}. More precisely, we will show that all the points in $OPT(c^*)$ with $c^*\in \opt_{pure}$ can be reassigned to the \emph{same} center $c'\in S$ without increasing too much their cost.  

For each remaining \emph{impure} center $c$, we identify a core set of clients in $\opt(c)$, the \emph{leaders},  which are \emph{sufficiently close} to $c$. We distinguish the impure clusters depending on whether the cost in $S$ of the leaders is small (\emph{core-cheap}) or large (\emph{core-expensive}).  The next step (see Section \ref{sec:leaders}) is to identify one leader in most of the core-expensive clusters of $\opt$. Let $\Lambda$ be the set of such leaders, and $\opt_{exp}$ be the core-expensive centers for which one leader was discovered. One can reassign all the clients $\opt(c^*)$ with $c^*\in \opt_{exp}$ to the corresponding leader while increasing their cost at most by a factor (roughly) $2$. Let $\opt_{cheap}$ be the remaining impure centers of $\opt$. We show that we can reassign all the clients $OPT(c^*)$ with $c^*\in \opt_{cheap}$ to the same center in $S$ while increasing their cost, again, by at most a factor (roughly) $2$. The {latter} two mentioned factors $2$ are {replaced by} $4$ in the analysis in \cite{CCGGLW2026kmeans}.

Consider next the solution that contains the leaders $\Lambda$ plus the centers in $S$ to which the points in the clusters of $\opt_{pure}\cup \opt_{cheap}$ are reassigned (which are at most $|\opt_{pure}|+|\opt_{cheap}|$ many). Notice that this is a feasible solution since $|\Lambda|=|\opt_{exp}|$. Furthermore, it is  (roughly) $2$-approximate by the previous discussion. Our final goal is to identify such a solution, or some good enough approximation of it. More precisely, we start with the solution $\Lambda\cup S$ (which is infeasible), and remove a set $S'_{disc}\subseteq S$ of $|\Lambda|$ many carefully chosen centers. This removal is done in two steps. First (see Section \ref{sec:removalofExpensive_eucl}), we remove a set $\calQ\subseteq S$ of centers which are \emph{expensive} in $S$. Then (see Section \ref{sec:removalofCheap_eucl}), we remove other $\bcalU$ centers from $S$, with $|\bcalU|=|\Lambda|-|\calQ|$.

\subsection{Locally Optimal Solution and Pure Clusters}
\label{sec:localSearchAnalysis_eucl}

Let $S$ be a locally optimal solution output by 
the following standard local search algorithm. Here, for technical reasons and differently from \cite{CharikarCGG98}, we use a truncated version of local search: this way {we do not need to assume that distances are integers}.
\begin{mdframed}[hidealllines=true, backgroundcolor=gray!15]
\vspace{-5mm}
\paragraph{$localSearch()$}\ \\
\vspace{-3mm}
\begin{algorithmic}[1]
\State $S \gets $ Arbitrary solution for $k$-Means
\While{There exists a solution $S'$ such that $|S \Delta S'|\le 2$ and $\clcost(S') \leq (1-\eps)\clcost(S)$}
\State $S \gets S'$
\EndWhile
\State \textbf{return} $S$
\end{algorithmic}
\end{mdframed}
We remark that the running time of the above procedure is polynomial.
It is standard fact that local search achieves a $\apxLSkmeans$ approximation for
$k$-Means, see \cite{GT08} (as we will see, the specific constant in not very relevant for us).
\begin{lemma}[\cite{GT08}] \label{lemma:localsearch}
The procedure $localSearch()$ runs in polynomial time and returns a feasible solution $S$ with $\clcost(S) \leq \apxLSkmeans\cdot \sopt$. 
\end{lemma}

We next fix $S$, and recall that we also fixed a reference optimal solution $\opt$. We identify in $\opt$ a set of pure clusters as follows. We say that a cluster $C^*$ of $\opt$ is \emph{pure} 
(w.r.t. $S$) if there exists a cluster $C'$ of $S$ such that $|C' \Delta C^*| \le {3\eps} \min(|C^*|,|C'|)$. In which
case we also say that $C'$ is pure w.r.t. $\opt$. We say that $C'$ and $C^*$ are associated. By $\opt_{pure}\subseteq \opt$ we denote the set of the centers of pure clusters (that we also call \emph{pure}). The centers (resp, clusters) which are not pure are \emph{impure}. The impure centers of $\opt$ are denoted by $\opt_{imp}$. We use a similar notation for the centers and clusters of $S$.
The following lemma follows from the fact that the input is $\beta$-stable
for $\beta := \eps/\log n$ and that the solution $S$  is a $\apxLSkmeans$-approximation by \Cref{lemma:localsearch}. Given a client or center $a$ and a distance $r$, we let $B(a,r)$ be the set of clients and centers at distance at most $r$ from $a$. We will use the following Lemma from \cite{charikar2025kmeans} (for technical reasons the original proof uses $25$ as the approximation ratio of the local search solution, however their proof works exactly in the same way with $\apxLSkmeans$).
\begin{lemma}[\cite{charikar2025kmeans}]
\label{lem:numnonpure_eucl}
$|\opt_{imp}|\leq \frac{\log n}{\eps^3}$.
\end{lemma}

We next define a replacement cost $r(p)$ for each client $p\in D$. Let $C^*$ be a pure cluster of $\opt$ and $C'$ be the associated cluster of $S$. Let
$c^*$ be the center of $C^*$ and ${c'}$ be the center of ${C'}$. We define $t(c^*) := {c'}$.
For any client
$p \in C^*$, let $r(p) := d^2(p, {c'}) = \dist^2(p, t(c^*))$. For any impure cluster $C^*$ with center $c^*$, we let $t(c^*)$ be the center of $S$ closest to ${\mu(C^*)}$,
and define $r(p)=\dist^2(p,t(c^*))$ for each $p\in C^*$.

The following lemma is the only one
that requires that the solution $S$ is obtained via local search (the proof of the previous lemma only used that it was a constant-factor approximation). In words, it says that $S$ approximates the connection cost of clients belonging to pure clusters of the optimal solution almost perfectly. We will use the following Lemma from \cite{charikar2025kmeans}.
\begin{lemma}[\cite{charikar2025kmeans}]
\label{lem:purecost_eucl}
$        
\sum_{c^*\in \opt_{pure}}\sum_{p\in \opt(c^*)} r(p) \leq \sum_{c^*\in \opt_{pure}}\sum_{p\in \opt(c^*)} \opt_p + O(\sqrt{\eps}) \cdot \sopt\,.
$
\end{lemma}

\subsection{\texorpdfstring{Leaders' Sampling of Leader-Expensive Clusters of $\opt$}{Leaders' Sampling of Leader-Expensive Clusters of OPT}}
\label{sec:leaders}

Let $\opt_{imp}=\opt-\opt_{pure}$ be the \emph{impure} centers of $\opt$. We next partition $\opt_{imp}$ according to the following definition. Consider any $c^*\in \opt_{imp}$ and {set $C^*=\opt(c^*)$}. We
let $\avg(c^*)$ be the squared distance from ${\mu(C^*)}$ to the $\eps |C^*|$-th closest client of $C^*$ to ${\mu(C^*)}$.
We further let $\opt_{lead}(c^*)$ be the set of clients of $C^*$ at a squared distance at most
$\avg(c^*)$ from ${\mu(C^*)}$: we call such clients \emph{leaders} of the cluster. So $\avg(c^*)$ is the \emph{maximum squared distance} from a leader to ${\mu(C^*)}$. Also note that we have, $|\opt_{lead}(c^*)|\geq \eps |C^*|$. 
We thus have that $\avg(c^*)$ {is roughly upper bounded by the average optimal cost of $C^*$, namely}
\begin{equation}\label{eqn:boundavg_eucl}
\avg(c^*)\leq {\frac{1}{(1-\eps)|C^*|}\sum_{p\in C^*} \dist^2(p, \mu(C^*))}\leq \frac{1}{(1-\eps)|C^*|}\sum_{p\in C^*} \opt_p.     
\end{equation}
At the same time $\opt_{lead}(c^*)$ contains a constant fraction of the clients in $\opt(c^*)$. We say that an impure center $c^*\in \opt_{imp}$ is \emph{leader-cheap} if the total cost in 
$S$ of the clients in $\opt_{lead}(c^*)$ is less than $\eps^{5}\sopt/\log n$, i.e., $\sum_{p\in \opt_{\avg}(c^*)}S_p\leq \frac{\eps^5}{\log n}\sopt$. The remaining impure centers are called leader-expensive. {We name similarly the respective clusters}. Let $\opt_{lcheap}$ and $\opt_{lexp}$ be the {leader}-cheap and {leader}-expensive centers, resp. The next lemma shows that leader-cheap clusters have a small replacement cost. There is a similar lemma in \cite{CCGGLW2026kmeans} where the factor $2$ is replaced by $4$.
\begin{lemma}
\label{lem:costLeaderCheap}
$ 
\sum_{c^*\in \opt_{lcheap}} \sum_{p\in \opt(c^*)} r(p) \leq \sum_{c^*\in \opt_{lcheap}} \sum_{p\in \opt(c^*)} 2\,\opt_p + O(\sqrt{\eps}) \sopt.
$
\end{lemma}
\begin{proof}
Consider any $c^*\in \opt_{lcheap}$, and  let $C^*:=\opt(c^*)$ and $C^*_{lead}:=\opt_{lead}(c^*)$. One has
$$
\sum_{p\in C^*}d^2(p,t(c^*))\overset{\text{Lem. }\ref{lem:centerMass}}{\leq} |C^*|d^2({\mu(C^*)},t(c^*))+\sum_{p\in C^*}\opt_p.
$$
Consider any $\ell\in C^*_{lead}$ and let $c(\ell)$ be the center closest to $\ell$ in $S$. Using \Cref{lem:apxTriangleInequality2} with $\gamma=1+\sqrt{\eps}$,
\begin{align*}
d^2({\mu(C^*)},t(c^*)) & \leq d^2({\mu(C^*)},c(\ell))\overset{\text{Lem. }\ref{lem:apxTriangleInequality2}}{\leq} (1+\sqrt{\eps})d^2({\mu(C^*)},\ell)+\frac{2}{\sqrt{\eps}}d^2(\ell,c(\ell))\\
& \leq (1+\sqrt{\eps})\avg(c^*)+\frac{2}{\sqrt{\eps}}S_\ell.
\end{align*}
Averaging over $\ell\in C^*_{lead}$,
$$
d^2({\mu(C^*)},t(c^*))\leq (1+\sqrt{\eps})avg(c^*)+\frac{2}{\sqrt{\eps}|C^*_{lead}|}\sum_{\ell\in C^*_{lead}}S_\ell\leq (1+\sqrt{\eps})avg(c^*)+\frac{2}{\eps^{1.5}|C^*|}\frac{\eps^5}{\log n}\sopt,
$$
where in the last inequality we used that $|C^*_{lead}|\geq \eps |C^*|$ and that $C^*$ is leader-cheap. Recall that $\avg(c^*)\leq \frac{1}{(1-\eps)|C^*|}\sum_{p\in C^*}\opt_p$ by \eqref{eqn:boundavg_eucl}. Altogether
\begin{align*}
\sum_{p\in C^*}d^2(p,t(c^*))\leq \sum_{p\in C^*}\opt_p+\frac{1+\sqrt{\eps}}{1-\eps}\sum_{p\in C^*}\opt_p+\frac{2\eps^{3.5}}{\log n}\sopt.    
\end{align*}
The claim follows by summing over $c^*\in \opt_{lcheap}$ and recalling that $|\opt_{lcheap}|\leq |\opt_{imp}|\leq \frac{\log n}{\eps^3}$ by Lemma \ref{lem:numnonpure_eucl}.
\end{proof}

Our next goal is to identify one leader for each  leader-expensive center. We will actually miss each one of the latter centers with some small probability, however this is not a problem as we will see. The idea is to sample each client $p$ with probability proportional to its cost $S_p$ for a sufficiently large number of times: this way it is unlikely to not hit each set $\opt_{lead}(c^*)$ of a leader-expensive center $c^*$. Then one extracts a proper set of leaders by brute force. Here we critically exploit the fact that there is only a logarithmic number of impure centers (by \Cref{lem:numnonpure_eucl}), hence of leader-expensive centers. We will use the following result in \cite{CCGGLW2026kmeans} that works when $S$ is $O(1)$ approximate.

\begin{lemma}\label{lem:costMiss}\cite{CCGGLW2026kmeans}
There is a polynomial-time procedure to compute a collection $\calL_{lead}$ of subsets of clients such that, with probability at least $1-\eps$, at least one $\Lambda\in \calL_{lead}$ is \emph{valid}, i.e., it satisfies the following property. Let $\opt_{exp}\subseteq \opt_{lexp}$ be the centers with exactly one leader in $\Lambda$ and $\opt_{miss}=\opt_{lexp}-\opt_{exp}$. Then  
$$
\sum_{c^*\in \opt_{miss}}\sum_{p\in \opt(c^*)}r(p)\leq  4\eps\, \sopt.
$$
\end{lemma}

We next assume that the event in the above lemma happens, and focus on the corresponding value of $\Lambda$. For $c^*\in \opt_{exp}$, we let $lead(c^*)$ be the leader in $\opt(c^*)\cap \Lambda$. Let us define $\opt_{exp}$ and $\opt_{miss}$ as in the above lemma. Let also $\opt_{cheap}=\opt_{lcheap}\cup \opt_{miss}$. We call \emph{expensive} and \emph{cheap} the centers in $\opt_{exp}$ and $\opt_{cheap}$, resp. Similarly for the respective clusters. The next lemma upper bounds the distance of the points in expensive clusters to the corresponding leaders. There is an analogous lemma in \cite{CCGGLW2026kmeans} where the $2$ is replaced by a $4$.

\begin{lemma}\label{lem:costExpensive}
Assume that $\Lambda$ is valid. One has
$$
\sum_{c^*\in \opt_{exp}}\sum_{p\in \opt(c^*)}\dist^2(p,\Lambda)\leq \sum_{c^*\in \opt_{exp}}\sum_{p\in \opt(c^*)}2\opt_p+O(\eps)\sopt
$$.
\end{lemma}
\begin{proof}
Consider any $c^*\in \opt_{exp}$. Let $C^*:=\opt(c^*)$ and $\ell{=lead(c^*)}\in \opt_{lead}(c^*)\cap \Lambda$. Notice that $\ell$ is well-defined by the definition of $\opt_{exp}$. Then
\begin{align*}
\sum_{p\in C^*}d^2(p,\Lambda) & \overset{\text{Lem. }\ref{lem:centerMass}}{\leq} |C^*|d^2({\mu(C^*)},\ell)+\sum_{p\in C^*}\opt_p  \leq |C^*|\avg(c^*)+\sum_{p\in C^*}\opt_p \\ 
& \overset{\eqref{eqn:boundavg_eucl}}{\leq} (1+O(\eps))\sum_{p\in C^*}\opt_p.    
\end{align*}
\end{proof}

The next lemma shows that there exists a choice of $|\opt_{exp}|$ many centers $S_{disc}\subseteq S$, such that, replacing $S_{disc}$ with $\Lambda$ in $S$ leads to a roughly $2$ approximate solution. 
Let $D_{exp}=\cup_{c^*\in \opt_{exp}}\opt(c^*)$ and $D_{cheap}=D-D_{exp}$. The proof of the following lemma is almost identical to the one of an analogous lemma in \cite{CCGGLW2026kmeans} (where the $2$ is replaced by a $4$).
\begin{lemma}
\label{lem:structSminusSdisc}
Suppose that $\Lambda$ is valid. Then there exists a set $S_{disc}\subseteq S$ that satisfies the following properties:
  \begin{enumerate}
  \item $|S_{disc}| = |\opt_{exp}|=|\Lambda| \le \frac{\log n}{\eps^3}$;
  \item for every $c \in S_{disc}$, there is no $c^* \in \opt-\opt_{exp}$ such that $t(c^*) = c$;
  \item $\clcost(S - S_{disc} \cup \Lambda) \le
  2 \sum_{p \in D_{exp}}\opt_p + \sum_{p \in D_{cheap}}
  r(p)  \le (2+O(\sqrt{\eps}))\sopt$.
  \end{enumerate}
  \label{lemma:S0properties_eucl}
\end{lemma}
\begin{proof}
By definition, the centers in $\opt_{exp}$ are not pure, hence $|\Lambda|=|\opt_{exp}|\leq \frac{\log n}{\eps^3}$ by \Cref{lem:numnonpure_eucl}.
For each center $c^*\in \opt_{pure}\cup \opt_{cheap}$, there is one associated center $t(c^*)$ in $S$, hence the total number of the latter centers $S'\subseteq S$ is at most $|\opt_{pure}|+|\opt_{cheap}|=k-|\opt_{exp}|$. Let us set $S_{disc}$ to any subset of $|\opt_{exp}|$ centers in $S- S'$. The first two bullets follow. 

Let us turn to the last bullet. By \Cref{lem:costExpensive}
$$
\sum_{p\in D_{exp}}d^2(p,S\cup \Lambda-S_{disc})\leq \sum_{p\in D_{exp}}d^2(p,\Lambda) \leq \sum_{p\in D_{exp}}2\opt_p +O(\eps) \sopt. 
$$
From the second bullet, for 
each client $p$ in a cluster of the optimum solution
whose center is $c^* \in \opt_{cheap}$, we have
that $t(c^*) \in S-S_{disc}$ and so its cost
is at most $r(p)$. Thus, by Lemmas \ref{lem:costLeaderCheap} and \ref{lem:costMiss},
$$
\sum_{p\in D_{cheap}}d^2(p,S\cup \Lambda-S_{disc})\leq \sum_{p\in D_{cheap}}r(p)\leq \sum_{p\in D_{cheap}}2\opt_p+O(\sqrt{\eps})\sopt.
$$
The last bullet follows.
\end{proof}

\subsection{\texorpdfstring{Removal of Expensive Centers of $S_{disc}$}{Removal of Expensive Centers of Sdisc}}
\label{sec:removalofExpensive_eucl}

Let us start with the tentative solution $S\cup \Lambda$ (which is infeasible since it contains more than $k$ centers). Our goal is to remove a proper subset $S'_{disc}$ of $|\Lambda|$ many centers from $S$, so that $S\cup \Lambda -S'_{disc}$, which is a feasible solution, is  roughly $2$-approximate. Ideally we would like to choose $S'_{disc}=S_{disc}$, so that Lemma \ref{lem:structSminusSdisc} directly gives what we need. We will rather find some $S'_{disc}$ which is comparably good. For a subset $A\subseteq S$, let $S_{A}:=S\cup \Lambda-A$. As usual, we let $S_A(c)$ be the cluster in $S_A$ associated with the center $c\in S_A$.

Consider any cluster $C$ of $S\cup \Lambda$ with center $c\in S_{disc}$. For reasons that will become clearer later, 
if we remove $c$ (which is then added to $S'_{disc}$), we would like to assign all the clients in $C$ to the same center $c'$ in $S\cup \Lambda-\{c\}$. Similarly for the later steps. However, this might be too expensive when the cost of $C$ in $S\cup \Lambda$ is large. Our next goal is to identify a subset $\calQ \subseteq S_{disc}$ of centers such that, after removing $\calQ$ from $S\cup \Lambda$, we enforce in some sense the above property. More precisely, we desire that the cost of the clusters with center in $S_{disc}-\calQ$ in the solution $S_{\calQ}:=S\cup \Lambda-\calQ$ is very small compared with $\sopt$.
This is done via the following lemma in \cite{CharikarCGG98} that is based on the fact that $\clcost(\dummyset\cup S-S_{disc})\leq O(1)\cdot\sopt$.

\begin{lemma}[\cite{CCGGLW2026kmeans}]\label{lem:successguessprocess_eucl}
There is a procedure that runs in time $n^{1/\eps^{O(1)}}$ and produces a collection $\calLexp$ of at most $n^{1/\eps^{O(1)}}$ subsets of $S$ such that with probability at least $1-1/n$ there exists $\calQ\in \calLexp$ satisfying
    \begin{enumerate}
    \item $\calQ \subseteq S_{disc}$, and
    \item The total cost in $S_{\calQ} $ of the clusters with centers in  $\calU:=S_{disc} - \calQ$ is at most $\eps \cdot \sopt$, namely $\sum_{c\in \calU}\sum_{p\in S_{\calQ}(c)}\dist^2(p,c)\leq \eps \cdot \sopt$.
    \end{enumerate}
\end{lemma}

\subsection{\texorpdfstring{Removal of Cheap Centers from $S$}{Removal of Cheap Centers from S}}
\label{sec:removalofCheap_eucl}

Our algorithm considers each choice of $\calQ\in \calL_{exp}$. We next assume that there exists at least one such $Q$ that satisfies the properties in Lemma \ref{lem:successguessprocess_eucl} (event which happens with high probability), and focus on the execution of the algorithm corresponding to the latter $\calQ$. Consider the (possibly infeasible) solution $S_{\calQ}$. Our target reference solution is $S_{S_{disc}}$, which is roughly $2$-approximate. Let $\calU:=S_{disc}-\calQ$. Ideally we would like to remove $\calU$ from $S_{\calQ}$, hence getting the desired solution $S_{\calQ\cup \calU}=S_{S_{disc}}$. We will rather remove a subset $\calU'$ of the same cardinality, whose removal leads to a solution of cost only marginally larger than the cost of $S_{S_{disc}}$. In order to simplify our argument, for each candidate set of centers $\calU'\subseteq S-\calQ$ with $|\calU'|=|\calU|$, we will restrict our attention to suboptimal assignments $\mu_{\calU'}$ of points to centers in $S_{\calQ\cup \calU'}$ such that, for each $c\in \calU'$, all the points $p\in S_{\calQ}(c)$ are assigned to the closest center $t(c,\calU')\in S_{\calQ\cup \calU'}$ to $c$. The remaining points are assigned by $\mu_{\calU'}$ to the same center as in $S_{\calQ}$. We call \emph{consistent} an assignment $\mu_{\calU'}$ of this type, and we denote the \emph{consistent cost} of the corresponding solution by 
$$
cost'(S_{\calQ\cup \calU'})=\sum_{p\in \clients}\dist^2(p,\mu_{\calU'}(p)).
$$
Obviously $cost'(S_{\calQ\cup \calU'})$ is an upper bound on the actual cost $cost(S_{\calQ\cup \calU'})$ of the solution $S_{\calQ\cup \calU'}$. The next lemma shows that, for $\calU'=\calU$, the consistent cost is small enough. The first inequality in the following lemma follows directly from an analogous lemma in \cite{CCGGLW2026kmeans}, while the second from \Cref{lem:structSminusSdisc}.3.
\begin{lemma}\label{lem:consistentCost}
$
cost'(S_{\calQ\cup \calU})\leq cost(S_{\calQ\cup \calU})+O(\sqrt{\eps}) \sopt
\leq (2+O(\sqrt{\eps}))\sopt.
$
\end{lemma}

Our next goal is to compute a set $\bcalU$ of cardinality $|\calU|$ such that  
$$
cost'(S_{\calQ\cup \bcalU})\leq cost'(S_{\calQ\cup \calU})+O(\sqrt{\eps})\sopt.
$$
This is done via the following lemma from \cite{CCGGLW2026kmeans}.
\begin{lemma}[\cite{CCGGLW2026kmeans}]\label{lem:successcheapremove_eucl}
Assume that $\Lambda$ and $\calQ$ satisfy the properties mentioned before. Then there is a polynomial-time procedure that computes a polynomial size collection $\calL_{cheap}$ of subsets of $S-\calQ$, each of size $|\Lambda|-|\calQ|$, such that at least one  $\bcalU\in \calL_{cheap}$ satisfies $cost'(S_{\calQ\cup \bcalU})\leq cost'(S_{\calQ\cup \calU})+O(\sqrt{\eps})\sopt$.
\end{lemma}

\subsection{\texorpdfstring{Putting Everything Together 
-- Proof of \Cref{thr:mainEuclideanStable}}{Putting Everything Together 
-- Proof of thr:mainEuclideanStable}}
\label{sec:stable:everythingtogether_eucl}

We next summarize the overall algorithm:
\begin{mdframed}[hidealllines=true, backgroundcolor=gray!15]
\vspace{-5mm}
\paragraph{$apxStableEuclideankMeans()$
\vspace{-2mm}
}\ \\
\begin{algorithmic}[1]
\State $\calL_{apx}\leftarrow \emptyset$
\State $S \gets localSearch()$ (see \Cref{sec:localSearchAnalysis_eucl})
\For{$\frac{1}{2\eps}\ln n$ many times}
\State  Compute $\calL_{lead}$ according to Lemma \ref{lem:costMiss} (see \Cref{sec:leaders})
\For{all $\Lambda\in \calL_{lead}$} \label{line:apxStablekMeans:forLambda}
\State Compute $\calL_{exp}$ according to Lemma \ref{lem:successguessprocess_eucl} (see \Cref{sec:removalofExpensive_eucl})
\For{all $\calQ\in \calL_{exp}$} \label{line:apxStablekMeans:forQ}
\State Compute ${\cal L}_{cheap}$ according to Lemma \ref{lem:successcheapremove_eucl} (see \Cref{sec:removalofCheap_eucl})
\For{all $\bcalU\in \calL_{cheap}$} \label{line:apxStablekMeans:forU}
\State $\calL_{apx}\leftarrow \calL_{axp}\cup \{S\cup \Lambda - \calQ -\bcalU\}$
\EndFor
\EndFor
\EndFor
\EndFor
\State \textbf{return} the cheapest solution $APX\in \calL_{apx}$
\end{algorithmic}
\end{mdframed}

We are now ready to prove Theorem \ref{thr:mainEuclideanStable}.
\begin{proof}[Proof of Theorem \ref{thr:mainEuclideanStable}]
Consider the algorithm $apxStableEuclideankMeans()$. Its running time is polynomial since the called procedures run in polynomial time (see Lemmas \ref{lemma:localsearch},  \ref{lem:costMiss}, \ref{lem:successguessprocess_eucl}, and \ref{lem:successcheapremove_eucl}), which also implies that each considered set $\calL_{lead}$, $\calL_{exp}$ and $\calL_{cheap}$ has polynomial size. 

Notice that all the solutions in $\calL_{apx}$ (including $APX$) contain at most $k$ centers, hence they are deterministically feasible. Consider one execution of the outer for loop. Suppose that the event from Lemma \ref{lem:costMiss} holds for some $\Lambda\in \calL_{lead}$, which happens with probability at least $1-\eps$, and focus on that choice of $\Lambda$ in Line \ref{line:apxStablekMeans:forLambda}. Notice that under this condition Lemmas \ref{lem:costExpensive} and \ref{lem:structSminusSdisc} hold. Suppose also that the event from Lemma \ref{lem:successguessprocess_eucl} holds for some $\calQ\in \calL_{exp}$, which happens with probability at least $1-1/n$, and focus on that choice of $\calQ$ in Line \ref{line:apxStablekMeans:forQ}. Under these assumptions, which hold with probability at least $1-2\eps$ altogether, there exists $\bcalU\in \calL_{cheap}$ which satisfies the conditions of Lemma \ref{lem:successcheapremove_eucl}: let us focus on the execution of Line \ref{line:apxStablekMeans:forU} for this value of $\bcalU$, and consider the corresponding solution $APX=S\cup \Lambda-\calQ-\bcalU$.  Observe that, under the above assumptions, Lemma \ref{lem:consistentCost} also holds. Thus
$$
cost(APX)\leq cost'(APX) \overset{Lem. \ref{lem:successcheapremove_eucl}}{\leq} cost'(S_{\calQ\cup \calU})+O(\sqrt{\eps})\sopt \overset{Lem. \ref{lem:consistentCost}}{\leq} (2+O(\sqrt{\eps}))\sopt.
$$
Since the outer for loop is executed $\frac{1}{2\eps}\ln n$ many times, the probability that no $APX\in \calL_{apx}$ is $2+O(\sqrt{\eps})$ approximate is at most $1/n$. The claim follows.
\end{proof}

\section*{Acknowledgement}
The authors developed the new algorithm, found evidence that it might improve upon the previous work, and established the $(3+\ln 2+\eps)$-approximation for the Euclidean case without any help of AI. 

For the metric case, the authors developed a $(5+\epsilon)$-analysis without any help of AI. The better $(4.9+\epsilon)$-analysis was obtained by GPT 5.2/5.4 Pro/Thinking after several rounds of communication between the authors and it. The writeup for the $(4.9+\eps)$-analysis has been verified and revised by the authors. The two analyses follow quite different routes within the spectral analysis framework and were found by different authors simultaneously.

Aditya Anand, Euiwoong Lee, and Amatya Sharma were supported in part by NSF grant 2236669. 
Ruiquan Gao was supported by NSF CCF-2112824 and Aviad Rubinstein's David and Lucile Packard Fellowship. 
Fabrizio Grandoni was partially supported by the SNF Grants 200021-200731 and 200021-236706.
Ernest van Wijland was supported in part by the French PEPR integrated projects EPIQ (ANR-22-PETQ-0007).

\printbibliography

\appendix

\section{Proof of Theorems \ref{thm:main_euclidean} and \ref{thm:main_metric} } \label{sec:finalproof}

\maineuclidean*
\mainmetric*
\begin{proof}[Proofs of \Cref{thm:main_metric,thm:main_euclidean}]
We compute a set of feasible solutions, and return the cheapest one. Let $\Delta$ be the maximum extra number of centers computed by the algorithm from Theorem \ref{thr:mainBicriteria} (this number is independent from $k$). Let 
$k'=\max\{1,k-\Delta\}$. Obtain a solution by computing the optimum solution with one center if $k'=1$, and otherwise by running
the algorithm from Theorem \ref{thr:mainBicriteria} with target number of centers being $k'$ (notice that it returns a solution with at most $k$ centers, and is hence feasible).

 We run the algorithm from either of \Cref{thr:mainMetricStable,thr:mainEuclideanStable} for every integer $k''\in (k',k]$ (thus obtaining solutions with $k''\leq k$ centers, hence feasible).

If $\opt_{k'}\leq {(1+\eps)}\opt_k$, 
the first solution is  ${(1+\eps)}\Gamma$ approximate. Otherwise, ${\opt}_{k'} > (1+\eps)\opt_k.$ Thus there exists an integer $k''\in (k',k]$ such that ${\opt}_{k''}\leq (1+\eps){\opt}_{k}$ and ${\opt}_{k''-1}\geq (1+\beta){\opt}_{k''}$ for $\beta = \Omega(\eps/\Delta)=\Omega(\eps^4/\log n)$. %
For that value of $k''$ the corresponding instance of Metric $k$-Means is $\beta$-stable, hence we can compute a solution with cost at most $(4+O({\sqrt{\eps}})){\opt}_{k''}\leq (1+\eps)(4+O({\sqrt{\eps}})){\opt}_k$ using~\Cref{thr:mainMetricStable}, which yields a $(1 + \eps)(4 + \sqrt{\epsilon})$ approximation. Similarly, in the Euclidean case, we get a $(1 + \eps)(2 + O(\sqrt{\epsilon}))$ approximation using~\Cref{thr:mainEuclideanStable}. The proof is then complete after suitably rescaling $\eps$. 
\end{proof}

\section{\texorpdfstring{Missing Proofs in \cref{sec:euclidean-lmp}}{error rendering title}}
\subsection{\texorpdfstring{Proof of \cref{lem:stoc-key-lemma}}{Proof of lem-stoc-key-lemma}}
\label{app:euclidean-lmp}
In this appendix, we prove \cref{lem:stoc-key-lemma} for the sake of completeness. 
\STOCKeyLemma*
\noindent Our proof works for the more general case where $x_i\in \mathbb{R}^{\di}$ for some $\di\geq 1$.

\subsubsection{\texorpdfstring{Preliminaries on weighted $k$-means}{Preliminaries on weighted k-means}}
We define the weighted versions of Euclidean $k$-Means that will be useful later in the proof.

\begin{definition}[weighted discrete Euclidean $k$-Means]
The input consists of a set of clients $D$, a set of facilities $\calF$ and their locations $x:D\cup\calF\to \mathbb{R}^{\di}$. In addition there is a weight function $w:D\to \mathbb{R}_{>0}$ mapping each client to a positive weight. The goal of the discrete Euclidean k-Means problem is to find a set of $k$ centers $S \subseteq \calF$ such that $|S|=k$ and the following objective is minimized:
\begin{align*}
    \sum_{j\in D} w(j) \min_{i\in S} \|x_i-x_j\|_2^2~.
\end{align*}
\end{definition}

\begin{definition}[weighted continuous Euclidean $k$-Means]
    \label{def:weighted-continuous-euclid-kmeans}
    The input consists of a set of clients $D$ and their locations $x:D \to \mathbb{R}^{\di}$. In addition there is a weight function $w:D\to \mathbb{R}_{>0}$ mapping each client to a positive weight. The goal of the weighted continuous Euclidean $k$-Means problem is to find a set of $k$ centers $S\subseteq\mathbb{R}^{\di}$ such that $|S|=k$ and the following objective is minimized:
    \begin{align*}
        \sum_{j\in D} w(j) \cdot \min_{c\in S} \|c-x_j\|_2^2~.
    \end{align*}
    In particular, an instance is unweighted if $w(j)=1$ for any client $j\in D$.
\end{definition}

One technical lemma we will use later is a characterization of the optimal cost of a cluster consisting of a set of clients among all possible choices of the center. 
\begin{lemma}[single-cluster cost of weighted continuous Euclidean $k$-Means]
    \label{lem:opt-single-cluster}
    Suppose the set $C\subseteq D$ forms a cluster. Suppose that $W=\sum_{j\in C} w(j)$. 
    The optimal center is the weighted average of the points $\sum_{j \in C} \frac{w(j)}{W}  x_j$, and the cost is 
    \begin{align*}
        \mathsf{cost}_w(C) = \frac{1}{W} \cdot \sum_{j, \ell \in \binom{C}{2}} w(j) w(l) \|x_j-x_\ell\|_2^2
    \end{align*}
    In particular, if the instance is unweighted, 
    we will use the notation $\cost(C)$ for simplicity.
\end{lemma}

\begin{proof}
    The optimal center is given by optimizing the following the total weighted cost
    \begin{alignat*}{2}
        f(c) = \sum_{j\in C} w(j) \cdot \|c-x_j\|^2_2 &=  \sum_{j\in C} w(j) \cdot (\|c\|_2^2 - 2\langle x_j, c\rangle + \|x_j\|_2^2)
        \\
        &= W\|c\|_2^2 
         - 2 \cdot \sum_{j\in C} \langle w(j) \cdot x_j, c\rangle
        + \sum_{j\in C} w(j) \cdot \|x_j \|_2^2~.
    \end{alignat*}
    It is clear that $f(c)$ is convex and the gradient is 
    $$\nabla f(c) = 2 W c - 2\cdot \sum_{j\in C} w(j) \cdot x_j~.$$
    Therefore, the optimal center is $c^* = \frac{\sum_{j\in C} w(j)\cdot x_j}{W}$. In particular, the total weighted cost:
    \begin{align*}
        f(c^*) &= - \frac{1}{W} \cdot  \left\| \sum_{j\in C} w(j) \cdot x_j\right\|_2^2 + \sum_{j\in C} w(j) \cdot \|x_j \|_2^2\\
        &= \frac{1}{W} \cdot \sum_{j,\ell\in C} w(j) \cdot w(\ell) \cdot (\|x_j\|_2^2 - \langle x_j, x_\ell\rangle)
        \\
        &= \frac{1}{W} \cdot \sum_{j,\ell\in \binom{C}{2}} w(j) \cdot w(\ell) \cdot (\|x_j\|_2^2 - 2\langle x_j, x_\ell\rangle + \|x_\ell\|_2^2)
        \\
        &= 
        \frac{1}{W} \cdot \sum_{j,\ell\in \binom{C}{2}} w(j) \cdot w(\ell) \cdot \|x_j-x_\ell\|_2^2~.
    \end{align*}
    This completes the proof.
\end{proof}

\begin{lemma}
    \label{lem:kmeans-cost-monotone}
    For any sets $C_1\subseteq C_2$ of clients and any weight function $w$, we have $\cost_w(C_1) \leq \cost_w(C_2)$.
\end{lemma}
\begin{proof}
    Suppose $c_2$ is the optimal center for $C_2$. Then, we have
    \begin{align*}
        \cost_w(C_1) \leq \sum_{j\in C_1} w(j) \cdot \|x_j-c_2\|_2^2 \leq \sum_{j\in C_2} w(j) \cdot \|x_j-c_2\|_2^2 = \cost_w(C_2)~.&\hfill \qedhere
    \end{align*}
\end{proof}

\begin{lemma}
    \label{lem:kmeans-cost-additive}
    Consider any set $C$ of clients, any client $j\notin C$ and any weight function $w$. Let $W=\sum_{\ell\in C} w(\ell)$. Then, we have $$\cost_w(C\cup \{j\}) = \frac{W}{W+w(j)} \cost_w(C) + \frac{w(j)}{W+w(j)} \cdot \sum_{\ell\in C} w(\ell)\|x_j-x_\ell\|_2^2~.$$
\end{lemma}
\begin{proof}
    Using \cref{lem:opt-single-cluster}, we have
    \begin{align*}
        \cost_w(C\cup \{j\}) &= \frac{1}{W+w(j)} \cdot \sum_{\ell, \ell'\in \binom{C}{2}} w(\ell) w(\ell') \|x_\ell-x_{\ell'}\|_2^2 + \frac{w(j)}{W+w(j)} \cdot \sum_{\ell\in C} w(\ell)\|x_j-x_\ell\|_2^2\\
        &= \frac{W}{W+w(j)} \cost_w(C) + \frac{w(j)}{W+w(j)} \cdot \sum_{\ell\in C} w(\ell) \|x_j-x_\ell\|_2^2~.
        &\hfill \qedhere
    \end{align*}
\end{proof}

\subsubsection{The proof}
Let $a\in [s-1]$ be the smallest integer such that $a\in DC^t$ for some $t\in [s]$. If no such $a$ exists, we set $a=s$ (note that otherwise $a<s$). 
We will prove by induction on $a$. 

\paragraph{Inductive hypothesis.} For any $s'\in [s]$ and any client $j\in [s]$, we let $w_{s'}(j)=\sum_{t\in [s']} \beta_t\eta_{tj}$. The inductive hypothesis is then: for any $s'\in [s]$, we have 
\begin{align}
    \label{eqn:stoc-inductive-hypothesis}
    \sum_{t\in [s']} \beta_t \sum_{j\in DC^t} w(j) \sum_{\ell\in IC^t\cup \{t\}} p_{tj\ell} \cdot \|x_j-x_\ell\|^2 \leq \cost_{w_{s'}}([a:s])~.
\end{align}
Because of recurrence relation of $\beta_t$ (\cref{eqn:Euclidean-beta-recurrence}), we have $w_s(j)=w(j)$ for each $j\in [s]$.
We also have $P_{j\ell} = \sum_{t\in [s]: j\in DC^t} \beta_t p_{tj\ell}$. Therefore, the LHS of \cref{eqn:stoc-inductive-hypothesis} equals $\sum_{j\in [s]} w(j) \sum_{\ell\in [s]} P_{j\ell} \cdot \|x_j-x_\ell\|^2$.
\cref{lem:stoc-key-lemma} then follows from the inductive hypothesis for $s'=s$:
\begin{align*}
    \sum_{j<\ell} P_{j\ell} \cdot \|x_j-x_\ell\|^2 \leq \cost_w([a:s]) \leq \cost_w([s]) \leq \sum_{j\in [s]} w(j) \|x_j\|^2~,
\end{align*}
where the last inequality follows from the fact that $\sum_{j\in [s]} w(j) \|x_j\|^2$ is the total weighted cost of opening the center at $0^{\di}$.

\paragraph{Base case.} The base case is when $a=s$. 
In this case, because $DC^t=\emptyset$ for any $t\in [s]$, we have $p_{tj\ell}=0$ for any $t,j,\ell\in [s]$, $j\in DC^t$. 
The proof of the base case is then straightforward because the LHS of \cref{eqn:stoc-inductive-hypothesis} equals $0$.

\paragraph{Inductive step.} We assume that the inductive hypothesis holds for any $a'>a$ and we will prove that it also holds for $a$.
Let $t_{a}$ the smallest index $t$ such that $a\in DC^t$.
Note that we have $a<t_a\leq s$ because $DC^a\subseteq [a-1]$ and $a\in DC^t$ for some $t\in [s]$.

\newcommand{\tiDC}{\widetilde{DC}}
\newcommand{\tiIC}{\widetilde{IC}}
\newcommand{\tip}{\tilde{p}}
\newcommand{\tieta}{\tilde{\eta}}
\newcommand{\tibeta}{\tilde{\beta}}
\newcommand{\tia}{\tilde{a}}
\newcommand{\tiw}{\tilde{w}}
\newcommand{\tipi}{\tilde{\pi}}
\newcommand{\tisigma}{\tilde{\sigma}}
Consider the case $(\tiDC, \tiIC, w)$, where the weights are the same, but for each $t\in [s]$, we have $\tiDC^t = DC^t\setminus \{a\}$ and $\tiIC^t = IC^t\cup \{a\}$.
We will use $\tip_{tj\ell}, \tieta_{tj}, \tibeta_t, \tia$ to denote the corresponding $p_{tj\ell}, \eta_{tj}, \beta_t, a$ in the case $(\tiDC, \tiIC, w)$.
In this case, we have $\tia>a$.
Let $\tiw_{s'}(j) = \sum_{t\in [s']} \tibeta_t \tieta_{tj}$ for any $s'\in [s]$ and any client $j\in [s]$.
Hence, according to the inductive hypothesis, we have
\begin{align*}
    \sum_{t\in [s']} \tibeta_t \sum_{j\in \tiDC^t} w(j) \sum_{\ell\in \tiIC^t\cup \{t\}} \tip_{tj\ell} \cdot \|x_j-x_\ell\|^2 \leq \cost_{\tiw_{s'}}([\tia:s])
\end{align*}
Next, we compare the parameters in the case $(DC, IC, w)$ and the case $(\tiDC, \tiIC, w)$.
\begin{itemize}
    \item For $t<t_a$, we have $\tiDC^t = DC^t$ and $\tiIC^t = IC^t$. Therefore, we have $\tip_{tj\ell} = p_{tj\ell}$ for any $j,\ell\in [s]$.
    \item For $t\geq t_a$, we have $DC^t = \tiDC^t\cup \{a\}$ and $\tiIC^t = IC^t\setminus \{a\}$. Therefore, we have the following properties for such $t$.
    \begin{itemize}
        \item We have $\pi_t(j) = \frac{\suf(a)}{\suf(a+1)} \cdot \tipi_{t}(j)$ for any $j\in [a+1:s]$ and $\pi_t(j)=\tipi_t(j)=1$ for $j\in [a]$. 
        \item We have $\eta_{tj} = \frac{\suf(a)}{\suf(a+1)} \cdot \tieta_{tj}$ for any $j\in [a+1:s]$. We have $\eta_{ta} = 0$, $\tieta_{ta}=1$. We have $\eta_{tj}=\tieta_{tj}=1$ for any $j\in [a-1]$. 
        \item We have $\sigma_t(j) = \frac{\suf(a)}{\suf(a+1)} \cdot \tisigma_t(j)$ for any $j\in [a+1:s]$ and $\sigma_t(j)=\tisigma_t(j)$ for $j\in [a]$.
        In particular, we have $\sigma_t(a) = \suf(a)$. 
        \item We have $p_{tj\ell} = \tip_{tj\ell}$ for any $j\in [a+1:s]\cap DC^t$ and $\ell\in IC^t\cup \{t\}$. We have $p_{ta\ell}=\eta_{t\ell}/\suf(a) = \tieta_{t\ell}/\suf(a+1)$ and $\tip_{ta\ell} = 0$ for any $\ell\in [a+1:s]\cap IC^t$. We have $p_{tj\ell}=\tip_{tj\ell}=0$ for all other cases. %
    \end{itemize}
\end{itemize}

Using the above properties, we have $\beta_t = \frac{\suf(a+1)}{\suf(a)} \cdot \tibeta_t$ for $t\geq t_a$ and $\beta_t=\tibeta_t$ for $a<t<t_a$. In addition, we have $\beta_a=1-\sum_{j>a} \beta_j$. 
It can be verified by the recurrence relation of $\beta_t$ (\cref{eqn:Euclidean-beta-recurrence}).
Therefore, we have $\beta_t \eta_{tj} = \tibeta_t \tieta_{tj}$ for any $t\in [s]$, $j\in [a+1:s]$. This implies that $w_{s'}(j) = \tiw_{s'}(j)$ for any $s'\in [s]$ and any $j\in [a+1:s]$.

\paragraph{case 1: $s'<t_a$.} In this case, we have $DC^t=\tiDC^t$ for any $t\in [s']$, and we have $\beta_t=\tibeta_t$ for any $t\in [s']$ except for $t=a$ (if $s'\geq a$). Because $DC^a=\emptyset$ and \cref{lem:kmeans-cost-monotone}, we have
\begin{align*}
    \sum_{t\in [s']} \beta_t \sum_{j\in DC^t} w(j) \sum_{\ell\in IC^t\cup \{t\}} p_{tj\ell} \|x_j-x_\ell\|^2 &= \sum_{t\in [s']} \tibeta_t \sum_{j\in \tiDC^t} w(j) \sum_{\ell\in \tiIC^t\cup \{t\}} \tip_{tj\ell} \|x_j-x_\ell\|^2\\ &\leq \cost_{\tiw_{s'}}([\tia:s]) = \cost_{w_{s'}}([\tia:s]) \leq \cost_{w_{s'}}([a:s]).
\end{align*}

\paragraph{case 2: $s'\geq t_a$.} In this case, we have $w_{s'}(a) = w(a)$ because $\eta_{ta}=0$ for any $t>s'$. Using the above properties, we have
\begin{align*}
    &\sum_{t\in [s']} \beta_t \sum_{j\in DC^t} w(j) \sum_{\ell\in IC^t\cup \{t\}} p_{tj\ell} \|x_j-x_\ell\|^2 \\
    =\ & \sum_{t\in [t_a-1]} \beta_t \sum_{j\in DC^t} w(j) \sum_{\ell\in IC^t\cup \{t\}} p_{tj\ell} \|x_j-x_\ell\|^2 + \sum_{t\in [t_a:s']} \beta_t \sum_{j\in DC^t} w(j) \sum_{\ell\in IC^t\cup \{t\}} p_{tj\ell} \|x_j-x_\ell\|^2 \\
    =\ & \sum_{t\in [t_a-1]} \tibeta_t \sum_{j\in \tiDC^t} w(j) \sum_{\ell\in \tiIC^t\cup \{t\}} \tip_{tj\ell} \|x_j-x_\ell\|^2 + \sum_{t\in [t_a:s']} \beta_t w(a) \sum_{\ell\in IC^t\cup \{t\}} p_{ta\ell} \|x_a-x_\ell\|^2 \\ & +  \sum_{t\in [t_a:s']} \frac{\suf(a+1)}{\suf(a)}\tibeta_t  \sum_{j\in \tiDC^t} w(j) \sum_{\ell\in \tiIC^t\cup \{t\}} \tip_{tj\ell} \|x_j-x_\ell\|^2 \tag{$\tip_{tja}=0$}\\
    =\ & \frac{w(a)}{\suf(a)} \sum_{t\in [t_a-1]} \tibeta_t \sum_{j\in \tiDC^t} w(j) \sum_{\ell\in \tiIC^t\cup \{t\}} \tip_{tj\ell} \|x_j-x_\ell\|^2 + \sum_{t\in [t_a:s']} \beta_t w(a) \sum_{\ell\in IC^t\cup \{t\}} p_{ta\ell} \|x_a-x_\ell\|^2 \\ & \frac{\suf(a+1)}{\suf(a)}  \sum_{t\in [s']} \tibeta_t  \sum_{j\in \tiDC^t} w(j) \sum_{\ell\in \tiIC^t\cup \{t\}} \tip_{tj\ell} \|x_j-x_\ell\|^2\\
    \leq\ & \frac{w(a)}{\suf(a)} \cdot \cost_{w_{t_a-1}}([\tia:s]) + \frac{\suf(a+1)}{\suf(a)} \cdot \cost_{w_{s'}}([\tia:s]) + \underbrace{\sum_{t\in [t_a:s']} \beta_t w(a) \sum_{\ell\in IC^t\cup \{t\}} p_{ta\ell} \|x_a-x_\ell\|^2}_{S_3}
    \tag{inductive hypothesis for $\tia$ and $w_{\cdot}(j)=\tiw_{\cdot}(j)$ for $j\in [a+1:s]$}
\end{align*}
Next, we consider the new term $S_3$. We have
\begin{align*}
    S_3 &= \sum_{t\in [t_a:s']} \beta_t w(a) \sum_{\ell\in IC^t\cup \{t\}} p_{ta\ell} \|x_a-x_\ell\|^2 \\
    &= \frac{\suf(a+1) w(a)}{\suf(a)} \sum_{t\in [t_a:s']} \tibeta_t \sum_{\ell>a:\ell\in IC^t\cup \{t\}} \frac{\tieta_{t\ell}}{\suf(a+1)} \|x_a-x_\ell\|^2~.
    \\
    &= \frac{w(a)}{\suf(a)} \sum_{t\in [t_a:s']} \tibeta_t \sum_{\ell>a} \tieta_{t\ell} \|x_a-x_\ell\|^2 \tag{$\tieta_{t\ell}=0$ for $\ell\in DC^t$}
    \\
    &= \frac{w(a)}{\suf(a)} \sum_{\ell>a} \|x_a-x_\ell\|^2 \cdot \sum_{t\in [t_a:s']}\tibeta_t \tieta_{t\ell} 
    \\
    &=
    \frac{w(a)}{\suf(a)} \sum_{\ell>a} \|x_a-x_\ell\|^2 \cdot \left(\sum_{t\in [s']}\tibeta_t \tieta_{t\ell} - \sum_{t\in [t_a-1]}\tibeta_t \tieta_{t\ell} \right)
    \\
    &= \frac{w(a)}{\suf(a)} \sum_{\ell>a} \|x_a-x_\ell\|^2 \cdot \left(\tiw_{s'}(\ell) - \tiw_{t_a-1}(\ell) \right)
    \\
    &= \frac{w_{s'}(a)}{\suf(a)} \sum_{\ell>a} w_{s'}(\ell) \|x_a-x_\ell\|^2  - \frac{w(a)}{\suf(a)} \sum_{\ell>a} w_{t_a-1}(\ell) \|x_a-x_\ell\|^2~.
\end{align*}
According to \cref{lem:kmeans-cost-additive,lem:kmeans-cost-monotone}, we have 
\begin{align*}
    &\frac{w_{s'}(a)}{\suf(a)} \sum_{\ell>a} w_{s'}(\ell) \|x_a-x_\ell\|^2  + \frac{\suf(a+1)}{\suf(a)} \cdot \cost_{w_{s'}}([\tia:s]) \\
    &\quad \leq \frac{w_{s'}(a)}{\suf(a)} \sum_{\ell>a} w_{s'}(\ell) \|x_a-x_\ell\|^2  + \frac{\suf(a+1)}{\suf(a)} \cdot \cost_{w_{s'}}([a+1:s]) = \cost_{w_{s'}}([a:s])~.
\end{align*}
Note that $\sum_{t>a} w_{t_a-1}(t) \|x_a-x_t\|^2$ is the cost of the cluster $[a+1:s]$ with the center at $x_a$. Therefore, we have $\sum_{t>a} w_{t_a-1}(t) \|x_a-x_t\|^2 \geq \cost_{w_{t_a-1}}([a+1:s])$. 
Therefore, putting $S_3$ back to the earlier inequality, we have completed the inductive step:
\begin{align*}
    \sum_{t\in [s']} \beta_t \sum_{j\in DC^t} w(j) \sum_{\ell\in IC^t\cup \{t\}} p_{tj\ell} \|x_j-x_\ell\|^2 \leq \cost_{w_s'}([a:s])~.
\end{align*}

\subsection{\texorpdfstring{Proof of \cref{lem:dual-feasibility-simpler-euclidean-special}}{Proof of lem-dual-feasibility-simpler-euclidean-special}}
\label{app:dual-feasibility-simpler-euclidean-special}

In the rest of this subsection, we will prove the above lemma.
We will use the same set of non-negative parameters $p_{tj\ell}$ and $\beta_t$ as in the proof of \cref{lem:dual-feasibility-simpler-euclidean} (in terms of $(w(t))_{t\in [s]}$, $(DC^t)_{t\in [s]}$ and $(IC^t)_{t\in [s]}$). 
One property we will use about the $\beta$-parameters is the following:
\begin{lemma}
  \label{lem:beta-zero-euclidean-special}
  For any $t\in [a+1:s-1]$, we have $\beta_t=0$.
\end{lemma}
\begin{proof}
  For any $t\in [a+1:s-1]$, we have $t\in IC^s$. Therefore, according to our definition of $\pi_t(\ell)$, we have $\pi_t(\ell)=\pi_t(s)$ for any $\ell\in [a+1:s]$. Further, $\eta_{s\ell} = w(\ell)\pi_t(s)$ for any $\ell\in [a+1:s]$. 
  Note that $\beta_s$ is defined by $\beta_s \eta_{ss} = w(s)$. We have $\beta_s = 1/\pi_t(s)$. Therefore, for any $\ell\in [a+1:s-1]$, we have $\beta_s\eta_{s\ell} = w(\ell)$. Using the recurrence relation in \cref{eqn:Euclidean-beta-recurrence}, we have $\beta_t \eta_{t\ell} = 0$ for any $\ell\in [a+1:s-1]$. Because $\eta_{tt}>0$, we have $\beta_t=0$ for any $t\in [a+1:s-1]$.
\end{proof}
We also consider adding one term that is no greater than $0$ to the LHS of the third constraint \ref{constr:euclidean-special-third} for each $t\in [a]\cup\{s\}$, using the same parameters $p_{tj\ell}$ as in the proof of \cref{lem:dual-feasibility-simpler-euclidean}. More specifically, we will consider the term on the LHS of \cref{eqn:adding-to-LHS-euclidean-special} in the following lemma. This is only slightly different from the term we considered in the proof of \cref{lem:dual-feasibility-simpler-euclidean}, with a $(1+2\eps)$ factor before each $(r_jt+d(j,\ell))^2$.
\begin{lemma}
  For each $t\in [a]\cup \{s\}$, we have
\begin{align}
  \label{eqn:adding-to-LHS-euclidean-special}
  \sum_{j\in DC^t} w(j) \cdot \sum_{\ell\in IC^t\cup\{t\}} p_{tj\ell}(\alpha^*_\ell - (1+2\epsilon)(r_{jt}+d(j,\ell))^2) \leq 0~.
\end{align}
\end{lemma}
\begin{proof}
  Because the second constraint \ref{constr:euclidean-special-second} holds for $t\in [a]$, it is straightforward to verify that the above term is upper bounded by $0$ for $t\in [a]$. Next, we will consider the case when $t=s$.

  Let $W^I=\sum_{\ell\in [a+1:s]} w(\ell)$ and $W^L=\sum_{\ell\in L} w(\ell)$. According to the first bullet of the fourth constraint \ref{constr:euclidean-special-second-approximate}, we have $W^L \leq \epsilon \cdot W^I$. Using the second and the third bullet of the fourth constraint, we have
  \begin{align}
    \begin{split}
  0\geq \sum_{j\in DC^s} w(j)& \cdot \sum_{\ell\in (IC^s\cup\{s\})\setminus L} p_{sj\ell}(\alpha^*_\ell - (1+\epsilon)(r_{js}+d(j,\ell))^2) \\
  + &\sum_{j\in DC^s} w(j) \sum_{\ell\in L} p_{sj\ell}\left(\alpha^*_\ell - \sum_{\ell'\in [a+1:s]} \frac{w(\ell')}{W^I} \cdot (r_{js}+d(j,\ell'))^2\right)
    \end{split}
    \label{eqn:additive-term-euclidean-special}
  \end{align}
  For the second part of the RHS of the above inequality, we have
  \begin{align*}
  \sum_{j\in DC^s} &w(j) \sum_{\ell\in L} p_{sj\ell}\left(\alpha^*_\ell - \sum_{\ell'\in [a+1:s]} \frac{w(\ell')}{W^I} \cdot (r_{js}+d(j,\ell'))^2\right) 
  \\
  &= \sum_{j\in DC^s} w(j) \sum_{\ell\in L} p_{sj\ell} \alpha^*_\ell - \sum_{j\in DC^s} w(j) \sum_{\ell\in L} p_{sj\ell} \sum_{\ell'\in [a+1:s]} \frac{w(\ell')}{W^I} \cdot (r_{js}+d(j,\ell'))^2
  \\
  &= \sum_{j\in DC^s} w(j) \sum_{\ell\in L} p_{sj\ell} \alpha^*_\ell - \sum_{j\in DC^s} w(j) \sum_{\ell'\in [a+1:s]} \left(\sum_{\ell\in L} p_{sj\ell}\right) \cdot \frac{w(\ell')}{W^I} \cdot (r_{js}+d(j,\ell'))^2~.
\end{align*}
Note that if $\ell\in [a], \ell'\in [a+1:s]$, we have $p_{sj\ell} \leq \frac{w(\ell)}{w(\ell')} \cdot p_{sj\ell'}$ according to our definition of $p_{tj\ell}$ (see \cref{eqn:euclidean-eta-tl,eqn:euclidean-p-tjl}). Therefore, we have 
\begin{align*}
  \sum_{j\in DC^s} &w(j) \sum_{\ell\in L} p_{sj\ell}\left(\alpha^*_\ell - \sum_{\ell'\in [a+1:s]} \frac{w(\ell')}{W^I} \cdot (r_{js}+d(j,\ell'))^2\right) 
  \\
  &\geq \sum_{j\in DC^s} w(j) \sum_{\ell\in L} p_{sj\ell} \alpha^*_\ell - \sum_{j\in DC^s} w(j) \sum_{\ell'\in [a+1:s]} p_{sj\ell'} \left(\sum_{\ell\in L} \frac{w(\ell)}{w(\ell')}\right) \cdot \frac{w(\ell')}{W^I} \cdot (r_{js}+d(j,\ell'))^2
  \\
  &= \sum_{j\in DC^s} w(j) \sum_{\ell\in L} p_{sj\ell} \alpha^*_\ell - \sum_{j\in DC^s} w(j) \sum_{\ell'\in [a+1:s]} p_{sj\ell'} \cdot \frac{W^L}{W^I} \cdot (r_{js}+d(j,\ell'))^2
  \\
  &\geq
  \sum_{j\in DC^s} w(j) \sum_{\ell\in L} p_{sj\ell} \alpha^*_\ell - \sum_{j\in DC^s} w(j) \sum_{\ell'\in [a+1:s]} \epsilon \cdot p_{sj\ell'} (r_{js}+d(j,\ell'))^2
\end{align*}
Putting this back to \cref{eqn:additive-term-euclidean-special}, we have
\begin{align*}
  0&\geq \sum_{j\in DC^s} w(j) \cdot \sum_{\ell\in (IC^s\cup\{s\})\setminus L} p_{sj\ell}(\alpha^*_\ell - (1+\epsilon)(r_{js}+d(j,\ell))^2) \\
  & \hspace{5em} + \sum_{j\in DC^s} w(j) \sum_{\ell\in L} p_{sj\ell} \alpha^*_\ell - \sum_{j\in DC^s} w(j) \sum_{\ell'\in [a+1:s]} \epsilon \cdot p_{sj\ell'} (r_{js}+d(j,\ell'))^2
  \\
      & \geq \sum_{j\in DC^s} w(j) \cdot \sum_{\ell\in (IC^s\cup\{s\})} p_{sj\ell}(\alpha^*_\ell - (1+2\epsilon)(r_{js}+d(j,\ell))^2)
  \end{align*}
  Hence, we have completed the proof of this lemma.
\end{proof}

By adding the LHS of \cref{eqn:adding-to-LHS-euclidean-special} to the LHS of the third constraint \ref{constr:euclidean-special-third} for each $t\in [a]\cup\{s\}$, we have the following inequality (for each $t\in [a]\cup\{s\}$):
\begin{align*}
  &\text{LHS of constraint \ref{constr:euclidean-special-third}} \\
  &\geq \sum_{\ell\in[t:s]} w(\ell) \alpha^*_t  + \sum_{\ell \in IC^t} w(\ell)\alpha^*_\ell + \sum_{j\in DC^t} w(j) \left(r_{jt}^2 + \sum_{\ell\in IC^t\cup\{t\}} p_{tj\ell}(\alpha^*_\ell - (1+2\epsilon)(r_{jt}+d(j,\ell))^2)\right)
  \\
  &= \sum_{\ell\in[t:s]} w(\ell) \alpha^*_t  + \sum_{\ell \in IC^t} w(\ell)\alpha^*_\ell + \sum_{j\in DC^t} w(j) \sum_{\ell\in IC^t\cup \{t\}} p_{tj\ell}\alpha^*_\ell \\
  &\hspace{3em} - \sum_{j\in DC^t} w(j) \left(2\epsilon r_{jt}^2 + \sum_{\ell\in IC^t\cup\{t\}} p_{tj\ell} (1+2\epsilon)(2r_{jt}d(j,\ell)+d^2(j,\ell))\right) \tag{$\sum_{\ell} p_{tj\ell} = 1$}
  \\
  &= \sum_{\ell\in IC^t\cup\{t\}} \eta_{t\ell} \alpha^*_\ell - \sum_{j\in DC^t} w(j) \left(2\epsilon r_{jt}^2 + \sum_{\ell\in IC^t\cup\{t\}} p_{tj\ell} (1+2\epsilon)(2r_{jt}d(j,\ell)+d^2(j,\ell))\right)
  \tag{\cref{lem:justify-beta-p-tjl}}
  \\
  &\geq \sum_{\ell\in IC^t\cup\{t\}} \eta_{t\ell} \alpha^*_\ell - \sum_{j\in DC^t} w(j) \left(2\epsilon R_j^2 + \sum_{\ell\in IC^t\cup\{t\}} p_{tj\ell} (1+2\epsilon)(2R_jd(j,\ell)+d^2(j,\ell))\right)
  \tag{$r_{jt}\leq R_j$}
\end{align*}

{Despite not having} the above inequality of $t\in [a+1:s-1]$, because we have $\beta_t=0$ for any $t\in [a+1:s-1]$ (\cref{lem:beta-zero-euclidean-special}), we have the following inequality for any $t\in [s]$: 
\begin{align*}
\beta_t\cdot \left[\sum_{\ell\in IC^t\cup\{t\}} \eta_{t\ell} \alpha^*_\ell - \sum_{j\in DC^t} w(j) \left(2\epsilon R_j^2 + \sum_{\ell\in IC^t\cup\{t\}} p_{tj\ell} (1+2\epsilon)(2R_jd(j,\ell)+d^2(j,\ell))\right)\right]
\\
\leq \beta_t \cdot \left(\hat{f} + \sum_{\ell \in IC^t \cup [t:s]} \rho w(\ell)d^2(\ell,i) + \sum_{j\in DC^t} w(j) d^2(j,i)\right)~.
\end{align*}
Summing over all $t\in [s]$, because of \cref{obv:beta-eta-sum}, $\rho=2$, $\Phi=\sum_{\ell} w(\ell)d^2(j,\ell)$, and $\sum_{t\in [s]} \beta_t \leq 1$ (\cref{cor:beta-sum-euclidean}), we have 
\begin{align*}
  \sum_{\ell\in [s]} w(\ell)\alpha^*_\ell - \sum_{t\in [s]} \beta_t \cdot \sum_{j\in DC^t} w(j) \left(2\epsilon R_j^2 + \sum_{\ell\in IC^t\cup\{t\}} p_{tj\ell} (1+2\epsilon)(2R_jd(j,\ell)+d^2(j,\ell))\right)
  \\
  \leq \hat f + 2\Phi - \sum_{t\in [s]} \beta_t \sum_{j\in DC^t} w(j) d^2(j,i)
  = \hat f + 2\Phi - \sum_{j\in [s]} w(j) d^+_j   d^2(j,i)~. \tag{\cref{obv:dplus-vs.-sum-beta}}
\end{align*}
Therefore, we have
\begin{align*}
&\sum_{j\in [s]} w(j) \cdot (\alpha^*_j - (\rho-1)R_j^2) - \hat{f} - 2\Phi\leq  - \sum_{j\in [s]} w(j) R_j^2 - \sum_{j\in [s]} w(j) d^+_j d^2(j,i) \\
& \hspace{5em} + \sum_{t\in [s]} \beta_t \sum_{j\in DC^t} w(j) \left(2\epsilon R_j^2 + \sum_{\ell\in IC^t\cup\{t\}} p_{tj\ell} (1+2\epsilon)(2R_jd(j,\ell)+d^2(j,\ell))\right)
\\
&\leq \sum_{j\in [s]} w(j) \left(-(1-2\eps) R_j^2 - d^+_j d^2(j,i) + \sum_{t\in [s]:j\in DC^t} \beta_t \sum_{\ell \in IC^t\cup\{t\}} p_{tj\ell} (1+2\epsilon)(2R_jd(j,\ell)+d^2(j,\ell))\right) 
\tag{$\sum_{t\in [s]:j\in DC^t} \beta_t =d^+_j \leq 1$, by \cref{obv:dplus-vs.-sum-beta,cor:dplus-upperbound}}
\\
&= \sum_{j\in [s]} w(j) \left(-(1-2\eps) R_j^2 - d^+_j d^2(j,i) + \sum_{\ell>j} P_{j\ell} (1+2\epsilon)(2R_jd(j,\ell)+d^2(j,\ell))\right) 
\tag{definition of $P_{j\ell}$}
\\
&\leq \sum_{j\in [s]} w(j) \left(- d^+_j d^2(j,i) + (1+2\epsilon)\sum_{\ell>j} P_{j\ell} d^2(j,\ell) + \frac{(1+2\eps)^2}{1-2\eps} \cdot \left(\sum_{\ell>j} P_{j\ell} d(j,\ell)\right)^2 \right) 
\\
&\leq (1+2\eps)\Phi + \sum_{j\in [s]} w(j) \cdot \left(-d^+_j d^2(j,i) + \frac{(1+2\eps)^2}{1-2\eps} \cdot \left(\sum_{\ell>j} P_{j\ell} d(j,\ell)\right)^2\right)
\tag{\cref{lem:stoc-key-lemma}}
\\
&\leq (1+2\eps)\Phi + \sum_{j\in [s]} w(j) \cdot \left(-d^+_j d^2(j,i) + (1+7\eps) \cdot \sum_{\ell>j} d^+_j P_{j\ell} d^2(j,\ell)\right)
\tag{$\eps$ is sufficiently small, Cauchy-Schwarz inequality}
\\
&\leq (1+9\eps)\Phi + \sum_{j<\ell} w(j) d^+_j P_{j\ell} d^2(j,\ell) - \sum_{j\in [s]} w(j) d^+_j d^2(j,i)
\tag{$d^+_j \leq 1$, \cref{lem:stoc-key-lemma}}
\\
& \leq (1+\ln(2)+9\eps)\Phi~. \tag{according to \cref{subsubsec:euclidean-lmp-final-proof}}
\end{align*}

\section{\texorpdfstring{Missing Proofs in \cref{sec:metric-lmp}}{error rendering title}}
\subsection{\texorpdfstring{Proof of \Cref{lem:functionupperbound}}{error rendering title}}
\label{sec:prooffunctionupperbound}

Define
\[
Q(x)\defeq 288x^4 - 235\tau x^3 - 1071x - 792 + 1128\tau.
\]
Differentiating \eqref{eq:function} gives
\begin{equation}
\label{eq:FQ-derivatives}
F'(x)=\frac{x}{2178}Q(x),\qquad
Q'(x)=1152x^3-705\tau x^2-1071,\qquad
Q''(x)=x(3456x-1410\tau).
\end{equation}
We shall use the rational bounds
\begin{equation}
\label{eq:xi-rational-bounds}
\frac{22894}{10000}<\tau<\frac{228943}{100000},\qquad
\Bigl(\frac{22894}{10000}\Bigr)^3<12<
\Bigl(\frac{228943}{100000}\Bigr)^3.
\end{equation}

We split the interval into three pieces.

\smallskip
\noindent{\bf Step 1: the interval $[1,\frac43]$.}
From \eqref{eq:xi-rational-bounds},
\[
3456-1410\tau
>
3456 - 1410\cdot \frac{228943}{100000}
=
\frac{2279037}{10000}
>0.
\]
Hence $Q''(x)>0$ on $[1,\tau]$, so $Q'$ is increasing. Therefore on $[1,\frac43]$,
\[
Q'(x)\le Q'\Bigl(\frac43\Bigr)
=
\frac{4979-3760\tau}{3}
<
\frac{4979-3760\cdot (22894/10000)}{3}
=
-\frac{453643}{375}
<0.
\]
Thus $Q$ is strictly decreasing on $[1,\frac43]$.

At the endpoints,
\[
Q(1)=893\tau-1575
>
893\cdot \frac{22894}{10000} - 1575
=
\frac{2347171}{5000}
>0,
\]
while
\[
Q\Bigl(\frac43\Bigr)
=
\frac{15416\tau-35364}{27}
<
\frac{15416\cdot (228943/100000)-35364}{27}
=
-\frac{876839}{337500}
<0.
\]
Therefore there is a unique $r\in (1,\frac43)$ such that $Q(r)=0$. Since $F'(x)$ has the sign of
$Q(x)$ by \eqref{eq:FQ-derivatives}, the function $F$ increases on $[1,r]$ and decreases on
$[r,\frac43]$. Hence
\[
\max_{1\le x\le 4/3} F(x)=F(r).
\]

We now compare $F(r)$ with $F(\frac43)$. By the mean value theorem, there exists
$\zeta\in (r,\frac43)$ such that
\[
Q\Bigl(\frac43\Bigr)-Q(r)=Q'(\zeta)\Bigl(\frac43-r\Bigr).
\]
Since $Q(r)=0$ and $Q'(\zeta)\le -453643/375$, while
\[
-Q\Bigl(\frac43\Bigr)
=
\frac{35364-15416\tau}{27}
<
\frac{35364-15416\cdot (22894/10000)}{27}
=
\frac{44131}{16875},
\]
we get
\[
\frac43-r
=
\frac{-Q(\frac43)}{-Q'(\zeta)}
<
\frac{44131/16875}{453643/375}.
\]
Because $Q$ is decreasing on $[r,\frac43]$ and vanishes at $r$, one has
\[
0\le -Q(x)< \frac{44131}{16875}\qquad (r\le x\le 4/3).
\]
Therefore
\begin{align*}
0\le F(r)-F\Bigl(\frac43\Bigr)
&=
\int_r^{4/3} -F'(x)\,dx
=
\int_r^{4/3}\frac{x(-Q(x))}{2178}\,dx
\\
&\le
\frac{4/3}{2178}\cdot \frac{44131}{16875}\cdot \Bigl(\frac43-r\Bigr)
\\
&<
\frac{4/3}{2178}\cdot \frac{44131}{16875}\cdot
\frac{44131/16875}{453643/375}
\\
&=
\frac{3895090322}{1125432995259375}
<4\times 10^{-6}.
\end{align*}
Next,
\[
F\Bigl(\frac43\Bigr)
=
\frac{97760\tau}{264627} + \frac{822427}{529254}
<
\frac{97760\cdot (228943/100000)}{264627} + \frac{822427}{529254}
=
\frac{793785221}{330783750}
<2.39971045.
\]
Hence
\[
\max_{1\le x\le 4/3} F(x) < 2.39971045 + 4\times 10^{-6} < \frac{12}{5}.
\]

\smallskip
\noindent{\bf Step 2: the interval $[\frac43,2]$.}
Since $Q''>0$, the function $Q$ is convex. We already know that $Q(\frac43)<0$. Also,
\[
Q(2)=1674-752\tau
<
1674-752\cdot \frac{22894}{10000}
=
-\frac{29768}{625}
<0.
\]
A convex function that is negative at both endpoints of an interval is negative throughout that
interval, so $Q(x)<0$ on $[\frac43,2]$. By \eqref{eq:FQ-derivatives}, the function $F$ is therefore
strictly decreasing on $[\frac43,2]$, and hence
\[
\max_{4/3\le x\le 2}F(x)=F\Bigl(\frac43\Bigr)<\frac{12}{5}.
\]

\smallskip
\noindent{\bf Step 3: the interval $[2,\tau]$.}
Again $Q''>0$, so $Q'$ is increasing. Moreover,
\[
Q'(2)=8145-2820\tau
>
8145-2820\cdot \frac{228943}{100000}
=
\frac{8444037}{5000}
>0.
\]
Thus $Q'(x)>0$ on $[2,\tau]$, so $Q$ is increasing there. Since also
\[
Q(2)=1674-752\tau
<
1674-752\cdot \frac{22894}{10000}
=
-\frac{29768}{625}
<0,
\]
the function $Q$ can cross zero at most once on $[2,\tau]$, and if it does, it crosses from
negative to positive. Therefore $F'(x)=xQ(x)/2178$ can change sign at most once, from negative to
positive, so the maximum of $F$ on $[2,\tau]$ is attained at one of the endpoints.

At $x=2$,
\[
F(2)=\frac{752\tau+3297}{2178}
<
\frac{752\cdot (228943/100000)+3297}{2178}
=
\frac{31366571}{13612500}
<\frac{12}{5}.
\]
At $x=\tau$, using $\tau^3=12$ and $\tau^6=144$, we obtain
\[
F(\tau)
=
\frac{48\tau^6-47\tau^6-357\tau^3+(564\tau-396)\tau^2+4665}{2178}
=
\frac{\tau^6+207\tau^3-396\tau^2+4665}{2178}
=
\frac{221}{66}-\frac{2}{11}\tau^2.
\]
Hence, from the lower bound in \eqref{eq:xi-rational-bounds},
\[
F(\tau)
<
\frac{221}{66} - \frac{2}{11}\Bigl(\frac{22894}{10000}\Bigr)^2
=
\frac{988148573}{412500000}
<\frac{12}{5}.
\]
So $F(x)<\frac{12}{5}$ throughout $[2,\tau]$ as well.

Combining the three steps proves the lemma.

\subsection{Weighted: Full Proof}
    \label{sec:dual-feasibility-weighted-full}

\providecommand{\sdist}{a}
\providecommand{\bcoef}{\eta}
\providecommand{\wcoeff}{\beta}
\providecommand{\swt}{W}

In this subsection, we prove \Cref{lem:dual-feasibility-simpler-metric}. %
Define
\[
    \sdist_t \defeq d(t,i),
    \qquad
    W(S)\defeq \sum_{j\in S} w(j) \quad (S\subseteq [s]),
    \qquad
    \swt_t \defeq \sum_{u=t}^s w(u).
\]

As briefly discussed in the overview in {\Cref{sec:spectral-overview}}, the proof has four steps. First, from the inequalities at time $\alpha_t-\eps$ given in condition (3) of \Cref{lem:dual-feasibility-simpler-metric}, we derive a family of constraints $\eqref{eq:Pk-weighted}$ by using the triangle inequality to lower bound the terms involving $r_{jt}$. Second, we choose coefficients $(\wcoeff)_{t\in[s]}$ so that, in the weighted combination of these constraints, the coefficient of each $\alpha_t^*$ is exactly $w(t)$. Third, after collecting terms, we obtain a quadratic form in the distances to the facility $i$. After rescaling by $\sqrt{w(t)}$, the quadratic form is represented by an ordinary symmetric nonnegative matrix. Finally, we bound the top eigenvalue of the corresponding matrix by the Collatz--Wielandt Formula, which yields the claimed constant.

\paragraph{Step 1: From $(I_t^w)$ to $(P_t^w)$.}
For each $t\in[s]$, condition~(3) gives
\begin{equation}
    \label{eq:Ik-weighted}
    \tag{$I_t^w$}
    \sum_{j\in DC^t} w(j)(r_{jt}^2-\sdist_j^2)
    +
    \sum_{j\in IC^t} w(j)(\alpha_j^*-\rho\sdist_j^2)
    +
    \sum_{j\ge t} w(j)(\alpha_t^*-\rho\sdist_j^2)
    \le \hat f.
\end{equation}
Fix $t\in[s]$. For each $j\in DC^t$, conditions~(1)--(2) and the triangle inequality imply
\[
    \alpha_t^*
    \le (r_{jt}+d(j,t))^2
    \le (r_{jt}+\sdist_j+\sdist_t)^2
    \le r_{jt}^2 + (\sdist_j+\sdist_t)^2 + 2(\sdist_j+\sdist_t)R_j.
\]
Rearranging and using $2\sdist_j\sdist_t\le \sdist_j^2+\sdist_t^2$, we obtain
\[
    r_{jt}^2-\sdist_j^2
    \ge
    \alpha_t^* - 2\sdist_t^2 - 3\sdist_j^2 - 2(\sdist_j+\sdist_t)R_j.
\]
Substituting into \Cref{eq:Ik-weighted}, we obtain
\begin{align}
    &(W(DC^t)+\swt_t)\alpha_t^*
    +
    \sum_{j\in IC^t} w(j)\alpha_j^*
    - \hat f
    \notag\\
    &\qquad\le
    2W(DC^t)\sdist_t^2
    + \sum_{j\in DC^t} w(j)\bigl(3\sdist_j^2+2(\sdist_t+\sdist_j)R_j\bigr)
    + \sum_{j\in IC^t}\rho w(j)\sdist_j^2
    + \sum_{j\ge t}\rho w(j)\sdist_j^2.
    \label{eq:Pk-weighted}
    \tag{$P_t^w$}
\end{align}

\paragraph{Step 2: Choice of the coefficients.}
We now choose nonnegative coefficients $\wcoeff_1,\dots,\wcoeff_s$ and multiply each inequality $\eqref{eq:Pk-weighted}$ with $\wcoeff_t$ and sum over all ${t\in[s]}$ such that the coefficient of each $\alpha_t^*$ is exactly $w(t)$.

\begin{lemma}
    \label{lem:focsweights-weighted}
    For each $t\in[s]$, define
    \[
        M_t\defeq W(DC^t),
        \qquad
        c_t\defeq \frac{\swt_t+M_t}{w(t)},
        \qquad
        g_t\defeq \sum_{u\ge t} \wcoeff_u,
    \]
    \[
        \bcoef_t\defeq \sum_{u>t:\,t\in DC^u} \wcoeff_u,
        \qquad
        A_t\defeq \frac{M_t\wcoeff_t}{w(t)}.
    \]
    Then there exist nonnegative coefficients $\wcoeff_1,\dots,\wcoeff_s$ such that, for every $t\in[s]$,
    \begin{equation}
        \label{eq:alphatcoeff-weighted}
        \wcoeff_t c_t + \sum_{u>t:\,t\in IC^u} \wcoeff_u = 1,
    \end{equation}
    and moreover,
    \[
        0\le \bcoef_t\le g_t,
        \qquad
        \sum_{t=1}^s \wcoeff_t = 1,
        \qquad
        A_t+g_t \le 1.
    \]
\end{lemma}

\begin{proof}
    Define the coefficients recursively, for $t=s,s-1,\dots,1$, by
    \begin{equation}
        \label{eq:focsreccur-weighted}
        \wcoeff_t=
        \frac{1-\sum_{u>t:\,t\in IC^u} \wcoeff_u}{c_t}.
    \end{equation}
    Since \Cref{eq:alphatcoeff-weighted} is triangular, this is the unique solution of the system. We verify that the resulting coefficients are nonnegative.

    Fix $t\in\{0\}\cup[s]$ and consider the suffix $\{\alpha^*_{t+1},\dots,\alpha_s^*\}$. In the final weighted combination, its total coefficient is $\swt_{t+1}$, where by convention $\swt_{s+1}=0$. On the other hand, for each $u>t$, the contribution of the $u$th inequality \Cref{eq:Pk-weighted} to this suffix is
    \[
        \swt_u + W(DC^u) + W(IC^u\cap [t+1,u-1]).
    \]
    Since $IC^u$ and $DC^u$ partition $[u-1]$, we have
    \begin{align*}
        &\swt_u + W(DC^u) + W(IC^u\cap [t+1,u-1]) \\
        &\qquad = \swt_u + W(DC^u) + W([t+1,u-1]) - W(DC^u\cap [t+1,u-1]) \\
        &\qquad = \swt_{t+1} + W(DC^u\cap [1,t]) \\
        &\qquad \ge \swt_{t+1}+M_t,
    \end{align*}
    where we used $DC^t\subseteq DC^u$ and $DC^t\subseteq [t-1]$. Summing over $u>t$ gives
    \[
        (\swt_{t+1}+M_t)\sum_{u>t} \wcoeff_u \le \swt_{t+1},
    \]
    and hence
    \[
        \sum_{u>t} \wcoeff_u \le \frac{\swt_{t+1}}{\swt_{t+1}+M_t} \le 1.
    \]
    Therefore,
    \[
        \sum_{u>t:\,t\in IC^u} \wcoeff_u \le \sum_{u>t} \wcoeff_u \le 1,
    \]
    so the numerator in \Cref{eq:focsreccur-weighted} is nonnegative, and hence $\wcoeff_t\ge 0$.

    The inequality $0\le \bcoef_t\le g_t$ is immediate from the definitions.

    To prove $\sum_{t=1}^s \wcoeff_t=1$, sum the coefficients of all $\alpha^*$-variables in the weighted combination of \Cref{eq:Pk-weighted}. Each inequality contributes total coefficient $\swt_1$, while by construction each $\alpha_u^*$ has coefficient $w(u)$. Hence
    \[
        \swt_1\sum_{t=1}^s \wcoeff_t = \swt_1,
    \]
    so $\sum_{t=1}^s \wcoeff_t=1$, and in particular $g_t\le 1$.

    Finally, fix $t\in[s]$ and consider the suffix $\{\alpha_t^*,\dots,\alpha_s^*\}$. Its total coefficient in the final weighted combination is $\swt_t$. For each $u\ge t$, the contribution of the $u$th inequality to this suffix is
    \[
        \swt_u + W(DC^u) + W(IC^u\cap [t,u-1]).
    \]
    Again using that $IC^u$ and $DC^u$ partition $[u-1]$, we obtain
    \begin{align*}
        &\swt_u + W(DC^u) + W(IC^u\cap [t,u-1]) \\
        &\qquad = \swt_u + W(DC^u) + W([t,u-1]) - W(DC^u\cap [t,u-1]) \\
        &\qquad = \swt_t + W(DC^u\cap [1,t-1]) \\
        &\qquad \ge \swt_t + M_t = c_t w(t).
    \end{align*}
    Summing over $u\ge t$ yields $c_t w(t) g_t\le \swt_t$, so
    \[
        g_t\le \frac{\swt_t}{c_t w(t)}.
    \]
    Since \Cref{eq:alphatcoeff-weighted} implies $c_t\wcoeff_t\le 1$, we also have $\wcoeff_t\le 1/c_t$. Therefore,
    \[
        A_t+g_t
        = \frac{M_t\wcoeff_t}{w(t)} + g_t
        \le \frac{M_t}{c_t w(t)} + \frac{\swt_t}{c_t w(t)}
        =1.
    \]
\end{proof}

\paragraph{Step 3: The weighted quadratic form.}
We now compute the quadratic form.

\begin{lemma}
    \label{lem:quadraticform-weighted}
    For each $t\in[s]$ and each $x\in\mathbb{R}^s$, define
    \[
        U_t(x) \defeq \sum_{u:\,t\in DC^u} \wcoeff_u x_u.
    \]
    Then the weighted sum of all inequalities \eqref{eq:Pk-weighted} over $t\in[s]$ implies
    \[
        \sum_{t\in[s]} w(t)\bigl(\alpha_t^*-(\rho-1)R_t^2\bigr) - \hat f
        \le
        \rho \sum_{t\in[s]} w(t)\sdist_t^2 + Q_\rho^w(\sdist),
    \]
    where, for every $x\in\mathbb{R}^s$,
    \begin{equation}
        \label{eq:Qrho-weighted}
        Q_\rho^w(x)
        =
        \sum_{t\in[s]} w(t)\bigl(2A_t+(3-\rho)\bcoef_t\bigr)x_t^2
        +
        \frac{1}{\rho-1}\sum_{t\in[s]} w(t)\bigl(U_t(x)+\bcoef_t x_t\bigr)^2.
    \end{equation}
\end{lemma}

\begin{proof}
    By \Cref{eq:alphatcoeff-weighted}, the weighted sum of the left-hand sides of $\eqref{eq:Pk-weighted}_{t\in[s]}$ is $\sum_t w(t)\alpha_t^*-\hat f$. Let $\mathcal R$ denote the weighted sum of the right-hand sides. Then
    \[
        \sum_{t\in[s]} w(t)\alpha_t^*-\hat f \le \mathcal R.
    \]
    We now collect the terms in $\mathcal R$.

    First,
    \[
        \sum_{t\in[s]} \wcoeff_t\, 2W(DC^t)\sdist_t^2
        = \sum_{t\in[s]} 2w(t)A_t\sdist_t^2.
    \]
    Second,
    \[
        \sum_{t\in[s]} \wcoeff_t\sum_{j\in DC^t} 3w(j)\sdist_j^2
        = \sum_{t\in[s]} 3w(t)\bcoef_t\sdist_t^2.
    \]
    Third,
    \[
        \sum_{t\in[s]} \wcoeff_t\sum_{j\in DC^t} 2w(j)(\sdist_t+\sdist_j)R_j
        = 2\sum_{t\in[s]} w(t)R_t\bigl(U_t(\sdist)+\bcoef_t\sdist_t\bigr).
    \]
    Finally, for each fixed $t$, the sets
    \[
        \{u\in[s]: t\in IC^u\},
        \qquad
        \{u\in[s]: t\in DC^u\},
        \qquad
        \{u\in[s]: u\le t\}
    \]
    partition $[s]$. Since $\sum_{u=1}^s \wcoeff_u=1$ by \Cref{lem:focsweights-weighted}, it follows that
    \[
        \sum_{u:\,t\in IC^u} \wcoeff_u + \sum_{u\le t} \wcoeff_u = 1-\bcoef_t,
    \]
    and hence
    \[
        \sum_{u\in[s]} \wcoeff_u\left(\sum_{j\in IC^u}\rho w(j)\sdist_j^2 + \sum_{j\ge u}\rho w(j)\sdist_j^2\right)
        = \rho\sum_{t\in[s]} w(t)(1-\bcoef_t)\sdist_t^2.
    \]
    Therefore,
    \[
        \mathcal R
        =
        \rho\sum_{t\in[s]} w(t)\sdist_t^2
        + \sum_{t\in[s]} w(t)\bigl(2A_t+(3-\rho)\bcoef_t\bigr)\sdist_t^2
        + 2\sum_{t\in[s]} w(t)R_t\bigl(U_t(\sdist)+\bcoef_t\sdist_t\bigr).
    \]
    Hence
    \begin{align*}
        &\sum_{t\in[s]} w(t)\bigl(\alpha_t^*-(\rho-1)R_t^2\bigr) - \hat f \\
        &\qquad\le
        \rho\sum_{t\in[s]} w(t)\sdist_t^2
        + \sum_{t\in[s]} w(t)\bigl(2A_t+(3-\rho)\bcoef_t\bigr)\sdist_t^2 \\
        &\qquad\qquad + \sum_{t\in[s]} w(t)\Bigl(2R_t\bigl(U_t(\sdist)+\bcoef_t\sdist_t\bigr)- (\rho-1)R_t^2\Bigr).
    \end{align*}
    As in the warm-up, we absorb the mixed term using
    \[
        2Rx-(\rho-1)R^2 \le \frac{x^2}{\rho-1} \qquad (R,x\in\mathbb{R}).
    \]
    Applying this with $x=U_t(\sdist)+\bcoef_t\sdist_t$ for each $t$, we obtain exactly the claimed inequality.
\end{proof}

\paragraph{Step 4: Weighted matrix form.}
It remains to bound $Q_\rho^w(x)$ by a multiple of
\[
    \sum_{t\in[s]} w(t)x_t^2.
\]
For $x\in\mathbb{R}^s$, write
\[
    \widetilde x \defeq \bigl(\sqrt{w(1)}\,x_1,\dots,\sqrt{w(s)}\,x_s\bigr).
\]

\begin{lemma}
    \label{lem:matrix-weighted}
    Let $B$ be the unique symmetric matrix such that
    \[
        Q_\rho^w(x)=\widetilde x^\top B\,\widetilde x
        \qquad\text{for all }x\in\mathbb{R}^s.
    \]
    Then, for every $t\in[s]$ and every $x\in\mathbb{R}_+^s$,
    \[
        (B\widetilde x)_t
        =
        \sqrt{w(t)}\Biggl[
        \Bigl(2A_t+(3-\rho)\bcoef_t+\frac{\bcoef_t^2}{\rho-1}\Bigr)x_t
        + \frac{\bcoef_t}{\rho-1}U_t(x)
        + \frac{\wcoeff_t}{(\rho-1)w(t)}\sum_{u\in DC^t} w(u)\bigl(U_u(x)+\bcoef_u x_u\bigr)
        \Biggr].
    \]
    Moreover, $B$ is entrywise nonnegative for all $1<\rho\le 3$. Consequently,
    \[
        Q_\rho^w(x)\le \lambda_{\max}(B)\sum_{t\in[s]} w(t)x_t^2
        \qquad\text{for all }x\in\mathbb{R}_+^s.
    \]
\end{lemma}

\begin{proof}
    Since $Q_\rho^w(x)=\widetilde x^\top B\,\widetilde x$, we have
    \[
        \frac{\partial Q_\rho^w(x)}{\partial x_t}=2\sqrt{w(t)}\,(B\widetilde x)_t.
    \]
    Differentiating \Cref{eq:Qrho-weighted} gives
    \begin{align*}
        \frac{(B\widetilde x)_t}{\sqrt{w(t)}}
        &= \bigl(2A_t+(3-\rho)\bcoef_t\bigr)x_t
        + \frac{\bcoef_t}{\rho-1}\bigl(U_t(x)+\bcoef_t x_t\bigr)
        + \frac{\wcoeff_t}{(\rho-1)w(t)}\sum_{u\in DC^t} w(u)\bigl(U_u(x)+\bcoef_u x_u\bigr) \\
        &= \Bigl(2A_t+(3-\rho)\bcoef_t+\frac{\bcoef_t^2}{\rho-1}\Bigr)x_t
        + \frac{\bcoef_t}{\rho-1}U_t(x)
        + \frac{\wcoeff_t}{(\rho-1)w(t)}\sum_{u\in DC^t} w(u)\bigl(U_u(x)+\bcoef_u x_u\bigr).
    \end{align*}
    Every coefficient in this expression is nonnegative when $1<\rho\le 3$, so $B$ is entrywise nonnegative. The final claim follows from the variational characterization of $\lambda_{\max}(B)$.
\end{proof}

We now fix $\rho=5/2$ and replace $\bcoef_t$ by $g_t$, since $\bcoef_t\le g_t$ by \Cref{lem:focsweights-weighted}. For every $t\in[s]$ and every $x\in\mathbb{R}_+^s$, \Cref{lem:matrix-weighted} gives
\begin{equation}
    \label{eq:finalkentry-weighted}
    \frac{(B\widetilde x)_t}{\sqrt{w(t)}}
    \le
    \Bigl(2A_t + \frac12 g_t + \frac23 g_t^2\Bigr)x_t
    + \frac23 g_t U_t(x)
    + \frac{2\wcoeff_t}{3w(t)}\sum_{u\in DC^t} w(u)U_u(x)
    + \frac{2\wcoeff_t}{3w(t)}\sum_{u\in DC^t} w(u)g_u x_u.
\end{equation}

\begin{lemma}
    \label{lem:function-weighted}
    Fix $\tau=\sqrt[3]{12}$. Then
    \[
        \lambda_{\max}(B)\le \max_{1\le x\le \tau} F(x),
    \]
    where
    \begin{equation}
        \label{eq:function-weighted}
        F(x)=
        \frac{48x^6-47\tau x^5-357x^3+(564\tau-396)x^2+4665}{2178}.
    \end{equation}
\end{lemma}

\begin{proof}
    Choose
    \[
        h_t=(1+11g_t)^{-2/3} \qquad (t\in[s]),
        \qquad
        \widetilde h_t=\sqrt{w(t)}\,h_t.
    \]
    Since $B$ is symmetric and entrywise nonnegative, the Collatz--Wielandt Formula \cite[Theorem 8.3.3]{meyer2000matrix} for the spectral radius $\rho(B)$ of nonnegative matrix $B\geq 0$ gives
    \[
        \lambda_{\max}(B) = \rho(B)= \min_{x\in \mathbb{R}_+^s}\max_{t\in[s], x_t \neq 0} \frac{(B\widetilde x)_t}{\widetilde x_t}\le \max_{t\in[s]} \frac{(B\widetilde h)_t}{\widetilde h_t}.
    \]

    We claim that, for every $t\in[s]$,
    \begin{align}
        \frac{2g_tU_t(h)}{3h_t}
        &\le
        \frac{2}{11}g_t\Bigl((1+11g_t)-(1+11g_t)^{2/3}\Bigr),\label{eq:lambda1-weighted}\\
        \frac{2\wcoeff_t}{3w(t)h_t}\sum_{u\in DC^t} w(u)g_uh_u
        &\le
        A_t\cdot \frac{2}{3}(1+11g_t)^{2/3}\tau^{-2},\label{eq:lambda2-weighted}\\
        \frac{2\wcoeff_t}{3w(t)h_t}\sum_{u\in DC^t} w(u)U_u(h)
        &\le
        A_t(1+11g_t)^{2/3}\cdot \frac{2}{11}(\tau-1).\label{eq:lambda3-weighted}
    \end{align}

    To prove \Cref{eq:lambda1-weighted}, note that the sets $DC^u$ are monotone in $u$, so the set $\{u\in[s]: t\in DC^u\}$ is a suffix of $\{t+1,\dots,s\}$. Let $u_t$ be its first element, with the convention $u_t=s+1$ if $t\notin DC^u$ for every $u$. Then
    \[
        U_t(h)=\sum_{u\ge u_t} \wcoeff_u h_u = \sum_{u\ge u_t} (g_u-g_{u+1})(1+11g_u)^{-2/3}.
    \]
    Since the function $z\mapsto (1+11z)^{-2/3}$ is decreasing, this sum is bounded by the corresponding integral:
    \begin{equation}
        \label{eq:ukh-weighted}
        U_t(h)
        \le \int_0^{\bcoef_t} (1+11z)^{-2/3}\,dz
        \le \frac{3}{11}\Bigl((1+11\bcoef_t)^{1/3}-1\Bigr).
    \end{equation}
    Together with $\bcoef_t\le g_t$, this gives \Cref{eq:lambda1-weighted}.

    For \Cref{eq:lambda2-weighted}, the function $g\mapsto g(1+11g)^{-2/3}$ is increasing on $[0,1]$, so
    \[
        g_u h_u = g_u(1+11g_u)^{-2/3} \le 12^{-2/3}=\tau^{-2}.
    \]
    Hence
    \[
        \sum_{u\in DC^t} w(u)g_uh_u \le W(DC^t)\tau^{-2}=M_t\tau^{-2},
    \]
    and therefore
    \[
        \frac{2\wcoeff_t}{3w(t)h_t}\sum_{u\in DC^t} w(u)g_uh_u
        \le
        \frac{2\wcoeff_tM_t}{3w(t)}(1+11g_t)^{2/3}\tau^{-2}
        =
        A_t\cdot \frac{2}{3}(1+11g_t)^{2/3}\tau^{-2}.
    \]

    For \Cref{eq:lambda3-weighted}, \Cref{eq:ukh-weighted} implies
    \[
        U_u(h)
        \le
        \frac{3}{11}\Bigl((1+11\bcoef_u)^{1/3}-1\Bigr)
        \le
        \frac{3}{11}(\tau-1),
    \]
    since $\bcoef_u\in[0,1]$. Therefore,
    \[
        \frac{2\wcoeff_t}{3w(t)h_t}\sum_{u\in DC^t} w(u)U_u(h)
        \le
        \frac{2\wcoeff_t}{3w(t)}(1+11g_t)^{2/3}\,M_t\cdot \frac{3}{11}(\tau-1)
        =
        A_t(1+11g_t)^{2/3}\cdot \frac{2}{11}(\tau-1).
    \]

    Dividing \Cref{eq:finalkentry-weighted} by $h_t$ and combining \Cref{eq:lambda1-weighted,eq:lambda2-weighted,eq:lambda3-weighted}, we obtain
    \begin{align*}
        \frac{(B\widetilde h)_t}{\widetilde h_t}
        &\le
        \Bigl(2A_t+\frac12 g_t+\frac23 g_t^2\Bigr)
        + \frac{2g_tU_t(h)}{3h_t}
        + \frac{2\wcoeff_t}{3w(t)h_t}\sum_{u\in DC^t} w(u)U_u(h)
        + \frac{2\wcoeff_t}{3w(t)h_t}\sum_{u\in DC^t} w(u)g_uh_u \\
        &\le
        2A_t+\frac12 g_t+\frac23 g_t^2
        + \frac{2}{11}g_t\Bigl((1+11g_t)-(1+11g_t)^{2/3}\Bigr) \\
        &\qquad + A_t(1+11g_t)^{2/3}\cdot \frac{2}{11}(\tau-1)
        + A_t\cdot \frac{2}{3}(1+11g_t)^{2/3}\tau^{-2} \\
        &\le G(g_t),
    \end{align*}
    where
    \begin{align}
        G(g)
        \defeq{}&
        2(1-g)+\frac12 g+\frac23 g^2
        + \frac23(1-g)(1+11g)^{2/3}\tau^{-2}
        \notag\\
        &+ \frac{2}{11}(1-g)(1+11g)^{2/3}(\tau-1)
        + \frac{2}{11}g\Bigl((1+11g)-(1+11g)^{2/3}\Bigr).
        \label{eq:scalar-majorant-g-weighted}
    \end{align}
    The last inequality uses $A_t+g_t\le 1$ from \Cref{lem:focsweights-weighted}.

    Substituting $g=(x^3-1)/11$ and $\tau^{-2}=\tau/12$ and simplifying yields
    \[
        \frac{(B\widetilde h)_t}{\widetilde h_t}\le G(g) = F(x)
        \qquad\text{for some }x\in[1,\tau].
    \]
    This proves the claim.
\end{proof}

{Since $F(x)$ is identical to its unweighted counterpart, from \Cref{lem:functionupperbound}, we know that}
\begin{equation}\label{eq:functionupperbound-weighted}
    \max_{x\in[1,\tau]} F(x) < \frac{12}{5}.
\end{equation}

We can now finish the proof of \Cref{lem:dual-feasibility-simpler-metric}. Using \Cref{lem:quadraticform-weighted} with $\rho=5/2$, we obtain
\begin{align*}
    &\sum_{t\in[s]} w(t)\bigl(\alpha_t^*-(\rho-1)R_t^2\bigr) - \hat f \\
    &\qquad\le \frac52 \sum_{t\in[s]} w(t)\sdist_t^2 + Q_{5/2}^w(\sdist) \tag{\Cref{lem:quadraticform-weighted}} \\
    &\qquad\le \frac52 \sum_{t\in[s]} w(t)\sdist_t^2 + \lambda_{\max}(B)\sum_{t\in[s]} w(t)\sdist_t^2 \tag{\Cref{lem:matrix-weighted}} \\
    &\qquad\le \frac52 \sum_{t\in[s]} w(t)\sdist_t^2 + \left(\max_{x\in[1,\tau]}F(x)\right)\sum_{t\in[s]} w(t)\sdist_t^2 \tag{\Cref{lem:function-weighted}} \\
    &\qquad\le \frac52 \sum_{t\in[s]} w(t)\sdist_t^2 + \frac{12}{5}\sum_{t\in[s]} w(t)\sdist_t^2 \tag{\Cref{eq:functionupperbound-weighted}} \\
    &\qquad= 4.9\sum_{t\in[s]} w(t)\sdist_t^2.
\end{align*}
This proves \Cref{lem:dual-feasibility-simpler-metric}.

\end{document}